\documentclass[a4paper,11pt]{article}
\usepackage{jheppub}
\usepackage[T1]{fontenc}
\usepackage[utf8]{inputenc}
\usepackage{amsmath}
\usepackage{amsfonts}
\usepackage{amssymb}
\usepackage{graphicx}
\usepackage{relsize}
\usepackage{pstricks}
\usepackage{color}
\usepackage{tikz}
\usepackage{booktabs}
\usepackage{csquotes}
\usepackage{blindtext}
\usepackage{imakeidx}
\indexsetup{firstpagestyle=empty}
\makeindex[columns=2,title=Alphabetical Index,intoc]

\newcommand{\dx}{\mathrm{d}}
\newcommand{\eps}{\varepsilon}
\newcommand{\ghy}{\phantom{}_2F_1}
\newcommand{\appell}{F_1}
\newcommand{\appellthree}{F_3}
\newcommand{\lauricella}{F_B^{(3)}}
\newcommand{\dilog}{\mathrm{Li}_2}
\newcommand{\Lv}[1]{L_{\mathbf{k}}\left(\mathbf{#1}\right)}
\newcommand{\Lvd}[1]{L^2_{\mathbf{k}}\left(\mathbf{#1}\right)}
\newcommand{\prop}[1]{\Delta_{\mathbf{k}}\left(\mathbf{#1}\right)}
\newcommand{\propd}[2]{\Delta_{\mathbf{k}}\left(\mathbf{#1},\mathbf{#2}\right)}
\newcommand{\propt}[3]{\Delta_{\mathbf{k}}\left(\mathbf{#1},\mathbf{#2},\mathbf{#3}\right)}
\newcommand{\propN}[3]{\Delta_{\mathbf{k}}\left(\mathbf{#1},\mathbf{#2},\cdots,\mathbf{#3}\right)}
\newcommand{\invprop}[1]{\Delta^{-1}_{\mathbf{k}}\left(\mathbf{#1}\right)}
\newcommand{\powprop}[2]{\Delta^{#2}_{\mathbf{k}}\left(\mathbf{#1}\right)}
\newcommand{\pochhammer}[2]{\left( #1\right)_{#2}}
\newcommand{\shuffle}{\sqcup\mathchoice{\mkern-8mu}{\mkern-8mu}{\mkern-3.5mu}{\mkern-3.5mu}\sqcup}

\newcommand{\seq}{\begin{subequations}}
\newcommand{\sen}{\end{subequations}}
\newcommand{\eq}{\begin{eqnarray}}
\newcommand{\en}{\end{eqnarray}}

\begin{document}

\title{\boldmath New ideas for handling of loop and 
angular integrals in D-dimensions in QCD}

\author[a,b,c,1]{Valery E. Lyubovitskij,\note{Corresponding author}}
\author[a]{Fabian Wunder,}
\author[d,c,e]{Alexey S. Zhevlakov}

\affiliation[a]{Institut f\"ur Theoretische Physik, Universit\"at T\"ubingen, \\
	Kepler Center for Astro and Particle Physics, \\ 
	Auf der Morgenstelle 14, D-72076 T\"ubingen, Germany}
\affiliation[b]{Departamento de F\'\i sica y Centro Cient\'\i fico
	Tecnol\'ogico de Valpara\'\i so-CCTVal, \\ 
        Universidad T\'ecnica Federico Santa Mar\'\i a, Casilla 110-V, Valpara\'\i so, Chile}
\affiliation[c]{Department of Physics, Tomsk State University, 634050 Tomsk, Russia}
\affiliation[d]{Bogoliubov Laboratory of Theoretical Physics,
Joint Institute for Nuclear Research, \\ 141980 Dubna, Russia} 
\affiliation[e]{Matrosov Institute for System Dynamics and
	Control Theory SB RAS,  \\ Lermontov str. 134, 664033, Irkutsk, Russia}

\emailAdd{valeri.lyubovitskij@uni-tuebingen.de}
\emailAdd{fabian.wunder@student.uni-tuebingen.de} 
\emailAdd{zhevlakov@theor.jinr.ru}

\abstract{
        We discuss new ideas for consideration of 
	loop diagrams and angular 
	integrals in $D$-dimensions in QCD. 
        In case of loop diagrams,   
	we propose the covariant formalism of expansion 
	of tensorial loop integrals into the 
	orthogonal basis of linear combinations of 
	external momenta. It gives a very simple representation 
	for the final results and is more convenient for 
        calculations on computer algebra systems. 
	In case of angular integrals we demonstrate 
        how to simplify the integration of 
	differential cross sections over 
	polar angles. Also we derive the recursion relations, which allow 
        to reduce all occurring angular integrals to a short set of 
        basic scalar integrals. All order $\eps$-expansion is given 
        for all angular integrals with up to two denominators based 
        on the expansion of the basic integrals and using recursion relations.  
        A geometric picture for partial fractioning is developed which provides 
        a new rotational invariant algorithm to reduce the number of denominators. 
        }


\flushbottom
            
\maketitle

\pagenumbering{roman}
\setcounter{page}{2}
\clearpage
\pagenumbering{arabic}
\setcounter{page}{1}
                       
\section{Introduction}

Evaluation of loop and angular integrals is one of the basic tasks in 
quantum field theory. During last five decades huge  progress 
has been achieved in analytical handling of loop integrals in QCD 
(see, e.g., Refs.~\cite{tHooft:1972tcz}-\cite{Kalmykov:2020cqz}).  
While many different methods for evaluating loop diagrams have been 
developed, they mainly use the following techniques: 

\begin{itemize}
\item Reduction of tensor structure of loop diagrams using 
the Passarino-Veltman (PV) method~\cite{Passarino:1978jh} 
based on Lorentz covariance of matrix elements and fractioning 
of the denominators in loop integrals. For extensions of the alternative 
PV methods up to six external legs see Ref.~\cite{Denner:2005nn};  
\item Reduction of numerators containing virtual momenta by means 
of derivatives acting on linear combinations of external 
momenta~\cite{Davydychev:1991va,Tarasov:1996br,Fleischer:1999hq,%
Smirnov:2006ry,Faessler:2006ft,Gutsche:2018msz,Zhevlakov:2018rwo};
\item Reduction of tensor 
to scalar integrals with shifted dimension~\cite{Davydychev:1991va}, 
which further reduce to the loop integrals 
in generic dimension~\cite{Tarasov:1996br,Fleischer:1999hq};
\item using helicity methods in evaluating loop integrals~\cite{Korner:1990sj,Korner:1982vg}; 
\item Method based on integration by parts (IBP) reduction 
is widely used for handling two and higher loops integrals~\cite{Laporta:2001dd}; 
\item Methods based on decomposition of Lorentz vectors 
in a parallel and orthogonal space in the context of one-loop
and multi-loop integrand reduction, and generalized 
unitarity~\cite{Bern:1994zx,Bern:1994cg,Britto:2004nc,Ellis:2007br,%
Ossola:2006us,Forde:2007mi,Giele:2008ve,Peraro:2020sfm}; 
\item Development of analytical and numerical methods for $\eps$-expansion of 
loop integrals~(see, e.g., Refs.~\cite{Kalmykov:2006pu,Kalmykov:2020cqz}). 
\end{itemize} 

In particular, the PV method~\cite{Passarino:1978jh} is based 
on expanding integrals, which contain the loop momentum in the numerator, 
in terms of a set of scalar functions (form factors) multiplying
a basis of tensors constructed from external momenta. 
The form factors can be expressed by the readily calculated scalar bubble, 
triangle and box one-loop integrals. The scalar form factors of tensorial 
loop integrals are obtained algebraically by solving a system of linear 
equations. This leads to the introduction of the Gram determinant composed of
scalar products of external momenta. As correctly stressed 
in Ref.~\cite{Fleischer:1999hq} the usage of the  PV method is best suited 
for diagrams with four or less external legs, otherwise the tensor structure 
of the diagrams with multi-legs becomes more complicated due to solving systems
of many algebraic equations and due to vanishing Gram determinants at some 
kinematic cases.

Success of the PV method~\cite{Passarino:1978jh} stimulated further development
of the idea to use the combinations of external momenta for the construction
of basis for the expansion of tensorial loop integrals. In particular,
the decomposition of Lorentz vectors in a parallel and
orthogonal space has already been extensively used in the context of one-loop
and multi-loop integrand reduction, generalized unitarity and recent developments
in the decomposition of amplitudes into tensors and form factors (see, 
e.g., discussion in Refs.~\cite{Ossola:2006us,Forde:2007mi,Peraro:2020sfm}). 
In Ref.~\cite{Denner:2005nn} the PV method was extended using alternative 
techniques up to six external legs. 

In the present paper we introduce the following improvement of the PV method. 
Instead of expanding loop integrals in the trivial basis of occurring external 
momenta we perform the expansion in the basis of orthogonal linear combinations 
of the external momenta. It allows us to exclude the stage of algebraically solving 
a system of equations in order to pin down the scalar functions in which we expand tensor 
loop diagrams. Instead we derive these scalar functions straightforwardly using 
fractioning relations involving inverse denominators in loop integrals. 

We consider only loop integrals without referring to specific type 
of the particles propagating in the loop. The proposed formalism can be adapted to full 
amplitudes.

Our method is very useful for a convenient analytic treatment of processes with massless particles, e.g., Drell-Yan (DY) 
and SIDIS processes, prompt photon and heavy flavor production, etc. However, it is important to notice 
that the proposed method can be generalized to arbitrary number of legs in one-loop  
diagrams and to massive particles.

While for many cutting-edge perturbative studies, like inclusive DY processes at $N^3$LO 
(see, e.g., Refs.~\cite{Duhr:2020seh,Duhr:2020sdp}), multi-loop techniques are key, we would like to stress that studies where a refinement of one-loop techniques can provide considerable improvement are still an active area of research.

For understanding angular distributions, spin effects and for access 
to the parton distributions (PDFs and TMDs), improving on the existing tools in producing analytical 
expressions as simply and economically as possible is beneficial.
Besides the importance of analytical methods for phenomenological study of QCD processes 
in the fixed target regime (DY, SIDIS, etc.), handy analytical expressions are also important 
for performing resummation of large logarithms, verifying of different identities (like Lam-Tung 
relation) involving hadronic structure functions, behavior of the hadronic amplitudes at different 
kinematical limits and their expansions in small parameters (like small transverse momentum of 
photon, etc.). See, e.g., discussion in Refs.~\cite{Boer:2006eq,Berger:2007jw}. 

In the second part of our paper we discuss new ideas concerning angular 
integration in QCD~\cite{vanNeerven:1985xr}-\cite{Anastasiou:2002yz}. 
We extend the known closed results for massless and single massive two denominator 
integrals in $D$ dimensions to the double massive case. 
All order $\eps$-expansion is given for all angular integrals with up to 
two denominators. A geometric picture for partial fractioning is developed which provides 
a new rotational invariant algorithm to reduce the number of denominators.

For the calculation of real emission corrections in higher order perturbative calculations 
the phase space integration (PSI) constitutes the most challenging part. When massless external 
particles are involved, the integrals are singular, thus numerical integration is not feasible. 
To regularize the divergent integrals, analytic integration methods are necessary. 
A prominent class of divergences are those of collinear type. 
They appear if two external massless particles propagate in the same direction making 
their scalar product vanish. These singularities appear in the angular PSI. 
Henceforth, analytic methods for their calculation are essential.

Since its introduction, the dimensional regularization (DR)~\cite{tHooft:1972tcz} by t'Hooft 
and Veltman is the preferred regularization method in gauge theories. In particular, 
it gives the straightforward and consistent recipe to perform QCD calculations in $D$ 
dimensions and take the physical limit $D\rightarrow 4$ only after the cancellation of 
all divergences. Throughout the paper we choose the most common definition of $D=4-2\eps$. 
Note that some of the references use $D=4 \pm \eps$ instead.  
The collinear divergences appearing in angular PSI manifest themselves 
as poles in the Laurent expansion about $\eps=0$.

Angular integrals appear in the calculation of the two, three and four particle PSIs 
and thus appear in a plethora of perturbative calculations. QCD examples include 
processes such as 
Drell-Yan~\cite{Schellekens:1981kq,matsuura1989calculation,matsuura1990contribution,%
hamberg1991complete,Mirkes:1992hu,Bahjat-Abbas:2018hpv}, 
deep inelastic scattering (DIS)~\cite{duke1982quantum}, 
semi-inclusive DIS (SIDIS)~\cite{Anderle:2016kwa,Wang:2019bvb}, 
prompt photon production~\cite{Gordon:1993qc}, 
hadron-hadron scattering~\cite{ellis1980large}, and 
heavy quark production~\cite{Beenakker:1988bq,Bojak:2000eu}.                              

In the literature those angular integrals were mostly considered in terms 
of the integrals, which we call in our paper the 
\textit{Neerven integrals}\index{Neerven integral}~\cite{vanNeerven:1985xr,Beenakker:1988bq}:   
\begin{align}
I^{j,l}_D=\int\dx\Omega_{k_1 k_2}
\frac{1}{(a+b\cos\theta_1)^j(A+B\cos\theta_1+C\sin\theta_1\cos\theta_2)^l}\,,
\label{eq:Van Neerven integral}
\end{align}
with 
\begin{align}
\int\dx\Omega_{k_1 k_2}=\int_0^\pi\dx\theta_1 \sin^{D-3}\theta_1 
\int_0^\pi\dx\theta_2 \sin^{D-4}\theta_2.
\end{align}
They are divided in four classes depending on whether the conditions $a^2=b^2$ and $A^2=B^2+C^2$ hold. 
We will call $I^{(j,l)}_D$ \textit{massless}, if both equations are satisfied and \textit{single massive},  
if one of them holds and else \textit{double massive}. The latter are finite for $D=4$, the other 
two involve singularities.

We will start our discussion with an overview of how those integral were used in the older 
literature. From their introduction in the early 80s onwards they were an important ingredient 
for calculations at the NNLO level. After providing context on the development of the knowledge 
about angular integrals, we discuss how the use of angular integrals compares to the method 
of \textit{reversed unitarity}, which is in frequent use today. This will come along with a review 
of modern application of angular integrals, which demonstrates that, thus known for four decades 
in QCD calculations, angular integrals are still in use today and a more systematic study of these 
was long overdue. 

First application of DR for treatment of divergences in the angular PSIs 
was performed by Ellis et al. in Ref.~\cite{ellis1980large}.  
In particular, angular integrals $\int\dx\Omega_{k_1 k_2}$ as part of 
the phase space $\dx\mathrm{PS}_3$ were considered and 
the massless integrals $I^{(1,1)}_D$ were introduced, to which all appearing PSIs could be 
reduced \enquote{by suitable rotation and partial fractioning}.
In a similar fashion Ref.~\cite{duke1982quantum} treated the three particle PSI. 
In Ref.~\cite{duke1982quantum} partial fractioning was introduced 
using identities involving the Mandelstam variables in order to reduce the PSIs 
to the massless $I^{(1,1)}_D$ and $I^{(1,0)}_D$.
Ref.~\cite{devoto1984analytic} extended the calculation of angular integrals to the single 
massive $I^{(1,1)}_D$, while a list of the double massive integrals 
$I^{(j,l)}_4$ for $-2\leq j,l\leq 2$ restricted to the four-dimensional case 
has been derived in Ref.~\cite{Schellekens:1981kq}.  

In this vein, major achievement was Neerven's analytic calculation of 
the massless integrals $I^{(j,l)}_D$ in
Ref.~\cite{vanNeerven:1985xr}. By calculating them as the discontinuity of a box graph and 
employing the optical theorem he heavily influenced the approach towards PSIs.
For a long time this constituted the only published detailed calculation of angular integrals.
Other papers were heavily based on the results of Ref.~\cite{vanNeerven:1985xr}.  
The most commonly used reference for the angular integrals for the last three decades 
is~\cite{Beenakker:1988bq}. Ref.~\cite{Beenakker:1988bq} gave the angular integral 
in the form of Eq.~(\ref{eq:Van Neerven integral}) using the set of parameters 
$\{a, b, A, B, C\}$ and produced a comprehensive, often cited list of angular integrals. 
The double massive integral $I^{(j,l)}_4$ was given for the case $|j|,|l|\leq 2$. 
The divergent integrals were expanded up to order $\eps$, the massless integral 
was given for $j=1,-2\leq l\leq 1$, and the single massive 
for $l=1$, $j=-2,-1,1,2$ and $l=2$, $j=0,1,2$. The single massive integrals were given in two groups 
without making the symmetry between $a^2\neq b^2$ and $A^2\neq B^2+C^2$ explicit. 
The list of the integrals was based on the calculation method put forward 
in Refs.~\cite{vanNeerven:1985xr,Schellekens:1981kq,devoto1984analytic}.
A similar list was compiled in Ref.~\cite{Smith:1989xz}.  
Some early use of those lists can be found in Ref.~\cite{matsuura1989calculation}. 
Ref.~\cite{matsuura1989calculation} mentioned the massless and single massive integrals $I^{(j,l)}_D$ 
and stated that the single massive integral \enquote{cannot be written in as elegant a form as} 
the massless integral, hence \enquote{brute force} methods were used. 
Ref.~\cite{matsuura1989calculation} derived an early version of two-point 
partial fractioning discussed in more details in Appendix~\ref{sec:Partial Fractioning} 
of the present paper. Ref.~\cite{matsuura1990contribution} considered the angular 
integration as part of the three-particle phase space. Again Eq.~(\ref{eq:Van Neerven integral}) 
appeared for the massless case. Note that only two-point partial fractioning was mentioned in 
Ref.~\cite{matsuura1989calculation} and there was 
no three-point partial fractioning~(see Appendix~\ref{sec:Partial Fractioning}).  
So seven integrals from the list of Ref.~\cite{matsuura1989calculation},  
including those containing the massive propagator, were said to be integrated 
by \enquote{brute force methods}. In the follow-up paper~\cite{hamberg1991complete} 
the Neerven integral came with a reference to the 1989 paper~\cite{Beenakker:1988bq}. 
The single massive integral was described as \enquote{very cumbersome}. About the double massive case they 
write \enquote{fortunately [the double massive integral] can be avoided by choosing an appropriate frame.} 
As a benchmark, their partial fractioning algorithm led to 217 different  3-body PSI. 

Later, in Ref.~\cite{Mirkes:1992hu} the Neerven integral\index{Neerven integral} 
was considered as part of a four-body phase space by Mirkes. 
The three classes of angular integrals he lists correspond to massless and single-massive 
integrals. Some of the single massive integrals were expanded up to order $\eps$. 
The recursion given for his class III integrals without reference 
fail for $\eps\neq 0$. We correct this shortcoming in Sec.~\ref{sec:Mirkes Type III}. 
The partial fractioning was described as \enquote{very involved}. 
In Ref.~\cite{Gordon:1993qc} the Neerven integral together with \enquote{extensive use of relations 
between Mandelstam variables} were employed for partial fractioning 
to calculate the 3-body phase space of prompt photon production. Additional phase space factors 
were accommodated for by dimensional shift $\eps\rightarrow \eps-1$. Such applications motivate to publish 
not only $\eps$-expansions but also general results permitting for expansion about other values of $\eps$. 
We consider algebraic identities to perform dimensional shifts in Sec.~\ref{sec:PDI}. 
Further consideration of the Neerven type integrals in the context of the three particle phase space has 
been considered in Ref.~\cite{Bojak:2000eu}. The \enquote{extensive partial fractioning} based on Mandelstam 
identities \enquote{though computerized [\dots] has the disadvantage of often yielding unnecessarily complicated 
expressions}. It leads to massless, single-massive and double-massive Neerven integrals. 
The symmetry between the two classes of single-massive integrals was recognized and relations between 
the double-massive integrals via differentiating with respect to $a$ and $A$ were given 
(compare~\ref{sec:remarks about massive Van Neerven integral}). 

In the decade following Neervens paper, there was not much improvement on the treatment 
of angular integrals. In 2001 Anastasiou and Melnikov~\cite{Anastasiou:2002yz} proposed
a new method called reverse unitarity for calculations of phase space
integrals. Expressing these integrals in terms of loop integrals made them accessible 
through the sophisticated mathematical methods developed for handling of loop integrals such as   
integration-by-parts relations, differential equations, etc. 
This approach computes cut diagrams as solutions of differential 
equations~\cite{Argeri:2007up,Moch:2007jk}. Technically the latter are needed to be augmented with
boundary conditions coming from a separate calculation.
Analytic results are obtained in terms of harmonic polylogarithms order by order in $\varepsilon$. 
During the last twenty years these methods received an evolution~\cite{Mitov:2005ps,Bonocore:2016doa}. 
The method of reverse unitarity has certain advantages.
It is applicable to rather complicated PSIs since it takes full advantage of the knowledge 
about loop integrals. Since every delta function in the phase space translates to additional 
propagators, the technique is especially suited for the treatment of inclusive PSIs.
Limitations are induced by the
development of technical methods 
for calculation of loop integrals. Difficulties in these approaches grow with the inclusion of 
more particles in the finals state, multiple scales and massive particles. 

The success of methods based on reversed unitarity provides an explanation of the scarcity 
of improvements regarding angular integrals. However, the direct approach towards angular 
integrals, rather than treating all PSIs by reversed unitarity still has its merits for certain 
applications. It is conceptually simple, since it makes straightforward use of their properties 
rather than taking a detour to loop integrals. One might compare the calculation of the massless Neerven integral
in Ref.~\cite{vanNeerven:1985xr} to ours in Sec.~\ref{sec:Massless integral with two denominators} as an example. By considering a specific class of PSIs, 
in the present paper angular integrals with two denominators, analytic results can be obtained 
in closed form and expanded to all orders in $\varepsilon$. Furthermore, recursion relations 
can explicitly be solved. This yields a completely solved building block which can be plugged 
into calculations, without the need to run an IBP reduction and to solve differential equations 
order by order. The expansion presented in Sec.~\ref{sec:All order expansion} has the benefit 
to be compatible with commonly used methods since the result is expressed in terms of multiple 
polylogarithms. Of course, employing the angular integral discussed here is only applicable for 
observables with phase space integrals of appropriate form and is not simply generalizable for 
other PSIs. However, the ideas presented should also be of value for related PSIs like angular integrals with 3 and 
more denominators. Such integrals appear, e.g., as master integrals in the reversed unitarity 
method (see Ref.~\cite{Anastasiou:2013srw}).

After introducing the angular integral in its historical context, we will look 
at modern developments and usage in the following. 
In 2011 Somogyi~\cite{Somogyi:2011ir} was the first who employed the property of rotational invariance for 
handling angular integrals. He did the first systematic calculation of the angular integrals with 
a parametrization independent of the choice of coordinate axis. Furthermore, his approach is much more 
straightforward than Neerven's method. Ref.~\cite{Somogyi:2011ir} gave the first closed result for the 
single-massive integral in terms of the Appell function. However, for the double massive integral 
Ref.~\cite{Somogyi:2011ir} can only give a Mellin-Barnes representation and no closed formula 
in terms of hypergeometric functions. {Ref.~\cite{Somogyi:2011ir} lists 
the massless integrals up to order $\eps$, the single massive and double massive 
to finite order. For the double massive integral the order $\eps$ term could not be derived using 
the Mellin-Barnes methods since it involved complicated triple sums. Unpublished work towards 
the order $\eps$ by other means is cited to also \enquote{not have an analytic 
answer}~\cite{SmithUnpublished}. In the present paper we will present this so far missing result in 
Sec.~\ref{sec:Double massive integral with two denominators}.

The closed analytic result for the single massive integral was first used 
in Ref.~\cite{Bolzoni:2010bt} for the calculation of subtraction terms in QCD jet cross sections 
at NNLO and more recently in Ref.~\cite{Lionetti:2018gko} for subtraction terms in Higgs production. 
The single massive integral found further application in Ref.~\cite{Specchia:2018uyj} to pin down 
phase space master integrals. Also the N$^3$LO study of Higgs production 
in Ref.~\cite{Anastasiou:2013srw} employed Somogyi's new results. A brought discussion of techniques 
for phase-space calculations, emphasizing the benefits of a rotational invariant approach based on 
Somogyi's methods, is given in Ref.~\cite{Hoschele:2015whb}. Ref.~\cite{Bahjat-Abbas:2018hpv} 
on the Drell-Yan process at N$^3$LO involves both the Neerven integral and Somogyis findings 
in the determination of phase space integrals in the soft region. Furthermore angular integrals 
in the Somogyi parametrization make their appearance in the determination of beam and soft functions 
(see Ref.~\cite{Baranowski:2020xlp}). 

However, many recent investigations are still based on the rather old Neerven list. 
The angular integral in Neerven parametrization appears in~\cite{Ringer:2015chp} 
and~\cite{Ringer:2015oaa} as part of the three particle phase space of single spin asymmetry 
in $W$-boson production. Ref.~\cite{Schlegel:2012ve} employed it for similar purpose. 
In Ref.~\cite{Anderle:2016kwa} the angular integral is used in Neerven form and some additional 
integrals wih higher indices are provided. Ref.~\cite{Kotlarski:2016zhv} employs normalized 
Neerven integrals for a study of supersymmetry. Further examples of recent uses of Neervens integral 
list are such diverse topics as the investigation of twist-3 single-spin 
asymmetry~\cite{Hinderer:2017bya}, heavy flavour production~\cite{Hekhorn:2019nlf}, 
or the study of Kaluza-Klein gluons in Ref.~\cite{Lillard:2016jxz}. 
The present paper is aimed at providing a long overdue update to the still frequently used 
tool of angular integration for future studies similar to those mentioned here.
 
The following three perturbative studies are further examples of recent practice where the presented 
method could have been conveniently applied. Ref.~\cite{Wang:2019bvb} considers the Neerven integral 
as part of the three-particle phase space of NNLO real corrections to SIDIS at high transverse momentum. 
The symmetry of single massive integrals is noticed by change of frame, a closed form for the massless 
integral is given however it is stated that \enquote{the integral no longer has a closed form} 
in the single massive case. They give a detailed description of their partial fractioning algorithm 
based on Mandelstam identities and provide an extensive list of integrals.
A recent paper on $e^+e^-$-annihilation gives the double massive integral $I^{(j,l)}_4$ for 
$1\leq j,l\leq 2$ and states \enquote{we were not able to find a closed form 
in $d$ dimensions}~\cite{Blumlein:2020jrf}. Their Eq.~(151)-(153) constitute two-point partial 
fractioning, eq. (154)-(159) is a version of three-point partial fractioning. 
In a recent publication on 
QED corrections in exclusive 
$\overline{B}\to \overline{K}{\mathrm{\ell}}^{+}{\mathrm{\ell}}^{-}$ 
decay~\cite{Isidori:2020acz}, the order $\eps$ of the double massive angular integral was explicitly needed to extract collinear logarithms from soft integrals. Private communication with Somogyi is cited to obtain a result. This showcases the need to improve on the existing literature on angular integrals.
 
The broad range of applications of the methods for dealing with angular integrals in QCD processes 
motivates a fresh study of this problem to supplement the toolbox for PSI.
In particular, one of the main purposes of our paper is to present the angular integrals in a systematic 
fashion based on the Somogyi framework using relativistically invariant variables 
(scalar products of four momenta/velocities)~\cite{Somogyi:2011ir}. 

Since the appearance of the angular integrals over a small number of propagators is intimately connected to 
partial fractioning, we also provide an algorithm which directly leads to the angular integrals in the form 
introduced in Sec.~\ref{sec:Somogyi integrals}. This has the benefit that partial fractioning becomes 
manifestly invariant under rotation and leads to considerably short expressions without the need to choose 
suitable coordinate axis, simplifying computer implementation.

The systematic approach offers a cleaner picture on the possible types of integrals appearing. 
Furthermore, we naturally obtain an expression for the double massive integral in $D$ dimensions with 
integer coefficients $j$, $l$ in terms of single massive integrals, not known in the existing literature. 
The general Neerven integral 
with integer valued $j$ and $l$ is calculated in $D$ dimensions and expanded to all orders in $\eps$ 
for the first time. The whole calculation is presented in a pedagogical form to encourage further studies 
and supplement the existing extensive lists of integrals with straightforward derivations. 

The paper is structured as follows.
In Sec.~\ref{sec:Loop_integrals} we discuss novel ideas for handling of 
loop tensorial integrals in $D$ dimensions using the basis of orthogonal external momenta. 
In Sec.~\ref{sec:Angular_integrals} we discuss  new ideas concerning angular integration 
in $D$ dimensions. Sec.~\ref{Sec:conclusion} contains our conclusions and summary of the main results.
Technical details of calculations, a discussion of parial fractioning in PSIs, and Tables of angular integrals 
are placed in Appendices. 

\clearpage 

\section{New formalism for reduction of tensor loop integrals}  
\label{sec:Loop_integrals} 

In this section we discuss our newly developed formalism for 
handling of loop tensorial integrals in $D$ dimensions
using an orthogonal basis of external momenta. 
As we stressed in Introduction we develop analytic method for 
evaluation of one-loop diagrams with its further application to QCD 
processes at fixed target regime (DY and SIDIS processes, prompt photon and 
heavy flavor production, etc.).  Our method can be extended to arbitrary number of legs in one-loop
diagrams and for both massless and massive particles. Development of analytical methods for study of 
QCD processes is very important for understanding of their different aspects: angular distributions, 
spin effects, access to the parton distributions. It is also important for performing resummation of large
logarithms, verifying of different identities (like Lam-Tung relation) involving hadronic structure functions,
behavior of the hadronic amplitudes at different kinematical limits and their expansions in small parameters
(like small transverse momentum of photon, etc.). 
See, e.g., discussion in Refs.~\cite{Boer:2006eq,Berger:2007jw}.

In present paper we restrict to consideration of 
one-loop integral with 2, 3, and 4 external legs and 
with specific kinematic relevant for QCD processes at fixed target 
regime. Typical loop diagrams 
with 2, 3, and 4 external legs are shown in Fig.~\ref{fig:box4d}.

\begin{figure}[htb]
  \begin{center}
		\epsfig{figure=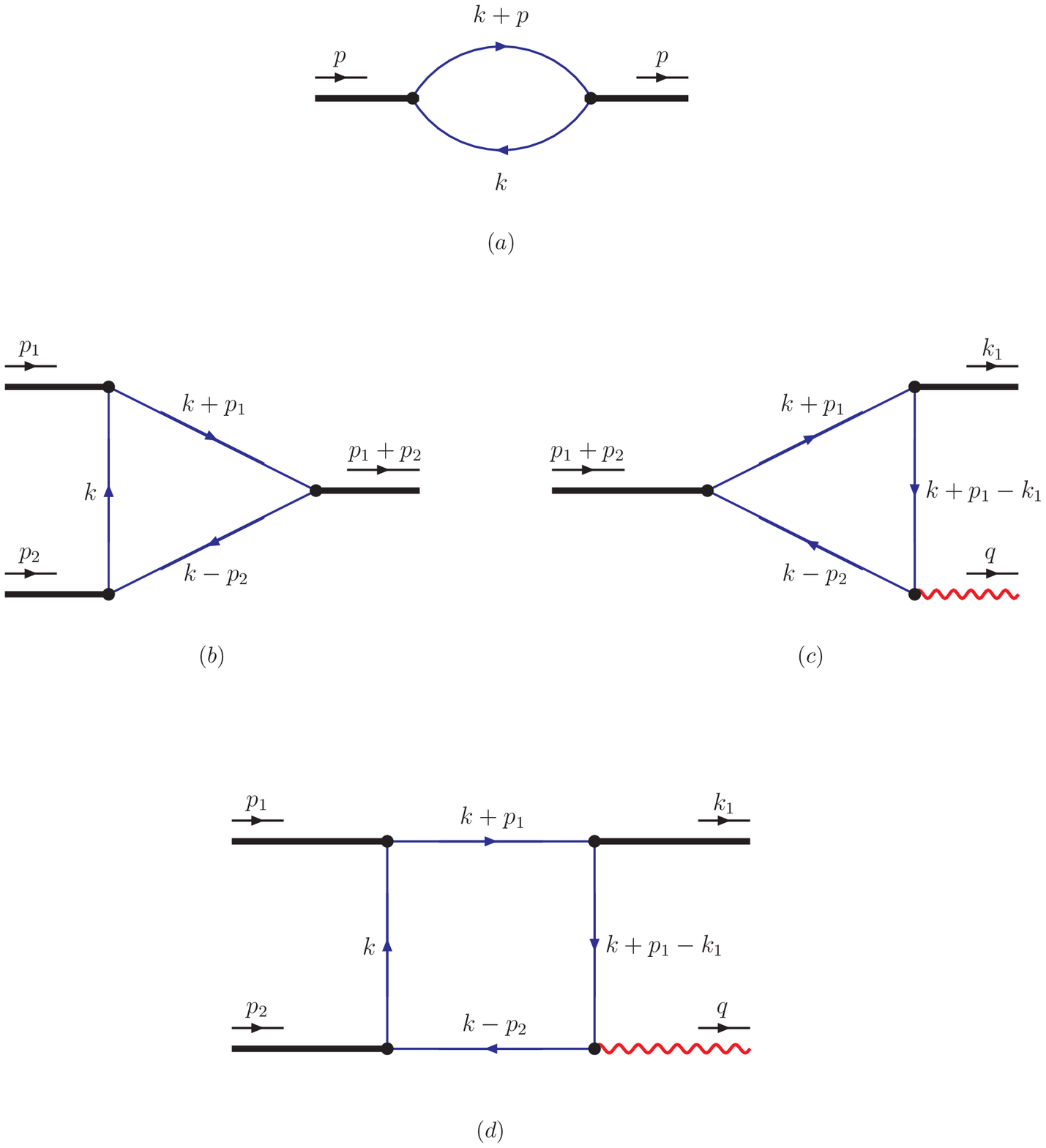,scale=.68} 
\caption{Loop diagrams: bubble (a), triangle (b) and (c), and box (d).} 
\label{fig:box4d}
\end{center}
\end{figure}

First, we define the 9 possible denominators occurring in these 
loop integrals: 
\eq 
& &
\Delta_0 = k^2          \,, \quad 
\Delta_1 = (k+p_1)^2    \,, \quad  
\Delta_2 = (k-p_2)^2    \,, \quad 
\Delta_3 = (k-k_1)^2    \,, \quad \nonumber\\
& &
\Delta_4 = (k-q)^2      \,, \quad 
\Delta_5 = (k+p_1-k_1)^2\,, \quad  
\Delta_6 = (k+p_1-q)^2  \,, \quad \\
& &
\Delta_7 = (k+p_1+p_2)^2\,, \quad  
\Delta_8 = (k+q)^2      \,, \quad  
\Delta_9 = (k+k_1)^2    \,. \nonumber
\en 

Our formalism is based on the PV technique.  
We improve it by expanding the tensor loop integrals in an} orthogonal
basis of vectors instead of plain external momenta. 
In case of four external particles the orthogonal basis is 
specified as following. Let $p_1$ and $p_2$ be the momenta of incoming 
massless particles with $p_1^2 = p_2^2 = 0$ 
and let $k_1$ and $q$ be the momenta of outcoming 
particles with massless $k_1^2 = 0$
and massive/virtual $q^2 = Q^2$. They obey 
four-momentum conservation $p_1 + p_2 = k_1 + q$.
The set of Mandelstam variables is specified as:
\eq\label{kinematics1} 
& &s = (p_1 + p_2)^2 = (k_1 + q)^2\,, \nonumber\\
& &t = (p_1 - q)^2 = (p_2 - k_1)^2\,, \nonumber\\
& &u = (p_1 - k_1)^2 = (p_2 - q)^2\,, \nonumber\\
& &s + t + u = q^2 = Q^2 \,. 
\en 
Now we define the basis of 3 orthogonal momenta as 
\eq\label{basis_3perp_V}
P^\mu = (p_1 + p_2)^\mu\,, 
\quad 
R^\mu = (p_1 - p_2)^\mu\,,  
\quad 
T^\mu = k_1^\mu 
- P^\mu \, \frac{Pk_1}{P^2}
- R^\mu \, \frac{Rk_1}{R^2} \,, 
\en 
where 
$P^2 = - R^2 = s$, $T^2 = - \dfrac{ut}{s}$ and 
$P \cdot R = P \cdot T = R \cdot T = 0$. 
Also we introduce the perpendicular metric 
tensor $g^{\mu\nu}_\perp$, 
which is manifestly orthogonal to all basis momenta: 
\eq\label{kinematics2}  
g^{\mu\nu}_\perp = g^{\mu\nu} 
- \frac{P^\mu P^\nu}{P^2}
- \frac{R^\mu R^\nu}{R^2}
- \frac{T^\mu T^\nu}{T^2}  \,, 
\en 
which satisfies 
$g^{\mu\nu}_\perp \, g_{\mu\nu}^\perp = D - 3$, 
$g^{\mu\nu}_\perp \, P_\mu = 
g^{\mu\nu}_\perp \, R_\mu = 
g^{\mu\nu}_\perp \, T_\mu = 0$. 
All above formulas can be extended to the case of a massive 
particle with momentum $k_1$. The only difference is 
that $T^2 =  \dfrac{Q^2k_1^2-ut}{s}$. 
The use of such basis considerably simplifies obtaining 
all scalar functions parameterizing the loop integrals. 

Note, that an extension to an arbitrary number of external and massive 
particles is straightforward. In particular, if we consider the scattering of two 
particles with momenta $p_1$ and $p_2$ producing $n$ particles 
in the final state with momenta $k_1, \ldots, k_n$, then 
the orthogonal basis involving $(n+1)$ momenta $\{P_1, \ldots P_{n+1}\}$ is 
specified as  
\eq 
P_1^\mu &=& P^\mu\,, \qquad \hspace*{1.75cm} P = p_1 + p_2\,,\nonumber\\
P_2^\mu &=& R^\mu - P^\mu \, \frac{PR}{P^2}\,, \qquad R = p_1 - p_2\,, \nonumber\\
P_3^\mu &=& k_1^\mu - P_1^\mu \, \frac{P_1k_1}{P_1^2}  
                    - P_2^\mu \, \frac{P_2k_2}{P_2^2} \,, \nonumber\\[2mm] 
\ldots&\ldots&\ldots\ldots\ldots\ldots\ldots\ldots\ldots\ldots\ldots    
 \nonumber\\[2mm] 
P_{n+1}^\mu &=& k_{n-1}^\mu  
              - P_1^\mu \, \frac{P_1k_{n-1}}{P_1^2}  
              - P_2^\mu \, \frac{P_2k_{n-1}}{P_2^2} 
              - \ldots 
              - P_n^\mu \, \frac{P_nk_{n-1}}{P_n^2} \,.
\en 
For the loop integrals with $n$ propagators 
and with a product of $m$ external momenta in numerator we use the notation: 
\eq
I^{\alpha_1 \ldots \alpha_{m}}_{i_1 \ldots i_n} &=& 
(i\pi^2)^{-1} \, \mu^{4-D}
\int d^Dk \, \frac{k^{\alpha_1} \ldots k^{\alpha_m}}
{\Delta_{i_1} \ldots \Delta_{i_n}} \,,
\en 
where $\mu$ is the renormalization scale. 
{Below we list the results for bubble, triangle, and box integrals 
with kinematics [see Eqs.~(\ref{kinematics1})-(\ref{kinematics2})]  
specific for the QCD processes at fixed target regime at NNLO.} 

\subsection{Bubble integrals} 

All bubble integrals are expressed through the scalar 
bubble PV function $B_0(p^2)$ 
\eq\label{PV_B0}
B_0(p^2) &=& (i \pi^2)^{-1} \, \mu^{4-D} \, 
\int d^Dk \frac{1}{k^2 (k+p)^2} \nonumber\\
&=& 
\pi^{\frac{D}{2}-2} \, \frac{\Gamma\Big(2-\frac{D}{2}\Big) 
	\, \Gamma^2\Big(\frac{D}{2}-1\Big)}{\Gamma\Big(D-2\Big)} \, 
\biggl(-\frac{p^2}{\mu^2}\biggr)^{\frac{D}{2}-2}  \,, 
\en 
where $\Gamma(x)$ is the gamma function\index{Gamma function}. 
The vector bubble integrals are given by: 
\eq 
B_1^\mu(p) 
&=& (i \pi^2)^{-1} \, \mu^{4-D} \, 
\int d^Dk \frac{k^\mu}{k^2 (k+p)^2} 
= - \frac{p^\mu}{2} \, B_0(p^2) \,,
\label{B1int}\\
B_2^{\mu\nu}(p) 
&=& (i \pi^2)^{-1} \, \mu^{4-D} \,  
\int d^Dk \frac{k^\mu \, k^\nu}{k^2 (k+p)^2} 
= \frac{1}{4 (D-1)} \, 
\biggl( p^\mu p^\nu \, D - g^{\mu\nu} p^2 \biggr) \, B_0(p^2) 
\nonumber\\
&=& \frac{1}{4} \, 
\biggl[ 
p^\mu p^\nu  - g^{\mu\nu}_{\perp;2} \frac{p^2}{D-1} \biggr] \, B_0(p^2) 
\,,
\label{B2int}
\en 
where 
$g^{\mu\nu}_{\perp;2} = g^{\mu\nu} - p^\mu p^\nu/p^2$ 
is the perpendicular metric tensor for the bubble diagram 
with the property: 
\eq 
g^{\mu\nu}_{\perp;2} \, 
g_{\perp;2; \mu\nu} = D - 1 \,.  
\en 
Here and in the following, we introduce three orthogonal metric 
tensors, one for each specific diagram --- bubble, triangle, and box ---  
which will be labeled by the indices 2, 3, and 4, respectively.
Also, we will see that the following normalization 
condition for the metric tensors holds: 
\eq 
g^{\mu\nu}_{\perp;n} \, 
g_{\perp;n; \mu\nu} = D + 1 - n \,.
\en 

For specific choices of momenta the scalar bubble integrals 
$I_{ij} \equiv B_0^{ij}$ read: 
\eq 
& &B_0(0)   = B_0^{01} 
            = B_0^{02} 
            = B_0^{03} 
            = B_0^{09} 
\,, \nonumber\\
& &B_0(u)   = B_0^{05}\,, \quad 
   B_0(t)   = B_0^{06}\,, \quad 
   B_0(s)   = B_0^{07}\,, \quad 
   B_0(Q^2) = B_0^{04} 
            = B_0^{08}\,. 
\en
Tensorial bubble integrals for specific momenta follow from 
Eqs.~(\ref{B1int}) and~(\ref{B2int}). Note that in case of bubble 
diagrams we work with a single transverse vector. Next, 
in case of triangle and box diagrams we will use 
two and three transverse vectors. However, these perpendicular 
vectors can be always expanded in a linear combination 
of the vectors of the basis~(\ref{basis_3perp_V}). 

\subsection{Triangle integrals} 

For the triangle integrals (scalar, vector, tensor rank-2) 
we use the following definitions, respectively:  
\eq
I_{ijk} &=& (i\pi^2)^{-1} \, \mu^{4-D} \, 
\int d^Dk \,\frac{1}{\Delta_i \,\Delta_j \,\Delta_k}           \,,\nonumber\\
I_{ijk}^{\mu} &=& (i\pi^2)^{-1} \, \mu^{4-D} \, 
\int d^Dk \,\frac{k^\mu}{\Delta_i \,\Delta_j \,\Delta_k}       \,,\\ 
I_{ijk}^{\mu\nu} &=& (i\pi^2)^{-1} \, \mu^{4-D} \,  
\int d^Dk \,\frac{k^\mu k^\nu}{\Delta_i \,\Delta_j \,\Delta_k} \,.\nonumber 
\en
The scalar integrals $I_{ijk} \equiv C_0^{ijk}$ obey the following relations 
and are expressed through the PV triangle scalar integrals $C_0(s_1,s_2)$ as:
\eq
& &C_0(s) = C_0^{012} = C_0^{017}           
\,, \nonumber\\
& &C_0(t) = C_0^{023} = C_0^{026} = C_0^{069} 
\,, \nonumber\\
& &C_0(u) = C_0^{015} = C_0^{019} 
\,, \\
& &C_0(s,Q^2)  = C_0^{038} = C_0^{049} = C_0^{078} = C_0^{079} 
\,, \nonumber\\
& &C_0(t,Q^2) = C_0^{016} = C_0^{018} = C_0^{046} 
\,, \nonumber\\
& &C_0(u,Q^2) = C_0^{024} = C_0^{025} = C_0^{035} = C_0^{058} 
\,. \nonumber 
\en 
Here, the integral $C_0(s_1,s_2,s_3)$ is 
defined as 
\eq\label{C0_def}
C_0(s_1,s_2,s_3) 
= (i\pi^2)^{-1} \, \mu^{4-D} \, 
\int d^Dk \, \frac{1}{k^2 \, (k+l_1)^2 \, (k+l_2)^2} \,,  
\en 
where $s_1 = l_1^2$, $s_2 = l_2^2$, $s_3 = l_3^2 = (l_1 - l_2)^2$,
and $2 l_1 l_2 = s_1 + s_2 - s_3$.

It is fully symmetric under all permutations of its arguments
\eq 
C_0(s_1,s_2,s_3) = 
C_0(s_1,s_3,s_2) &=& 
C_0(s_2,s_3,s_1) = 
C_0(s_2,s_1,s_3)   \nonumber\\
=
C_0(s_3,s_1,s_2) &=& 
C_0(s_3,s_2,s_1) \,. 
\en 
In specific limit, when one of the arguments is equal to zero,  
it is expressed through the bubble PV integral $B_0(s)$ as: 
\eq 
C_0(s_1,s_2,0) = 
\frac{2 \, (D-3)}{4-D} \ \frac{B_0(s_1) - B_0(s_2)}{s_1 - s_2} \,. 
\en 
Now we consider tensorial triangle integrals. 
In contrast to the bubble diagram case, here we introduce 
a ``perpendicular basis'' for the triangle master integral containing 
two vectors. These two can be related 
to our ``orthogonal basis'' containing the momenta 
$P$, $R$, $T$ used to parametrize all considered 
types of loop diagrams --- bubble, triangle, box. 

The master vector triangle integral reads 
\eq 
C_1^\mu(s_1,s_2,s_3) &=& (i\pi^2)^{-1} \, \mu^{4-D} \,  
\int d^Dk \,\frac{k^\mu}{k^2 \, (k+l_1)^2 \, (k+l_2)^2} \,. 
\en
For handling triangle diagrams the 
perpendicular basis reads: 
\eq 
l_{1\perp}^\mu &=& l_1^\mu - l_2^\mu \, \frac{l_1^2}{l_1 \,l_2}\,,
\nonumber\\
l_{2\perp}^\mu &=& l_2^\mu - l_1^\mu \, \frac{l_2^2}{l_1 \,l_2}  \,,
\en 
with the following properties:
\eq 
& &l_{i\perp} \cdot l_i = 0\,, \quad i=1,2 \,, \nonumber\\
& &l_{i\perp}^2 = - \frac{l_i^2}{4 \, (l_1 \, l_2)^2} \, 
\lambda(l_1^2,l_2^2,l_3^2) \, \, , \quad i=1,2 \,, \nonumber\\
& &l_{1\perp} \cdot l_{2\perp} = \frac{\lambda(l_1^2,l_2^2,l_3^2)}
{4 l_1 l_2} , \nonumber\\
& &(l_{1\perp} \cdot l_{2\perp})^2 = 
\frac{(l_1 l_2)^2}{l_1^2 l_2^2} \ 
l_{1\perp}^2 \, l_{2\perp}^2 \,, 
\en 
where 
\eq\label{Kallen_f}
\lambda(x,y,z) = x^2 + y^2 + z^2 - 2xy - 2yz - 2xz \, ,
\en 
is the kinematic triangle K\"allen function. 
In addition, to accompany the perpendicular  basis,  
we introduce orthogonal metric tensor 
$g^{\mu\nu}_{\perp;3}$, which is defined 
as 
\eq 
g^{\mu\nu}_{\perp;3} = 
g^{\mu\nu} 
- \frac{l_{1\perp}^\mu l_{1\perp}^\nu l_2^2}{(l_{1 \perp} l_{2 \perp})^2} 
- \frac{l_{2\perp}^\mu l_{2\perp}^\nu l_1^2}{(l_{1 \perp} l_{2 \perp})^2} 
- \frac{(l_{1\perp}^\mu l_{2\perp}^\nu + 
	l_{2\perp}^\mu l_{1\perp}^\nu) l_1l_2}{(l_{1 \perp} l_{2 \perp})^2}   \, ,
\en 
and has the following properties 
\eq 
g^{\mu\nu}_{\perp;3} \cdot l_{i\perp; \mu} = 
g^{\mu\nu}_{\perp;3} \cdot l_{i\perp; \nu} = 
g^{\mu\nu}_{\perp;3} \cdot l_{i; \mu} = 
g^{\mu\nu}_{\perp;3} \cdot l_{i; \nu} = 0 \,, \quad i = 1,2   \, ,
\en 
and 
\eq 
g^{\mu\nu}_{\perp;3} \cdot g_{\perp;3; \mu\nu} = D-2 \,. 
\en 
Note, that such choice of basis is free of soft 
singularities occurring in the limit $l_1^2$ and/or $l_2^2=0$. 
In particular, in this limit $l_{i\perp} \to l_i$. 
Also it gives the straightforward results for the scalar 
functions in the expansion of tensorial triangle diagrams, 
i.e.\,without the need of solving a system of equations like 
in the original PV method. 

The expansion of the integral 
$C_1^\mu(s_1,s_2,s_3)$ in the basis reads 
\eq 
C_1^\mu(s_1,s_2,s_3) = l_{1\perp}^\mu \, C_{1;l1} \, + \, 
l_{2\perp}^\mu \, C_{1;l2}  \,, 
\en
where the scalar functions $C_{1;l1}$ and $C_{1;l2}$ are fixed 
by contraction of our loop integral with $l_2$ and $l_1$, 
respectively: 
\eq 
C_{1;l1}(s_1,s_2,s_3) 
= \frac{1}{l_{1\perp} l_{2\perp}} \, 
C_1^\mu(s_1,s_2,s_3) \, l_{2\mu} \,, \nonumber\\
C_{1;l2}(s_1,s_2,s_3) 
= \frac{1}{l_{1\perp} l_{2\perp}} \, 
C_1^\mu(s_1,s_2,s_3) \, l_{1\mu} \,. 
\en 
Analytic expressions for the functions $C_{1;l1}$ and $C_{1;l2}$ read 
\eq 
C_{1;l1}(s_1,s_2,s_3) 
&=& C_1(s_1,s_2,s_3) = 
\frac{2 l_1 l_2}{\lambda} \, 
\biggl[ B_0(s_1) - B_0(s_3) - s_2 \, C_0(s_1,s_2,s_3)
\biggr]\,, \nonumber\\
C_{1;l2}(s_1,s_2,s_3)  
&=& C_1(s_2,s_1,s_3) = 
\frac{2 l_1 l_2}{\lambda} \, 
\biggl[ B_0(s_2) - B_0(s_3) - s_1 \, C_0(s_1,s_2,s_3)
\biggr] \,.
\label{C1_expression}
\en 
Note, that $C_{1;l1}$ and $C_{1;l2}$ are related via exchange of 
momenta $l_1 \leftrightarrow l_2$ or $s_1 \leftrightarrow s_2$.

By analogy we pin down triangle tensor rank-2 integral 
\eq 
C_2^{\mu\nu}(s_1,s_2,s_3) 
&=& (i\pi^2)^{-1} \, \mu^{4-D} \,  
\int d^Dk \,\frac{k^\mu k^\nu}{k^2 \, (k+l_1)^2 \, (k+l_2)^2} 
\\
&=& g^{\mu\nu}_{\perp;3} \, C_{2;g}  \nonumber
+ 
 l_{1\perp}^\mu l_{1\perp}^\nu  \, C_{2;l_1l_1} + 
 l_{2\perp}^\mu l_{2\perp}^\nu  \, C_{2;l_2l_2} + 
(l_{1\perp}^\mu l_{2\perp}^\nu + 
 l_{2\perp}^\mu l_{1\perp}^\nu) \, C_{2;l_1l_2} \,. 
\en
Again, we can easily and directly extract the scalar functions 
$C_{2;l1l1}$, $C_{2;l2l2}$, and 
$C_{2;l1l2}$ using the formulas 
\eq
C_{2;l_1l_1} 
&=& \frac{1}{(l_{1\perp} l_{2\perp})^2}\, 
C_2^{\mu\nu}(s_1,s_2,s_3) \, l_{2\mu} l_{2\nu}\,, \nonumber\\
C_{2;l_2l_2} &=& 
\frac{1}{(l_{1\perp} l_{2\perp})^2}\, 
C_2^{\mu\nu}(s_1,s_2,s_3) \, l_{1\mu} l_{1\nu}\,, \nonumber\\
C_{2;l_1l_2} 
&=& \frac{1}{(l_{1\perp} l_{2\perp})^2}\, 
C_2^{\mu\nu}(s_1,s_2,s_3) \, l_{1\mu} l_{2\nu}\,, \nonumber\\
C_{2;g} &=& \frac{1}{D-2}\, 
C_2^{\mu\nu}(s_1,s_2,s_3) \, g_{\perp;3;\mu\nu} \,. 
\en 
Analytic expressions for these functions read 
\eq\label{C2_functions} 
C_{2;l_1l_1}(s_1,s_2,s_3) &=& \frac{4 (l_1 l_2)^2}{\lambda^2} \, 
\biggl[ 
- l_1l_2  \, B_0(s_1) + (s_2 + l_1 l_2) B_0(s_3) 
- \frac{\lambda \, s_2}{2 l_1 l_2} 
\, C_{1;l_1}(s_1,s_2,s_3)
\biggr] \,,\nonumber\\
C_{2;l_2l_2}(s_1,s_2,s_3) &=& \frac{4 (l_1 l_2)^2}{\lambda^2}  \, 
\biggl[ 
- l_1l_2  \, B_0(s_2) + (s_1 + l_1 l_2) B_0(s_3) 
- \frac{\lambda \, s_1}{2 l_1 l_2} 
\, C_{1;l_2}(s_1,s_2,s_3)
\biggr] \,,\nonumber\\
C_{2;l_1l_2}(s_1,s_2,s_3) &=& \frac{2 (l_1 l_2)^2}{\lambda^2} \, 
\biggl[ 
- s_1  \, B_0(s_1)  
- s_2  \, B_0(s_2) 
+ (s_1 + s_2 + 2 l_1 l_2)  \, B_0(s_3) \nonumber\\
&-& \frac{\lambda}{2 l_1 l_2} 
\, \Big( 
  s_1 C_{1;l_1}(s_1,s_2,s_3) 
+ s_2 C_{1;l_2}(s_1,s_2,s_3) \Big) 
\biggr] \,,\nonumber\\
C_{2;g}(s_1,s_2,s_3) &=& \frac{1}{D-2} \, \biggl[ 
B_0(s_3) + \frac{\lambda}{4 (l_1 l_2)^2} \, 
\Big(s_1 C_{2;l_1l_1}(s_1,s_2,s_3) 
   + s_2 C_{2;l_2l_2}(s_1,s_2,s_3) 
\Big) 
\nonumber\\
&-& \frac{\lambda}{2 l_1 l_2} \, C_{2;l_1l_2}(s_1,s_2,s_3) 
\biggr] \,.
\en 
Note, that the functions $C_{2;l_1l_1}(s_1,s_2,s_3)$ 
and $C_{2;l_1l_1}(s_1,s_2,s_3)$ 
are symmetric under exchange of first two arguments 
$s_1 \leftrightarrow s_2$.  
In Appendix~\ref{Triangle_details} we present the solutions for 
the scalar functions $C_{2;g}$, 
$C_{2;l_1l_1}$, $C_{2;l_2l_2}$, and $C_{2;l_1l_2}$ for 
specific choice of momenta (including soft limits). 
Finally, we note that the perpendicular basis vector $l_{i\perp}$ 
can be easily expressed through the basis vectors $P$, $R$, and $T$.  

Finally we present analytic results for triangle tensor rank-3 integral 
\eq 
C_3^{\mu\nu\alpha}(s_1,s_2,s_3) 
&=& (i\pi^2)^{-1} \, \mu^{4-D} \,  
\int d^Dk \,\frac{k^\mu k^\nu k^\alpha}{k^2 \, (k+l_1)^2 \, (k+l_2)^2} 
\nonumber\\
&=& g^{\mu\nu\alpha}_{\perp;l_{1\perp}} \, C_{3;gl_1}  
 +  g^{\mu\nu\alpha}_{\perp;l_{2\perp}} \, C_{3;gl_2}  
\nonumber\\
&+&  \{ l_{1\perp} l_{1\perp} l_{1\perp} \}^{\mu\nu\alpha} \, C_{3;l_1l_1l_1}   
 +  \{ l_{2\perp} l_{2\perp} l_{2\perp} \}^{\mu\nu\alpha} \, C_{3;l_2l_2l_2}   
\nonumber\\
&+&  \{ l_{1\perp} l_{1\perp} l_{2\perp} \}^{\mu\nu\alpha} \, C_{3;l_1l_1l_2}   
 +   \{ l_{2\perp} l_{2\perp} l_{1\perp} \}^{\mu\nu\alpha} \, C_{3;l_2l_2l_1}   
\,. 
\en
where 
\eq 
g^{\mu\nu\alpha}_{\perp;l_{i\perp}} &=& 
g^{\mu\nu}_{\perp;3} l_{i\perp}^\alpha + 
g^{\mu\alpha}_{\perp;3} l_{i\perp}^\beta + 
g^{\mu\beta}_{\perp;3} l_{i\perp}^\alpha  \,, \nonumber\\
\{ A A A \}^{\mu\nu\alpha} &=& 
A^\mu A^\nu A^\alpha \,, \nonumber\\
\{ A A B \}^{\mu\nu\alpha} &=& 
A^\mu A^\nu B^\alpha + A^\nu A^\alpha B^\mu + A^\alpha A^\mu B^\nu 
\,.
\en 
The scalar functions $C_{3;gl_1}$, 
$C_{3;gl_2}$, 
$C_{3;l_1l_1l_1}$, $C_{3;l_2l_2l_2}$, $C_{3;l_1l_1l_2}$, and 
$C_{3;l_2l_2l_1}$ are given by 
\eq 
C_{3;gl_1} &=& \frac{4 l_1l_2}{(D-2) \, \lambda(l_1^2,l_2^2,l_3^2)} \, 
g_{\mu\nu;3} \, l_{2\alpha} \, C_3^{\mu\nu\alpha}(s_1,s_2,s_3) 
\nonumber\\
&=& - \frac{l_1l_2}{(D-2) \, \lambda(l_1^2,l_2^2,l_3^2)} \, 
\frac{B_0(l_3^2)}{2} \, l_2 (2l_2 + l_3) \,, \\
C_{3;l_1l_1l_1} &=& \frac{64 (l_1l_2)^3}{\lambda^3(l_1^2,l_2^2,l_3^2)} \, 
l_{2\mu} \, l_{2\nu} \, l_{2\alpha} \, 
C_3^{\mu\nu\alpha}(s_1,s_2,s_3) 
\nonumber\\
&=& \frac{8 (l_1l_2)^3}{\lambda^3(l_1^2,l_2^2,l_3^2)} \, 
\biggl[ 
B_0(l_2^2) \, \frac{D (l_1l_2)^2 - l_1^2l_2^2}{D-1} 
\,-\, 
B_0(l_3^2) \, \frac{D (l_1l_3)^2 - l_1^2l_3^2}{D-1} \, 
\nonumber\\
&-& B_0(l_3^2)  \Big(l_1 (2l_1-l_3)\Big)^2
\,-\, C_{2;l_2l_2} \, \frac{l_1^2 \, \lambda^2(l_1^2,l_2^2,l_3^2)}{4 (l_1l_2)^2} 
\biggr] \,, \\
C_{3;l_1l_1l_2} &=& \frac{64 (l_1l_2)^3}{\lambda^3(l_1^2,l_2^2,l_3^2)} \, 
l_{2\mu} \, l_{2\nu} \, l_{1\alpha} \, 
C_3^{\mu\nu\alpha}(s_1,s_2,s_3) 
\nonumber\\
&=& \frac{8 (l_1l_2)^3}{\lambda^3(l_1^2,l_2^2,l_3^2)} \, 
\biggl[ 
B_0(l_2^2) \, l_2^4 
\,-\, 
B_0(l_3^2) \, \Big(l_2 (2l_2+l3)\Big)^2 
\nonumber\\
&-& C_{2;l_1l_1} \, 
\frac{l_1^2  \, \lambda^2(l_1^2,l_2^2,l_3^2)}{4 (l_1l_2)^2} 
\biggr] \,. 
\en 
Here functions 
$C_{3;gl_2}$, $C_{3;l_2l_2l_2}$, $C_{3;l_2l_2l_1}$ are 
obtainded from functions 
$C_{3;gl_1}$, $C_{3;l_1l_1l_1}$, $C_{3;l_1l_1l_2}$ 
under replacements $l_1 \to l_2$, $l_2 \to l_1$, $l_3 \to - l_3$. 
As before, the perpendicular basis vector $l_{i\perp}$ 
is easily expressed through the basis vectors $P$, $R$, and $T$.  

\subsection{Box integrals} 

A typical box integral is shown in Fig.~\ref{fig:box4d}(d) 
[the others are obtained by permutation of external momenta 
in the propagators]. 

We start with the calculation of the scalar integral 
\eq 
D_0(s,u,Q^2) \equiv I_{0125} 
= (i\pi^2)^{-1} \, \mu^{4-D} \, 
\int d^Dk \,\frac{1}{\Delta_0 \,\Delta_1 \,\Delta_2 \, \Delta_5} \,.          
\en 
To pin down this integral we use Feynman $\alpha$-parametrization 
(for calculation of fully massless integral with $Q^2 = 0$ see 
Ref.~\cite{Smirnov:2006ry}). The final results can be written 
in terms of the Gauss hypergeometric function\index{Gauss hypergeometric function} 
$_2F_1(a,b,c,z)$ 
[see details of calculation in Appendix~\ref{Box_calculation}].  
The expression for the box integral reads 
(see also Refs.~\cite{Duplancic:2000sk,Cachazo:2004zb,Ellis:2007qk}):
\eq 
D_0(s,u,Q^2) &=& - \frac{1}{su} \  
\Biggl[\pi^{\frac{D}{2}-2} \frac{\Gamma\Big(2-\frac{D}{2}\Big) \, 
	\Gamma^2\Big(\frac{D}{2}-2\Big)}{\Gamma(D-4)}\Biggr] \nonumber\\ 
&\times& 
\Biggl[ \Big(-\frac{s}{\mu^2}\Big)^{\frac{D}{2}-2} 
\  _2F_1\Big(1,\frac{D}{2}-2,\frac{D}{2}-1,-\frac{t}{u}\Big) \nonumber\\
&+&     \Big(-\frac{u}{\mu^2}\Big)^{\frac{D}{2}-2} 
\  _2F_1\Big(1,\frac{D}{2}-2,\frac{D}{2}-1,-\frac{t}{s}\Big) \nonumber\\
&-&     \Big(-\frac{Q^2}{\mu^2}\Big)^{\frac{D}{2}-2} 
\ _2F_1\Big(1,\frac{D}{2}-2,\frac{D}{2}-1,-\frac{Q^2 t}{s u}\Big) 
\Biggr] 
\nonumber\\
&=& - \frac{4}{su} \ \Biggl[\frac{D-3}{D-4}\Biggr] \  
\Biggl[ B_0(s) \  _2F_1\Big(1,\frac{D}{2}-2,\frac{D}{2}-1,-\frac{t}{u}\Big) 
\nonumber\\ 
&+& B_0(u) \  _2F_1\Big(1,\frac{D}{2}-2,\frac{D}{2}-1,-\frac{t}{s}\Big) 
\nonumber\\
&-& B_0(Q^2) \ _2F_1\Big(1,\frac{D}{2}-2,\frac{D}{2}-1,-\frac{Q^2 t}{s u}\Big) 
\Biggr]
\,. 
\en 
Note that in the on-shell limit $Q^2 = s + t + u = 0$ we 
reproduce the result for the scalar box given in Ref.~\cite{Smirnov:2006ry}. 
$\eps$ expansion of the scalar box diagram is discussed in 
detail in Appendix~\ref{Box_calculation}. The results 
for the other scalar box integrals are obtained via permutation 
of Mandelstam variables $(s,t,u)$. 
Here we display important symmetry properties of the scalar box integral:
\eq 
D_0(s,u,Q^2) &=& D_0(u,s,Q^2) = I_{0125}\,, \   \nonumber\\
D_0(s,t,Q^2) &=& D_0(t,s,Q^2) = I_{0126}\,, \   \nonumber\\
D_0(u,t,Q^2) &=& D_0(t,u,Q^2) = I_{0235}\,.  
\en 

Now we are in the position to derive the expressions for the tensorial 
box integrals (with one, two, and three loop momentum vectors in numerator) 
using Lorentz covariance in the orthogonal basis. The loop integrals 
will be expanded in terms of scalar bubble and triangle PV functions. 

We start with vector integrals: 
\eq
D_1^\mu(s,u,Q^2) &=& (i\pi^2)^{-1} \, \mu^{4-D} \, 
\int d^Dk \,\frac{k^\mu}{\Delta_0 \,\Delta_1 \,\Delta_2 \, \Delta_5}     
\nonumber\\      
&=& P^\mu D_{1P}^{0125} 
+ R^\mu D_{1R}^{0125} 
+ T^\mu D_{1T}^{0125}\,, \label{Box_Vector1} \\ 
D_1^\mu(s,t,Q^2) &=& (i\pi^2)^{-1} \, \mu^{4-D} \, 
\int d^Dk \,\frac{k^\mu}{\Delta_0 \,\Delta_1 \,\Delta_2 \, \Delta_6}      \nonumber\\    
&=& P^\mu D_{1P}^{0126} 
+ R^\mu D_{1R}^{0126} 
+ T^\mu D_{1T}^{0126}\,, \label{Box_Vector2} \\ 
D_1^\mu(t,u,Q^2) &=& (i\pi^2)^{-1} \, \mu^{4-D} \, 
\int d^Dk \,\frac{k^\mu}{\Delta_0 \,\Delta_2 \,\Delta_3 \, \Delta_5}        \nonumber\\  
&=& P^\mu D_{1P}^{0235} 
+ R^\mu D_{1R}^{0235} 
+ T^\mu D_{1T}^{0235}\,, \label{Box_Vector3} 
\en  
where $D_{1F}^{ijkl}$ are the scalar functions, which are obtained 
by contraction of the vector integrals $D_1^\mu(s_1,s_2,Q^2)$ 
with corresponding Lorentz structure.  
In contrast to the original PV method, there remains no system 
of equations to be solved.
E.g., the function $D_{1P}^{0125}$ is simply fixed as 
\eq 
D_{1P}^{0125} = \frac{1}{P^2} \, P_\mu \cdot D_1^\mu(s,u,Q^2) \,. 
\en 
Then we use the fractioning identity 
\eq\label{2kP_frac_ID} 
2kP = \Delta_1 - \Delta_2  \,. 
\en 
to reduce the number of factors 
in the denominator of $D_1^\mu(s,u,Q^2)$. 
Using Eq.~(\ref{2kP_frac_ID}) we get an expression for 
$D_{1P}^{0125}$ as linear combination of known scalar triangle 
integrals:  
\eq 
D_{1P}^{0125} = \frac{1}{2} \, 
\biggl[C_0^{025} - C_0^{015}\biggr] 
= \frac{1}{2} \, 
\biggl[C_0(u,Q^2) - C_0(u) \biggr]  \,.
\en 
By analogy we pin down all scalar functions 
occurring in Eqs.~(\ref{Box_Vector1})-(\ref{Box_Vector3}), 
which are listed in Appendix~\ref{Vector_Box_Functions}. 
Also there we give the set of the fractional identities 
needed to simplify the box diagrams. 

The tensor rank-2 box integrals read: 
\eq
D_2^{\mu\nu}(s,u,Q^2) &=& (i\pi^2)^{-1} \, \mu^{4-D} \, 
\int d^Dk \,\frac{k^\mu k^\nu}
{\Delta_0 \, \Delta_1 \, \Delta_2 \, \Delta_5}         \nonumber\\ 
&=& g^{\mu\nu}_\perp \, D_{2g}^{0125} 
+ \{P P\}^{\mu\nu}   \, D_{2PP}^{0125} 
+ \{R R\}^{\mu\nu}   \, D_{2RR}^{0125} 
+ \{T T\}^{\mu\nu}   \, D_{2TT}^{0125} \nonumber\\[2mm]
&+& 
  \{P R\}^{\mu\nu} \, D_{2PR}^{0125} 
+ \{P T\}^{\mu\nu} \, D_{2PT}^{0125} 
+ \{R T\}^{\mu\nu} \, D_{2RT}^{0125} 
\,, \label{Box_Tensor2_1} \\[4mm] 
D_2^{\mu\nu}(s,t,Q^2) &=& (i\pi^2)^{-1} \, \mu^{4-D} \, 
\int d^Dk \,\frac{k^\mu k^\nu}
{\Delta_0 \,\Delta_1 \,\Delta_2 \, \Delta_6}    \nonumber\\      
&=& g^{\mu\nu}_\perp \, D_{2g}^{0126} 
+ \{P P\}^{\mu\nu}   \, D_{2PP}^{0126} 
+ \{R R\}^{\mu\nu}   \, D_{2RR}^{0126} 
+ \{T T\}^{\mu\nu}   \, D_{2TT}^{0126} \nonumber\\[2mm]
&+& 
\{P R\}^{\mu\nu}    \, D_{2PR}^{0126} 
+ \{P T\}^{\mu\nu}  \, D_{2PT}^{0126} 
+ \{R T\}^{\mu\nu}  \, D_{2RT}^{0126} 
\,, \label{Box_Tensor2_2} \\[4mm] 
D_2^{\mu\nu}(t,u,Q^2) &=& (i\pi^2)^{-1} \, \mu^{4-D} \, 
\int d^Dk \,\frac{k^\mu k^\nu}
{\Delta_0 \,\Delta_2 \,\Delta_3 \, \Delta_5}  \nonumber\\        
&=& g^{\mu\nu}_\perp \, D_{2g}^{0235} 
+ \{P P\}^{\mu\nu}   \, D_{2PP}^{0235} 
+ \{R R\}^{\mu\nu}   \, D_{2RR}^{0235} 
+ \{T T\}^{\mu\nu}   \, D_{2TT}^{0235} \nonumber\\[2mm]
&+& 
  \{P R\}^{\mu\nu}   \, D_{2PR}^{0235} 
+ \{P T\}^{\mu\nu}   \, D_{2PT}^{0235} 
+ \{R T\}^{\mu\nu}   \, D_{2RT}^{0235} 
\,, \label{Box_Tensor2_3} 
\en 
where 
$\{A A\}^{\mu\nu} = A^\mu A^\nu$ and 
$\{A B\}^{\mu\nu} = A^\mu B^\nu + A^\nu B^\mu$. 
We handle tensor rank-2 box integrals analogous to  vector 
integrals. In particular, we fix the scalar functions 
$D_{2F}^{ijkl}$ occurring in the  
expansions~(\ref{Box_Tensor2_1})-(\ref{Box_Tensor2_3}) 
by multiplying the corresponding loop integral with the corresponding 
Lorenz structures $g^{\mu\nu}_\perp$, $P^\mu P^\nu$, $R^\mu R^\nu$, 
and $T^\mu T^\nu$, respectively. 
We do not need to solve any system of equations as 
in the original PV  method. Again, as for vectorial 
box integrals we use the fractioning identities 
(see Appendix~\ref{Vector_Box_Functions}). 
In particular, the results for the $D_{2g}^{0125}$, 
$D_{2PP}^{0125}$, $D_{2PR}^{0125}$, and $D_{2TT}^{0125}$ 
functions read: 
\eq 
D_{2PP}^{0125} &=& 
 \frac{1}{P^4} \, P_\mu P_\nu \cdot D_2^{\mu\nu}(s,u,Q^2)  
= \frac{1}{2s} \, \biggl[C_{1P}^{025} - C_{1P}^{015}\biggr]\,, 
\label{D2PP_res}\\
D_{2RR}^{0125} &=& 
 \frac{1}{R^4} \, R_\mu R_\nu \cdot D_2^{\mu\nu}(s,u,Q^2)  
= \frac{1}{2s} \, \biggl[C_{1R}^{025} + C_{1R}^{015} 
- 2 C_{1R}^{078} + C_0^{025} \biggr]\,, 
\label{D2RR_res}\\
D_{2TT}^{0125} &=& 
 \frac{1}{T^4} \, T_\mu T_\nu \cdot D_2^{\mu\nu}(s,u,Q^2)  = - \frac{s}{2ut} \, \biggl[ u D_{1T}^{0125} 
- C_{1T}^{015} \, \frac{u}{s} \nonumber\\
&+& C_{1T}^{025} \, \frac{s+t}{s} 
 +  C_{1T}^{078} \, \frac{u-t}{s} 
\biggr]\,, 
\label{D2TT_res}\\
D_{2g}^{0125} &=& 
\frac{1}{D-3} \, g_{\mu\nu}^\perp \cdot D_2^{\mu\nu}(s,u,Q^2)  
= \frac{1}{D-3} \, \biggl[C_0^{078} 
- s \, D_{2PP}^{0125}    \nonumber\\
&+& s \, D_{2RR}^{0125}
 + \frac{ut}{s} \, D_{2TT}^{0125} 
\biggr]
\label{D2g_res} \,.
\en 

The tensorial rank-3 box integrals read: 
\eq\label{Box_Tensor3_1} 
\hspace*{-.5cm}
D_3^{\mu\nu\alpha}(s,u,Q^2) &=& (i\pi^2)^{-1} \, \mu^{4-D} \, 
\int d^Dk \,\frac{k^\mu k^\nu k^\alpha}
{\Delta_0 \,\Delta_1 \,\Delta_2 \, \Delta_5}          
= g^{\mu\nu\alpha}_{\perp P} \, D_{3gP}^{0125} 
+ g^{\mu\nu\alpha}_{\perp R} \, D_{3gR}^{0125} 
+ g^{\mu\nu\alpha}_{\perp T} \, D_{3gT}^{0125} 
\nonumber\\[2mm]
\hspace*{-.5cm}
&+&\{P P P\}^{\mu\nu\alpha}  \, D_{3PPP}^{0125} 
+  \{R R R\}^{\mu\nu\alpha}  \, D_{3RRR}^{0125} 
+  \{T T T\}^{\mu\nu\alpha}  \, D_{3TTT}^{0125} 
\nonumber\\[2mm]
\hspace*{-.5cm}
&+&\{P P R\}^{\mu\nu\alpha}  \, D_{3PPR}^{0125} 
+  \{P P T\}^{\mu\nu\alpha}  \, D_{3PPT}^{0125}
+  \{R R P\}^{\mu\nu\alpha}  \, D_{3RRP}^{0125} 
\nonumber\\[2mm]
\hspace*{-.5cm}
&+&\{R R T\}^{\mu\nu\alpha}  \, D_{3RRT}^{0125} 
+  \{T T P\}^{\mu\nu\alpha}  \, D_{3TTP}^{0125} 
+  \{T T R\}^{\mu\nu\alpha}  \, D_{3TTR}^{0125} 
\nonumber\\[2mm]
\hspace*{-.5cm}
&+&\{P R T\}^{\mu\nu\alpha}  \, D_{3PRT}^{0125} 
\,, 
\en 
\eq\label{Box_Tensor3_2}  
\hspace*{-.5cm}
D_3^{\mu\nu\alpha}(s,t,Q^2) &=& (i\pi^2)^{-1} \, \mu^{4-D} \, 
\int d^Dk \,\frac{k^\mu k^\nu k^\alpha}
{\Delta_0 \,\Delta_1 \,\Delta_2 \, \Delta_6}          
= g^{\mu\nu\alpha}_{\perp P} \, D_{3gP}^{0126} 
+ g^{\mu\nu\alpha}_{\perp R} \, D_{3gR}^{0126} 
+ g^{\mu\nu\alpha}_{\perp T} \, D_{3gT}^{0126} 
\nonumber\\[2mm]
\hspace*{-.5cm}
&+&\{P P P\}^{\mu\nu\alpha}  \, D_{3PPP}^{0126} 
+  \{R R R\}^{\mu\nu\alpha}  \, D_{3RRR}^{0126} 
+  \{T T T\}^{\mu\nu\alpha}  \, D_{3TTT}^{0126} 
\nonumber\\[2mm]
\hspace*{-.5cm}
&+&\{P P R\}^{\mu\nu\alpha}  \, D_{3PPR}^{0126} 
+  \{P P T\}^{\mu\nu\alpha}  \, D_{3PPT}^{0126}
+  \{R R P\}^{\mu\nu\alpha}  \, D_{3RRP}^{0126} 
\nonumber\\[2mm]
\hspace*{-.5cm}
&+&\{R R T\}^{\mu\nu\alpha}  \, D_{3RRT}^{0126} 
+  \{T T P\}^{\mu\nu\alpha}  \, D_{3TTP}^{0126} 
+  \{T T R\}^{\mu\nu\alpha}  \, D_{3TTR}^{0126} 
\nonumber\\[2mm]
&+&\{P R T\}^{\mu\nu\alpha}  \, D_{3PRT}^{0126} 
\,, 
\en
\eq\label{Box_Tensor3_3}   
\hspace*{-.5cm}
D_3^{\mu\nu\alpha}(t,u,Q^2) &=& (i\pi^2)^{-1} \, \mu^{4-D} \, 
\int d^Dk \,\frac{k^\mu k^\nu k^\alpha}
{\Delta_0 \,\Delta_2 \,\Delta_3 \, \Delta_5}          
= g^{\mu\nu\alpha}_{\perp P} \, D_{3gP}^{0235} 
+ g^{\mu\nu\alpha}_{\perp R} \, D_{3gR}^{0235} 
+ g^{\mu\nu\alpha}_{\perp T} \, D_{3gT}^{0235} 
\nonumber\\[2mm]
\hspace*{-.5cm}
&+&\{P P P\}^{\mu\nu\alpha}  \, D_{3PPP}^{0235} 
+  \{R R R\}^{\mu\nu\alpha}  \, D_{3RRR}^{0235} 
+  \{T T T\}^{\mu\nu\alpha}  \, D_{3TTT}^{0235} 
\nonumber\\[2mm]
\hspace*{-.5cm}
&+&\{P P R\}^{\mu\nu\alpha}  \, D_{3PPR}^{0235} 
+  \{P P T\}^{\mu\nu\alpha}  \, D_{3PPT}^{0235}
+  \{R R P\}^{\mu\nu\alpha}  \, D_{3RRP}^{0126} 
\nonumber\\[2mm]
\hspace*{-.5cm}
&+&  \{R R T\}^{\mu\nu\alpha}  \, D_{3RRT}^{0235} 
+  \{T T P\}^{\mu\nu\alpha}  \, D_{3TTP}^{0235} 
+  \{T T R\}^{\mu\nu\alpha}  \, D_{3TTR}^{0235} 
\nonumber\\[2mm]
&+&\{P R T\}^{\mu\nu\alpha}  \, D_{3PRT}^{0235} 
\,, 
\en 
where 
\eq
g^{\mu\nu\alpha}_{\perp A} &=& 
  g^{\mu\nu}_{\perp} A^\alpha 
+ g^{\mu\alpha}_{\perp} A^\nu
+ g^{\nu\alpha}_{\perp} A^\mu\,, 
\nonumber\\
\{A A A\}^{\mu\nu\alpha} &=& 
A^\mu A^\nu A^\alpha \,, \nonumber\\
\{A A B\}^{\mu\nu\alpha} &=& 
  A^\mu A^\nu    B^\alpha 
+ A^\mu A^\alpha B^\nu 
+ A^\nu A^\alpha B^\mu \,, \\
\{A B C\}^{\mu\nu\alpha} &=& 
  A^\mu    B^\nu    C^\alpha 
+ A^\mu    B^\alpha C^\nu 
+ A^\nu    B^\mu    C^\alpha
+ A^\nu    B^\alpha C^\mu 
+ A^\alpha B^\mu    C^\nu 
+ A^\alpha B^\nu    C^\mu 
\,. \nonumber
\en
As before the scalar functions defining expansion of rank-3 box integrals 
are simply fixed by contraction of the integral 
$D_3^{\mu\nu\alpha}$ with corresponding Lorentz structure: 
\eq 
D_{3AAA} &=& 
\frac{A_\mu A_\nu A_\alpha}{A^6} \, 
D_3^{\mu\nu\alpha}\,, 
\\[1mm]
D_{3AAB} &=& 
\frac{A_\mu A_\nu B_\alpha}{A^4 B^2} \, 
D_3^{\mu\nu\alpha}\,, 
\\[1mm]
D_{3ABC} &=& 
\frac{A_\mu B_\nu C_\alpha}{A^2 B^2 C^2} \, 
D_3^{\mu\nu\alpha} 
\\[1mm]
D_{3gA} &=& 
\frac{g_{\mu\nu}^\perp A_\alpha}{(D-3) A^2} \, 
D_3^{\mu\nu\alpha}
\,, \label{D3list} 
\en  
where $A, B, C = P, R, T$. 
We list all scalar functions occurring in the expansion of tensorial 
rank-2 and rank-3 box integrals in Appendix~\ref{Tensor_rank23_Box}. 

Extension on tensorial box integral with higher rank is straightforward. 
One should just specify the Lorentz structure of corresponding 
integral and then fix the scalar functions on which the integral is 
expanded. In particular, decomposition of the rank-4 box inegral 
with denominators $\Delta_0$, $\Delta_1$, $\Delta_2$, and 
$\Delta_5$ reads: 
\eq 
D_4^{\mu\nu\alpha\beta}(s,u,Q^2) &=& (i\pi^2)^{-1} \, \mu^{4-D} \, 
\int d^Dk \,\frac{k^\mu k^\nu k^\alpha k^\beta}
{\Delta_0 \, \Delta_1 \, \Delta_2 \, \Delta_5}         \nonumber\\ 
&=& g^{\mu\nu\alpha\beta}_\perp \, D_{4gg}^{0125} 
+ g^{\mu\nu\alpha\beta}_{\perp; PP} \, D_{4g; PP}^{0125} 
+ g^{\mu\nu\alpha\beta}_{\perp; RR} \, D_{4g; RR}^{0125} 
+ g^{\mu\nu\alpha\beta}_{\perp; TT} \, D_{4g; TT}^{0125} 
\nonumber\\
&+& g^{\mu\nu\alpha\beta}_{\perp; PR} \, D_{4g; PR}^{0125} 
+   g^{\mu\nu\alpha\beta}_{\perp; PT} \, D_{4g; PT}^{0125} 
+   g^{\mu\nu\alpha\beta}_{\perp; RT} \, D_{4g; RT}^{0125} 
\nonumber\\ 
&+& \{P P P P\}^{\mu\nu\alpha\beta}   \, D_{4PPPP}^{0125} 
+   \{R R R R\}^{\mu\nu\alpha\beta}   \, D_{4RRRR}^{0125} 
+   \{T T T T\}^{\mu\nu\alpha\beta}   \, D_{4TTTT}^{0125}
\nonumber\\ 
&+& 
   \{P P P R\}^{\mu\nu\alpha\beta}   \, D_{4PPPR}^{0125} 
+  \{P P P T\}^{\mu\nu\alpha\beta}   \, D_{4PPPT}^{0125}  
+  \{R R R P\}^{\mu\nu\alpha\beta}   \, D_{4RRRP}^{0125}  
\nonumber\\ 
&+&
   \{R R R T\}^{\mu\nu\alpha\beta}   \, D_{4RRRT}^{0125} 
+  \{T T T P\}^{\mu\nu\alpha\beta}   \, D_{4TTTP}^{0125}  
+  \{T T T R\}^{\mu\nu\alpha\beta}   \, D_{4TTTR}^{0125}  
\nonumber\\ 
&+& 
   \{P P R R\}^{\mu\nu\alpha\beta}   \, D_{4PPRR}^{0125} 
+  \{P P T T\}^{\mu\nu\alpha\beta}   \, D_{4PPTT}^{0125}  
+  \{P P R T\}^{\mu\nu\alpha\beta}   \, D_{4PPRT}^{0125}  
\nonumber\\ 
&+& 
   \{R R T T\}^{\mu\nu\alpha\beta}   \, D_{4RRTT}^{0125} 
+  \{R R P T\}^{\mu\nu\alpha\beta}   \, D_{4RRPT}^{0125}  
+  \{T T P R\}^{\mu\nu\alpha\beta}   \, D_{4TTPR}^{0125}  
\,, \nonumber\\
\label{Box_Tensor4_1} 
\en 
where 
\eq 
g^{\mu\nu\alpha\beta}_\perp &=& 
g^{\mu\nu}_\perp \, g^{\alpha\beta}_\perp 
\,+\, 
g^{\mu\alpha}_\perp \, g^{\nu\beta}_\perp 
\,+\, 
g^{\mu\beta}_\perp \, g^{\nu\alpha}_\perp 
\,, \\ 
g^{\mu\nu\alpha\beta}_{\perp; AA} &=& 
g^{\mu\nu}_{\perp} \, \{AA\}^{\alpha\beta} 
\,+\, 
g^{\mu\alpha}_{\perp} \, \{AA\}^{\nu\beta} 
\,+\, 
g^{\mu\beta}_{\perp} \, \{AA\}^{\nu\alpha} 
\,, \\
g^{\mu\nu\alpha\beta}_{\perp; AB} &=& 
g^{\mu\nu}_{\perp} \, \{AB\}^{\alpha\beta} 
\,+\, 
g^{\mu\alpha}_{\perp} \, \{AB\}^{\nu\beta} 
\,+\, 
g^{\mu\beta}_{\perp} \, \{AB\}^{\nu\alpha} 
\nonumber\\
&+& 
g^{\alpha\beta}_{\perp} \, \{AB\}^{\mu\nu} 
\,+\, 
g^{\nu\beta}_{\perp} \, \{AB\}^{\mu\alpha} 
\,+\, 
g^{\nu\alpha}_{\perp} \, \{AB\}^{\nu\beta} \,, \\
\{A A A A\}^{\mu\nu\alpha\beta} &=& 
A^\mu A^\nu A^\alpha A^\beta\,, \nonumber\\
\{A A A B\}^{\mu\nu\alpha} &=& 
  A^\mu A^\nu A^\alpha B^\beta 
+ A^\nu A^\alpha A^\beta B^\mu 
+ A^\alpha A^\beta A^\mu B^\nu 
+ A^\beta A^\mu A^\nu B^\alpha 
\,, \\
\{A A B C\}^{\mu\nu\alpha} &=& 
  A^\mu A^\nu B^\alpha C^\beta 
+ A^\mu A^\nu B^\beta  C^\alpha 
+ A^\nu A^\alpha B^\beta C^\mu 
+ A^\nu A^\alpha B^\mu   C^\beta 
\nonumber\\
&+& A^\alpha A^\beta B^\mu C^\nu 
+ A^\alpha A^\beta B^\nu C^\mu 
\,. \nonumber
\en 
Again, all scalar functions defining expansion of rank-4 box integrals 
are simply fixed by contraction of the integral 
$D_4^{\mu\nu\alpha\beta}$ with corresponding Lorentz structure: 
\eq
D_{4gg} &=& \frac{1}{(D-3) (D-1)} \, 
g_{\mu\nu}^\perp \, g_{\alpha\beta}^\perp  \cdot  
D_4^{\mu\nu\alpha\beta} \,, \\ 
D_{4gg;AA} &=& \frac{1}{(D-3) A^4} \, 
g_{\mu\nu}^\perp \, A_\alpha A_\beta \cdot  
D_4^{\mu\nu\alpha\beta} \,, \\ 
D_{4gg;AB} &=& \frac{1}{(D-3) A^2 B^2} \, 
g_{\mu\nu}^\perp \, A_\alpha B_\beta \cdot  
D_4^{\mu\nu\alpha\beta} \,, \\ 
D_{4AAAA} &=& 
\frac{A_\mu A_\nu A_\alpha A^\beta}{A^8} \, 
D_4^{\mu\nu\alpha\beta}\,, 
\\[1mm]
D_{4AAAB} &=& 
\frac{A_\mu A_\nu A_\alpha B^\beta}{A^6 B^2} \, 
D_4^{\mu\nu\alpha\beta}\,, 
\\[1mm]
D_{4AABC} &=& 
\frac{A_\mu A_\nu B_\alpha C^\beta}{A^4 B^2 C^2} \, 
D_4^{\mu\nu\alpha\beta}
\,. \label{D4list} 
\en

\section{Angular integrals in \texorpdfstring{$D$}{TEXT} dimensions}  
\label{sec:Angular_integrals}

In this section we discuss new ideas concerning angular integration in QCD. 
We extend the known closed results for massless and single massive two denominator 
integrals in $D$ dimensions derived 
in Refs.~\cite{vanNeerven:1985xr,Beenakker:1988bq,Somogyi:2011ir} 
to the double massive case. 
Furthermore we investigate novel analytic and algebraic properties of angular integrals. 
An all-order $\eps$-expansion is given for all angular integrals with up to
two denominators. A supplementary geometric picture for partial fractioning is developed 
in Appendix~\ref{sec:Partial Fractioning}
which provides a new rotational invariant algorithm to reduce the number of denominators 
and also proves to be useful for the double massive integral. 

\subsection{Generalized two particle phase space in \texorpdfstring{$D$}{TEXT} dimensions}

To set the stage, we consider the Lorentz invariant phase space integral (PSI) 
\begin{align}
\int\dx\mathrm{PS}_{2,P}=\int\frac{\dx^{D-1} k_1}{(2\pi)^{D-1} 2k_{1}^0}
\int\frac{\dx^{D-1} k_2}{(2\pi)^{D-1}2k_{2}^0}\,(2\pi)^D \delta^D\left(P-k_1-k_2\right)\,.
\label{eq:2bodyPS}
\end{align}
where $P=p_1+p_2+\dots+ p_n$ is the sum of some fixed momenta. 
One might think of this PSI  
potentially as a part of some larger PSI. 
$P$ shall be time-like and future-pointing such that there exists a rest frame of $P$ 
with $P^0>0$. The $k_1$ and $k_2$ are assumed as momenta of massless particles, 
i.e.\,$k_{1}^0=|{\bf k_1}|$ and $k_{2}^0=|{\bf k_2}|$. 
But we should stress that, in general, 
final particles are not neccessary massless and we can deal with 
massive particles (see also Ref.~\cite{Mirkes:1992hu}). 

Applying the identity
\begin{align}
\int\dx^D k \, \delta(k^2)  \, \theta(E_k) \, f(k)=\int\frac{\dx^{D-1} k}{2 k^0} \, f(k) \,,
\end{align}
for the integral over $k_2$ and evaluating it using the $D$ dimensional $\delta$-function, 
the phase space is written as 
\begin{align}
\int\dx\mathrm{PS}_{2,P}=\int\frac{\dx^{D-1} k_1}{(2\pi)^{D-2}|{\bf k_1}|}\,
\delta\left((P_1-k_1)^2\right)\theta(P^0-k_1^0)\,.
\end{align}
By choosing the rest frame of $P$ we have 
$P=(\sqrt{P^2},\mathbf{0})$ and $k_1=(|\mathbf{k}|,\mathbf{k})$ 
leading to $(P-k_1)^2=2\sqrt{P^2}\left(\frac{\sqrt{P^2}}{2}-|\mathbf{k}|\right)$\,.
Employing spherical coordinates in this frame and evaluating the radial integral, 
we get
\begin{align}
\int\dx\mathrm{PS}_{2,P}&=
\frac{1}{2(2\pi)^{D-2} \sqrt{P^2}} \, \int_0^\infty \dx |\mathbf{k}| \, |\mathbf{k}|^{D-3}
\int\dx\Omega_{D-2}\,\delta\left(|\mathbf{k}| - \frac{\sqrt{P^2}}{2}\right) 
\theta\left(\sqrt{P^2}-|\mathbf{k}|\right)\,
\nonumber\\
&=\frac{(P^2)^\frac{D-4}{2}}{2(4\pi)^{D-2}}\int\dx\Omega_{D-2}\,.
\end{align}
Notice that the $\theta$-function is trivially satisfied since $P^0>0$. 
If we are in a situation, where the function we want to integrate over the phase space, 
does only depend on two angles, the remaining $D-4$ dimensional spherical integral 
can be integrated out (see 
details of spherical integration in dimensional regularization in Appendix~\ref{app:spher_int}). 
Defining 
\begin{align}\label{dOmega}
\dx\Omega_{k_1 k_2}\equiv\dx\theta_1\sin^{1-2\eps}\theta_1\dx\theta_2\sin^{-2\eps}\theta_2\,
\end{align}
with $D=4-2\eps$ and using the area formula for $n$ dimensional spheres
\begin{align}
\Omega_n=\frac{2 \pi^\frac{n+1}{2}}{\Gamma\left(\frac{n+1}{2}\right)}=\frac{2^n\pi^\frac{n}{2}
\Gamma\left(\frac{n}{2}\right)}{\Gamma(n)}=
\frac{2 (4\pi)^\frac{n}{2}\Gamma\left(1+\frac{n}{2}\right)}{\Gamma(1+n)}
\end{align}
the angular integral becomes
\begin{align}
\int\dx\Omega_{D-2}=\int\dx\Omega_{D-4}\int\dx\Omega_{k_1 k_2}\longrightarrow \Omega_{D-4}
\int\dx\Omega_{k_1 k_2}=\frac{2\, (4\pi)^{-\eps}\Gamma(1-\eps)}{\Gamma(1-2\eps)}\int\dx\Omega_{k_1 k_2}\,.
\end{align}
Therefore, 
we find the well known formula for the two-body phase space in $D=4-2\eps$ dimensions 
in the rest frame of $P$
\begin{align}
\int\dx\mathrm{PS}_{2,P}=\frac{\Gamma(1-\eps)}{ (4\pi)^{2-\eps}
\,\Gamma(1-2\eps)}\left(P^2\right)^{-\eps}
\int\dx\Omega_{k_1 k_2}\,.
\end{align}

If the integrand of the PSI is a rational function of the angular dependent 
Mandelstam variables, then it can be further simplified and 
reduced to a set of master integrals by partial fraction decomposition. 
In Appendix~\ref{sec:Partial Fractioning} we discuss in details a novel geometric approach 
developed for this task. There, we considered the double real corrections
in Drell-Yan process as an illustration of how the partial fractioning works (see Appendix~\ref{DY_example}).

\subsection{Analytic evaluation of angular integrals in \texorpdfstring{$D$}{TEXT}  dimension}
\label{sec:Somogyi integrals}

We introduce a set of basis integrals to which all of the Neerven integrals can be reduced, 
but that are characterized by properties that are invariant under rotation rather than 
the parameters $a$, $b$, $A$, $B$, and $C$ from Eq.~(\ref{eq:Van Neerven integral}).
They will now depend on the variables $v_{ij} = v_i\, \cdot\, v_j$,  
which are all invariant under change of coordinates.\footnote{Note that this more natural normalization 
is different than that used by Somogyi~\cite{Somogyi:2011ir}, which is taylored 
to the Mellin-Barnes representations.}
We define 

\noindent 
Zero-denominator integral:  
\eq
I^{(0)}(\eps)&:=&\int\dx\Omega_{k_1k_2} \,.
\en
One-denominator massless integral:  
\eq
I^{(0)}_j(\eps)&:=&\int\dx\Omega_{k_1k_2} \ \dfrac{1}{(v_1\cdot k)^j}\,, 
\qquad v_{11}=0 \,. 
\en
One-denominator massive integral:  
\eq
I^{(1)}_j(v_{11};\eps)&:=\int\dx\Omega_{k_1k_2} \  
\dfrac{1}{(v_1\cdot k)^j}\,, \qquad v_{11}\neq 0 \,. 
\en
Two-denominator massless integral:  
\eq
I^{(0)}_{j,l}(v_{12};\eps)&:=\int\dx\Omega_{k_1k_2} \  
\dfrac{1}{(v_1\cdot k)^j \, (v_2\cdot k)^l}\,, \qquad v_{11}=v_{22}=0 \,.
\en
Two-denominator single-massive integral: 
\eq 
I^{(1)}_{j,l}(v_{12},v_{11};\eps)&:=\int\dx\Omega_{k_1k_2} \  
\dfrac{1}{(v_1\cdot k)^j \, (v_2\cdot k)^l}\,, \qquad v_{11}\neq 0,\, v_{22}=0 \,.
\en 
Two-denominator double-massive integral: 
\eq 
I^{(2)}_{j,l}(v_{12},v_{11},v_{22};\eps)&:=\int\dx\Omega_{k_1k_2} \ 
\dfrac{1}{(v_1\cdot k)^j \, (v_2\cdot k)^l}\,, \qquad v_{11}\neq 0,\, v_{22}\neq 0 \,.
\en
Here 
\begin{align}
\int\dx\Omega_{k_1k_2}=\int_0^\pi\dx\theta_1\sin^{1-2\eps}\theta_1\int_0^\pi\dx\theta_2\sin^{-2\eps}\theta_2 \, ,
\end{align}
is the integration measure
in a frame with $k=(1,\dots,\cos\theta_2\sin\theta_1,\cos\theta_1)$ 
[the dots represent irrelevant angles since $\mathbf{v_1}$ and $\mathbf{v_2}$ lie in the $(x_{D-1},x_D)$-plane]. 
The upper number in parenthesis gives the number of masses involved and the number of lower indices gives 
the number of different denominators. All appearing $D$-vectors are normalized such that their 0-component 
is equal to 1.

We will calculate the integrals one by one which has the benefit that we gradually increase the difficulty of 
the calculation and can introduce the employed methods one by one.

\subsubsection{Integral without denominator}

We start with the basic integral $I^{(0)}(\eps)$ without any denominator
\begin{align}
I^{(0)}(\eps)=\int\dx\Omega_{k_1k_2}=\int_0^\pi\dx\theta_1\sin^{1-2\eps}\theta_1
\int_0^\pi\dx\theta_2\sin^{-2\eps}\theta_2\,.
\end{align}
Changing the variables in both integrals as $\cos\theta_i=1-2t_i$ with $d\theta_i \, \sin\theta_i = 2 dt_i$ and 
$\sin\theta_i = 2 \sqrt{t_i \, (1-t_i)}$ we rewrite the integral $I^{(0)}(\eps)$ in the form 
\begin{align}
I^{(0)}(\eps)=2^{1-4\eps}\int_0^1\dx t_1\,t_1^{-\eps}(1-t_1)^{-\eps} \, 
                         \int_0^1\dx t_2\,t_2^{-\frac{1}{2}-\eps}(1-t_2)^{-\frac{1}{2}-\eps} \,,
\end{align}
containing the product of two integral representations of the beta function\index{Beta function} 
\eq\label{beta_Int} 
B(x,y) = \frac{\Gamma(x)\Gamma(y)}{\Gamma(x+y)} = 
\int\limits_0^1 \dx t \, t^{x-1} \, (1-t)^{y-1} \,.
\en 
Taking into account Eq.~(\ref{beta_Int}) we finally get:
\eq 
I^{(0)}(\eps)=2^{1-4\eps}\frac{\Gamma^2(1-\eps)}{\Gamma(2-2\eps)}
\frac{\Gamma^2\left(\frac{1}{2}-\eps\right)}{\Gamma(1-2\eps)}\,.
\en 
Using the Legendre duplication formula\index{Legendre duplication formula}  
\begin{align}
\Gamma\left(\frac{1}{2}-\eps\right)\Gamma(1-\eps)=\frac{\sqrt{\pi}}{2^{-2\eps}}\Gamma(1-2\eps) \,,
\end{align}
$I^{(0)}(\eps)$ is further simplified to 
\begin{align}
I^{(0)}(\eps)=2\pi\,\frac{\Gamma(1-2\eps)}{\Gamma(2-2\eps)}
=\frac{2\pi}{1-2\eps}\,.
\end{align}
The latter factor will occur as universal multiplier in all other angular integrals.

\subsubsection{Massless integral with one denominator}

In the calculations of the angular integrals we follow ideas proposed and 
developed in Ref.~\cite{Somogyi:2011ir}.  In particular, 
for the calculation of the massless one denominator integral $I^{(0)}_j(\eps)$ 
we rotate the frame such that 
the unit vector $\mathbf{v_1}$ points in the $x_D$-direction, i.e.\,$v_1=(1,\mathbf{0}_{D-3},0,1)$, 
$k=(1,\mathbf{0}_{D-3},\sin\theta_1\cos\theta_2,\cos\theta_1)$, 
and $v_1\cdot k = 1-\cos\theta_1$. Therefore, we have
\begin{align}
I^{(0)}_j(\eps)&=\int\dx\Omega_{k_1k_2}\frac{1}{(v_1\cdot k)^j}
\nonumber\\
&=\int_0^\pi\dx\theta_1\frac{\sin^{1-2\eps}\theta_1}{(1-\cos\theta_1)^j}
\int_0^\pi\dx\theta_2\sin^{-2\eps}\theta_2\,.
\end{align}
Next we proceed by analogy with the integral $I^{(0)}(\eps)$: employing the same change of variables 
and using the integral representation~(\ref{beta_Int}) for the beta function\index{Beta function}. 
We find:  
\begin{align}
I^{(0)}_j(\eps)&=2^{1-j-4\eps}\int_0^1\dx t_1\,t_1^{-j-\eps}(1-t_1)^{-\eps} \, 
                              \int_0^1\dx t_2\,t_2^{-\frac{1}{2}-\eps}(1-t_2)^{-\frac{1}{2}-\eps}
\nonumber\\
&=2^{1-j-4\eps}\frac{\Gamma(1-j-\eps)\Gamma(1-\eps)}{\Gamma(2-j-2\eps)}
\frac{\Gamma^2\left(\frac{1}{2}-\eps\right)}{\Gamma(1-2\eps)}\,.
\end{align}
Separating the factor $I^{(0)}(\eps)$ and using the 
Pochhammer symbol\index{Pochhammer symbol} 
\eq
(x)_n = \frac{\Gamma(x+n)}{\Gamma(x)} 
\en
yields 
\begin{align}
I^{(0)}_j(\eps)&=\frac{2\pi}{1-2\eps} 2^{-j}
\frac{\Gamma(1-j-\eps)\Gamma(2-2\eps)}{\Gamma(1-\eps)\Gamma(2-j-2\eps)}
=I^{(0)}(\eps)\,\frac{(2-j-2\eps)_j}{2^j (1-j-\eps)_j}\,.
\end{align}
In this form $I^{(0)}_j(\eps)$ 
is a manifestly rational function
of $\eps$ if $j\in\mathbb{Z}$, such that apparently 
no polygamma constants will clutter the $\eps$-expansion.

\subsubsection{Massive integral with one denominator}

We choose the same frame as in the massless case with $\mathbf{v}_1$ in $x_D$ direction:  
$v=(1,\mathbf{0}_{D-3},0,\beta)$ and $v_1\cdot k = 1-\beta\cos\theta_1$ 
with $v_{11}\neq 0$ and $\beta\equiv \sqrt{1-v_{11}}\neq 1$.  
Therefore, 
\begin{align}
\hspace*{-.2cm}
I^{(1)}_j(v_{11};\eps)&=\int\dx\Omega_{k_1k_2}\frac{1}{(v_1\cdot k)^j}
\nonumber\\
\hspace*{-.2cm}
&=\int_0^\pi\dx\theta_1\frac{\sin^{1-2\eps}\theta_1}{(1-\beta\cos\theta_1)^j}
\int_0^\pi\dx\theta_2\sin^{-2\eps}\theta_2\,.
\end{align}
The $\cos\theta_1=1-2t$ substitution now gives 
\begin{align}
\hspace*{-.2cm}
1-\beta\cos\theta_1=(1-\beta)\left(1+\frac{2\beta}{1-\beta}\,t\right)\,,
\end{align}
thus
\begin{align}
\hspace*{-.2cm}
I^{(1)}_j(v_{11};\eps)&=\frac{2^{1-4\eps}}{(1-\beta)^{j}}
\int_0^1\dx t_1\,t_1^{-\eps}(1-t_1)^{-\eps}
\left(1+\frac{2\beta}{1-\beta}\,t\right)^{-j}
\int_0^1\dx t_2\,t_2^{-\frac{1}{2}-\eps}(1-t_2)^{-\frac{1}{2}-\eps}\,.
\end{align}
Next, using the integral representation for the 
Gauss hypergeometric function\index{Gauss hypergeometric function} 
(Euler formula\index{Euler formula}) 
\eq 
\ghy(a,b,c,z) = \frac{\Gamma(c)}{\Gamma(b) \Gamma(c-b)} \,
\int_0^1 \dx t \, t^{b-1} \, (1-t)^{c-b-1} \, (1-z t)^{-a} \,,
\en 
in case of the integral over $t_1$ 
and isolating the factor $I^{(0)}(\eps)$ one gets: 
\begin{align}
I^{(1)}_j(v_{11};\eps)&=\frac{2^{1-4\eps}}{(1-\beta)^{j}}
\frac{\Gamma^2(1-\eps)}{\Gamma(2-2\eps)}\,\ghy\left(j,1-\eps,2-2\eps,-\frac{2\beta}{1-\beta}\right)
\frac{\Gamma^2\left(\frac{1}{2}-\eps\right)}{\Gamma(1-2\eps)}
\nonumber\\
&=\frac{I^{(0)}(\eps)}{(1-\beta)^j}\,\ghy\left(j,1-\eps,2-2\eps,-\frac{2\beta}{1-\beta}\right)\,.
\label{eq:Massive integral with one denominator as Gauss hypergeometric function with integer coefficients}
\end{align} 
Here we should make two important remarks on the 
Gauss function\index{Gauss hypergeometric function}: 
(1) the Gauss function with integer coefficients occurring in 
Eq.~(\ref{eq:Massive integral with one denominator as Gauss hypergeometric function with integer coefficients}) 
is well suited for the $\eps-$expansion; 
(2) the Gauss function can be defined by the power series using 
Pochhammer symbol\index{Pochhammer symbol} 
as: 
\eq\label{Gauss_Pochhamer} 
\ghy(a,b,c,z) = \sum\limits_{n=0}^\infty \, \frac{(a)_n \, (b)_n}{(c)_n} \, 
\frac{z^n}{n!} \,,
\en
which is manifestly symmetric under interchange of the first and second arguments $a \leftrightarrow b$. 

The integral $I^{(1)}_j(v_{11};\eps)$ will be a cornerstone for consideration of further integrals. 
For this purpose it is feasible to deduce a Mellin-Barnes representation of the integral,  
which has a particularly simple dependence on $v_{11}$. To this end we first employ 
the quadratic hypergeometric transformation
\begin{align}
\ghy(a,b,2b,x)=\left(1-\frac{x}{2}\right)^{-a}\,\ghy\left(\frac{a}{2},\frac{a+1}{2},b+\frac{1}{2},
\left(\frac{x}{2-x}\right)^2\right) \,,
\end{align}
to convert $I^{(1)}_j(v_{11};\eps)$ into
\begin{align}
I^{(1)}_j(v_{11};\eps)=I^{(0)}(\eps)\,\ghy\left(\frac{j}{2},\frac{j+1}{2},\frac{3}{2}-\eps,1-v_{11}\right)\,.
\label{eq:OneDenominatorMassiveQuadratic}
\end{align}
In a second step we use the Mellin-Barnes representation
\begin{align}\label{MBR}
\ghy(a,b,c,x)&=\frac{\Gamma(c)}{\Gamma(a)\Gamma(b)\Gamma(c-a)\Gamma(c-b)}
\nonumber\\
&\times\int_{-i\infty}^{i\infty}\frac{\dx z}{2\pi i}\,\Gamma(a+z)\Gamma(b+z)\Gamma(c-a-b-z)\Gamma(-z)(1-x)^z\,.
\end{align}
This results in
\begin{align}
I^{(1)}_j(v_{11};\eps)&=I^{(0)}(\eps)\,\frac{\Gamma\left(\frac{3}{2}-\eps\right)}{\Gamma\left(\frac{j}{2}\right)
\Gamma\left(\frac{j+1}{2}\right)\Gamma\left(\frac{3-j}{2}-\eps\right)\Gamma\left(\frac{2-j}{2}-\eps\right)}
\nonumber\\
&\times\int_{-i\infty}^{i\infty}\frac{\dx z}{2\pi i}\Gamma\left(\frac{j}{2}+z\right)
\Gamma\left(\frac{j+1}{2}+z\right)\Gamma\left(1-j-\eps-z\right)\Gamma\left(-z\right)v_{11}^z\,.
\end{align}
Applying the Legendre duplication formula\index{Legendre duplication formula}   
on each gamma function that contains a half integer, results 
in the Mellin-Barnes representation\index{Mellin-Barnes integral} 
\begin{align}
\hspace*{-.5cm}
I^{(1)}_j(v_{11};\eps)&=\frac{I^{(0)}(\eps)}{\Gamma(1-\eps)}\frac{(2-j-2\eps)_j}{2^j\Gamma(j)}
\int_{-i\infty}^{i\infty}\frac{\dx z}{2\pi i}\Gamma(j+2z)
\Gamma(1-j-\eps-z)\Gamma(-z)\left(\frac{v_{11}}{4}\right)^z\,.
\label{eq:one denominator massive MB representation}
\end{align}
This form will prove valuable for further calculations due to its simple dependence on $v_{11}$.

As a crosscheck of our calculations we are able to demonstrate that applying the limit $v_{11}\rightarrow 0$ 
we get  
$I^{(1)}_j(v_{11},\eps)\overset{v_{11}\rightarrow 0}{\longrightarrow} I^{(0)}_j(\eps)$. For this purpose 
the form (\ref{eq:OneDenominatorMassiveQuadratic}) is well suited. In particular, using the 
Gauss theorem\index{Gauss theorem} 
\begin{align}\label{Gauss_theorem}
\ghy(a,b,c,1)=\frac{\Gamma(c)\Gamma(c-a-b)}{\Gamma(c-a)\Gamma(c-b)} 
\quad\text{for }\mathrm{Re}(c)>\mathrm{Re}(a+b) \,,
\end{align} 
and the Legendre duplication formula one gets 
\begin{align}
I^{(1)}_j(0,\eps)&=I^{(0)}(\eps)\,\ghy\left(\frac{j}{2},\frac{j+1}{2},\frac{3}{2}-\eps,1\right)
\nonumber\\
&=I^{(0)}(\eps)\,\frac{\Gamma\left(\frac{3}{2}-\eps\right)\Gamma(1-j-\eps)}{\Gamma\left(\frac{3-j}{2}-\eps\right)
\Gamma\left(1-\frac{j}{2}-\eps\right)}
\nonumber\\
&=
I^{(0)}(\eps) \,\frac{(2-j-2\eps)_j}{2^j \, (1-j-\eps)_j}=I^{(0)}_j(\eps)\,.
\end{align}

\subsubsection{Massless integral with two denominators}
\label{sec:Massless integral with two denominators}

For the two denominator case there is the issue that the $\theta_1$ and $\theta_2$ integrals do no 
longer factorize. Therefore, it is feasible to disregard a direct $t$-substitution and 
favor an approach based on Feynman parametrization to combine the denominators. 
For the moment we assume $j,l\notin\mathbb{Z}_{\leq 0}$.
Taking advantage of the linearity of the propagator this results in
\begin{align}
I^{(0)}_{j,l}(v_{12};\eps)&=\int\dx\Omega_{k_1k_2}\frac{1}{(v_1\cdot k)^j (v_2\cdot k)^l}
\nonumber\\
&=\frac{1}{B(j,l)} \, \int_0^1\dx x_1\,x_1^{j-1}\int_0^1\dx x_2\, x_2^{l-1} \, \delta(1-x_1-x_2)
\nonumber\\
&\times
\int\dx\Omega_{k_1k_2}\frac{1}{((x_1 v_1+x_2 v_2)\cdot k)^{j+l}}\,.
\end{align}
We introduce the vector $v\equiv x_1 v_1+x_2 v_2$ and rotate the coordinate system such that 
$\mathbf{v}$ points in the $x_D$-direction. 
In this frame it holds $v=(1,\mathbf{0}_{D-3},0,\sqrt{1-2x_1x_2 v_{12}})$, 
since in the massless case we have $v_{11}=v_{22}=0$ such that 
$v^2=(x_1 v_1+x_2 v_2)^2=2x_1 x_2 v_{12}$.
Thus, we can write
\begin{align}
I^{(0)}_{j,l}(v_{12};\eps)&=\frac{\Gamma(j+l)}{\Gamma(j)\Gamma(l)}\int_0^1\dx x_1\,x_1^{j-1}
\int_0^1\dx x_2\, x_2^{l-1}\delta(1-x_1-x_2) I^{(1)}_{j+l}(2x_1x_2v_{12})\,.
\end{align}
Employing the Mellin-Barnes representation of $I^{(1)}_j(v_{11};\eps)$ from 
Eq.~(\ref{eq:one denominator massive MB representation}) brings this into the form
\begin{align}
I^{(0)}_{j,l}(v_{12};\eps)&=\frac{I^{(0)}(\eps)}{\Gamma(1-\eps)}
\frac{(2-j-l-2\eps)_{j+l}}{2^{j+l}\Gamma(j)\Gamma(l)}\int_0^1\dx x_1\,x_1^{j-1}
\int_0^1\dx x_2\, x_2^{l-1}\delta(1-x_1-x_2)
\nonumber\\
&\times\int_{-i\infty}^{i\infty}\frac{\dx z}{2\pi i}\Gamma(j+l+2z)\Gamma(1-j-l-\eps-z)
\Gamma(-z)\left(\frac{x_1x_2v_{12}}{2}\right)^z\,.
\label{eq:MB representation of massless two denominator integral}
\end{align}
Now the Feynman parameter integrals are factorized and 
can be evaluated in terms of beta function\index{Beta function} as 
\begin{align}\label{Beta_Function}
\int_0^1\dx x_1\,x_1^{j+z-1}\int_0^1\dx x_2\, x_2^{l+z-1}\delta(1-x_1-x_2)=
B(j+z,l+z) \,. 
\end{align}
Using the Mellin-Barnes representation\index{Mellin-Barnes integral} 
for the Gauss hypergeometric function\index{Gauss hypergeometric function}~(\ref{MBR}) 
and Eq.~(\ref{Beta_Function}) we get 
\begin{align}\label{Ijl_v12_eps}
I^{(0)}_{j,l}(v_{12};\eps)&=
\frac{I^{(0)}(\eps)}{\Gamma(1-\eps)}\frac{(2-j-l-2\eps)_{j+l}}{2^{j+l}\Gamma(j)\Gamma(l)}
\nonumber\\
&\times\int_{-i\infty}^{i\infty}\frac{\dx z}{2\pi i}\Gamma(j+z)\Gamma(l+z)\Gamma(1-j-l-\eps-z)\Gamma(-z)
\left(\frac{v_{12}}{2}\right)^z
\nonumber\\
&=I^{(0)}(\eps)\,\frac{(2-j-l-2\eps)_{j+l}}{2^{j+l}}\frac{\Gamma(1-j-\eps)
\Gamma(1-l-\eps)}{\Gamma^2(1-\eps)}\,\ghy\left(j,l,1-\eps,1-\frac{v_{12}}{2}\right)
\nonumber\\
&=I^{(0)}(\eps)\,\frac{(2-j-l-2\eps)_{j+l}}{2^{j+l}(1-j-\eps)_j(1-l-\eps)_l}\,
\ghy\left(j,l,1-\eps,1-\frac{v_{12}}{2}\right) \,,
\end{align}
or alternatively by factorizing out the integral $I^{(0)}_{j+l}(\eps)$: 
\begin{align}
I^{(0)}_{j,l}(v_{12};\eps)&=I^{(0)}_{j+l}(\eps)\,\frac{(1-j-l-\eps)_{j}}{(1-j-\eps)_j}
\,\ghy\left(j,l,1-\eps,1-\frac{v_{12}}{2}\right)\,.
\label{eq:Massless two denominator integral}
\end{align}
This result also holds for the previously excluded cases $j,l\in\mathbb{Z}_{\leq 0}$.
As for the integral with one massive denominator we obtain a hypergeometric function 
to be expanded in $\eps$ about integer parameters.

Using the Gauss theorem\index{Gauss theorem}~(\ref{Gauss_theorem}) 
we obtain in the limit $v_{12}\rightarrow 0$
\begin{align}
I^{(0)}_{j,l}(0;\eps)&=I^{(0)}_{j+l}(\eps)\,\frac{(1-j-l-\eps)_{j}}{(1-j-\eps)_j}
\,\ghy\left(j,l,1-\eps,1\right)
\nonumber\\
&=I^{(0)}_{j+l}(\eps)\,\frac{(1-j-l-\eps)_{j}}{(1-j-\eps)_j}
\frac{\Gamma(1-\eps)\Gamma(1-j-l-\eps)}{\Gamma(1-j-\eps)
\Gamma(1-l-\eps)}=I^{(0)}_{j+l}(\eps)\,.
\end{align} 
One can see that the expressions~(\ref{Ijl_v12_eps}) 
and~(\ref{eq:Massless two denominator integral}) 
are manifestly symmetric under interchange $j \leftrightarrow l$, hence 
$I^{(0)}_{j,l}(v_{12};\eps)=I^{(0)}_{l,j}(v_{12};\eps)$. 
From Eq.(\ref{eq:Massless two denominator integral}) it is evident that
$I^{(0)}_{j,0}(v_{12};\eps)=I^{(0)}_{j}(\eps)$.

\subsubsection{Single massive integral with two denominators}

Now we consider the single massive case
\begin{align}
I^{(1)}_{j,l}(v_{12},v_{11};\eps)&=\int\dx\Omega_{k_1k_2}\frac{1}{(v_1\cdot k)^j (v_2\cdot k)^l}\,, 
\qquad v_{11}\neq 0,\,\,v_{22}=0\,.
\end{align}
We can start our consideration after Feynman parametrization of the two denominator integral. 
The vector $v$ has now norm $v^2=x_1^2 v_{11}+2 x_1 x_2 v_{12}$. Employing again the Mellin-Barnes 
representation of $I^{(1)}_{j+l}(v^2)$ gives 
[compare with Eq.~(\ref{eq:MB representation of massless two denominator integral})]  
\begin{align}
I^{(1)}_{j,l}(v_{12},v_{11};\eps)
&=
\frac{I^{(0)}(\eps)}{\Gamma(1-\eps)}\frac{(2-j-l-2\eps)_{j+l}}{2^{j+l}
\Gamma(j)\Gamma(l)}\int_0^1\dx x_1\,x_1^{j-1}\int_0^1\dx x_2\, x_2^{l-1}\delta(1-x_1-x_2)
\nonumber\\
&\times\int_{-i\infty}^{i\infty}\frac{\dx z}{2\pi i}\Gamma(j+l+2z)
\Gamma(1-j-l-\eps-z)\Gamma(-z)\left(\frac{x_1^2 v_{11}}{4}+\frac{x_1x_2v_{12}}{2}\right)^z\,.
\end{align}
In order to evaluate the Feynman parameter integral following our strategy in 
Sec.~\ref{sec:Massless integral with two denominators}, a factorization of 
$\left(\frac{x_1^2 v_{11}}{4}+\frac{x_1x_2v_{12}}{2}\right)^z$ is required. Applying the 
Binomi-Mellin-Newton integral\index{Binomi-Mellin-Newton integral} 
representation~\footnote{This integral representation does not have 
	a name in the literature 
	(see, e.g., Eq.~(5.1) in Ref.~\cite{Smirnov:2012gma}, where it is just 
	referred to as a simple formula, however due to its frequent occurrence 
	and usefulness, it deserves, at least in the authors' opinion, 
	a name analogous to the well-established Cahen-Mellin integral. 
	As~(\ref{eq: Binomi-Newton-Mellin integral}) is the Mellin integral 
	version of Newton's Binomial Theorem this name suggests itself.}
\begin{align}
(a+b)^{z_1}=\frac{1}{\Gamma(-z_1)}\int_{-i\infty}^{i\infty}\frac{\dx z_2}{2\pi i}a^{z_1-z_2} b^{z_2} 
\Gamma(-z_2)\Gamma(-z_1+z_2) \,,
\end{align}
on this term, results in
\begin{align}
I^{(1)}_{j,l}(v_{12},v_{11};\eps)&=\frac{I^{(0)}(\eps)}{\Gamma(1-\eps)}\frac{(2-j-l-2\eps)_{j+l}}{2^{j+l}
\Gamma(j)\Gamma(l)}
\int_{-i\infty}^{i\infty}\frac{\dx z_1}{2\pi i}\Gamma(j+l+2z_1)\Gamma(1-j-l-\eps-z_1)
\nonumber\\
&\times
\int_{-i\infty}^{i\infty}\frac{\dx z_2}{2\pi i}\left(\frac{v_{11}}{4}\right)^{z_1-z_2}
\left(\frac{v_{12}}{2}\right)^{z_2}\Gamma(-z_2)\Gamma(-z_1+z_2)
\nonumber\\
&\times
\int_0^1\dx x_1\,x_1^{j+2z_1-z_2-1}\int_0^1\dx x_2\, x_2^{l+z_2-1}\delta(1-x_1-x_2)\,.
\end{align} 
Evaluating the Feynman parameter integral again cancels the $\Gamma(j+l+2z_1)$ and yields
\begin{align}
I^{(1)}_{j,l}(v_{12},v_{11};\eps)&=\frac{I^{(0)}(\eps)}{\Gamma(1-\eps)}
\frac{(2-j-l-2\eps)_{j+l}}{2^{j+l}\Gamma(j)\Gamma(l)}
\int_{-i\infty}^{i\infty}\frac{\dx z_1}{2\pi i}\int_{-i\infty}^{i\infty}
\frac{\dx z_2}{2\pi i}\left(\frac{v_{11}}{4}\right)^{z_1}\left(\frac{2v_{12}}{v_{11}}\right)^{z_2}
\nonumber\\
&\times
\Gamma(1-j-l-\eps-z_1)\Gamma(-z_1+z_2)\Gamma(j+2z_1-z_2)\Gamma(-z_2)\Gamma(l+z_2)\,.
\end{align}
Performing the substitutions $z_1\rightarrow 1-j-l-\eps+z_1$ and $z_2\rightarrow 1-j-l-\eps+z_1-z_2$ 
results in the two-fold Mellin-Barnes representation
\begin{align}\label{I1jl_expression}
I^{(1)}_{j,l}(v_{12},v_{11};\eps)&=\frac{I^{(0)}(\eps)}{\Gamma(1-\eps)}
\frac{(2-j-l-2\eps)_{j+l}}{2^{j+l}\Gamma(j)\Gamma(l)} 
\int_{-i\infty}^{i\infty}\frac{\dx z_1}{2\pi i}\int_{-i\infty}^{i\infty}
\frac{\dx z_2}{2\pi i}\left(\frac{v_{11}}{4}\right)^{1-j-l-\eps+z_1}
\nonumber\\
&\times
\left(\frac{2v_{12}}{v_{11}}\right)^{1-j-l-\eps+z_1-z_2}
\Gamma(-z_1)\Gamma(-z_2)\Gamma(1-l-\eps+z_1+z_2)
\nonumber\\
&\times\Gamma(-1+j+l+\eps-z_1+z_2)\Gamma(1-j-\eps+z_1-z_2)\,.
\end{align}
Replacing the product of the last two gamma functions in 
Eq.~(\ref{I1jl_expression}) by the corresponding beta function\index{Beta function} 
integral representation    
\begin{align}
&\Gamma(-1+j+l+\eps-z_1+z_2)\Gamma(1-j-\eps+z_1-z_2)
\nonumber\\
&=\Gamma(l)\int_0^1\dx t\,t^{-2+j+l+\eps-z_1+z_2}(1-t)^{-j-\eps+z_1-z_2} \,,
\end{align}
leads to
\begin{align}
I^{(1)}_{j,l}(v_{12},v_{11};\eps)&=\frac{I^{(0)}(\eps)}{\Gamma(1-\eps)}
\frac{(2-j-l-2\eps)_{j+l}}{2^{j+l}\Gamma(j)}
\left(\frac{v_{12}}{2}\right)^{1-j-l-\eps}\int_0^1\dx t\,t^{-2+j+l+\eps}(1-t)^{-j-\eps}
\nonumber\\
&\times
\int_{-i\infty}^{i\infty}\frac{\dx z_1}{2\pi i}\int_{-i\infty}^{i\infty}\frac{\dx z_2}{2\pi i}
\left(\frac{1-t}{t}\frac{v_{12}}{2}\right)^{z_1}
\left(\frac{t}{1-t}\frac{v_{11}}{2v_{12}}\right)^{z_2}
\nonumber\\
&\times
\Gamma(-z_1)\Gamma(-z_2)\Gamma(1-l-\eps+z_1+z_2)\,.
\label{eq:single massive two denominator MB representation}
\end{align}
The appearing two-fold Mellin-Barnes integral\index{Mellin-Barnes integral} 
evaluates by the Binomi-Mellin-Newton integral\index{Binomi-Mellin-Newton integral} to
\begin{align}
&\int_{-i\infty}^{i\infty}\frac{\dx z_1}{2\pi i}\int_{-i\infty}^{i\infty}\frac{\dx z_2}{2\pi i}
\left(\frac{1-t}{t}\frac{v_{12}}{2}\right)^{z_1}
\left(\frac{t}{1-t}\frac{v_{11}}{2v_{12}}\right)^{z_2}
\Gamma(-z_1)\Gamma(-z_2)\Gamma(1-l-\eps+z_1+z_2)
\nonumber\\
&=
\Gamma(1-l-\eps)\left(1+\frac{1-t}{t}\frac{v_{12}}{2}+\frac{t}{1-t}\frac{v_{11}}{2 v_{12}}\right)^{l-1+\eps}
\nonumber\\
&=\Gamma(1-l-\eps)\left(\frac{v_{12}}{2t(1-t)}\right)^{l-1+\eps}\left[1-t\left(2-\frac{2}{v_{12}}\right)
+t^2\left(1-\frac{2}{v_{12}}+\frac{v_{11}}{v_{12}^2}\right)\right]^{l-1+\eps}\,.
\nonumber\\
&=\Gamma(1-l-\eps)\left(\frac{v_{12}}{2t(1-t)}\right)^{l-1+\eps}(1-\tau_+ t)^{l-1+\eps}(1-\tau_- t)^{l-1+\eps}\,,
\end{align}
where we introduced the notation 
$\tau_\pm=1-(1\pm\sqrt{1-v_{11}})/v_{12}$.  
Plugging this into integral (\ref{eq:single massive two denominator MB representation}) gives
\begin{align}
I^{(1)}_{j,l}(v_{12},v_{11};\eps)&=\frac{I^{(0)}(\eps)}{2^l v_{12}^j}
\,\frac{(2-j-l-2\eps)_{j+l}}{(1-l-\eps)_l\Gamma(j)}
\nonumber\\
&\times
\int_0^1\dx t\,t^{j-1}(1-t)^{1-j-l-2\eps}(1-\tau_+ t)^{l-1+\eps}(1-\tau_- t)^{l-1+\eps}\,.
\end{align}
The $t$-integral is the Euler integral representation of the 
Appell function\index{Appell function} $\appell(a,b,c,d,x,y)$:
\eq 
\appell(a,b,c,d,x,y) = \frac{\Gamma(c)}{\Gamma(a) \Gamma(d-a)} \, 
\int_0^1\dx t\,t^{a-1}(1-t)^{d-a-1}(1-xt)^{-b}(1-yt)^{-c}\,.
\en
Here for completeness we also present the definition of the 
Appell function\index{Appell function} 
in terms of the Pochhammer symbol\index{Pochhammer symbol}
\eq\label{Appell_Pochhamer} 
\appell(a,b,c,d,x,y) = \sum\limits_{m,n=0}^\infty \, 
\frac{(a)_{m+n} (b)_m (c)_n}{(d)_{m+n}} \, 
\frac{x^m y^n}{m! n!} \,. 
\en
Hence, we obtain
\begin{align}
I^{(1)}_{j,l}(v_{12},v_{11};\eps)&=\frac{I^{(0)}(\eps)}{2^l v_{12}^j}
\,\frac{(2-j-l-2\eps)_{j+l}}{(1-l-\eps)_l}
\frac{\Gamma(2-j-l-2\eps)}{\Gamma(2-l-2\eps)}
\nonumber\\
&\times\appell(j,1-l-\eps,1-l-\eps,2-l-2\eps,\tau_+,\tau_-)\,.
\nonumber\\
&=\frac{I^{(0)}(\eps)}{2^l v_{12}^j}\,
\frac{(2-l-2\eps)_{l}}{(1-l-\eps)_l}\appell(j,1-l-\eps,1-l-\eps,2-l-2\eps,\tau_+,\tau_-)
\nonumber\\
&=\frac{I^{(0)}_l(\eps)}{v_{12}^j}\appell(j,1-l-\eps,1-l-\eps,2-l-2\eps,\tau_+,\tau_-)\,.
\end{align}
As in Sec.~\ref{sec:Massless integral with two denominators} this result 
is valid for all $j$ and $l$ even though 
the derivation assumed that neither of them is a non-positive integer.
Employing the transformation
\begin{align}
\appell(a,b_1,b_2,c,x,y)=(1-x)^{-b_1}(1-y)^{-b_2}
\appell\left(c-a,b_1,b_2,c,\frac{x}{x-1},\frac{y}{y-1}\right) \,,
\end{align}
gives the alternative representation
\begin{align}\label{Ijl_1}
&I^{(1)}_{j,l}(v_{12},v_{11};\eps)=\frac{I^{(0)}_l(\eps)}{v_{12}^j}
\left(\frac{v_{11}}{v_{12}^2}\right)^{l-1+\eps}
\nonumber\\
&\times
\appell\left(2-j-l-2\eps,1-l-\eps,1-l-\eps,2-l-2\eps,
\frac{\tau_+}{\tau_+ - 1},\frac{\tau_-}{\tau_- - 1}\right)\,,
\end{align}
which is a suitable starting point for the $\eps$-expansion 
in the case of positive integer $j$ and $l$.

For $j\in\mathbb{Z}_{\leq 0}$ the hypergeometric series terminates at 
$m+n=-j$, since $(j)_{m+n}=0$ otherwise. 
Therefore, 
\begin{align}
\appell(j,1-l-\eps,1-l-\eps,2-l-2\eps,x,y)
=\sum_{m,n=0}^\infty \frac{(j)_{m+n}(1-l-\eps)_m (1-l-\eps)_n}{(2-l-2\eps)_{m+n} }\,
\frac{x^m}{m!}\frac{y^n}{n!}&
\nonumber\\
\overset{j\in\mathbb{Z}_{\leq 0}}{=}\sum_{m,n\leq -j} 
\frac{(j)_{m+n}(1-l-\eps)_m (1-l-\eps)_n}{(2-l-2\eps)_{m+n} }\,\frac{x^m}{m!}\frac{y^n}{n!}&\,,
\end{align} 
which makes the $\eps$-expansion of $I^{(1)}_{j,l}(v_{12},v_{11};\eps)$ trivial.

For $l\in\mathbb{Z}_{\leq 0}$ the Appell function 
reduces to a sum of hypergeometric functions. It holds
\begin{align}
&\appell(j,1-l-\eps,1-l-\eps,2-l-2\eps,x,y)
\nonumber\\
&\overset{l\in\mathbb{Z}_{\leq 0}}{=}\frac{(2-2l-2\eps)_l}{(1-y)^j}
\sum_{n=0}^{-l}\binom{-l}{n}(-1)^{l-n}\frac{(j)_n\,
\ghy\left(j+n,1-l-\eps,2-2l-2\eps,\frac{x-y}{1-y}\right)}{(1-y)^n (j+l-1+2\eps)_{l+n}}\,.
\label{eq:Single massive integral Gauss hypergeometric reduction}
\end{align}
This form allows for a much simpler $\eps-$expansion since we reduced the two-variable 
Appell function\index{Appell function}   
to a finite sum of genuinely one-variable functions.

\subsubsection{Double massive integral with two denominators}
\label{sec:Double massive integral with two denominators}

An approach similar to the massless and single-massive case yields a complicated three-fold 
Mellin-Barnes integral\index{Mellin-Barnes integral}  
(see Ref.~\cite{Somogyi:2011ir}). Even though this might be a good starting point for numerically 
establishing $\eps$-expansions, there has been 
no successful attempt to obtain a closed analytic expression similar 
to the single-massive case~\cite{Somogyi:2011ir}. 
We will close this gap in Sec.~\ref{sec:Hypergeometric representation of angular integrals}

For the purpose of $\eps$-expansion, 
we propose the following promising new method. By restricting ourselves to $j,l\in\mathbb{N}$ we obtain 
a closed expression for the double massive integral for arbitrary $D$ by employing 
the generalized two-mass splitting 
formula\index{Two-mass splitting formula}~(\ref{eq:Generalized Two-mass splitting}). 
This reduces the double massive integral to a sum of $j+l$ single-massive integrals. 
For $j,l\in\mathbb{Z}_{\leq 0}$ we use Eq.~(\ref{eq:PropagatorNegativePowerBinomial}) 
for reduction to single massive integrals.

Using the generalized two-point splitting lemma\index{Two-point splitting lemma} 
[see details in Appendix~\ref{sec:Partial Fractioning}] 
we get 
\begin{align}
I_{j,l}^{(2)}(v_{12},v_{11},v_{22};\eps)&=\int\dx\Omega_{k_1 k_2}\frac{1}{(v_1\cdot k)^j(v_2\cdot k)^l}
=\int\dx\Omega_{k_1 k_2}\powprop{v_1}{j}\powprop{v_2}{l}
\nonumber\\
&=\int\dx\Omega_{k_1 k_2}\left[\sum_{n=0}^{j-1}\binom{l-1+n}{l-1}\lambda_\pm^l(1-\lambda_\pm)^n
\powprop{v_1}{j-n}\powprop{v^\pm_3}{l+n}\right.
\nonumber\\
&\left.\hphantom{\,\,\,\int\dx\Omega_{k_1 k_2}}+\sum_{n=0}^{l-1}\binom{j-1+n}{j-1}
\lambda_\pm^n(1-\lambda_\pm)^j\powprop{v_2}{l-n}\powprop{v^\pm_3}{j+n}\right]
\nonumber\\
&=\sum_{n=0}^{j-1}\binom{l-1+n}{l-1}\lambda_\pm^l(1-\lambda_\pm)^n
\int\dx\Omega_{k_1 k_2}\powprop{v_1}{j-n}\powprop{v^\pm_3}{l+n}
\nonumber\\
&+\sum_{n=0}^{l-1}\binom{j-1+n}{j-1}\lambda_\pm^n(1-\lambda_\pm)^j
\int\dx\Omega_{k_1 k_2}\powprop{v_2}{l-n}\powprop{v^\pm_3}{j+n}\,,
\end{align}
where $\left(v_3^\pm\right)^2=0$, therefore the double massive integral 
with $j,l\in\mathbb{N}$ can be evaluated for general $\eps$ as
\begin{align}
I_{j,l}^{(2)}(v_{12},v_{11},v_{22};\eps)&=\sum_{n=0}^{j-1}
\binom{l-1+n}{l-1}\lambda_\pm^l(1-\lambda_\pm)^n I^{(1)}_{j-n,l+n}(v_{13}^\pm,v_{11};\eps)
\nonumber\\
&+\sum_{n=0}^{l-1}\binom{j-1+n}{j-1}\lambda_\pm^n(1-\lambda_\pm)^j I^{(1)}_{l-n,j+n}(v_{23}^\pm,v_{22};\eps)\,.
\label{eq:Two mass integral general result}
\end{align}
where
\begin{align}
\lambda_{\pm}&=\frac{v_{12}-v_{11}\pm\sqrt{v_{12}^2-v_{11}v_{22}}}{2v_{12}-v_{11}-v_{22}}\,,
\nonumber\\
v_{13}^\pm &=(1-\lambda_\pm)v_{11}+\lambda_\pm v_{12}=\frac{v_{11} 
\left(v_{22}\pm\sqrt{v_{12}^2-v_{11} v_{22}}\right)-v_{12} 
\left(v_{12}\pm\sqrt{v_{12}^2-v_{11} v_{22}}\right)}{v_{11}+v_{22}-2 v_{12}}\,,
\nonumber\\
v_{23}^\pm &=(1-\lambda_\pm)v_{12}+\lambda_\pm v_{22}=
\frac{v_{22} \left(v_{11}\mp\sqrt{v_{12}^2-v_{11} v_{22}}\right)-v_{12}
\left(v_{12}\mp\sqrt{v_{12}^2-v_{11} v_{22}}\right)}{v_{11}+v_{22}-2 v_{12}}\,.
\end{align}
Formula~(\ref{eq:Two mass integral general result}) holds for both sign choices.

When the power of one of the massive propagators is negative one can derive a simpler expression for the corresponding angular integral. 
In particular, using Eqs.~(\ref{2M_to_1M_red_a}) and~(\ref{2M_to_1M_red_b}), it holds 
\begin{align}
I_{j,l}^{(2)}(v_{12},v_{11},v_{22};\eps)&=\sum_{n=0}^{-j}\binom{-j}{n}
\left(1-\sqrt{1-v_{11}}\,\right)^{-j-n} \left(1-v_{11}\right)^{n/2}
\nonumber\\
&\qquad\times I_{l,-n}^{(1)}\left(1-\frac{1-v_{12}}{\sqrt{1-v_{11}}},v_{22};\eps\right) \,,
\end{align} 
for $j\in\mathbb{Z}_{\leq 0}$ and equivalently 
\begin{align}
I_{j,l}^{(2)}(v_{12},v_{11},v_{22};\eps)&=\sum_{n=0}^{-l}\binom{-l}{n}  
\left(1-\sqrt{1-v_{22}}\,\right)^{-l-n} \left(1-v_{22}\right)^{n/2}
\nonumber\\
&\qquad\times  I_{j,-n}^{(1)}\left(1-\frac{1-v_{12}}{\sqrt{1-v_{22}}},v_{11};\eps\right) \,,
\end{align} 
for $l\in\mathbb{Z}_{\leq 0}$.  

\subsection{New properties of angular integrals} 
\label{sec:New properties of angular integrals}
In this section we review new properties of angular integrals proposed and 
developed in present paper for the first time. In particular, we 
derive: (1) hypergeometric representation of the general two denominator angular integral 
in $D$ dimensions; (2) differential, partial integration, and recursion relations 
for angular integrals. The main advantage of our findings is that it allow 
to reduce all known angular integrals to a small set of basis integrals. 

\subsubsection{Hypergeometric representation of angular integrals}
\label{sec:Hypergeometric representation of angular integrals}

It is known (see, e.g., detailed discussion in Ref.~\cite{Somogyi:2011ir}) 
that many angular integrals are represented in terms of hypergeometric (Gauss, Appell) functions.  
In this section we demonstrate how to derive the 
hypergeometric representation of the general angular integral having two denominators 
in terms of the Lauricella function $\lauricella$. 
The latter being a three-variable generalization of the Gauss and Appell function.     

For this purpose we consider generic angular integral in $D=4-2\eps$ dimensions
\begin{align}
I_{j,l}(v_{12},v_{11},v_{22};\eps)=\int\dx\Omega_{k_1 k_2}\, 
\frac{1}{(v_1\cdot k)^j(v_2\cdot k)^l}\,,
\end{align}
where as before 
$\dx\Omega_{k_1 k_2}=\dx\theta_1 \sin^{1-2\eps}\theta_1\,\dx\theta_2\sin^{-2\eps}\theta_2$ 
and $v_{ij}=v_i\cdot v_j$. 

Using Feynman parametrization one gets 
\begin{align}
I_{j,l}(v_{12},v_{11},v_{22};\eps) = B(j,l) 
\int_0^1\dx x_1\, 
x_1^{j-1}\int_0^1\dx x_2\, x_2^{l-1}\delta(1-x_1-x_2)\, 
I^{(1)}_{j+l}(w_{12};\eps)\,, 
\label{eq:Feynman parametrized Ijl}
\end{align}
where $w_{12}=(x_1v_1+x_2v_2)^2$ is the Feynman ``mass parameter''. 

The one denominator massive integral $I^{(1)}_{j+l}(v,\eps)$ is given by 
\begin{align}
I^{(1)}_{j+l}(v;\eps)=I^{(0)}(\eps)\,\ghy\left(\frac{j+l}{2},
\frac{j+l+1}{2},\frac{3}{2}-\eps,1-v\right)\,.
\end{align}
Due to the $\delta$-function 
it follows $2x_1 x_2=1-x_1^2-x_2^2$, and therefore $w_{12}$ 
can be presented in the form, where the variables $x_1$ and $x_2$ are separated,
\begin{align}
w_{12}=x_1^2v_{11}+2x_1x_2v_{12}+x_2^2v_{22}
=v_{12}-x_1^2 (v_{12}-v_{11})-x_2^2(v_{12}-v_{22})\,.
\label{eq:Feynman parametrized mass}
\end{align}
Representation~(\ref{eq:Feynman parametrized mass}) is very useful since it leads to 
a Mellin-Barnes integral, which is considerably simpler than that derived 
in Ref.~\cite{Somogyi:2011ir}. In addition, it can be expressed in terms of 
the known hypergeometric function $\lauricella$. Using Eq.~(\ref{eq:Feynman parametrized mass}), 
the Mellin-Barnes representation of the Gauss hypergeometric function
\begin{align}
\ghy(a,b,c,x)=\int_{-i\infty}^{i\infty}\frac{\dx z}{2\pi i}
\frac{\pochhammer{a}{z}\pochhammer{b}{z}}{\pochhammer{c}{z}}\Gamma(-z)\left(-x\right)^{z} \,,
\label{eq:MB rep of 2F1}
\end{align}
and the Newton-Binomi-Mellin integral 
\begin{align}
(a+b+c)^z=\frac{1}{\Gamma(-z)}\int_{-i\infty}^{i\infty}\frac{\dx z_1}{2\pi i}
\int_{-i\infty}^{i\infty}\frac{\dx z_2}{2\pi i} \, a^{z-z_1-z_2} \, b^{z_1} \, 
c^{z_2}\, \Gamma(-z_1)\Gamma(-z_2)\Gamma(-z+z_1+z_2) \,,
\end{align}
we obtain
\begin{align}\label{I_jl}
&I^{(1)}_{j+l}\left(w_{12};\eps\right)=
I^{(0)}(\eps)\int_{-i\infty}^{i\infty}\frac{\dx z}{2\pi i}\Gamma(-z)
\nonumber\\
&\qquad \times \left(-(1-v_{12})-x_1^2(v_{12}-v_{11})-x_2^2(v_{12}-v_{22})\right)^{z} \, 
\frac{\pochhammer{\frac{j+l}{2}}{z}\pochhammer{\frac{j+l+1}{2}}{z}}{\pochhammer{\frac{3}{2}-\eps}{z}}
\nonumber\\
&=I^{(0)}(\eps)\int_{-i\infty}^{i\infty}\frac{\dx z}{2\pi i}\int_{-i\infty}^{i\infty}
\frac{\dx z_1}{2\pi i}\int_{-i\infty}^{i\infty}\frac{\dx z_2}{2\pi i}
\Gamma(-z_1)\Gamma(-z_2)\Gamma(-z+z_1+z_2)
\nonumber\\
&\qquad\times(v_{11}-v_{12})^{z_1}(v_{22}-v_{12})^{z_2}(v_{12}-1)^{z-z_1-z_2}
\frac{\pochhammer{\frac{j+l}{2}}{z}
\pochhammer{\frac{j+l+1}{2}}{z}}{\pochhammer{\frac{3}{2}-\eps}{z}}\, x_1^{2z_1}x_2^{2z_2}
\nonumber\\
&=I^{(0)}(\eps)\int_{-i\infty}^{i\infty}\frac{\dx z_1}{2\pi i}\int_{-i\infty}^{i\infty}
\frac{\dx z_2}{2\pi i}\int_{-i\infty}^{i\infty}\frac{\dx z_3}{2\pi i}\Gamma(-z_1)\Gamma(-z_2)\Gamma(-z_3)
\nonumber\\
&\qquad\times(v_{11}-v_{12})^{z_1}(v_{22}-v_{12})^{z_2}(v_{12}-1)^{z_3}
\frac{\pochhammer{\frac{j+l}{2}}{z_1+z_2+z_3}\pochhammer{\frac{j+l+1}{2}}{z_1+z_2+z_3}}
{\pochhammer{\frac{3}{2}-\eps}{z_1+z_2+z_3}}\, x_1^{2z_1}x_2^{2z_2}\,. 
\end{align}
In the last step of derivation of Eq.~(\ref{I_jl}) we changed variable  $z \to z_3=z-z_1-z_2$. 
The combination of the Pochhammer symbols\index{Pochhammer symbol} 
can be presented in the following form 
\begin{align}
\frac{\pochhammer{\frac{j+l}{2}}{z_1+z_2+z_3}
\pochhammer{\frac{j+l+1}{2}}{z_1+z_2+z_3}}{\pochhammer{\frac{3}{2}-\eps}{z_1+z_2+z_3}}
&=\frac{4^{-z_1-z_2}\pochhammer{j+l}{2z_1+2z_2}}{\pochhammer{\frac{3}{2}-\eps}{z_1+z_2}}
\nonumber\\
&\times\frac{\pochhammer{\frac{j+l}{2}+z_1+z_2}{z_3}\pochhammer{\frac{j+l+1}{2}+z_1+z_2}{z_3}}
{\pochhammer{\frac{3}{2}-\eps+z_1+z_2}{z_3}}\,,
\end{align}
whereby the $z_3$ integral becomes 
a Gauss hypergeometric function\index{Gauss hypergeometric function}, 
\begin{align}
I^{(1)}_{j+l}\left(w_{12};\eps\right)
&=I^{(0)}(\eps)\int_{-i\infty}^{i\infty}\frac{\dx z_1}{2\pi i}
\int_{-i\infty}^{i\infty}\frac{\dx z_2}{2\pi i}\left(\frac{v_{11}-v_{12}}{4}\right)^{z_1}
\left(\frac{v_{22}-v_{12}}{4}\right)^{z_2}
\nonumber\\
&\times\Gamma(-z_1)\Gamma(-z_2)\frac{\pochhammer{j+l}{2z_1+2z_2}}
{\pochhammer{\frac{3}{2}-\eps}{z_1+z_2}}\,x_1^{2z_1}x_2^{2z_2}
\nonumber\\
&\times
\ghy\left(\frac{j+l}{2}+z_1+z_2,\frac{j+l+1}{2}+z_1+z_2,\frac{3}{2}-\eps+z_1+z_2,1-v_{12}\right)\,.
\end{align}
Next to reduce the number of occurrences of the integration variables $z_1$ and $z_2$ in the 
arguments of the Gauss hypergeometric function\index{Gauss hypergeometric function} 
we perform the Euler transformation\index{Euler transformation}   
\begin{align}
&\ghy\left(\frac{j+l}{2}+z_1+z_2,\frac{j+l+1}{2}+z_1+z_2,\frac{3}{2}-\eps+z_1+z_2,1-v_{12}\right)\nonumber\\
=\,\,&v_{12}^{1-j-l-\eps-z_1-z_2}
\, \ghy\left(\frac{3-j-l-2\eps}{2},\frac{2-j-l-2\eps}{2},\frac{3}{2}-\eps+z_1+z_2,1-v_{12}\right)\,,
\end{align}
Introducing again the Mellin-Barnes representation\index{Mellin-Barnes integral} 
of the 
Gauss hypergeometric function\index{Gauss hypergeometric function}~(\ref{eq:MB rep of 2F1}) 
and combining the Pochhammer symbols\index{Pochhammer symbol}
leads to
\eq 
\hspace*{-.25cm}
I^{(1)}_{j+l}\left((x_1v_1+x_2v_2)^2;\eps\right)
&=&I^{(0)}(\eps)v_{12}^{1-j-l-\eps}\int_{-i\infty}^{i\infty}\frac{\dx z_1}{2\pi i}
\int_{-i\infty}^{i\infty}\frac{\dx z_2}{2\pi i}\int_{-i\infty}^{i\infty}
\frac{\dx z_3}{2\pi i} 
\nonumber\\
\hspace*{-.25cm}
&\times& \Gamma(-z_1)\Gamma(-z_2)\Gamma(-z_3)
\left(\frac{v_{11}-v_{12}}{4v_{12}}\right)^{z_1}
\left(\frac{v_{22}-v_{12}}{4v_{12}}\right)^{z_2}\left(\frac{v_{12}-1}{4}\right)^{z_3} 
\nonumber\\
\hspace*{-.25cm}
&\times& 
x_1^{2z_1}x_2^{2z_2}\frac{\pochhammer{j+l}{2z_1+2z_2}\pochhammer{2-j-l-2\eps}{2z_3}}
{\pochhammer{\frac{3}{2}-\eps}{z_1+z_2+z_3}}\,.
\en
Plugging this into expression~(\ref{eq:Feynman parametrized Ijl}) for $I_{j,l}$ and 
evaluating the Feynman parameter integral as
\begin{align}
\frac{1}{B(j,l)} \, \int_0^1\dx x_1\, x_1^{j-1}\int_0^1\dx x_2\, 
x_2^{l-1}\delta(1-x_1-x_2)x_1^{2z_1}x_2^{2z_2}
=\frac{\pochhammer{j}{2z_1}\pochhammer{l}{2z_2}}{\pochhammer{j+l}{2z_1+2z_2}}\,,
\end{align}
yields the three-fold Mellin-Barnes representation\index{Mellin-Barnes integral}
\begin{align}
&I_{j,l}(v_{12},v_{11},v_{22};\eps)=I^{(0)}(\eps)\,v_{12}^{1-j-l-\eps}\int_{-i\infty}^{i\infty}
\frac{\dx z_1}{2\pi i}\int_{-i\infty}^{i\infty}\frac{\dx z_2}{2\pi i}\int_{-i\infty}^{i\infty}
\frac{\dx z_3}{2\pi i}\,\Gamma(-z_1)\Gamma(-z_2)\Gamma(-z_3)
\nonumber\\
&\qquad\qquad\quad\times\left(\frac{v_{11}-v_{12}}{4v_{12}}\right)^{z_1}
\left(\frac{v_{22}-v_{12}}{4v_{12}}\right)^{z_2}\left(\frac{v_{12}-1}{4}\right)^{z_3}
\frac{\pochhammer{j}{2z_1}\pochhammer{l}{2z_2}\pochhammer{2-j-l-2\eps}{2z_3}}
{\pochhammer{\frac{3}{2}-\eps}{z_1+z_2+z_3}}\,.
\end{align}
Upon employing the duplication identity 
$\pochhammer{2x}{2n}=2^{2n}\pochhammer{x}{n}\pochhammer{x+\frac{1}{2}}{n}$ 
on each Pochhammer symbol\index{Pochhammer symbol}
in the numerator we obtain the 
hypergeometric representation\index{Hypergeometric representation}   
of $I_{j,l}$ in term of the Lauricella function\index{Lauricella function}
$\lauricella$: 
\begin{align}
I_{j,l}(v_{12},v_{11},v_{22};\eps)=\frac{2\pi}{1-2\eps}\,v_{12}^{1-j-l-\eps}
\, \lauricella\left(a_1,a_2,a_3,b_1,b_2,b_3,c;x_1,x_2,x_3\right) \,, 
\label{eq:Hypergeometric representation of Ijl} 
\end{align}
where 
\eq 
\hspace*{-.25cm}
& &a_1 = \frac{j}{2}\,, \quad 
   a_2 = \frac{l}{2}\,, \quad 
   a_3 = \frac{3}{2} - a_1 - a_2 - \eps\,, \quad 
   b_1 = a_1 + \frac{1}{2}\,, \quad 
   b_2 = a_2 + \frac{1}{2}\,, \quad b_3 = a_3 - \frac{1}{2}\,,
\nonumber\\ 
\hspace*{-.25cm}
& & c   = a_1 + a_2 + a_3\,, \quad 
    x_1 = 1-\frac{v_{11}}{v_{12}}\,, \quad 
    x_2 = 1-\frac{v_{22}}{v_{12}}\,, \quad 
    x_3 = 1-v_{12}\,. 
\en 
Here the Lauricella function $\lauricella$ 
is equivalently defined as the Mellin-Barnes integral\index{Mellin-Barnes integral}    
\eq\label{Master_formula_LMB}
\lauricella\left(a_1,a_2,a_3,b_1,b_2,b_3,c,x_1,x_2,x_3\right)&=&
\int_{-i\infty}^{i\infty}\frac{\dx z_1}{2\pi i}
\int_{-i\infty}^{i\infty}\frac{\dx z_2}{2\pi i}
\int_{-i\infty}^{i\infty}\frac{\dx z_3}{2\pi i}
\,\Gamma(-z_1)\Gamma(-z_2)\Gamma(-z_3)
\nonumber\\
&\times& (-x_1)^{z_1} \, (-x_2)^{z_2} \, (-x_3)^{z_3} \,  
\nonumber\\
&\times&\frac{
\pochhammer{a_1}{z_1}\,  
\pochhammer{a_2}{z_2}\,  
\pochhammer{a_3}{z_3}\,
\pochhammer{b_1}{z_1}\,  
\pochhammer{b_2}{z_2}\,  
\pochhammer{b_3}{z_3}}
{\pochhammer{c}{z_1+z_2+z_3}}\,.
\en
or as the sum in terms of the Pochhammer symbols\index{Pochhammer symbol}
\eq\label{Master_formula_LPS}
\lauricella\left(a_1,a_2,a_3,b_1,b_2,b_3,c,x_1,x_2,x_3\right)
&=&\sum_{m,n,p=0}^\infty
\frac{\pochhammer{a_1}{m}
      \pochhammer{a_2}{n}
      \pochhammer{a_3}{p}
      \pochhammer{b_1}{m}
      \pochhammer{b_2}{n}
      \pochhammer{b_3}{p}}
     {\pochhammer{c}{m+n+p}}\,
\nonumber\\
&\times&\frac{x_1^m}{m!}\frac{x_2^n}{n!}\frac{x_3^p}{p!}\,.
\en

Eqs.~(\ref{eq:Hypergeometric representation of Ijl})-(\ref{Master_formula_LPS}) 
represent the main result of this section --- 
{\it Hypergeometric representation}\index{Hypergeometric representation}
for generic angular integral with two denominators 
without referring to their mass-shell properties (i.e.\,they are massless or massive). 
In Appendix~\ref{sec:Consistency of the hypergeometric representation of angular integrals} 
we explicitly demonstrate the consistency with known special cases.

\subsubsection{Partial differential identities\index{Partial differential identity}}
\label{sec:PDI}

In this section we derive differential identities involving angular integrals, 
which help to establish recurrence relations between the latter. Note, 
there were before some efforts to obtain differential relations between 
\index{Neerven integral}
Neerven-type integrals\index{Neerven integral} in literature (see, e.g., 
Ref.~\cite{Bojak:2000eu}). Here we perform a study of differential properties 
of angular integrals in a systematic way. 
For that purpose we consider the generic angular integral with two denominators 
\begin{align}
I_{j,l}(v_{12},v_{11},v_{22};\eps)=\int\dx\Omega_{k_1k_2}
\frac{1}{(v_1\cdot k)^j(v_2\cdot k)^l}=\int\dx\Omega_{k_1k_2}\powprop{v_1}{j}\powprop{v_2}{l} \,. 
\end{align}
It is convenient to choose coordinates such that
\begin{align}
\prop{v_1}&=\frac{1}{1-\mathbf{v_1}\cdot\mathbf{k}}=\frac{1}{1-\beta_1 \cos\theta_1} \,, 
\\
\prop{v_2}&=\frac{1}{1-\mathbf{v_2}\cdot\mathbf{k}}=
\frac{1}{1-\beta_2\cos\vartheta\cos\theta_1-\beta_2\sin\vartheta\sin\theta_1\cos\theta_2} \,,
\end{align}
with 
\eq\label{two_sets_variables} 
& &\beta_1=\sqrt{1-v_{11}}\,, \qquad \beta_2=\sqrt{1-v_{22}}    \,, 
\nonumber\\
& &\cos\vartheta=\frac{1-v_{12}}{\sqrt{1-v_{11}}\sqrt{1-v_{22}}} \,, 
\nonumber\\
& &\Delta_{12} = (\beta_1 \beta_2 \sin\vartheta)^2 
= (1-v_{11}) (1-v_{22}) - (1-v_{12})^2 \,. 
\en 
Taking partial derivatives of the propagators gives
\begin{align}
\frac{\partial}{\partial\beta_1}\prop{v_1}&=\cos\theta_1 \powprop{v_1}{2}\,,
\\
\frac{\partial}{\partial\beta_2}\prop{v_2}&=(\cos\vartheta\cos\theta_1
+\sin\vartheta\sin\theta_1\cos\theta_2)\powprop{v_2}{2}\,,
\\
\frac{\partial}{\partial\vartheta}\prop{v_2}&=\beta_2(-\sin\vartheta\cos\theta_1
+\cos\vartheta\sin\theta_1\cos\theta_2)\powprop{v_2}{2}\,.
\end{align}
Expressing appearing angular prefactors through the propagators yields
\begin{align}
\frac{\partial}{\partial\beta_i}\prop{v_i}&=\frac{1}{\beta_i} \, 
\Lv{v_i} \, \powprop{v_i}{2}\,, 
\\
\frac{\partial}{\partial\vartheta}\prop{v_2}&=\cot\vartheta\left[ \Lv{v_2} 
- \frac{\beta_2}{\beta_1\cos\vartheta} \, \Lv{v_1} \right] \, \powprop{v_2}{2}\,,
\end{align}
where $\Lv{v} = 1-\powprop{v}{-1}$. 
Using these differential identities we can calculate the derivatives of $I_{j,l}$ 
with respect to $\beta_1$, $\beta_2$, and $\vartheta$. We have
\begin{align}
\frac{\partial}{\partial\beta_1} I_{j,l}&=\int\dx\Omega_{k_1k_2} \, 
\frac{\partial}{\partial\beta_1}\powprop{v_1}{j}\powprop{v_2}{l}
\nonumber
\\
&=\frac{j}{\beta_1} \, 
\int\dx\Omega_{k_1k_2} \, \Lv{v_1} \, \powprop{v_1}{j-1}\powprop{v_2}{l} 
\powprop{v_1}{2}
\nonumber
\\
&=\frac{j}{\beta_1} \, \Big(I_{j+1,l}-I_{j,l}\Big) \,,
\end{align}
and by analogy 
\begin{align}
\frac{\partial}{\partial\beta_2} I_{j,l}&=\frac{l}{\beta_2}(I_{j,l+1}-I_{j,l})\,,
\label{Diff_beta2}\\
\frac{\partial}{\partial\vartheta} I_{j,l}&=l \cot\vartheta\left(
\left(1-\frac{\beta_2}{\beta_1\cos\vartheta}\right)I_{j,l+1}-I_{j,l}
+\frac{\beta_2}{\beta_1\cos\vartheta}I_{j-1,l+1}\right)\,.
\label{Diff_theta}
\end{align} 
Due to rotational invariance the angular integral $I_{j,l}$ does not 
change under replacement of orientation of vectors $v_1$ and $v_2$ 
with respect to integration momentum $k$. This helps to 
get the identity equivalent to Eq.~(\ref{Diff_theta}):  
\begin{align}
\frac{\partial}{\partial\vartheta} I_{j,l}&=j \cot\vartheta
\left(\left(1-\frac{\beta_1}{\beta_2\cos\vartheta}\right)I_{j+1,l}-I_{j,l}
+\frac{\beta_1}{\beta_2\cos\vartheta}I_{j+1,l-1}\right)\,.
\end{align}
Next, we derive the differential identities for angular integrals 
with respect to the set of variables $(v_{11},v_{22},v_{12})$ 
using relations~(\ref{two_sets_variables}) and 
\begin{align}
\frac{\partial}{\partial v_{12}}&=
\frac{1}{\beta_1\beta_2\sin\vartheta}\frac{\partial}{\partial \vartheta}\,,
\\
\frac{\partial}{\partial v_{11}}&=-\frac{1}{2\beta_1}\frac{\partial}{\partial \beta_1}
-\frac{\cot\vartheta}{2\beta_1^2}\frac{\partial}{\partial \vartheta}\,,
\\
\frac{\partial}{\partial v_{22}}&=-\frac{1}{2\beta_2}\frac{\partial}{\partial \beta_2}
-\frac{\cot\vartheta}{2\beta_2^2}\frac{\partial}{\partial \vartheta}\,.
\end{align}
One gets: 
\begin{align}
\frac{\partial}{\partial v_{12}}I_{j,l}&= \frac{l}{\Delta_{12}} \, 
\biggl[ (v_{22}-v_{12})I_{j,l+1}-(1-v_{12})I_{j,l}
+(1-v_{22})I_{j-1,l+1} \biggr] 
\label{eq: dv12 Ijl 1st}
\\
&=\frac{j}{\Delta_{12}} \, 
\biggl[ (v_{11}-v_{12})I_{j+1,l}-(1-v_{12})I_{j,l}+(1-v_{11})I_{j+1,l-1} \biggr] \,, 
\label{eq: dv12 Ijl 2nd}
\\
\frac{\partial}{\partial v_{11}}I_{j,l}&=\frac{j}{2 \Delta_{12}} \, 
\biggl[ (v_{22}-v_{12})I_{j+1,l}+(1-v_{22})I_{j,l}-(1-v_{12})I_{j+1,l-1} 
\biggr] \,, 
\label{eq: dv11 Ijl}
\\
\frac{\partial}{\partial v_{22}}I_{j,l}&=\frac{l}{2 \Delta_{12}} \, 
\biggl[ (v_{11}-v_{12})I_{j,l+1}
+(1-v_{11})I_{j,l}-(1-v_{12})I_{j-1,l+1} 
\biggr] \,.
\label{eq: dv22 Ijl}
\end{align}
Two independent ways of calculating $\frac{\partial}{\partial v_{12}}I_{j,l}$ 
[see Eqs.~(\ref{eq: dv12 Ijl 1st}) and~(\ref{eq: dv12 Ijl 2nd})]
lead to the consistency relation
\eq\label{Recur1} 
& &l(v_{22}-v_{12})I_{j,l+1}-l(1-v_{12})I_{j,l}+l(1-v_{22})I_{j-1,l+1}
\nonumber\\
&=&j(v_{11}-v_{12})I_{j+1,l}-j(1-v_{12})I_{j,l}+j(1-v_{11})I_{j+1,l-1}\,.
\en
This is a first algebraic (recurrence) relation between $I_{j,l}$ with different indices.

From Eqs.~(\ref{eq: dv12 Ijl 1st})-(\ref{eq: dv22 Ijl}) 
we deduce the following identities 
\begin{align}
2l\, \frac{\partial}{\partial v_{11}}I_{j,l+1}&=j\,\frac{\partial}{\partial v_{12}}I_{j+1,l}\,, 
\label{eq:diff relation v11 v12}
\\
2j\, \frac{\partial}{\partial v_{22}}I_{j+1,l}&=l\,\frac{\partial}{\partial v_{12}}I_{j,l+1}\,.
\label{eq:diff relation v22 v12}
\end{align}
Combining equations (\ref{eq:diff relation v11 v12}) and (\ref{eq:diff relation v22 v12}), 
we find that $I_{j,l}(v_{12},v_{11},v_{22};\eps)$ obeys the second order 
partial differential equation\index{Partial differential equation} (PDE) 
\begin{equation}
\left(\frac{\partial^2}{\partial v_{12}^2}-4\frac{\partial^2}{\partial v_{11}
\partial v_{22}}\right)I_{j,l}(v_{12},v_{11},v_{22};\eps)=0\,.
\end{equation}
Introducing the light-cone coordinates $v_\pm=\frac{1}{2}(v_{11}\pm v_{22})$ 
this PDE can be written in the standard form
\begin{align}
\left(\frac{\partial^2}{\partial v_{+}^2}-\frac{\partial^2}{\partial v_{-}^2}
-\frac{\partial^2}{\partial v_{12}^2}\right)I_{j,l}(v_{12},v_{+}+v_{-},v_{+}-v_{-};\eps)=0\,.
\end{align}
We see that $I_{j,l}(v_{12},v_{+}+v_{-},v_{+}-v_{-};\eps)$ satisfies 
a two-dimensional homogenous wave equation with ``time'' $v_+$ and ``speed of light'' $c=1$. 
The ``light-cone" is given by $0=v_+^2-v_-^2-v_{12}^2=v_{11}v_{22}-v_{12}^2$. 
It is the surface of vanishing symmetric rank-2 Gram determinant\index{Gram determinant} 
$G = - X = v_{11}v_{22}-v_{12}^2$. 
Since the PDE is independent of $\eps$, it must be fulfilled independently at all orders 
in the $\eps$-expansion. Therefore, it could serve as a consistency check 
for individual terms in the $\eps$-expansion. 

Up until now we only considered the dependence on the indices $j$ and $l$. 
However, we will see that the dependence of $I_{j,l}$ on $\eps$ is structurally quite similar. 
A good starting point is the representation
\begin{align}
I_{j,l}&=\frac{1}{B(j,l)} \, I^{(0)}(\eps) \, 
\int_0^1\dx x_1 x_1^{j-1} 
\int_0^1\dx x_2 x_2^{l-1}\delta(1-x_1-x_2)
\nonumber\\
&\times\ghy\left(\frac{j+l}{2},\frac{j+l+1}{2},\frac{3}{2}-\eps,1-x_1^2 
v_{11}-x_2^2 v_{22}-2x_1 x_2 v_{12}
\right)\,.
\label{eq:Ijl ghy representation}
\end{align}
Calculating the derivative with respect to $v_{12}$ using the identity
\begin{align}
\frac{\dx}{\dx x}\,\ghy(a,b,c,x)=\frac{a b}{c}\,\ghy(a+1,b+1,c+1,x) \,,
\end{align}
yields 
\begin{align}
\frac{\partial}{\partial v_{12}}I_{j,l}(\eps)=-\frac{j l}{1-2\eps}I_{j+1,l+1}(\eps-1)\,.
\end{align}
Using either Eq.~(\ref{eq: dv12 Ijl 1st}) or Eq.~(\ref{eq: dv12 Ijl 2nd}) 
to express $\partial I_{j,l}/\partial v_{12}$ 
results in an algebraic equation for $I_{j,l}(\eps-1)$ 
in terms of $I_{j^\prime,l^\prime}(\eps)$. 
Another way to obtain a formula of this kind is to start from the original angular integral 
and to express the additional factor $\sin^2\theta_1\sin^2\theta_2$ in terms of propagators: 
\eq\label{sin2_id} 
\sin^2\theta_1\sin^2\theta_2&=& 
1 - \frac{1}{\Delta_{12}} \, 
\biggl[ \beta_1^2 \, \Lvd{v_2}
      + \beta_2^2 \, \Lvd{v_1}
      - 2\beta_1\beta_2\cos\vartheta \, \Lv{v_1} \Lv{v_2} 
\biggr] \nonumber\\
&=& \frac{1}{\Delta_{12}} \, \biggl[ 
 v_{11}v_{22}-v_{12}^2
-(1-v_{11})\powprop{v_2}{-2}
-(1-v_{22})\powprop{v_1}{-2}
\nonumber\\
&+& 2(v_{12}-v_{11})\powprop{v_2}{-1}
 +  2(v_{12}-v_{22})\powprop{v_1}{-1}
\nonumber\\
&+&2(1-v_{12})\powprop{v_1}{-1}\powprop{v_2}{-1}
\biggr] \,.
\en
Next, we use Eq.~(\ref{sin2_id}) to obtain the  
{\it dimensional recurrence identity}\index{Dimensional recurrence identity} 
for angular integral $I_{j,l}(\eps)$: 
\eq\label{DRI1} 
\hspace*{-1.25cm}
I_{j,l}(\eps-1)&=&\int\dx\Omega_{k_1k_2}\sin^2\theta_1\sin^2\theta_2\, 
\powprop{v_1}{j}\powprop{v_2}{l}
\nonumber\\
\hspace*{-1.25cm}
&=&\frac{1}{\Delta_{12}} \, \Big[
(v_{11}v_{22}-v_{12}^2)I_{j,l}(\eps)
-(1-v_{11})I_{j,l-2}(\eps)-(1-v_{22})I_{j-2,l}(\eps)
\nonumber\\
\hspace*{-1.25cm}
&+&2(v_{12}-v_{11})I_{j,l-1}(\eps)
 + 2(v_{12}-v_{22})I_{j-1,l}(\eps)
 + 2(1-v_{12})I_{j-1,l-1}(\eps)
\Big] \,.
\en
Here we take into account that 
\eq
\dx\Omega_{k_1 k_2}(\eps) \, \sin^2\theta_1\sin^2\theta_2 \equiv \dx\Omega_{k_1 k_2}(\eps-1) \,. 
\en 
Now we are in the position to derive the second 
{\it dimensional recurrence identity}\index{Dimensional recurrence identity}, which complements 
the identity~(\ref{DRI1}).   
Doing this we start again with Eq.~(\ref{eq:Ijl ghy representation}) and shift 
dimension $\eps\rightarrow\eps+1$. Next,  
using the contiguos neighbours relation of the 
Gauss hypergeometric function\index{Gauss hypergeometric function} 
\begin{align}
\ghy(a,b,c-1,x)=\frac{1}{c-1}\left[a\,\ghy(a+1,b,c,x)-(a-c+1)\,\ghy(a,b,c,x)\right] \,,
\end{align}
we expand the angular integral $I_{j,l}(\eps+1)$ into two terms:  
\begin{align}\label{Ijl_eps_1}
I_{j,l}(v_{12},v_{11},v_{22};\eps+1)&= \frac{1}{B(j,l)} \, I^{(0)}(\eps+1)
\int_0^1\dx x_1 x_1^{j-1}\int_0^1\dx x_2 x_2^{l-1}\delta(1-x_1-x_2)
\nonumber\\
&\times\ghy\left(\frac{j+l}{2},\frac{j+l+1}{2},\frac{1}{2}-\eps,1-w_{12}\right)
\nonumber\\
&= \frac{1}{B(j,l)} \, 
I^{(0)}(\eps+1)\int_0^1\dx x_1 x_1^{j-1}
\int_0^1\dx x_2 x_2^{l-1}\delta(1-x_1-x_2)
\nonumber\\
&\times\frac{1}{1-2\eps}\left[(j+l)\,\ghy\left(\frac{j+l+1}{2},
\frac{j+l+2}{2},\frac{3}{2}-\eps,w_{12}\right)
\right.\nonumber\\
&\left.-(j+l-1+2\eps)\,\ghy\left(\frac{j+l}{2},\frac{j+l+1}{2},
\frac{3}{2}-\eps,w_{12}\right)\right]\,. 
\end{align}

The second term in Eq.~(\ref{Ijl_eps_1}) directly converts into 
integral $I_{j,l}(\epsilon)$. In case of the first term we apply 
the identity for the integrand 
\begin{align}
\delta(1-x_1-x_2) \, x_1^m x_2^n = 
\delta(1-x_1-x_2) \, (x_1^{m+1} x_2^n + x_1^m x_2^{n+1}) \,,
\end{align}
to obtain a sum of two angular integrals $I_{j+1,l}$ and $I_{j,l+1}$. 
Finally, using 
\begin{equation}
I^{(0)}(\eps+1)=\frac{2\pi}{1-2(\eps+1)}=-\frac{1-2\eps}{1+2\eps}I^{(0)}(\eps) \,,
\end{equation}
we get
\begin{align}
I_{j,l}(\eps+1)=\frac{j+l-1+2\eps}{1+2\eps}I_{j,l}(\eps)-\frac{j}{1+2\eps}I_{j+1,l}(\eps)
-\frac{l}{1+2\eps}I_{j,l+1}(\eps)\,.
\label{eq:dim recurrence eps to eps+1}
\end{align} 
The latter identity is the second 
{\it dimensional recurrence identity}\index{Dimensional recurrence identity}, which is more 
compact than the first one~(\ref{DRI1}).  

We finalize this section with the derivation of the differential equations for index  
and dimension rising operators. In particular, 
using the set of identities~(\ref{eq: dv12 Ijl 1st})-(\ref{eq: dv22 Ijl}) 
we introduce 
the {\it index $j/l$ rising differential operators}\index{Index rising differential operator} 
at fixed dimension:  
\eq 
\hat{D}_{1,j} &=& 
-2 (1-v_{11}) \frac{\partial}{\partial v_{11}}-(1-v_{12})\frac{\partial}{\partial v_{12}} + j
\,,\\
\hat{D}_{2,l} &=& 
-2 (1-v_{22}) \frac{\partial}{\partial v_{22}}-(1-v_{12})\frac{\partial}{\partial v_{12}} + l \,, 
\en 
which obey the following {\it index rising differential equations}
\index{Index rising differential equations}: 
\eq 
\hat{D}_{1,j} \, I_{j,l}(\eps) &=& j\,I_{j+1,l}(\eps)\,,
\label{IRDE1}\\ 
\hat{D}_{2,l} \, I_{j,l}(\eps) &=& l\,I_{j,l+1}(\eps)\,.
\label{IRDE2}
\en 
Next, we can sum up Eqs.~(\ref{IRDE1}) and~(\ref{IRDE2}), where r.h.s. 
is further simplified using identity~(\ref{eq:dim recurrence eps to eps+1}): 
\eq\label{Id4} 
j\,I_{j+1,l}(\eps) + l\,I_{j,l+1}(\eps) = (j+l-1+2\eps)\,I_{j,l}(\eps) 
- (1+2\eps)\,I_{j,l}(\eps+1) \,. 
\en  
Now it is convenient to define 
the {\it dimension rising differential operator}\index{Dimension rising differential operator} 
at fixed indices:  
\eq 
\hat{D}_{\eps} = \sum\limits_{v=v_{11},v_{12},v_{22}} \!\!\!\!\! 
2(1-v) \, \frac{\partial}{\partial v} \,-\, 1+2\eps \,, 
\en 
which obeys the following 
{\it dimension rising differential equation}\index{Dimension rising differential equation}: 
\eq 
\hat{D}_{\eps} \, I_{j,l}(\eps) = (1+2\eps)\, I_{j,l}(\eps+1)\,. 
\en

\subsubsection{Partial integration identities\index{Partial integration identity}}
\label{sec:PII}

Partial integration identities between angular integrals are another and complementary 
possibility to differential identities derived in previous section. 
We will perform the partial integration with respect to $\theta_1$. We will need
\begin{align}
\frac{\partial}{\partial\theta_1}\prop{v_1}&=-\beta_1\sin\theta_1\, \powprop{v_1}{2}\,, 
\\
\frac{\partial}{\partial\theta_1}\prop{v_2}&=-\beta_2 \, 
\Big(\cos\vartheta\sin\theta_1
-\sin\vartheta\cos\theta_1\cos\theta_2\Big)\, \powprop{v_2}{2}\,.
\end{align}
Using
\begin{align}
\frac{\dx}{\dx\theta_1}\frac{\sin^{2-2\eps}\theta_1}{2-2\eps}=\sin^{1-2\eps}\theta_1\cos\theta_1 \, ,
\end{align}
and 
\eq 
1 \equiv  (1-\beta_1\cos\theta_1) \, \prop{v_1} \, ,
\en 
it holds
\begin{align}
I_{j,l}&=\int\dx\Omega_{k_1 k_2}\powprop{v_1}{j}\powprop{v_2}{l}
\nonumber\\
&=I_{j+1,l}+\frac{\beta_1}{2-2\eps}\int_0^\pi\dx\theta_2\sin^{-2\eps}\theta_2\int_0^\pi\dx\theta_1
\sin^{2-2\eps}\theta_1\frac{\partial}{\partial\theta_1}\left[\powprop{v_1}{j+1}\powprop{v_2}{l}\right]  
\,. 
\label{eq:PI part 1}
\end{align}
Making the partial derivative 
\eq 
\hspace*{-1.25cm}
\beta_1\sin\theta_1\frac{\partial}{\partial\theta_1}\left[\powprop{v_1}{j+1}\powprop{v_2}{l}\right]
&=&l \, \powprop{v_1}{j+1} \powprop{v_2}{l+1} 
\Big[\Lv{v_1} \Lv{v_2} - \beta_1\beta_2\cos\vartheta\biggr]  
\nonumber\\
\hspace*{-1.25cm}
&+&(j+1) \, \powprop{v_1}{j+2} \powprop{v_2}{l} 
\Big[\Lvd{v_1} - \beta_1^2\Big] \, ,
\en
and plugging it into Eq.~(\ref{eq:PI part 1}) yields
\begin{align}
I_{j,l}=I_{j+1,l}+\frac{1}{2-2\eps}&\biggl[\Big(j+1\Big) 
\Big((1-\beta_1^2)I_{j+2,l}-2I_{j+1,l}+I_{j,l}\Big)
\nonumber\\
&
+l\Big((1-\beta_1\beta_2\cos\vartheta)I_{j+1,l+1}-I_{j,l+1}-I_{j+1,l}+I_{j,l}\Big)\biggr]\,.
\end{align}
Collecting $I_{j,l}$ with same indices and using $1-\beta_1^2=v_{11}$ and 
$1-\beta_1\beta_2\cos\vartheta=v_{12}$ results in the partial integration identity
\eq\label{Recur2}
0&=&(j+l-1+2\eps) \, I_{j,l}
    -(2j+l+2\eps) \, I_{j+1,l} 
+ v_{11} \, (j+1) \, I_{j+2,l}  \nonumber\\
&-&             l \, I_{j,l+1} 
    + l \, v_{12} \, I_{j+1,l+1} \,.
\en
Interchanging $\mathbf{v_1}$ and $\mathbf{v_2}$ yields the second complementary relation
\eq\label{Recur3}
0&=&(j+l-1+2\eps) \, I_{j,l}
    -(2l+j+2\eps) \, I_{j,l+1}
+ v_{22} \, (l+1) \, I_{j,l+2} \nonumber\\
&-&             j \, I_{j+1,l}
     +j \, v_{12} \, I_{j+1,l+1} \,.
\en

\subsubsection{Recursion relations\index{Recursion relation}}
\label{sec:recursion_relations}

In Sec.~\ref{sec:PDI} and~\ref{sec:PII} we derived three important 
relations~(\ref{Recur1})-(\ref{Recur3}) based on differential and integration properties 
of angular integrals:
\eq
0&=&(j-l) \, (1-v_{12}) \, I_{j,l}
       -j \, (1-v_{11}) \, I_{j+1,l-1}
       +l \, (1-v_{22}) \, I_{j-1,l+1}   \nonumber\\
&+&j \, (v_{12}-v_{11}) \, I_{j+1,l} 
 - l \, (v_{12}-v_{22}) \, I_{j,l+1}
\label{eq:R1} \,,
\\[3mm]
0&=&(j+l-1+2\eps)   \, I_{j,l}
    -(2j+l+2\eps)   \, I_{j+1,l}   
  +v_{11} \, (j+1)  \, I_{j+2,l}   \nonumber\\
&-&              l  \, I_{j,l+1}
      +l  \, v_{12} \, I_{j+1,l+1}
\label{eq:R2} \,,
\\[3mm]
0&=&(j+l-1+2\eps)   \, I_{j,l}
    -(2l+j+2\eps)   \, I_{j,l+1}  
  +v_{22} \, (l+1)  \, I_{j,l+2}   \nonumber\\
&-&              j  \, I_{j+1,l}
      +j  \, v_{12} \, I_{j+1,l+1} 
\label{eq:R3} \,.
\en
The first relation is symmetric under exchange $(j,v_{11}) \leftrightarrow (l,v_{22})$, 
while the second and the third are equivalent under such exchange. 
Using these relations one can establish recursion relations to deduce all 
angular integrals $I_{j,l}$ with $j,l\in \mathbb{Z}$. As starting point of the recursion 
procedure we use the set of three basic integrals: with no denominators 
$I_{0,0} = I^{(0)} = 2\pi/(1-2\eps)$, 
with one massive denominator $I_{1,0} = I_{1,0}^{(1)}(v_{11};\eps)$,  
and with two massive denominators $I_{1,1} = I_{1,1}^{(2)}(v_{12},v_{11},v_{22})$.   

We are now in the position to provide the complete set of recursive relations involving angular 
integrals $I_{j,l}$ for all $j,l\in\mathbb{Z}$. We consider five cases depending on signs of 
the indices $j$ and $l$. 

\vspace*{.25cm}
{\bf First case:} $l=0$ and $j>0$ (or $j=0$ and $l>0$).  
\vspace*{.25cm}

We start with trivial case, when one of the indices is zero and the other is positive. 
In case of $l=0$ and $j>0$ we take the identity~(\ref{eq:R2}) put $l=0$ and shift index $j$ 
as $j \to j-2$. We get the required recursion relation 
\begin{align}\label{recursion_1mass}
I_{j,0}=\frac{1}{v_{11}(j-1)} \, \left[2 (j-2+\eps)I_{j-1,0}-(j-3+2\eps)I_{j-2,0}\right] \, ,
\end{align}
valid for $j \ge 2$. The starting input for the recursion are two basic (parent) integrals 
$I_{0,0}$ and $I_{1,0}$ corresponding to $j=0$ and $j=1$, respectively. 

Next three integrals for this case are determined by the relations:
\eq 
I_{2,0}&=&\frac{1}{v_{11}} \, \left[2\eps \, I_{1,0} + (1-2\eps) \, I_{0,0}\right]\,, 
\nonumber\\
I_{3,0}&=&\frac{1}{v_{11}} \, \left[ (1+\eps) \, I_{2,0}  - \eps \, I_{1,0}\right]\,, 
\nonumber\\
I_{4,0}&=&\frac{1}{3v_{11}}\, \left[2 (2+\eps) \, I_{3,0} - (1+2\eps) \, I_{2,0}\right]\,. 
\en  
The recursion procedure for $j=0$, $l>0$ is completely analogous with $(j,v_{11})$ and $(l,v_{22})$ 
interchanged. 

For massless case $v_{11}=0$ at $l=0$ and $j>0$ the recursion is again derived from 
identity~(\ref{eq:R2}) by shifting index $j$ as $j \to j-1$: 
\begin{align}\label{Rec_Massless_1}
I_{j,0}=\frac{j-2+2\eps}{2j-2+2\eps}\,I_{j-1,0} \, ,
\end{align}
valid for $j \ge 1$ and is starting with basic integral $I_{0,0}$. 

In this case the next four integrals are fixed from: 
\eq 
I_{1,0}&=&I_{0,0} \, \frac{-1+2\eps}{2\eps} \,, 
\nonumber\\
I_{2,0}&=&I_{1,0} \, \frac{\eps}{1+\eps} \,, 
\nonumber\\
I_{3,0}&=&I_{2,0} \, \frac{1+2\eps}{4+2\eps} \,, 
\nonumber\\
I_{4,0}&=&I_{3,0} \, \frac{1+\eps}{3+\eps} \,. 
\en 
The recursion for $j=0$, $l>0$ is completely analogous with $j$ and $l$ 
interchanged. 

In Appendix~\ref{sec:tables_AI} we present explicit results for the angular integrals 
and perform the $\eps$ expansion. 

\vspace*{.25cm}
{\bf Second case:} $l=0$ and $j<0$ (or $j=0$ and $l<0$).  
\vspace*{.25cm}

In case $l=0$ and $j<0$ we again use the identity~(\ref{eq:R2}) 
and resort the terms there to get the following 
recursion relation: 
\begin{align}\label{recursion_11mass} 
I_{j,0}=\frac{1}{1-j-2\eps} \, 
\left[ - 2 (j+\eps) \, I_{j+1,0} + v_{11} (j+1) \, I_{j+2,0}\right] \, ,
\end{align}
starting with $I_{0,0}$ and valid for $j \le -1$. 
In particular, the next four integrals are: 
\eq 
I_{-1,0}&\equiv& I_{0,0} \,, \nonumber\\
I_{-2,0}&=& I_{0,0}  \, \frac{4-v_{11}-2\eps}{3-2\eps}\,, 
\nonumber\\
I_{-3,0}&=& \frac{1}{2-\eps}  \Big[ - v_{11} I_{0,0}  + (3-\eps) I_{-2,0} \Big]\,,
\nonumber\\
I_{-4,0}&=& \frac{1}{5-2\eps} \Big[ - 3 v_{11} I_{-2,0}  + 2 (4-\eps) I_{-3,0} \Big]\,. 
\en  
The recursion for $j=0$, $l<0$ is completely analogous 
with $(j,v_{11})$ and $(l,v_{22})$ interchanged. 

In the massless case ($j<0$, $l=0$, $v_{11}=0$) 
the recursion relation is further simplified to 
\begin{align}\label{recursion_11massless}  
I_{j,0}=-\frac{2 (j+\eps)}{1-j-2\eps}\,I_{j+1,0}\,, 
\end{align} 

In particular, the results for the next three integrals read 
\eq 
I_{-2,0} &=& I_{-1,0} \, \frac{4-2\eps}{3-2\eps}\,, 
\nonumber\\
I_{-3,0} &=& I_{-2,0} \, \frac{3-\eps}{2-\eps}\,, 
\nonumber\\
I_{-4,0} &=& I_{-3,0} \, \frac{8-2\eps}{5-2\eps}\,.
\en 
The case ($j=0$, $l<0$, $v_{22}=0$) is completely analogous with $j$ and $l$ interchanged.

\vspace*{.25cm}
{\bf Third case:} $j < 0$ and $l < 0$. 
\vspace*{.25cm}

In the case of no actual denominators ($j < 0$ and $l < 0$) 
we can give a closed polynomial formula for $I_{j,l}$.  
Using Eqs.~(\ref{Gauss_Pochhamer}), (\ref{Ijl_v12_eps}),  
(\ref{2M_to_1M_red_a}), and~(\ref{2M_to_1M_red_b}) 
we get 
\begin{align}
I_{j,l}(v_{12},v_{11},v_{22};\eps)&\overset{j,l\leq 0}{=}\pi\sum_{m=0}^{-j}
\binom{-j}{m} \, (1-\sqrt{1-v_{11}})^{-j-m} \, (1-v_{11})^{m/2}
\nonumber\\
&\times\sum_{n=0}^{-l}\binom{-l}{n} \, 
(1-\sqrt{1-v_{22}})^{-l-n} \, (1-v_{22})^{n/2} 
\nonumber\\
&\times 2^{m+n+1} \, 
\frac{\pochhammer{1-\eps}{m}\pochhammer{1-\eps}{n}}
{\pochhammer{1-2\eps}{m+n+1}}
\,\ghy\left(-m,-n,1-\eps,1-\frac{\hat{v}_{12}}{2}\right)
\nonumber\\
&=\pi\sum_{m=0}^{-j} \, \sum_{n=0}^{-l}
\sum_{p=0}^{\mathrm{min}(m,n)}\binom{-j}{m}\binom{-l}{n}
(1-\sqrt{1-v_{11}})^{-j-m} \, (1-v_{11})^{m/2} 
\nonumber\\
&\times(1-\sqrt{1-v_{22}})^{-l-n} \, (1-v_{22})^{n/2} \, 
2^{m+n+1} \, 
\frac{\pochhammer{1-\eps}{m}\pochhammer{1-\eps}{n}}
{\pochhammer{1-2\eps}{m+n+1}}
\nonumber\\
&\times
\frac{\pochhammer{-m}{p}\pochhammer{-n}{p}}{\pochhammer{1-\eps}{p}p!} \, 
\left(1-\frac{\hat{v}_{12}}{2}\right)^p\,,
\end{align}
where 
\eq 
\hat{v}_{12} = \frac{1}{2}
\left(1+\frac{1-v_{12}}{\sqrt{1-v_{11}}\sqrt{1-v_{22}}}\right) \,.
\en 
Here the sum over $p$ terminates at $\mathrm{min}(m,n)$, because 
$\pochhammer{-m}{p} \, \pochhammer{-n}{p} = 0$ at 
$p > \mathrm{min}(m,n)$. 

To complement the exact result for both negative indices we derive 
the recursion relation. It is derived summing two basic 
formulas~(\ref{eq:R2}) and~(\ref{eq:R3}). One gets: 
\eq\label{recursions_negative} 
I_{j,l} &=& \frac{1}{2 (j+l-1+2\eps)} \, 
\biggl[ 
(3l + j + 2\eps) \, I_{j,l+1} \,+\, 
(3j + l + 2\eps) \, I_{j+1,l} \nonumber\\
&-&
v_{11} \, (j+1) \, I_{j+2,l} \,-\, 
v_{22} \, (l+1) \, I_{j,l+2} \,-\, 
v_{12} \, (j+l)  \, I_{j+1,l+1} 
\biggr] \,. 
\en 
This recursion relation is valid for all $j, l \le -1$ and 
manifestly symmetric under interchange of two sets 
$(j,v_{11})$ and $(l,v_{22})$. E.g., the expressions for 
determination of a few first integrals read: 
\eq 
I_{-1,-1} &=& \frac{1}{3-2\eps} \, \biggl[ (2-\eps) \, \Big(I_{-1,0}+I_{0,-1}\Big) 
- v_{12} I_{0,0} \biggr] = 
I_{0,0} \, \frac{4-2\eps-v_{12}}{3-2\eps}         
\,,  \nonumber\\[2mm] 
I_{-1,-2} &=& \frac{1}{4 (2-\eps)} \, \biggl[ 
(7-2\eps) I_{-1,-1} \,+\, 
(5-2\eps) I_{0,-2}  \,-\, 
v_{22} \, I_{-1,0} \,-\, 
3v_{12} \, I_{0,-1} \biggr] \nonumber\\
&=& I_{0,0} \ \frac{6-2\eps-2v_{12}-v_{22}}{3-2\eps} 
\,,  \nonumber\\[2mm]  
I_{-2,-2} &=& \frac{1}{5-2\eps} \, \biggl[ 
(4-\eps) \, \Big(I_{-2,-1} + I_{-1,-2}\Big) \,-\, 
\frac{v_{11}}{2} \, I_{0,-2} \,-\, 
\frac{v_{22}}{2} \, I_{-2,0} \,-\,
2v_{12} \, I_{-1,-1}\bigg] \nonumber\\ 
&=& \frac{I_{0,0}}{(3-2\eps) (5-2\eps)} \, 
\biggl[ (3-\eps) \Big(4-\eps - \frac{v_{11}+v_{22}+4v_{22}}{2} \Big) 
+ v_{11} v_{22} + 2 v_{12}^2 \biggr] \,.
\en 
In Appendix~\ref{sec:tables_AI} we display the integrals for $-4 \le j,l \le -1$ 
and perform their $\eps$ expansion up to order $\eps$. 
Note, that the limit of massless particles $v_{11}=0$ or/and 
$v_{22}=0$ is straightforward in this case. 

\vspace*{.25cm}
{\bf Fourth case:} $j>0$ and $l<0$ (or $l>0$, $j<0$).  
\vspace*{.25cm}

For the case, that the indices of the angular integral have opposite signs 
$j>0$ and $l<0$ (or $l>0$, $j<0$) we involve all three 
master identities~(\ref{eq:R1})-(\ref{eq:R3}). In particular, 
in case $j>0$ and $l<0$ we do the following: 
(1) first, we shift indices $j \to j-1$ in (\ref{eq:R2}) 
and $l \to l-1$ in (\ref{eq:R3}); (2) sum three equations 
multiplying each of them with $-1$, $-(1-v_{12})$, and $(1-v_{11})$, 
respectively. After shifting index $l \to l+1$ in the resulting expression 
we get the recursive relation 
\eq\label{Req_jl1} 
I_{j,l} &=& \frac{1-v_{12}}{1-v_{11}} \, I_{j-1,l+1} 
\,+\, 
\frac{1}{(1-v_{11}) \, (j+l-1+2\eps)} \, 
\biggl[ 
I_{j-1,l+2} \, (l+1) \, (v_{12}-v_{22}) \nonumber\\
&-&I_{j,l+2}   \, (l+1) \, X \,+\, 
I_{j,l+1} \, (j+2l+2\eps) \, (v_{12}-v_{11}) 
\biggr] \,,
\en 
where $X = v_{12}^2 - v_{11} v_{22}$. 
Starting integrals for this recursion, which are known to us, 
are $I_{0,0}$, $I_{1,0}$, and $I_{0,1}$. The integral $I_{1,1}$ 
does not appear in this recursion since $(l+1) I_{j,l+2} = 0$ 
at $l=-1$. 

We present a few examples of the expressions for the angular 
integrals with $j>0$ and $l<0$: 
\eq
I_{1,-1} &=& 
\frac{1-v_{12}}{1-v_{11}} \, I_{0,0} \,+\, 
\frac{v_{12}-v_{11}}{1-v_{11}}  \, I_{1,0} \,,  
\nonumber\\ 
I_{1,-2} &=& 
\frac{1-v_{12}}{1-v_{11}} \, I_{0,-1} \,+\, 
\frac{1}{2 (1-v_{11}) (1-\eps)} \, 
\biggl[ (v_{12}-v_{22}) \, I_{0,0} \,-\, X \, I_{1,0} 
\nonumber\\
&+&  
(v_{12}-v_{11}) (3-2\eps) \, I_{1,-1} \biggr] \,, 
\nonumber\\ 
I_{2,-2} &=& 
\frac{1-v_{12}}{1-v_{11}} \, I_{1,-1} \,+\, 
\frac{1}{(1-v_{11}) (1-2\eps)} \, 
\biggl[ (v_{12}-v_{22}) \, I_{1,0} \,-\, X \, I_{2,0}
\nonumber\\ 
&+&  
2 (v_{12}-v_{11}) (1-\eps) \, I_{2,-1} \biggr] \,. 
\en 
By analogy we can derive the recursive relation 
for the case with $j<0$ and $l>0$ making 
interchange of two sets 
$(j,v_{11})$ and $(l,v_{22})$: 
\eq\label{Req_jl2}  
I_{j,l} &=& \frac{1-v_{12}}{1-v_{22}} \, I_{j+1,l-1} 
\,+\, 
\frac{1}{(1-v_{22}) \, (j+l-1+2\eps)} \, 
\biggl[ 
I_{j+2,l-1} \, (j+1) \, (v_{12}-v_{11}) \nonumber\\
&-&I_{j+2,l}   \, (j+1) \, X \,+\, 
I_{j+1,l} \, (2j+l+2\eps) \, (v_{12}-v_{22}) 
\biggr] \,. 
\en  

\vspace*{.25cm}
{\bf Fifth case:} $j>0$ and $l>0$. 
\vspace*{.25cm}

To derive the recursive relations for the 
$j>0$ and $l>0$ case we start with Eqs.~(\ref{Req_jl1}), (\ref{Req_jl2})  
and shift the indices there as 
$l \to l-2$ in Eq.~(\ref{Req_jl1}) and 
$j \to j-2$ in Eq.~(\ref{Req_jl2}). We get two 
recursive relations 
\eq 
I_{j,l} &=& \frac{1}{X (j-1)} \, 
\biggl[ 
I_{j-1,l}   \, (v_{12}-v_{22}) \, (2j+l-4+2\eps) 
\,-\, 
I_{j-2,l}   \, (1-v_{22}) \, (j+l-3+2\eps) 
\nonumber\\
&+& 
I_{j,l-1}   \, (j-1) \, (v_{12}-v_{11}) \,+\, 
I_{j-1,l-1} \, (j+l-3+2\eps) \, (1-v_{12}) 
\biggr] \,, \label{Recj2l1}\\
I_{j,l} &=& \frac{1}{X (l-1)} \, 
\biggl[ 
I_{j,l-1}   \, (v_{12}-v_{11}) \, (2l+j-4+2\eps) \,-\, 
I_{j,l-2}   \, (1-v_{11}) \, (j+l-3+2\eps) 
\nonumber\\
&+&
I_{j-1,l}   \, (l-1) \, (v_{12}-v_{22}) \,+\, 
I_{j-1,l-1} \, (j+l-3+2\eps) \, (1-v_{12}) 
\biggr] \, , \label{Recj1l2}
\en 
valid for $j \ge 2$, $l \ge 1$ and $j \ge 1$, $l \ge 2$ and starting 
with basic integral $I_{1,1}$. 
One can see that Eq.~(\ref{Recj2l1}) is not valid for $j=1, l=2$, 
while Eq.~(\ref{Recj1l2}) is not valid for $j=2, l=1$ because of a pole. 
To avoid the pole and to get the recursive relation, which is manifestly 
symmetric under interchange of $(j,v_{11})$ and $(l,v_{22})$ 
we simply multiply Eq.~(\ref{Recj2l1}) with factor $X (j-1)$ and Eq.~(\ref{Recj1l2}) 
with factor $X (l-1)$ and sum them up. As a result we 
get the recursive relation, which is more suitable for our purposes 
\eq\label{Main_Rec_5} 
I_{j,l} &=& \frac{j+l-3+2\eps}{X (j+l-2)} \, 
\biggl[ 
2 (1-v_{12}) \, I_{j-1,l-1} \,-\, 
  (1-v_{11}) \, I_{j,l-2}   \,-\,
  (1-v_{22}) \, I_{j-2,l} 
\biggr] \nonumber\\
&+& 
\frac{2j+2l-5+2\eps}{X (j+l-2)} \, 
\biggl[
  (v_{12}-v_{11}) \, I_{j,l-1} \,+\, 
  (v_{12}-v_{22}) \, I_{j-1,l} 
\biggr] \, ,
\en 
valid for all $j,l \ge 1$ and $j+l \neq 2$. Note, that $j+l \neq 2$ means that 
$j=l=1$ and this choice corresponds to the master integral $I_{1,1}$. 

Below we list few examples of integrals with positive $j$ and $l$: 
\eq 
I_{1,2} &=& \frac{1}{X} \, 
\biggl[ (v_{12} - v_{11}) (1+2\eps) \, I_{1,1} 
- 2 (1-v_{11}) \, \eps \,      I_{1,0} 
\nonumber\\
&+& 2 (1-v_{12})   \, \eps \,  I_{0,1} 
 + (v_{12}-v_{22}) \, I_{0,2} 
\biggr] \,, \nonumber\\
I_{2,1} &=& \frac{1}{X} \, 
\biggl[ (v_{12} - v_{22}) (1+2\eps) \, I_{1,1} 
- 2 (1-v_{22}) \, \eps \,      I_{0,1} 
\nonumber\\
&+& 2 (1-v_{12})   \, \eps \,  I_{1,0} 
 + (v_{12}-v_{11}) \,  I_{2,0} 
\biggr] \,, \nonumber\\
I_{2,2} &=& 
\frac{1+2\eps}{2X} \, 
\biggl[ 
2 (1-v_{12}) \, I_{1,1} 
- (1-v_{11}) \, I_{2,0} 
- (1-v_{22}) \, I_{0,2}
\biggr] 
\nonumber\\
&+& 
\frac{3+2\eps}{2X} \, 
\biggl[ 
  (v_{12}-v_{11}) \, I_{2,1} 
+ (v_{12}-v_{22}) \, I_{1,2} 
\biggr] \, .
\en 

Finally, we stress that the main result of this section is the derivation of the set 
of recursive relations [see Eqs.~(\ref{recursion_1mass}),     
(\ref{Rec_Massless_1}), 
(\ref{recursion_11mass}), 
(\ref{recursion_11massless}),  
(\ref{recursions_negative}), 
(\ref{Req_jl1}), 
(\ref{Req_jl2}), 
and  
(\ref{Main_Rec_5})], which together with three 
master integrals $I_{0,0}$, $I_{1,0}$, and $I_{1,1}$ can deduce any angular 
$I_{j,l}$ with arbitrary indices $j$ and $l$. 
Therefore, the $\eps$-expansion of any $I_{j,l}$ follows from expansion 
of master integrals $I_{0,0}$, $I_{1,0}$, and $I_{1,1}$. 
The recursion relations have been implemented together with the all-order expansions of the basis 
integrals in \textit{Mathematica}. 
This allows for fast calculation of any $I_{j,l}$ with $j,l\in\mathbb{Z}$ to arbitrary order 
in~$\eps$.

\subsection{All order \texorpdfstring{$\eps$}{TEXT}-expansion of angular integrals}
\label{sec:All order expansion}

The hypergeometric representations of the angular integrals derived 
in Sec.~\ref{sec:Somogyi integrals} allow for 
a systematic $\eps$-expansion up to desired order. In the cases where the angular integrals are expressible 
in terms of the Gauss hypergeometric function\index{Gauss hypergeometric function}  
there are publicly available algorithms for the $\eps$-expansion up to considerably 
high order. A well-suited Mathematica-based package is~\textit{HypExp}~\cite{Huber:2005yg}, which provides 
analytic expansions in terms of harmonic polylogarithms.

In this chapter we derive the analytical formalism allowing  for 
the all order $\eps$-expansions for the integrals up to two denominators (i.e.\,  
{\it the basic angular integrals}) in terms of the single and double 
Nielsen polylogarithms\index{Nielsen polylogarithm}~\cite{Kolbig:1969zza,Kolbig:1983qt}. 
As it was shown 
in Sec.~\ref{sec:Double massive integral with two denominators} for arbitrary dimension $D$
the double massive angular integral can be transformed to a finite sum of  
single-massive integrals based on the reduction formalism proposed by us. Therefore, 
$\eps$-expansion of the double massive integral for any $j,l\in\mathbb{Z}$ can be generated 
by use of the $\eps$-expansion of the single massive integral.
Using two mass splitting (see Sec.~\ref{sec:Double massive integral with two denominators}) 
and recursion relations (see Sec.~\ref{sec:recursion_relations}) 
any angular integral $I_{j,l}(v_{12},v_{11},v_{22};\eps)$ 
with $j,l\in\mathbb{Z}$ can be expressed in terms of the basic angular integrals  
mentioned above.
Note, that these results have never been discussed or derived before in literature. 
Only expansion up to a few first orders in $\eps$ were known before 
(see, e.g. results in Refs.~\cite{vanNeerven:1985xr,Beenakker:1988bq,Somogyi:2011ir}).  
Therefore, we think that our findings will be useful as input for the calculation 
of QCD processes. 

\subsubsection{Massive integral with one denominator}

We start with the hypergeometric representation for the massive integral 
with one denominator~(\ref{eq:Massive integral with one denominator 
as Gauss hypergeometric function with integer coefficients}) 
\begin{align}
I_1^{(1)}(v_{11};\eps)&=\frac{I^{(0)}(\eps)}{(1-\beta)}
\,\ghy\left(1,1-\eps,2-2\eps,-\frac{2\beta}{1-\beta}\right)
\nonumber\\
&=\frac{2\pi}{1-2\eps}\frac{1}{1-\sqrt{1-v_{11}}}
\,\ghy\left(1,1-\eps,2-2\eps,1-\frac{1+\sqrt{1-v_{11}}}{1-\sqrt{1-v_{11}}}\right)\,.
\end{align}
The $\eps$-expansion of the $\ghy(1,1-\eps,2-2\eps,x)$ is established the following way. 
First, we perform the Euler transformation and employ the Euler integral representation for 
the Gauss hypergeometric function\index{Gauss hypergeometric function} $\ghy$:
\begin{align}\label{2F1_Gauss}
\ghy(1,1-\eps,2-2\eps,x)&=(1-x)^{-\eps}\ghy(1-\eps,1-2\eps,2-2\eps,x)
\nonumber\\
&=(1-x)^{-\eps}(1-2\eps)\int_0^1\dx t\,t^{-2\eps}(1-xt)^{-1+\eps}\,.
\end{align}
Next using $\frac{\partial}{\partial x}(1-x t)^\eps=-\eps t (1-x t)^{-1+\eps}$ we get 
\begin{align}
\ghy(1,1-\eps,2-2\eps,x)&=-\frac{1-2\eps}{\eps}(1-x)^{-\eps}\frac{\partial}{\partial x}
\int_0^1\frac{\dx t}{t}\,t^{-2\eps}(1-xt)^\eps\,.
\end{align}
To separate the pole at $t=0$ in Eq.~(\ref{2F1_Gauss}) we use the decomposition 
\eq
(1-xt)^\eps = 1 + [(1-xt)^\eps - 1] 
\en
and drop the first constant term, 
since it vanishes when differentiated w.r.t.\,$x$. 
Then we perform the $\eps$ expansion of the remaining $(1-xt)^\eps - 1$. It yields 
\eq 
\ghy(1,1-\eps,2-2\eps,x)
&=&-\frac{1-2\eps}{\eps}(1-x)^{-\eps}\sum_{n=0}^\infty\sum_{m=1}^\infty\frac{(-2)^n}{n!m!}
\eps^{m+n} \nonumber\\
&\times& \frac{\partial}{\partial x}\int_0^1\frac{\dx t}{t}\,\log^n t\log^m(1-xt)\,. 
\en
The latter expansion can be written in a more compact form using  
the familiar~\textit{Nielsen polylogarithms}\index{Nielsen polylogarithm}~\cite{Kolbig:1969zza,Kolbig:1983qt}
\begin{align}
S_{n,p}(x)\equiv\frac{(-1)^{n+p-1}}{(n-1)! p!}\int_0^1\frac{\dx t}{t}\,\log^{n-1} t\log^p(1-xt)\,,
\end{align}
which satisfies the differential equation 
\begin{align}
\frac{\partial}{\partial x}S_{n,p}(x)=\frac{1}{x}S_{n-1,p}(x)\,.
\end{align}
Hence, we have
\begin{align}
\ghy(1,1-\eps,2-2\eps,x)&=-\frac{1-2\eps}{\eps}(1-x)^{-\eps}\sum_{n=0}^\infty\sum_{m=1}^\infty 
2^n (-1)^m\eps^{m+n}\frac{\partial}{\partial x}S_{n+1,m}(x)
\nonumber\\
&=-\frac{1-2\eps}{\eps}\frac{(1-x)^{-\eps}}{x}\sum_{n=0}^\infty\sum_{m=1}^\infty 
2^n (-1)^m\eps^{m+n}S_{n,m}(x)\,.
\end{align}
Next, rearranging the summation to make it over $N=m+n-1$ 
and shifting $m\rightarrow m+1$ yields 
\begin{align}
\ghy(1,1-\eps,2-2\eps,x)&=(1-2\eps)\frac{(1-x)^{-\eps}}{x}\sum_{N=0}^\infty
\sum_{m=0}^N 2^{N-m}(-1)^m  S_{N-m,m+1}(x)\, \eps^N\,.
\end{align}
Therefore, we have established the all order $\eps$-expansion of the 
massive integral with one denominator~(\ref{eq:Massive integral with one 
denominator as Gauss hypergeometric function with integer coefficients}) 
\begin{align}\label{expansion_1mass}
I_1^{(1)}(v_{11};\eps)&=\frac{\pi}{\sqrt{1-v_{11}}}
\left(\frac{1+\sqrt{1-v_{11}}}{1-\sqrt{1-v_{11}}}\right)^{-\eps}
\nonumber\\
&\times\sum_{N=0}^\infty\sum_{m=0}^N 2^{N-m} (-1)^{m+1}S_{N-m,m+1}
\left(1-\frac{1+\sqrt{1-v_{11}}}{1-\sqrt{1-v_{11}}}\right)\,\eps^N\,.
\end{align}
Using specific Nielsen polylogarithms 
$S_{0,p}(x)=(-1)^p \log^p(1-x)/p!$ and $S_{n,1}(x)=\mathrm{Li}_{n+1}(x)$ 
this explicitly gives up to order $\eps$
\begin{align}
I_1^{(1)}(v_{11};\eps)&=\frac{\pi}{\sqrt{1-v_{11}}}
\left(\frac{1+\sqrt{1-v_{11}}}{1-\sqrt{1-v_{11}}}\right)^{-\eps}
\left[\log\left(\frac{1+\sqrt{1-v_{11}}}{1-\sqrt{1-v_{11}}}\right)
\right.
\nonumber\\
&\left.
+\frac{\eps}{2}\log^2\left(\frac{1+\sqrt{1-v_{11}}}{1-\sqrt{1-v_{11}}}\right)-2\eps\,
\dilog\left(1-\frac{1+\sqrt{1-v_{11}}}{1-\sqrt{1-v_{11}}}\right)+\mathcal{O}(\eps^2)\right]\,.
\end{align}
The latter formula is exactly the same expansion as obtained in Ref.~\cite{Somogyi:2011ir}. 

Eq.~(\ref{expansion_1mass}) together with the recursion relation~(\ref{recursion_1mass}) 
completes all order $\eps$-expansion of the angular integrals $I^{(1)}_j(v_{11};\eps)$
with one massive denominator. 
In particular, starting from $I_0^{(1)}(v_{11};\eps) \equiv I^{(0)}(\eps) = 2\pi/(1-2\eps)$ 
and $I_1^{(1)}(v_{11};\eps)$ given by Eq.~(\ref{expansion_1mass}) we can derive 
analytical results for any power of massive denominator $j$. For example, 
the results for $j=2,3,4$ are given by the recursion identities: 
\eq 
I^{(1)}_2(v_{11};\eps) &=& \frac{1}{v_{11}} \, 
\biggl[ 2\eps I^{(1)}_1(v_{11};\eps) + (1-2\eps) I^{(0)}(\eps) \biggr]\,, 
\nonumber\\
I^{(1)}_3(v_{11};\eps) &=& \frac{1}{2v_{11}} \, 
\biggl[ 2(1+\eps) I^{(1)}_2(v_{11};\eps) - 2\eps I^{(1)}_1(v_{11};\eps) \biggr]\,, 
\\
I^{(1)}_4(v_{11};\eps) &=& \frac{1}{3v_{11}} \, 
\biggl[ 2(2+\eps) I^{(1)}_3(v_{11};\eps) - (1+2\eps) I^{(2)}_2(v_{11};\eps) \biggr] 
\nonumber \,.
\en 

\subsubsection{Massless integral with two denominators}

For the two denominator massless integral we start with the 
hypergeometric representation\index{Hypergeometric function}~(\ref{Ijl_v12_eps}) 
\begin{align}
I_{1,1}^{(0)}(v_{12};\eps)&=I^{(0)}(\eps)
\frac{\pochhammer{-2\eps}{2}}{2^2\pochhammer{-\eps}{1}\pochhammer{-\eps}{1}}
\,\ghy\left(1,1,1-\eps,1-\frac{v_{12}}{2}\right)
\nonumber\\
&=-\frac{\pi}{\eps}\,\ghy\left(1,1,1-\eps,1-\frac{v_{12}}{2}\right)\,.
\end{align}
The $\eps$-expansion of the $\ghy\left(1,1,1-\eps,x\right)$ runs analogous to the 
one denominator one mass case considered above. First, we perform the Euler transformation 
and employ the Euler integral\index{Euler formula} representation of the $\ghy$:
\begin{align}
\ghy(1,1,1-\eps,x)&=(1-x)^{-1-\eps}\ghy(-\eps,-\eps,1-\eps,x)
\nonumber\\
&=(1-x)^{-1-\eps}(-\eps)\int_0^1\dx t\,t^{-\eps-1}(1-xt)^{\eps}\,.
\end{align}
The pole at $t=0$ is taken care of by subtracting the numerator at the pole. 
Afterwards we can safely expand in $\eps$.
\begin{align}
\ghy(1,1,1-\eps,x)&=-\eps(1-x)^{-1-\eps}\left(\int_0^1\frac{\dx t}{t}
\,t^{-\eps}\left[(1-x t)^{\eps}-1\right]+\int_0^1\dx t\,t^{-\eps-1}\right)
\nonumber\\
&=-\eps(1-x)^{-1-\eps}\left(\int_0^1\frac{\dx t}{t}
\sum_{n=0}^\infty\frac{(-\eps)^n}{n!}\log^n t\sum_{m=1}^\infty\frac{\eps^m}{m!}
\log^m(1-xt)-\frac{1}{\eps}\right)\,.
\end{align}
Rearranging the terms and evaluating the logarithmic integrals in terms of the 
Nielsen polylogarithms\index{Nielsen polylogarithm} yields
\begin{align}
\ghy(1,1,1-\eps,x)&=\eps(1-x)^{-1-\eps}\left[1-\eps\sum_{n=0}^\infty\sum_{m=1}^\infty\frac{(-1)^n}{n!m!}
\eps^{m+n}\int_0^1\frac{\dx t}{t}\,\log^n t\log^m(1-xt)\right]
\nonumber\\
&=\eps(1-x)^{-1-\eps}\left[1-\eps\sum_{n=0}^\infty\sum_{m=1}^\infty(-1)^m\eps^{m+n}S_{n+1,m}(x)\right]\,.
\end{align}
Performing the summation over $N=m+n+1$ and $m$ gives
\begin{align}
\ghy(1,1,1-\eps,x)&=(1-x)^{-1-\eps}\left[1-\sum_{N=2}^\infty
\sum_{m=1}^{N-1}(-1)^m S_{N-m,m}(x)\,\eps^N\right]\,.
\end{align}
Therefore we have established the all order expansion
\begin{align}
I_{1,1}^{(0)}(v_{12},\eps)=\pi\left(\frac{v_{12}}{2}\right)^{-1-\eps}
\left[-\frac{1}{\eps}+\sum_{N=1}^\infty\sum_{m=1}^N(-1)^m S_{N-m+1,m}\left(1-\frac{v_{12}}{2}\right)
\,\eps^N\right]\,.
\end{align}
For checking purposes we explicitly write out the terms up to order $\eps$ and 
confirm that it exactly coincides with result derived 
before in Refs.~\cite{vanNeerven:1985xr,Beenakker:1988bq,Somogyi:2011ir}:
\begin{align}
I_{1,1}^{(0)}(v_{12},\eps)=\pi\left(\frac{v_{12}}{2}\right)^{-1-\eps}
\left[-\frac{1}{\eps}-\eps\,\dilog\left(1-\frac{v_{12}}{2}\right)+\mathcal{O}(\eps^2)\right]\,.
\end{align}

\subsubsection{Single massive integral with two denominators}

For the two denominator single massive integral we start with the 
hypergeometric representation\index{Hypergeometric representation}~(\ref{Ijl_1}) 
\eq 
I_{1,1}^{(1)}(v_{12},v_{11};\eps)&=&\frac{I_1^{(0)}(\eps)}{v_{12}}
\left(\frac{v_{11}}{v^2_{12}}\right)^\eps \appell\left(-2\eps,-\eps,-\eps,
1-2\eps,\omega_+,\omega_-\right)\,, \nonumber\\
\omega_\pm &=& \frac{\tau_\pm}{1-\tau_\pm} = 1-\frac{v_{12}}{1\pm\sqrt{1-v_{11}}} \,, 
\en
where $I_1^{(0)}(\eps) = - \pi/\eps$. 
The $\eps$-expansion of the Appell function\index{Appell function} $\appell(-2\eps,-\eps,-\eps,1-2\eps,x,y)$ 
is established similarly to the previous cases. Employing the 
Euler integral\index{Euler formula} representation and separating the pole gives
\begin{align}
\appell(-2\eps,-\eps,-\eps,1-2\eps,x,y)&=-2\eps\int_0^1\dx t\,t^{-1-2\eps}(1-x t)^\eps(1-y t)^\eps 
\nonumber\\
&=-2\eps\left(\int_0^1\frac{\dx t}{t}\, t^{-2\eps}\left[(1-xt)^\eps (1-yt)^\eps-1\right]
+\int_0^1\dx t\,t^{-1-2\eps}\right)\,.
\end{align}
Now we can safely expand in $\eps$ under the integral to obtain
\begin{align}
\appell(\dots)&=-2\eps\left(\int_0^1\frac{\dx t}{t}
\,\sum_{n=0}^\infty\frac{(-2\eps)^n}{n!}\log^n t\sum_{m=1}^\infty\frac{\eps^m}{m!}
\log^m\left((1-xt)(1-yt)\right)-\frac{1}{2\eps}\right)
\nonumber\\
&=1-2\eps\sum_{n=0}^\infty\sum_{m=1}^\infty\frac{(-2)^n}{n!m!}\eps^{m+n}\int_0^1
\frac{\dx t}{t}\log^n t\log^m\left((1-xt)(1-yt)\right)\,.
\end{align}
By binomial expansion the $\log^m$ term can be written in the form
\eq 
\log^m\left((1-xt)(1-yt)\right)&=&\left(\log(1-x t)+\log(1-y t)\right)^m
\nonumber\\ 
&=&\sum_{k=0}^m\binom{m}{k}\log^{m-k}(1-xt)\log^k(1-yt)\,.
\en
Hence we can write
\begin{align}
\appell(\dots)&=1-2\eps\sum_{n=0}^\infty\sum_{m=1}^\infty\frac{(-2)^n}{n!m!}
\eps^{m+n}\sum_{k=0}^m\binom{m}{k}\int_0^1\frac{\dx t}{t}\log^n t\log^{m-k}(1-xt)\log^k(1-yt)\,.
\end{align}
The logarithmic integral is the natural two-variable generalization of the Nielsen polylogarithm. 
Defining the \textit{double Nielsen polylogarithm}\index{Nielsen polylogarithm} 
[see discussion of their properties in Appendix~\ref{app:Nielsen}] 
\begin{align}
S_{n,p_1,p_2}(x,y)\equiv\frac{(-1)^{n+p_1+p_2-1}}{(n-1)!p_1!p_2!}\int_0^1\frac{\dx t}{t}
\,\log^{n-1} t\log^{p_1}(1-xt)\log^{p_2}(1-yt)\,,
\end{align}
we can evaluate  the integrals to
\begin{align}
\appell(\dots)&=1-2\eps\sum_{n=0}^\infty\sum_{m=1}^\infty (-1)^m 2^n\eps^{m+n}
\sum_{k=0}^m S_{n+1,m-k,k}(x,y)\,.
\end{align}
Changing summation to run over $N=m+n+1$ and $m$ yields the result
\begin{align}
\appell(-2\eps,-\eps,-\eps,1-2\eps,x,y)&=1-\sum_{N=2}^\infty\sum_{m=1}^{N-1} 
(-1)^m 2^{N-m}\sum_{k=0}^m S_{N-m,m-k,k}(x,y)\,\eps^N\,.
\end{align}
This establishes the all-order expansion of the single massive 
two denominator integral in the form
\eq 
I_{1,1}^{(1)}(v_{12},v_{11},\eps)&=&\frac{\pi}{v_{12}}
\biggl(\frac{v_{11}}{v^2_{12}}\biggr)^\eps\biggl[-\frac{1}{\eps}
\nonumber\\
&+&\sum_{N=1}^\infty\sum_{m=1}^{N} (-1)^m 2^{N-m+1}\sum_{k=0}^m 
S_{N-m+1,m-k,k}(\tau_+,\tau_-)\,\eps^N\biggr]\,. 
\en
Up to order $\eps$ this expression is
\eq 
I_{1,1}^{(1)}(v_{12},v_{11},\eps)&=&\frac{\pi}{v_{12}}
\biggl(\frac{v_{11}}{v^2_{12}}\biggr)^\eps \, \biggl[-\frac{1}{\eps}
-2\eps\biggl( \dilog\biggl(1-\frac{v_{12}}{1+\sqrt{1-v_{11}}}\biggr)
\nonumber\\
&+&\dilog\biggl(1-\frac{v_{12}}{1-\sqrt{1-v_{11}}}\biggr)\biggr)
+\,\mathcal{O}(\eps^2)\vphantom{\frac{1}{1}}\biggr]\,.
\en
Genuine double Nielsen polylogarithms start appearing at order $\eps^2$.

\subsubsection{Double massive integral with two denominators}
\label{sec:Double_massive}

As we showed in Sec.~(\ref{sec:Double massive integral with two denominators}) 
the double massive integral reduces to a sum of single massive 
integrals using two mass splitting 
[see details in Eqs.~(\ref{eq:Two mass integral general result}) 
and~(\ref{eq:Two mass integral general result})]. In particular, 
there are two fully equivalent decompositions of the double mass integral 
governed by the mixing parameter $\lambda$. Due to rotational invariance 
both solutions for the $\lambda$ parameter and correspondingly for the 
scalar products of the velocities $v_{13}$ and $v_{23}$ are equivalent 
to each other. Hence, we restrict to one of them, e.g. to 
\begin{align}
\lambda &=\frac{v_{11}-v_{12}-\sqrt{X}}{v_{11}+v_{22}-2v_{12}}\,,
\\
v_{13}&=\frac{v_{11}\left(v_{22}+\sqrt{X}\right)-v_{12}
\left(v_{12}+\sqrt{X}\right)}{v_{11}+v_{22}-2v_{12}}\,,
\\
v_{23}&=\frac{v_{22}\left(v_{11}-\sqrt{X}\right)-v_{12}
\left(v_{12}-\sqrt{X}\right)}{v_{11}+v_{22}-2v_{12}}\,,
\end{align} 
where $X=v_{12}^2-v_{11}v_{22}$. 

In this case the expression for the 
double massive integral $I_{1,1}^{(2)}(v_{12},v_{11},v_{22};\eps)$ 
reads: 
\begin{align}
I_{1,1}^{(2)}(v_{12},v_{11},v_{22};\eps)=\frac{1}{\sqrt{X}}
\left[v_{13}I_{1,1}^{(1)}(v_{13},v_{11};\eps)
     -v_{23}I_{1,1}^{(1)}(v_{23},v_{22};\eps)\right]\,.
\end{align}
Here we used the relations $\lambda=v_{13}/\sqrt{X}$ and $1-\lambda=-v_{23}/\sqrt{X}$.
After cancelling the pole and some rearrangement one gets 
\begin{align}\label{Eq_I11_2mass}
I_{1,1}^{(2)}(v_{12},v_{11},v_{22};\eps)&=\frac{\pi}{\sqrt{X}}
\left[\log\left(\frac{v_{12}+\sqrt{X}}{v_{12}-\sqrt{X}}\right)
-\sum_{n=1}^\infty\frac{\eps^n}{(n+1)!}
\left(\log^{n+1}\frac{v_{11}}{v^2_{13}}
     -\log^{n+1}\frac{v_{22}}{v^2_{23}}
\right)\right.
\nonumber\\
&+\left(\frac{v_{11}}{v^2_{13}}\right)^\eps\sum_{N=1}^\infty\sum_{m=1}^{N} 
(-1)^m 2^{N-m+1}\sum_{k=0}^m S_{N-m+1,m-k,k}(\tau_+^{13},\tau_-^{13})\,\eps^N
\nonumber\\
&\left.-\left(\frac{v_{22}}{v^2_{23}}\right)^\eps\sum_{N=1}^\infty\sum_{m=1}^{N} 
(-1)^m 2^{N-m+1}\sum_{k=0}^m S_{N-m+1,m-k,k}(\tau_+^{23},\tau_-^{23})\,\eps^N\right]\,,
\end{align}
with $\tau_\pm^{13}=1-{v_{13}}/({1\pm\sqrt{1-v_{11}}})$ and 
$\tau_\pm^{23}=1-{v_{23}}/({1\pm\sqrt{1-v_{11}}})$. 
In derivation of Eq.~(\ref{Eq_I11_2mass}) we use the identity: 
\eq 
\log\frac{v_{22}}{v^2_{23}} - \log\frac{v_{11}}{v^2_{13}} = 
\log\left(\frac{v_{12}+\sqrt{X}}{v_{12}-\sqrt{X}}\right) \,. 
\en 
Explicitly, the $\eps$ expansion up to second order reads: 
\begin{align}\label{dm_eps_2nd}
  I_{1,1}^{(2)}(v_{12},v_{11},v_{22};\eps)&=\frac{1}{\sqrt{X}}\left[
 v_{13}\,I^{(1)}_{1,1}(v_{13},v_{11};\eps)
-v_{23}\,I^{(1)}_{1,1}(v_{23},v_{22};\eps)
\right]
\nonumber\\
&=\frac{\pi}{\sqrt{X}}\left[
\log\left(\frac{v_{12}+\sqrt{X}}{v_{12}-\sqrt{X}}\right)
-\eps\left(\frac{1}{2}\log^2\frac{v_{11}}{v_{13}^2}-\frac{1}{2}\log^2\frac{v_{22}}{v_{23}^2}
\right.\right.
\nonumber\\
&\qquad\quad+2\,\dilog\left(1-\frac{v_{13}}{1-\sqrt{1-v_{11}}}\right)
+2\,\dilog\left(1-\frac{v_{13}}{1+\sqrt{1-v_{11}}}\right)
\nonumber\\
&\qquad\quad\left.\left.
-2\,\dilog\left(1-\frac{v_{23}}{1-\sqrt{1-v_{22}}}\right)
-2\,\dilog\left(1-\frac{v_{23}}{1+\sqrt{1-v_{22}}}\right)
\right)\right]
\nonumber\\
&\qquad\quad+\mathcal{O}(\eps^2)\,.
\end{align}

\subsection{Neerven integrals\index{Neerven integral}}

In this section we would like to discuss some aspects of the Neerven angular 
integrals: 
\begin{itemize}
\item[1.] connection between 
the Neerven~\cite{vanNeerven:1985xr,Beenakker:1988bq} 
and Somogyi~\cite{Somogyi:2011ir} parametrization of 
angular integrals;
\item[2.] some important remarks on massless and 
double massive Neerven integrals;
\item[3.] comment on Mirkes's recursive formula~\cite{Mirkes:1992hu} 
for a special class of single massive Neerven integrals.
\end{itemize}
\subsubsection{Connection between Neerven and Somogyi parametrization}
\label{sec:Connection between Neerven and Somogyi parametrization} 

Discussion of angular integrals in the literature revolves for the most part around results
for integrals of the form
\begin{align}
I^{(j,l)}_D(a,b,A,B,C)=\int\dx\Omega_{k_1k_2}\frac{1}{(a+b\cos\theta_1)^j
(A+B\cos\theta_1+C\sin\theta_1\cos\theta_2)^l}\,.
\end{align}
In this section we will review how these angular integrals presented in the Neerven form can 
be translated to those in the Somogyi form
\begin{align}
\int\dx\Omega_{k_1k_2}\frac{1}{(v_1\cdot k)^j (v_2\cdot k)^l}\,,
\end{align}
introduced in Sec.~\ref{sec:Somogyi integrals}.
Defining
\begin{align}
v_1=\left(1,\mathbf{0}_{D-3},0,-\frac{b}{a}\right)\, \, ,\qquad 
v_2=\left(1,\mathbf{0}_{D-3},-\frac{C}{A},-\frac{B}{A}\right) \, ,
\end{align}
we have
\begin{align}
  v_{11}=1-\frac{b^2}{a^2}     \,, \quad 
  v_{22}=1-\frac{B^2+C^2}{A^2} \,, \quad 
  v_{12}=1-\frac{b B}{a A}     \,.
\end{align}
With $k=(1,\dots,\sin\theta_1\cos\theta_2,\cos\theta_1)$ we can write the generic Neerven integral as
\begin{align}
I^{(j,l)}_{D}(a,b,A,B,C)=\frac{1}{a^j A^l}\int\dx\Omega_{k_1k_2}\frac{1}{(v_1\cdot k)^j (v_2\cdot k)^l}\,.
\end{align}
For the four classes of the Neerven integrals this implies
\paragraph{Massless case: $a^2=b^2$ and $A^2=B^2+C^2$}
\begin{align}
I^{(j,l)}_{4-2\eps}(a,b,A,B,C)=\frac{1}{a^jA^l}I^{(0)}_{j,l}\left(1-\frac{bB}{aA};\eps\right) \,.
\end{align}
\paragraph{Single massive case: $a^2=b^2$}
\begin{align}
I^{(j,l)}_{4-2\eps}(a,b,A,B,C)=\frac{1}{a^jA^l}I^{(1)}_{j,l}
\left(1-\frac{bB}{aA},1-\frac{B^2+C^2}{A^2};\eps\right) \,. 
\end{align}
\paragraph{Single massive case: $A^2=B^2+C^2$}
\begin{align}
I^{(j,l)}_{4-2\eps}(a,b,A,B,C)=\frac{1}{a^jA^l}I^{(1)}_{j,l}
\left(1-\frac{bB}{aA},1-\frac{b^2}{a^2};\eps\right) 
\,.
\end{align}

\paragraph{Double massive case: $a^2\neq b^2$ and $A^2\neq B^2+C^2$:}
\begin{align}
I^{(j,l)}_{4-2\eps}(a,b,A,B,C)=\frac{1}{a^jA^l}I^{(2)}_{j,l}
\left(1-\frac{bB}{aA},1-\frac{b^2}{a^2},1-\frac{B^2+C^2}{A^2};\eps\right)\,.
\end{align}
Note, that there are the following limits:
\begin{itemize}
\item[1.] $I^{(2)}_{j,l}$ reduces to $I^{(1)}_{j,l}$ at $v_{11}=0$ or $v_{22}=0$, 
\item[2.] $I^{(1)}_{j,l}$ reduces to $I^{(0)}_{j,l}$ at $v_{11}=0$. 
\end{itemize}
Therefore, the integral $I^{(2)}_{j,l}$ can be regarded as the 
basic Neerven integral~(\ref{eq:Van Neerven integral}) encoding information 
about the others. 
Applying the result from Sec.~\ref{sec:Hypergeometric representation of angular integrals} 
we can write down the general Neerven integral in hypergeometric form in terms of 
the Lauricella function \index{Lauricella function} $\lauricella$: 
\begin{align}
&I^{(j,l)}_{4-2\eps}(a,b,A,B,C)=\frac{1}{a^j A^l}\frac{2\pi}{1-2\eps}
\left(1-\frac{b B}{a A}\right)^{1-j-l-\eps}
\nonumber\\
&\times
\lauricella\left(\frac{j}{2},\frac{l}{2},\frac{3-j-l-2\eps}{2},\frac{j+1}{2},
\frac{l+1}{2},\frac{2-j-l-2\eps}{2},\frac{3}{2}-\eps;x,y,z\right)\,,
\end{align}
with the abbreviations $x=\frac{b(Ab-aB)}{a(aA-bB)}$, 
$y=\frac{a(B^2+C^2)-AbB}{A(aA-bB)}$,  
and $z=\frac{bB}{aA}$.

For $j,l\in\mathbb{N}_{>0}$ we also derive the closed formula for
the angular integral using results from Sec.~\ref{sec:Somogyi integrals} 
in terms of the Appell function\index{Appell function} $\appell$
\begin{align}
&I^{(j,l)}_{4-2\eps}(a,b,A,B,C)=\frac{1}{a^j A^l}\frac{2\pi}{1-2\eps}
\nonumber\\
&\times\biggl[\sum_{n=0}^{j-1}\binom{l-1+n}{l-1}\frac{\lambda_\pm^l(1-\lambda_\pm)^n 
(2-l-n-2\eps)_{l+n}}{(v_{13}^\pm)^{j-n}2^{l+n}
(1-l-n-\eps)_{l+n}}\left(\frac{v_{11}}{v_{13}^{\pm 2}}\right)^{l+n-1+\eps} 
\nonumber\\
&\times\appell\left(2-j-l-2\eps,1-l-n-\eps,1-l-n-\eps,2-l-n-2\eps,x_{13}^\pm,y_{13}^\pm\right)
\nonumber\\
&+\sum_{n=0}^{l-1}\binom{j-1+n}{j-1}\frac{\lambda_\pm^n(1-\lambda_\pm)^j 
(2-l-n-2\eps)_{j+n}}{(v_{23}^\pm)^{l-n}2^{j+n}
(1-j-n-\eps)_{j+n}}\left(\frac{v_{22}}{v_{23}^{\pm 2}}\right)^{j+n-1+\eps}
\nonumber\\
&\times\appell\Big(2-j-l-2\eps,1-j-n-\eps,1-j-n-\eps,2-j-n-2\eps,x_{23}^\pm,y_{23}^\pm\Big)
\vphantom{\frac{v}{v}^2}\biggr]\,,
\end{align}
where the following notations are used: 
\begin{align}
v_{11}&=1-\frac{b^2}{a^2}\,,\,\,
v_{22}=1-\frac{B^2+C^2}{A^2}\,,\,\,
v_{12}=1-\frac{bB}{aA}\,,\,\,
\nonumber\\
\lambda_\pm &=\frac{A \left(a-a b B+A b^2 \pm\sqrt{(A b-a B)^2+C^2 (a-b) (a+b)}\right)}{a^2 C^2+(A b-a B)^2}\,,
\nonumber\\
v_{13}^\pm &=(1-\lambda_\pm)v_{11}+\lambda_\pm v_{12}\,,\,\,
v_{23}^\pm=(1-\lambda_\pm)v_{12}+\lambda_\pm v_{22}\,,\,\,
\nonumber\\
&x_{13}^\pm=1-\frac{v_{13}^\pm}{1+\sqrt{1-v_{11}}}\,,\,\,
y_{13}^\pm=1-\frac{v_{13}^\pm}{1-\sqrt{1-v_{11}}}\,,\,\,
\nonumber\\
& x_{23}^\pm=1-\frac{v_{23}^\pm}{1+\sqrt{1-v_{22}}}\,,\,\,
y_{23}^\pm=1-\frac{v_{23}^\pm}{1-\sqrt{1-v_{22}}}\,.
\end{align}

\subsubsection{On massless integral in the Neerven parametrization}

For a check of consistency, we want to derive the massless 
Neerven integral in its most commonly used form. 
This concerns the integral (compare with Ref.~\cite{vanNeerven:1985xr})
\begin{align}
\int\dx\Omega_{k_1k_2}\, \frac{1}{(1-\cos\theta_1)^j
(1-\cos\vartheta\cos\theta_1-\sin\vartheta\sin\theta_1\cos\theta_2)^l}\,.
\label{eq:Massless Van Neerven integral}
\end{align}
In the notation we have established, we write Eq.~(\ref{eq:Massless Van Neerven integral}) as
\begin{align}
I^{(j,l)}_{4-2\eps}(1,-1,1,-\cos\vartheta,-\sin\vartheta)=
I^{(0)}_{j,l}(1-\cos\vartheta;\eps)\,.
\end{align}
employing the result~(\ref{eq:Massless two denominator integral}) 
and rewriting the cosine with the half-angle formula we obtain
\begin{align}
I^{(j,l)}_{4-2\eps}(1,-1,1,-\cos\vartheta,-\sin\vartheta)=I_{j+l}^{(0)}(\eps)
\frac{(1-j-l-\eps)_j}{(1-j-\eps)_j}\,\ghy\left(jl,1-\eps,\cos^2\frac{\vartheta}{2}\right)\,.
\end{align}
Writing down $I_{j+l}^{(0)}(\eps)$ and the 
Pochhammer symbols\index{Pochhammer symbol} in terms of 
gamma functions\index{Gamma function} we arrive 
at the well known result (see, e.g., Ref.~\cite{Gordon:1993qc})
\begin{align}
&\int\dx\Omega_{k_1k_2}\frac{1}{(1-\cos\theta_1)^j(1-\cos\vartheta\cos\theta_1
-\sin\vartheta\sin\theta_1\cos\theta_2)^l}
\nonumber\\
&=2^{1-j-l}\pi \, \frac{\Gamma(1-2\eps)}{\Gamma^2(1-\eps)} \, 
B(1-j-\eps,1-l-\eps) \,\ghy\left(j,l,1-\eps,\cos^2\frac{\vartheta}{2}\right)\,.
\end{align}

\subsubsection{On massive integral in the Neerven parametrization
\index{Neerven integral}} 
\label{sec:remarks about massive Van Neerven integral}

To derive the four dimensional result for the double massive integral the proposed method via 
two-mass splitting is of course unnecessarily involved. One can instead employ 
Weierstrass substitution\index{Weierstrass substitution} 
$\xi=\tan(\theta/2)$ on both integrals in Eq.~(\ref{eq:Van Neerven integral}) and obtain, 
by taking advantage of the trigonometric identities
\begin{align}
\cos x=\frac{1-\tan^2\frac{x}{2}}{1+\tan^2\frac{x}{2}}\,, \qquad 
\sin x=\frac{2\tan\frac{x}{2}}{1+\tan^2\frac{x}{2}}\,,
\end{align}
the integral representation
\eq
I^{(j,l)}_{4-2\eps}(a,b,A,B,C)&=&2^{3-4\eps}\int_0^\infty\dx\xi_1\int_0^\infty\dx\xi_2 
\nonumber\\[2mm]
&\times& 
\frac{(1+\xi_1^2)^{j+l-2+2\eps} (1+\xi_2^2)^{l-1+2\eps} \xi_1^{1-2\eps} \xi_2^{-2\eps}}
{\Delta^j(a,b,\xi_1) \Delta^l(A,B,C,\xi_1,\xi_2)}\,, 
\en 
where 
\eq 
\Delta(a,b,\xi_1) &=& a+b+(a-b) \xi_1^2  \,, \nonumber\\
\Delta(A,B,C,\xi_1,\xi_2) &=& 
\Big(A+B+(A-B)\xi_1^2\Big) (1+\xi_2^2) +2C\xi_1^2 (1-\xi_2^2)\,. 
\en 
For $\eps=0$ and $j=l=1$ this vastly simplifies and can be elementary integrated 
to yield (compare, e.g., with Ref.~\cite{Schellekens:1981kq})
\begin{align}
I^{(1,1)}_{4}(a,b,A,B,C)=\frac{\pi}{\sqrt{X}}
\log\left(\frac{aA-bB+\sqrt{X}}{aA-bB-\sqrt{X}}\right)\, \, ,
\end{align}
with $X=(aA-bB)^2-(A^2-B^2-C^2)(a^2-b^2)$.
Any higher values $j,l\in\mathbb{N}$ can be reached by the differential identities 
(see also Ref.~\cite{Bojak:2000eu})
\begin{align}
I^{(j+1,l)}_D=-\frac{1}{j}\frac{\partial}{\partial a}I^{(j,l)}_D\,, \qquad 
I^{(j,l+1)}_D=-\frac{1}{l}\frac{\partial}{\partial A}I^{(j,l+1)}_D\,,
\end{align}
the first requiring $a^2\neq b^2$ the second $A^2\neq B^2+C^2$.

\subsubsection{Recursive determination of Mirkes type III integrals}
\label{sec:Mirkes Type III}

In Sec.~\ref{sec:New properties of angular integrals} we developed the methods for the derivation of 
relations between angular integrals. One should stress that 
there have been attempts in this direction in the literature before.
For example, in Ref.~\cite{Mirkes:1992hu} Mirkes made an effort to
derive the recursive formula for a special kind of single-massive Neerven integrals.
We find that his formula is correct only in four dimensions
and here we discuss how to extend it to arbitrary dimension.
The subject of study is the single massive angular integral
\begin{align}\label{Mirkes1}
I^{j,l}_D(A,B,C) = 
\int\dx\Omega_{k_1k_2}
\frac{(-\cos\theta_1)^l}{\left(A+B\cos\theta_1+C\sin\theta_1\cos\theta_2\right)^j} \,, 
\qquad l \in \mathbb{N}\,.   
\end{align} 
In order to derive the recursive relation for the integral~(\ref{Mirkes1}) we write it down in the 
form of a double massive integral with two denominators and express it in terms, 
which occurred before in our calculations 
(see Sec.~\ref{sec:Double massive integral with two denominators}): 
\eq 
I^{j,l}_D(A,B,C) &=& \lim_{\beta \to \infty} \, 
I^{j,l}_D(A,B,C;\beta)\,,\text{with} \nonumber\\ 
I^{j,l}_D(A,B,C;\beta) &=& 
\int\dx\Omega_{k_1k_2} \, 
\frac{\Big(\frac{1}{\beta}-\cos\theta_1\Big)^l}
{\left(A+B\cos\theta_1+C\sin\theta_1\cos\theta_2\right)^j} 
\nonumber\\
&=& \frac{1}{A^j \beta^l} \, 
\int\dx\Omega_{k_1k_2} \, 
\frac{(1 - \beta \cos\theta_1)^l}
{\left(1+\frac{B}{A} \cos\theta_1+\frac{C}{A} \sin\theta_1\cos\theta_2\right)^j} \,, 
\en 
where the parameter $\beta$ should go to infinity after deriving the recursive relations. 
Here, the velocities of massive particles are chosen as
\eq
v_1=\Big(1,{\bf 0}_{D-3},-\frac{C}{A},-\frac{B}{A}\Big)\,, \qquad
v_2=\Big(1,{\bf 0}_{D-2},\beta\Big)\,
\en
The scalar products of the velocities read
\eq\label{Mirkes2}
& &v_{11}=v_1^2 = 1-\frac{B^2+C^2}{A^2}\,, \qquad
v_{22}=v_2^2 = 1-\beta^2\,, \qquad
v_{12}=v_1\cdot v_2 = 1 + \beta \frac{B}{A} \,, \nonumber\\
& &X=v_{12}^2 - v_{11}v_{22} = \frac{(A\beta+B)^2+C^2(1-\beta^2)}{A^2} \,.
\en
Using notations from~(\ref{Mirkes2}) the integral $I^{j,l}_D(A,B,C;\beta)$ can be
classified as a double massive integral with two denominators (with positive $j$
and negative $-l$ powers):
\eq
I^{j,l}_D(A,B,C;\beta) = \frac{1}{A^j \beta^l} \, I_{j,-l}^{(2)}(v_{12},v_{11},v_{22})
\en
Next we obtain the recursion relation for
$I^{j,l}_D(A,B,C;\beta)$ using the 4th type recursion relation~(\ref{Req_jl1})
derived for the integral $I_{j,l}^{(2)}(v_{12},v_{11},v_{22})$
at $j>0$ and $-l<0$. We have:
\eq
I^{j,l}_D(A,B,C;\beta) &=& \frac{1}{B^2+C^2} \,
\biggl[ - B \, I^{j-1,l-1}_D(A,B,C;\beta)
\nonumber\\
&+& \frac{1-l}{j-l-1+2\eps} \, \biggl(A + \frac{B}{\beta} \biggr)
\, I^{j-1,l-2}_D(A,B,C;\beta)
\nonumber\\
&-& \frac{1-l}{j-l-1+2\eps} \,
\biggr(A^2-C^2 + \frac{2 A B}{\beta} + \frac{B^2+C^2}{\beta^2}\biggr)
\, I^{j,l-2}_D(A,B,C;\beta) \,
\nonumber\\
&+& \frac{j-2l+2\eps}{j-l-1+2\eps} \, \biggl(A B-\frac{A^4}{\beta (B^2+C^2)}\biggr)
\, I^{j,l-1}_D(A,B,C;\beta)
\biggr]
\en
Finally taking the limit $\beta \to \infty$ we derive the recursion relation
for type III integrals for arbitrary dimension and valid
for $l \ge 1$ and $j > 0$:
\eq
I^{j,l}_D(A,B,C) &=&
\frac{1}{B^2+C^2} \, \biggl[ - B \, I^{j-1,l-1}_D(A,B,C)
\nonumber\\
&+& \frac{1-l}{j-l-1+2\eps} \,
\biggl(A \, I^{j-1,l-2}_D(A,B,C)
\,-\, (A^2-C^2) \, I^{j,l-2}_D(A,B,C) \biggr)
\nonumber\\
&+& \frac{j-2l+2\eps}{j-l-1+2\eps} \, A B
\, I^{j,l-1}_D(A,B,C)
\biggr] \,.
\en
This is the main result of this Section. Note, that in specific cases
$j=1$ and $j=2$ the recursions relation read
\eq
I^{1,l}_D(A,B,C) &=&
\frac{1}{l \, (B^2+C^2)} \,
\biggl[ \frac{1+(-1)^l}{2} \  \frac{l}{l-2\eps} \,
\dfrac{
 \Big(\frac{3}{2}\Big)_{\frac{l}{2}-1}}
{\Big(\frac{3}{2}-\eps\Big)_{\frac{l}{2}-1}}
\ A \, I^{(0)}
\nonumber\\
&-&   \dfrac{1-(-1)^l}{2} \
\dfrac{\Big(\frac{3}{2}\Big)_{\frac{l-1}{2}}}
{\Big(\frac{3}{2}-\eps\Big)_{\frac{l-1}{2}}}  \ B \, I^{(0)}
+ \frac{l \, (1-l)}{l-2\eps} \, (A^2-C^2) \, I^{1,l-2}_D(A,B,C)
\nonumber\\[3mm]
&-& \frac{l \, (1-2l+2\eps)}{l-2\eps} \, A B \, I^{1,l-1}_D(A,B,C)
\biggr] \,, \label{RecIII_1}\\[3mm]
I^{2,l}_D(A,B,C) &=& - \frac{\partial}{\partial A}I^{1,l}_D(A,B,C) =
- \frac{1}{l \, (B^2+C^2)} \
\biggl[ \dfrac{1+(-1)^l}{2} \frac{l}{l-2\eps} \,
\dfrac{
 \Big(\frac{3}{2}\Big)_{\frac{l}{2}-1}}
{\Big(\frac{3}{2}-\eps\Big)_{\frac{l}{2}-1}}
\, I^{(0)}
\nonumber\\
&+& \frac{l \, (1-l)}{l-2\eps} \,
\biggl( 2 A \, I^{1,l-2}_D(A,B,C)
\,-\, (A^2-C^2) \, I^{2,l-2}_D(A,B,C) \biggr)
\nonumber\\[3mm]
&-& \frac{l \, (1-2l+2\eps)}{l-2\eps} \,
B \, \Big( I^{1,l-1}_D(A,B,C;\beta) - A \, I^{2,l-1}_D(A,B,C;\beta) \Big)
\biggr] \,. \label{RecIII_2}
\en
To derive Eqs.~(\ref{RecIII_1}) and~(\ref{RecIII_2}),  we used
\eq
I^{0,l}_D =
\int\dx\Omega_{k_1k_2} \, (-\cos\theta_1)^l =
\frac{1+(-1)^l}{2 \, (1+l)} \
\frac{\Big(\frac{3}{2}\Big)_{\frac{l}{2}}}
{\Big(\frac{3}{2}-\eps\Big)_{\frac{l}{2}}} \ I^{(0)} \,.
\en
In the limit $\eps=0$ (i.e. $D=4$)\,we exactly reproduce the
recursion relations (D.23) and (D.24) in Ref.~\cite{Mirkes:1992hu}.
The latter are valid only at  $\eps= 0$ and fail for $\eps\neq 0$.

In case of $l = 0$ and $j \ge 2$ we use the recursion relation~(\ref{recursion_1mass}),
which is in notations of type III integrals reads:
\eq\label{RecIII_0}
I^{j,0}_D(A,B,C) &=& \frac{1}{A \, v_{11} \, (j-1)}
\, \biggl[2 \, (j-2+\eps) \, I^{j-1,0}_D(A,B,C) \nonumber\\
&-& \frac{1}{A} \, (j-3+2\eps) \, I^{j-2,0}_D(A,B,C)\biggr]\,,
\en
where $v_{11}$ is defined in Eq.~(\ref{Mirkes2}). As before,
the recursion~(\ref{RecIII_0}) starts with master integrals
$I^{0,0}_D(A,B,C) = I^{(0)}$ and
$I^{1,0}_D(A,B,C) = I^{(1)}_1/A$.
In Appendix~\ref{sec:tables_Mirkes} using Eqs.~(\ref{RecIII_1})-(\ref{RecIII_2})
we present analytical results for the integrals for $j=1,2$ and $l=0,\ldots,4$.

By analogy with integrals~(\ref{Mirkes_analogue})
one can pin down the similar integrals
\begin{align}\label{Mirkes_analogue}
J^{j,l}_D(A,B,C) =
\int\dx\Omega_{k_1k_2}
\frac{(-\sin\theta_1\cos\theta_2)^l}
{\left(A+B\cos\theta_1+C\sin\theta_1\cos\theta_2\right)^j} \,.
\end{align}
It is clear that in this case we can do the same trick --- introducing
the massive propagator $(1/\beta - \sin\theta_1\cos\theta_2)^l$ with $\beta\to\infty$
and choose the second velocity $v_2$ and therefore the scalar product $v_{12}$ as
$v_2=\Big(1,{\bf 0}_{D-3},\beta,0\Big)$ and $v_{12} = 1 + \beta \frac{C}{A}$. 
One can see that the two similar types of integrals
$I^{j,l}_D(A,B,C)$ and $J^{j,l}_D(A,B,C)$ are related upon interchange of two arguments
$B \leftrightarrow C$ as
\eq
J^{j,l}_D(A,B,C) \equiv I^{j,l}_D(A,C,B) \,.
\en

\section{Conclusions}
\label{Sec:conclusion}

In present paper we discussed and developed new ideas and methods for calculation of loop
and angular integrals in $D=4-2\varepsilon$ dimensions in QCD, which serve as input for the study of
perturbative matrix elements relevant for different processes such as Drell-Yan,
SIDIS, etc. In particular,  our paper contains two main topics --- handling of loop and angular
integrals. In the first case we further develop the Pasarino-Veltman (PV)
technique~\cite{Passarino:1978jh} for dealing with tensorial
integrals. We proposed a covariant formalism for consideration of loop integrals with rich Lorentz
structure (so-called  tensorial loop integrals) arising from decomposition of Feynman matrix elements.
The method is based on the use of the orthogonal basis~[see Eq.~(\ref{basis_3perp_V})], which is formed
by specific linear combinations of external momenta, which is quite simple and convenient
for making analytical and numerical calculations. Note that our choice of the orthogonal basis
is by construction free of soft
singularities occurring in the limit of vanishing momenta.
Also it gives the straightforward results for the scalar
functions in the expansion of tensor loop integrals,
i.e.\,without solving a system of equations like  in the original
PV method. All scalar functions are manifestly Lorentz covariant and depend on the set
of Mandelstam variables. In present paper we restricted ourselves to the consideration of
diagrams up to four external legs, but an extension of our method to more complicated diagrams
(e.g., with 5 or 6 legs) is straightforward and we commented on it.
All scalar functions are related via recurrence relations and finally presented in terms
of PV functions or in terms of hypergeometric functions.

In the second part of our paper we presented a detailed and systematic analysis of angular
integrals, which play an important role in the study of angular structure of QCD observables at
next-to-leading order in the expansion in strong coupling constant. In particular,
it is important to have complete and analytical set of these integrals for massive and massless particles
and perform their $\eps$ expansion up to desired order. Based on methods developed before in
Refs.~\cite{vanNeerven:1985xr}-\cite{Mirkes:1992hu} we implement new ideas for consistent
and analytical treatment of angular integral with arbitrary number of propagators and
independent of their power and kinematical properties. We explicitly discuss our method for
splitting of the product of arbitrary number of propagators allowing to reduce it to the
product of one or two propagators in Appendix~\ref{sec:Partial Fractioning}. 
Also we studied in detail the integro-differential properties
of angular integrals in order to derive partial differential and integral identities and
the recursion relations. For the first time in literature we derive a complete set of recursion relations, 
which allow to deduce any angular integral with two
denominators in arbitrary power and with arbitrary on-shell properties.
We consider all particular cases:
\begin{itemize}
\item[1.] the powers of the propagators are of the same sign and positive;
\item[2.] the powers of the propagators are of the same sign and negative;
\item[3.] the powers of the propagators have opposite signs; 
\item[4.]  the power of one denominator is simply zero.
\end{itemize}
Next important issue we considered is the systematic $\eps$ expansion of angular integrals up to desired order.
We demonstrated that this task can be solved for all types of integrals (i.e.\,for any
powers of denominators and for any kinematical situations). For some integrals, e.g.,
with negative powers for both propagators we derived a closed polynomial formula
for the angular integral. For other cases, based on the recursion relations it
is enough to perform the $\eps$-expansion of the three
master integrals $I_{0,0}$, $I_{1,0}$, and $I_{1,1}$. 
We developed a \textit{Mathematica} package to reduce angular integrals 
to master integrals and to perform the symbolic all-order $\eps$ expansion. 
This allows for fast calculation of any $I_{j,l}$ with arbitrary $j,l\in\mathbb{Z}$ and
up to arbitrary order in~$\eps$ expansion.
All technical details of our consideration of loop and angular integrals are placed in
the Appendices, including tabulated integrals with specific on-shell properties and
powers of the denominators.

\vspace*{.25cm}
Finally, we summarize our main results:

\begin{enumerate}

\item We developed new formalism for reduction of tensor loop integrals in $D$ dimensions
based on the PV technique~\cite{Passarino:1978jh}. Our formalism allows
to expand matrix elements through a set of Lorentz structures induced
by the basis of orthogonal linear combinations of external momenta.

\item We discovered new properties of angular integrals:
partial fraction decomposition (reduction of a number of denominators in the integrand);
integro-differential identities leading to derivation of recursion relations between
angular integrals with arbitrary indices; hypergeometric representation for all two-denominators
angular integrals including double-massive case was derived.

\item We formulated approach for all order $\eps$-expansion of angular integrals, 
which is based on expansion of the set of three parent integrals, while the expansion 
of the others is performed with the use of recursion relations. 
The recursion starts with the parent integrals.

\end{enumerate}

\begin{acknowledgments}

We thank Werner Vogelsang for stimulating and useful discussions.
This work was funded by BMBF ``Verbundprojekt 05P2018 - Ausbau von ALICE                                     
am LHC: Jets und partonische Struktur von Kernen''
(F\"orderkennzeichen: 05P18VTCA1),
by ANID PIA/APOYO AFB180002 (Chile) 
and by FONDECYT (Chile) under Grant No. 1191103.
This study was supported by Deutsche Forschungsgemeinschaft (DFG) through
the Research Unit FOR 2926 ``Next Generation pQCD for
Hadron Structure: Preparing for the EIC'' (Project number 40824754).

\end{acknowledgments}

\clearpage

\appendix

\section{Triangle integrals}
\label{Triangle_details}

Here, we present analytical results for the scalar functions occurring
in the expansion of the triangle vector and tensorial (rank-2)
loop integrals.

\subsection{Scalar and vector triangle integrals}

First, we discuss calculation of scalar triangle
integral $C_0(s_1,s_2)$ with two massive external legs.
Here, we follow derivation done in Ref.~\cite{Smirnov:2006ry}.
We start with the integral
\eq
C_0(s_1,s_2) &=&
(i \pi^2)^{-1} \mu^{4-D}
\, \int d^Dk \, \frac{1}{k^2 (k+l_1)^2 (k+l_2)^2} \,,
\en
where $s_1 = l_1^2$, $s_2 = l_2^2$, $s_3 = (l_1 - l_2)^2 = 0$.
Note that for $s_3 \neq 0$ the $C_0(s_1,s_2,s_3)$ is fully
symmetric under permutation of the arguments $s_1$, $s_2$, and $s_3$.
Therefore, result, which will be obtained for $s_3=0$ and $s_1, s_2 \neq 0$
is trivially extended to other two cases of one vanishing mass:
$s_1=0$, $s_2, s_3 \neq 0$, and $s_2=0$, $s_1, s_3 \neq 0$.

After $\alpha$ parametization and integration over loop
momentum, the integral $C_0(s_1,s_2)$ takes the form
\eq
\hspace*{-1cm}C_0(s_1,s_2) =
- (\pi\mu^{-2})^{\frac{D}{2}-2}
\, \Gamma\Big(3-\frac{D}{2}\Big)
\, \int\limits_0^\infty d\alpha_1 \ldots \int\limits_0^\infty d\alpha_3
\, \frac{\delta\biggl(1-\sum\limits_{i=1}^3 \alpha_i\biggr) \,
        \biggl(\sum\limits_{i=1}^3 \alpha_i\biggr)^{3-D}}
{\Big(  - s_1 \alpha_1 \alpha_3
        - s_2 \alpha_2 \alpha_3
\Big)^{3-\frac{D}{2}}} \,.
\en
Then we apply the change of variables ($\alpha$ parameters)
using the trick from~\cite{Smirnov:2006ry}:
\eq
\alpha_1 = \eta_1 \xi\,, \qquad
\alpha_2 = \eta_1 (1-\xi)\,, \qquad
\alpha_3 = \eta_2 \,.
\en
Subsequently we take into account that the Jacobian of such
the change of variables is $J = \eta_1$ and
integrate over $\eta_2$ using $\delta$ function.

One gets:
\eq
C_0(s_1,s_2) = -
(\pi\mu^{-2})^{\frac{D}{2}-2}
\, \Gamma\Big(3-\frac{D}{2}\Big)
\, \int\limits_0^1 d\xi \int\limits_0^1 d\eta \,
\frac{\eta^{\frac{D}{2}-2} \, (1-\eta)^{\frac{D}{2}-3}}
{\Big(-s_1 \xi - s_2 (1-\xi)\Big)^{3-\frac{D}{2}}} \,.
\en
Next we integrate over $\eta$ using the integral representation
for the beta function\index{Beta function}
\eq\label{Beta_int}
B(m,n) = \frac{\Gamma(m) \Gamma(n)}
{\Gamma(m+n)} = \int\limits_0^1 d\eta \, \eta^{m-1} \,
(1-\eta)^{n-1}\,,
\en
where $\Gamma(m)$ is the gamma function\index{Gamma function}.
Integration over $\xi$ proceeds using simple
table integral
\eq
\int\limits_0^1 \, \frac{d\xi}{(\xi + A)^{n+1}}
= \frac{A^{-n} - (1+A)^{-n}}{n} \,.
\en

After these tricks the scalar triangle integral takes the form
\eq
\hspace*{1cm}
C_0(s_1,s_2) = \pi^{\frac{D}{2}-2} \,
\frac{2 \Gamma\Big(2-\frac{D}{2}\Big) \, \Gamma^2\Big(\frac{D}{2}-1\Big)}
{(4-D) \, \Gamma(D-4)} \,
\frac{\Big(-\frac{s_1}{\mu^2}\Big)^{\frac{D}{2}-2} \,-\,
      \Big(-\frac{s_2}{\mu^2}\Big)^{\frac{D}{2}-2}}
{s_1 \,-\, s_2} \, .
\en
Next using the definition of the scalar bubble function
$B_0(s)$~(\ref{PV_B0}) we express $C_0(s_1,s_2)$ in
terms of $B_0(s)$ functions as:
\eq
C_0(s_1,s_2) =
\frac{2 \, (D-3)}{4-D} \ \frac{B_0(s_1) - B_0(s_2)}{s_1 - s_2}  \,,
\en
which is manifestly symmetric under change of variables $s_1$ and $s_2$.
For the $C_0$ function depending on one argument we introduce the following
notation
\eq
C_0(s) = C_0(s,0) = C_0(0,s) \,.
\en
Now we show results for the vector integrals:
\eq
C^{\mu;0ij}_1 = (i \pi^2)^{-1} \, \mu^{4-D} \,
\int d^Dk \, \frac{k^\mu}{\Delta_0 \, \Delta_i \, \Delta_j}
= P^\mu \, C_{1P}^{0ij}
+ R^\mu \, C_{1R}^{0ij}
+ T^\mu \, C_{1T}^{0ij} \,.
\en
All scalar functions $C_{1P}^{0ij}$, $C_{1R}^{0ij}$, and
$C_{1T}^{0ij}$ are expressed in terms
of scalar function~(\ref{C1_expression}) manifestly symmetric
under $s_1 \leftrightarrow s_2$ permutation:
\eq
C_1(s_1,s_2,s_3) =
\frac{s_1+s_2-s_3}{\lambda(s_1,s_2,s_3)} \,
\biggl[ B_0(s_1) - B_0(s_3) - s_2 \, C_0(s_1,s_2,s_3)
\biggr]\,,
\en
For specific limits it reduces to:

Limit I ($s_3 = 0$)
\eq
C_1(s_1,s_2,0)
&=& \frac{s_1+s_2}{(s_1-s_2)^2} \,
\biggl[ B_0(s_1) - B_0(0) - s_2 \, C_0(s_1,s_2) \biggr]
\nonumber\\
&=&
\frac{s_1+s_2}{(s_1-s_2)^2} \,
\biggl[ \frac{4-D}{2 (D-3)} \ s_1 \, C_0(s_1) - s_2 \, C_0(s_1,s_2)
\biggr] \,.
\en
Limit II ($s_1 = 0$)
\eq
C_1(0,s_2,s_3)
&=& \frac{1}{s_2-s_3} \,
\biggl[ B_0(0) - B_0(s_3) - s_2 \, C_0(s_2,s_3) \biggr]
\nonumber\\
&=&
- \frac{1}{s_2-s_3} \,
\biggl[ \frac{4-D}{2 (D-3)} \ s_3 \, C_0(s_3) + s_2 \, C_0(s_2,s_3)
\biggr] \,.
\en
Limit III ($s_2 = 0$)
\eq
C_1(s_1,0,s_3)
&=& \frac{1}{s_1-s_3} \,
\biggl[ B_0(s_1) - B_0(s_3) \biggr]
\nonumber\\
&=&
\frac{4-D}{2 (D-3)} \ C_0(s_1,s_3) \,.
\en
Limit IV ($s_1 = s_3 = 0$)
\eq
C_1(0,s_2,0) \equiv  C_1(0,s_2) = -C_0(s_2) \,.
\en
Limit V ($s_2 = s_3 = 0$)
\eq
C_1(s_1,0,0) \equiv C_1(s_1,0)
&=& \frac{B_0(s_1) - B_0(0)}{s_1}
\nonumber\\
&=& \frac{4-D}{2 (D-3)} \, C_0(s_1) \,.
\en
Limit VI ($s_1 = s_2 = 0$)
\eq
C_1(0,0,s_3) \equiv C_1(s_3,0)
&=& \frac{B_0(s_3) - B_0(0)}{s_3}
\nonumber\\
&=& \frac{4-D}{2 (D-3)} \, C_0(s_3) \,.
\en
One can see that Limits V and VI are specified by the same function
$C_1(s,0)$.

In the following it is convenient to use the notations:
\eq
F(s_1,s_2) &=& \frac{B_0(s_1)-B_0(s_2)}{s_1-s_2} = \frac{4-D}{2 (D-3)}
\ C_0(s_1,s_2) \,, \\
G(s_1,s_2) &=& \frac{B_0(s_1)+B_0(s_2)}{s_1-s_2} \,, \\
H(s_1,s_2) &=& G(s_1,s_2) + \frac{s_1+s_2}{s_1-s_2} \, F(s_1,s_2) \,.
\en
Functions $F(s_1,s_2)$ and $G(s_1,s_2)$ are  manifestly symmetric resp. antisymmetric
under exchange of variables $s_1 \leftrightarrow s_2$. Function $F(s_1,s_2)$ obeys the
useful identity:
\eq
F(s_1,s_2) = \frac{s_1 \, F(s_1,0) - s_2 \, F(s_2,0)}{s_1 - s_2} \,.
\en
Function $H(s_1,s_2)$ has the following property::
\eq
H(s,0) = - H(0,s) = \frac{2 B_0(s)}{s} \,.
\en
Furthermore we use the following relations:
\eq
& &C_0(s,0) = C_0(0,s) = - C_1(0,s) = 2 F(s,0) \, \frac{D-3}{4-D} \,,
\nonumber\\[1mm]
& &C_1(s,0) = F(s,0) \,.
\en
Also we introduce the 
functions $\alpha_\pm(s_1,s_2,s_3)$,
$\beta_\pm(s_1,s_2,s_3)$,
$\gamma_\pm(s_1,s_2,s_3)$, and 
$\sigma_\pm(s_1,s_2,s_3)$  
depending on three different
Mandelstam variables with $s_1 \neq s_2 \neq s_3$ and $s_1+s_2+s_3=Q^2$:
\eq
\alpha_\pm(s_1,s_2,s_3)  &=&
\frac{(Q^2-s_1)^2 - (Q^2s_2-s_1s_3)}{s (Q^2 \pm s_1)}\,,
\\
\beta_\pm(s_1,s_2,s_3)  &=&
\frac{(Q^2-s_1)^2 + (Q^2s_2-s_1s_3)}{s (Q^2 \pm s_1)}\,,
\\
\gamma_\pm(s_1,s_2,s_3) &=& \frac{(Q^2-s_1)^2 + s_1 (s_2-s_3)}
{s (Q^2 \pm s_1)}\,,
\\
\sigma_\pm(s_1,s_2,s_3) &=& \frac{(Q^2-s_1)^4 + s_1 (s_2-s_3) (Q^2s_3-s_1s_2)}
{s^2 (Q^2 \pm s_1)^2} \,.
\en
It is important to stress that the numerator of
functions $\alpha_\pm(s_1,s_2,s_3)$ is
manifestly symmetric under exchange of variables
$s_1 \leftrightarrow s_2$ and obeys the identities
\eq\label{Identity_s1s2s3}
& &(Q^2-s_1) (Q^2-s_2) - 2s_1s_2
\nonumber\\
&=& Q^2s_3 - s_1s_2
\nonumber\\
&=&(Q^2-s_1)^2 - Q^2s_2+s_1s_3
\nonumber\\
&=&(Q^2-s_2)^2 - Q^2s_1+s_2s_3 \,,
\en
Another interesting identity reads:
\eq
\gamma_\pm(s_1,s_2,s_3) \, \gamma_\pm(s_1,s_3,s_2)
= \frac{(s_2+s_3)^4 - s_1^2 (s_2-s_3)^2}{s^2 (Q^2 \pm s_1)^2} \,.
\en
In above formulas the set of variables $(s_1,s_2,s_3)$
is not fixed to a specific ordering of $(s,t,u)$ variables,
i.e.\,all possible 6 sets of arguments
due to permutations of $s_1$, $s_2$, and $s_3$ could occur.

Now we are in the position to list the scalar functions, which occur in
the expansion of vectors triangle integrals. In doing this we group the
scalar functions into pairs taking into account symmetric relations
between them. 
Most of the relations between the scalar functions are derived
using the change of the external momenta: $p_1 \to -p_2$,
$p_2 \to -p_1$,
$k_1 \to -k_1$, $q \to -q$ leading to the
following change of Mandelstam variables and
the momenta from orthogonal basis:
$s \to s\,, \quad u \to t\,, \quad t \to u\,, \quad
P \to - P\,, \quad R \to R\,, \quad T \to -T$. 
It concerns the pairs of the integrals
$(C^{015}_{1J},C^{026}_{1J})$,
$(C^{016}_{1J},C^{025}_{1J})$,
$(C^{018}_{1J},C^{024}_{1J})$,
$(C^{019}_{1J},C^{023}_{1J})$,
$(C^{035}_{1J},C^{069}_{1J})$,
$(C^{046}_{1J},C^{058}_{1J})$. 
The relations between the integrals
$C^{038}_{1J}$ and $C^{069}_{1K}$ are obtained 
by inverse of the loop momentum $k \to -k$.
The relations between the integrals
$C^{012}_{1J}$ and $C^{017}_{1J}$ are obtained 
by change of the loop momentum $k \to -(k+p_1)$,
while in the case of $C^{078}_{1J}$ and $C^{079}_{1J}$ integrals
we apply the following change of the loop momentum $k \to -(k+P)$.

We proceed using the following strategy:
\begin{itemize}
\item[1.] indicate the pairs of
loop integrals obeying specific symmetry relations;
\item[2.] display the sets of corresponding scalar functions and relations between them;
\item[3.] write down explicit results for the scalar functions.
\end{itemize}

\vspace*{0.25cm}
\noindent
\textbf{Loop integrals $C^{\mu;012}$ and $C^{\mu;017}$}

\noindent 
Sets of the scalar functions:  
\eq
& &C_{1P}^{012}(s)\,, C_{1R}^{017}(s)\,, C_{1T}^{012}(s) \,, \nonumber\\
& &C_{1P}^{017}(s)\,, C_{1R}^{017}(s)\,, C_{1T}^{017}(s) \,.
\en
\noindent 
Relations: 
\eq 
\left\{   
\begin{array}{lccl}
C_{1P}^{012}(s) &=& -& C_{1P}^{017}(s) - \frac{1}{2} C_0(s,0)\,, \nonumber\\
C_{1R}^{012}(s) &=& -& C_{1R}^{017}(s) - \frac{1}{2} C_0(s,0)\,, \nonumber\\
C_{1T}^{012}(s) &=& -& C_{1T}^{017}(s)\,.                        \nonumber
\end{array}
\right.
\en 
Explicit results:
\eq
C_{1P}^{012}(s) = C_{1T}^{012}(s) = 0 
\,, \quad C_{1R}^{012}(s) = C_1(s,0) = F(s,0) \,. 
\en 

\newpage  
\noindent 
\textbf{Loop integrals $C^{\mu;015}$ and $C^{\mu;026}$ }

\noindent 
Sets of the scalar functions: 
\eq
& &C_{1P}^{015}(s,u,Q^2)\,,
   C_{1R}^{015}(s,u,Q^2)\,,
   C_{1T}^{015}(u) \,, \nonumber\\
& &C_{1P}^{026}(s,t,Q^2)\,,
   C_{1R}^{026}(s,t,Q^2)\,,
   C_{1T}^{026}(t) \,.
\en
\noindent 
Relations:
\eq
\left\{
\begin{array}{lccl}
C_{1P}^{015}(s,u,Q^2) &=& -& C_{1P}^{026}(s,u,Q^2)\,, \nonumber\\[1mm]
C_{1R}^{015}(s,u,Q^2) &=& +& C_{1R}^{026}(s,u,Q^2)\,, \nonumber\\[1mm]
C_{1T}^{015}(u)       &=& -& C_{1T}^{026}(u)\,.     \nonumber
\end{array}
\right.
\en
Explicit results: 
\eq
C_{1P}^{015}(s,u,Q^2) &=&
\frac{C_1(0,u)}{2} +
\frac{C_1(u,0)}{2} \, \frac{Q^2-2s}{s}
= F(u,0) \,
\biggl[ \frac{1}{D-4} + \frac{Q^2}{2s} \biggr]
\,, \nonumber\\[1mm]
C_{1R}^{015}(s,u,Q^2) &=&
\frac{C_1(0,u)}{2} -
\frac{C_1(u,0)}{2} \, \frac{Q^2-2t}{s}
= C_{1P}^{015}(s,u,Q^2) - F(u,0) \, \frac{u}{s}
\,, \nonumber\\[1mm]
C_{1T}^{015}(u) &=& - C_1(u,0) = - F(u,0)
\,.
\en
\vspace*{.25cm}
\noindent 
\textbf{Loop integrals $C^{\mu;016}$ and $C^{\mu;025}$} 

\noindent 
Sets of the scalar functions: 
\eq
& &C_{1P}^{016}(s,t,Q^2)\,,
   C_{1R}^{016}(s,t,Q^2)\,,
   C_{1T}^{016}(t,Q^2)\,, \nonumber\\
& &C_{1P}^{025}(s,u,Q^2)\,,
   C_{1R}^{025}(s,u,Q^2)\,,
   C_{1T}^{025}(u,Q^2) \,.
\en
\noindent 
Relations: 
\eq 
\left\{   
\begin{array}{lccl}
C_{1P}^{016}(s,t,Q^2) &=& -& C_{1P}^{025}(s,t,Q^2)\,, \nonumber\\
C_{1R}^{016}(s,t,Q^2) &=& +& C_{1R}^{025}(s,t,Q^2)\,, \nonumber\\
C_{1T}^{016}(t,Q^2)   &=& -& C_{1T}^{025}(t,Q^2)\,. \nonumber
\end{array}
\right.
\en
\noindent 
Explicit results: 
\eq 
C_{1P}^{016}(s,t,Q^2) &=& \frac{C_1(0,t,Q^2)}{2} -
\frac{C_1(t,0,Q^2)}{2} \, \gamma_-(t,u,s) \nonumber\\[1mm]
&=& 
- F(Q^2,t) \, \biggl[ \frac{t}{(Q^2-t) (D-4)} 
+ \frac{Q^2}{2 s} \biggr] 
+ F(Q^2,0) \, \frac{Q^2}{2 (Q^2-t)} 
\,, \nonumber\\[2mm]
C_{1R}^{016}(s,t,Q^2) &=& \frac{C_1(0,t,Q^2)}{2} +
\frac{C_1(t,0,Q^2)}{2} \, \gamma_-(t,s,u) \nonumber\\[1mm]
&=& 
C_{1P}^{016}(s,t,Q^2) + F(Q^2,t) \, \frac{Q^2-t}{s}
\,, \nonumber\\[2mm]
C_{1T}^{016}(t,Q^2) &=& C_1(t,0,Q^2) = F(Q^2,t) 
\,.
\en  
\vspace*{.25cm}
\noindent 
\textbf{Loop integrals $C^{\mu;018}$ and $C^{\mu;024}$}

\noindent 
Sets of the scalar functions: 
\eq
& &C_{1P}^{018}(s,t,Q^2)\,,
   C_{1R}^{018}(s,t,Q^2)\,,
   C_{1T}^{018}(t,Q^2)\,, \nonumber\\
& &C_{1P}^{024}(s,u,Q^2)\,,
   C_{1R}^{024}(s,u,Q^2)\,,
   C_{1T}^{024}(u,Q^2) \,.
\en
\noindent 
Relations:
\eq 
\left\{   
\begin{array}{lccl}
C_{1P}^{018}(s,t,Q^2) &=& -& C_{1P}^{024}(s,t,Q^2)\,, \nonumber\\
C_{1R}^{018}(s,t,Q^2) &=& +& C_{1R}^{024}(s,t,Q^2)\,, \nonumber\\
C_{1T}^{018}(t,Q^2) &=& -& C_{1T}^{024}(t,Q^2)\,. \nonumber
\end{array}
\right.
\en 
Explicit results: 
\eq 
C_{1P}^{018}(s,t,Q^2) &=& 
  \frac{C_1(0,Q^2,t)}{2} 
+ \frac{C_1(Q^2,0,t)}{2} \, 
\alpha_-(t,s,u) \nonumber\\[2mm]
&=& 
F(Q^2,t) \, 
  \biggl[ \frac{Q^2}{(Q^2-t) \, (D-4)} 
+ \frac{Q^2+2s}{2 s} \biggr]
- F(Q^2,0) \, \frac{Q^2}{2 (Q^2-t)} 
\,, \nonumber\\[2mm]
C_{1R}^{018}(s,t,Q^2) &=&
  \frac{C_1(0,Q^2,t)}{2} 
- \frac{C_1(Q^2,0,t)}{2} \, 
  \beta_-(t,s,u)  \nonumber\\[2mm]
&=& 
C_{1P}^{018}(s,t,Q^2) - F(Q^2,t) \, \frac{Q^2-t}{s} 
\,, \nonumber\\[2mm]
C_{1T}^{018}(t,Q^2) &=& - C_1(Q^2,0,t) = - F(Q^2,t) 
\,. 
\en  
\vspace*{.25cm}
\noindent  
\textbf{Loop integrals $C^{\mu;019}$ and $C^{\mu;023}$}

\noindent 
Sets of the scalar functions: 
\eq
& &C_{1P}^{019}(s,u,Q^2)\,,
   C_{1R}^{019}(s,u,Q^2)\,,
   C_{1T}^{019}(u) \,, \nonumber\\
& &C_{1P}^{023}(s,t,Q^2)\,,
   C_{1R}^{023}(s,t,Q^2)\,,
   C_{1T}^{023}(t) \,.
\en
\noindent 
Relations:
\eq 
\left\{   
\begin{array}{lccl}
C_{1P}^{019}(s,u,Q^2) &=& -& C_{1P}^{023}(s,u,Q^2)\,, \nonumber\\
C_{1R}^{019}(s,u,Q^2) &=& +& C_{1R}^{023}(s,u,Q^2)\,, \nonumber\\
C_{1T}^{019}(u)       &=& -& C_{1T}^{023}(u)      \,. \nonumber
\end{array}
\right.
\en 
Explicit results: 
\eq 
C_{1P}^{019}(s,u,Q^2) &=&
- \frac{C_1(u,0)}{2 s} \, (Q^2 - 2s) = 
- F(u,0) \, \frac{Q^2-2s}{2 s} 
\,, \nonumber\\[1mm]
C_{1R}^{019}(s,u,Q^2) &=&
- \frac{C_1(u,0)}{2 s} \, (Q^2 - 2t) = 
- F(u,0) \, \frac{Q^2-2t}{2 s}  
\,, \nonumber\\[1mm]
C_{1T}^{019}(u) &=& C_1(u,0) = F(u,0) 
\,.
\en  
\vspace*{.25cm}
\noindent 
\textbf{Loop integrals $C^{\mu;035}$ and $C^{\mu;069}$}

\noindent 
Sets of the scalar functions:
\eq
& &C_{1P}^{035}(s,u,Q^2)\,,
   C_{1R}^{035}(s,u,Q^2)\,,
   C_{1T}^{035}(u) \,, \nonumber\\
& &C_{1P}^{069}(s,t,Q^2)\,,
   C_{1R}^{069}(s,t,Q^2)\,,
   C_{1T}^{069}(t) \,.
\en
\noindent 
Relations:
\eq 
\left\{   
\begin{array}{lccl}
C_{1P}^{035}(s,u,Q^2) &=& -& C_{1P}^{069}(s,u,Q^2)\,, \nonumber\\
C_{1R}^{035}(s,u,Q^2) &=& +& C_{1R}^{069}(s,u,Q^2)\,, \nonumber\\
C_{1T}^{035}(u)       &=& -& C_{1T}^{069}(u)    \,. \nonumber
\end{array}
\right.
\en 
Explicit results: 
\eq 
C_{1P}^{035}(s,u,Q^2) &=& \frac{C_1(0,u)}{2} \, \frac{Q^2-s}{s}
- \frac{C_1(u,0)}{2} \, \frac{Q^2-2s}{s} 
= F(u,0) \, \biggl[ \frac{Q^2-s}{s \, (D-4)} + \frac{Q^2}{2s} \biggr]
\,, \nonumber\\[1mm]
C_{1R}^{035}(s,u,Q^2) &=& \frac{C_1(0,u)}{2} \, \frac{t-u}{s}
+ \frac{C_1(u,0)}{2} \, \frac{Q^2-2t}{s} 
= C_{1P}^{035}(s,u,Q^2) -  \frac{F(u,0) \, u}{s} \, \frac{D-2}{D-4}  
\,, \nonumber\\[1mm]
C_{1T}^{035}(u) &=& - C_1(0,u) + C_1(u,0) = - F(u,0) \, \frac{D-2}{D-4} 
\,. 
\en 
\vspace*{.25cm}
\noindent 
\textbf{Loop integrals $C^{\mu;038}$ and $C^{\mu;049}$}

\noindent 
Sets of the scalar functions:
\eq
& &C_{1P}^{038}(s,Q^2)\,,
   C_{1R}^{038}(s,u,Q^2)\,,
   C_{1T}^{038}(s,Q^2) \,, \nonumber\\
& &C_{1P}^{049}(s,Q^2)\,,
   C_{1R}^{049}(s,u,Q^2)\,,
   C_{1T}^{049}(s,Q^2) \,.
\en
\noindent 
Relations:
\eq 
\left\{   
\begin{array}{lccl}
C_{1P}^{038}(s,Q^2)   &=& -& C_{1P}^{049}(s,Q^2)\,, \nonumber\\
C_{1R}^{038}(s,u,Q^2) &=& -& C_{1R}^{049}(s,u,Q^2)\,, \nonumber\\
C_{1T}^{038}(s,Q^2)   &=& -& C_{1T}^{049}(s,Q^2)\,. \nonumber
\end{array}
\right.
\en 
Explicit results: 
\eq 
C_{1P}^{038}(s,Q^2) &=& 
\frac{Q^2-s}{2s} \, \biggl[ C_1(0,Q^2,s) - C_1(Q^2,0,s) \biggr] 
\nonumber\\[1mm]
&=& \frac{Q^2}{s} \, \biggl[
        F(Q^2,s) \, \frac{D-3}{D-4} 
- \frac{F(Q^2,0)}{2}  \biggr] 
\,, \nonumber\\[1mm]
C_{1R}^{038}(s,u,Q^2) &=& 
\frac{u-t}{2s} \,  C_{1T}^{038}(s,Q^2) 
\,, \nonumber\\[1mm]
C_{1T}^{038}(s,Q^2) &=& 
- \biggl[C_1(0,Q^2,s) - 
         C_1(Q^2,0,s) \, \frac{Q^2+s}{Q^2-s} \biggr] 
\nonumber\\[1mm]
&=& - \frac{2s}{Q^2-s} \, 
\biggl[ C_{1P}^{038}(s,Q^2) - F(Q^2,s) \biggr]
\,. 
\en  
\vspace*{.25cm}
\noindent 
\textbf{Loop integrals $C^{\mu;046}$ and $C^{\mu;058}$} 

\noindent 
Sets of the scalar functions: 
\eq
& &C_{1P}^{046}(s,t,Q^2)\,,
   C_{1R}^{046}(s,t,Q^2)\,,
   C_{1T}^{046}(t,Q^2)\,, \nonumber\\
& &C_{1P}^{058}(s,u,Q^2)\,,
   C_{1R}^{058}(s,u,Q^2)\,,
   C_{1T}^{058}(u,Q^2) \,.
\en 
\noindent 
Relations:
\eq 
\left\{   
\begin{array}{lccl}
C_{1P}^{046}(s,t,Q^2) &=& -& C_{1P}^{058}(s,t,Q^2)\,, \nonumber\\
C_{1R}^{046}(s,t,Q^2) &=& +& C_{1R}^{058}(s,t,Q^2)\,, \nonumber\\
C_{1T}^{046}(t,Q^2) &=& -& C_{1T}^{058}(t,Q^2)\,. \nonumber
\end{array}
\right.
\en 
Explicit results:
\eq 
C_{1P}^{046}(s,t,Q^2) &=& \frac{C_1(Q^2,t,0)}{2} \, \alpha_+(t,s,u)
                      - \frac{C_1(t,Q^2,0)}{2} \, \gamma_+(t,u,s) 
\nonumber\\[1mm]
&=& 
- F(Q^2,t) \, \biggl[ \frac{(Q^2 - t)^2 + u t}{(Q^2-t) s \, (D-4)}
+ \frac{Q^2}{2 s} \biggr]
- F(Q^2,0) \, \frac{Q^2}{2 (Q^2-t)} 
\,, \nonumber\\[1mm]
C_{1R}^{046}(s,t,Q^2) &=& - \frac{C_1(Q^2,t,0)}{2} \,  \beta_+(t,s,u) 
                        + \frac{C_1(t,Q^2,0)}{2} \, \gamma_+(t,s,u) 
\nonumber\\[1mm]
&=&
C_{1P}^{046}(s,t,Q^2) + F(Q^2,t) \, \frac{Q^2-t}{s} \, \frac{D-2}{D-4} 
\,, \nonumber\\[1mm]
C_{1T}^{046}(t,Q^2) &=& - \frac{Q^2-t}{Q^2+t} \, 
\biggl[ C_1(Q^2,t,0) - C_1(t,Q^2,0) \biggr] 
\nonumber\\[1mm]
&=& F(Q^2,t) \, \frac{D-2}{D-4} 
\,. 
\en  
\hspace*{0.25cm}
\noindent 
\textbf{Loop integrals $C^{\mu;078}$ and $C^{\mu;079}$} 

\noindent 
Sets of the scalar functions: 
\eq
& &C_{1P}^{078}(s,Q^2)\,,
   C_{1R}^{046}(s,u,Q^2)\,,
   C_{1T}^{046}(s,Q^2)\,, \nonumber\\
& &C_{1P}^{079}(s,Q^2)\,,
   C_{1R}^{069}(s,t,Q^2)\,,
   C_{1T}^{058}(s,Q^2) \,.
\en
\noindent 
Relations:
\eq 
\left\{   
\begin{array}{lccl}
C_{1P}^{078}(s,Q^2)   &=& -& C_{1P}^{079}(s,Q^2) - C_0(Q^2,s)\,, \nonumber\\
C_{1R}^{078}(s,u,Q^2) &=& -& C_{1R}^{079}(s,u,Q^2)\,, \nonumber\\
C_{1T}^{078}(s,Q^2)    &=& -& C_{1T}^{079}(s,Q^2)\,. \nonumber
\end{array}
\right.
\en 
Explicit results:
\eq 
C_{1P}^{078}(s,Q^2) &=& 
\frac{C_1(Q^2,s,0)}{2} \, 
\frac{(Q^2-s)^2}{s (Q^2+s)}  =  
  F(Q^2,s) \, \frac{D-3}{D-4} 
+ F(Q^2,0) \, \frac{Q^2}{2 s} 
\,, \nonumber\\[1mm]
C_{1R}^{078}(s,u,Q^2) &=& 
\frac{u-t}{2s} \,  C_{1T}^{078}(s,Q^2) 
\,, \nonumber\\[1mm]
C_{1T}^{078}(s,Q^2) &=& C_1(s,Q^2,0) \, \frac{2s}{Q^2+s}
- C_1(Q^2,s,0) \nonumber\\[1mm]
&=& \frac{2s}{Q^2-s} \, \biggl[ 
\frac{F(Q^2,s)}{D-4} - 
F(Q^2,0) \, \frac{Q^2}{2 s} \biggr] 
\,.
\en  

\subsection{Tensor triangle integrals}

In this section we show results for the triangle tensor rank-2  integrals: 
\eq
C^{\mu\nu;0ij}_2 &=& (i \pi^2)^{-1} \, \mu^{4-D} \,
\int d^Dk \, \frac{k^\mu}{\Delta_0 \, \Delta_i \, \Delta_j}
= g^{\mu\nu;3}_\perp \, C_{2g}^{0ij}
+ P^\mu P^\nu \, C_{2PP}^{0ij}
+ R^\mu R^\nu \, C_{2RR}^{0ij}
\nonumber\\
&+& T^\mu T^\nu \, C_{2TT}^{0ij} 
  + \{P R\}^{\mu\nu} \, C_{2PR}^{0ij}
  + \{P T\}^{\mu\nu} \, C_{2PT}^{0ij}
  + \{R T\}^{\mu\nu} \, C_{2RT}^{0ij} \,,
\en
where $\{ A B\}^{\mu\nu} = A^\mu B^\nu + A^\nu B^\mu$.

All scalar functions are expressed in terms of three scalar functions 
derived in Eq.~(\ref{C2_functions}): 
\eq\label{C2_functions_app} 
C_{21}(s_1,s_2,s_3) &=& 
C_{2;g}(s_1,s_2,s_3) 
\nonumber\\
&=& 
\frac{1}{D-2} \, \biggl[ 
B_0(s_3) + 
\lambda \, \frac{s_1 \, C_{22}(s_1,s_2,s_3) + s_2 \, C_{22}(s_2,s_1,s_3)} 
{(s_1+s_2-s_3)^2} 
\nonumber\\
&-& \frac{\lambda \, C_{23}(s_1,s_2,s_3)}{s_1+s_2-s_3} 
\biggr] \,, 
\nonumber\\
C_{22}(s_1,s_2,s_3) &=& 
C_{2;l_1l_1}(s_1,s_2,s_3) = 
C_{2;l_2l_2}(s_2,s_1,s_3) \nonumber\\
&=& 
\frac{(s_1+s_2-s_3)^2}{2 \lambda^2} \, 
\biggl[ 
- (s_1+s_2-s_3)  \, B_0(s_1) 
+ (s_1+3s_2-s_3) \, B_0(s_3) 
\nonumber\\
&-& \frac{2s_2 \, \lambda}{s_1+s_2-s_3} \, 
C_1(s_1,s_2,s_3) 
\biggr] \,,\nonumber\\
C_{23}(s_1,s_2,s_3) &=& 
C_{2;l_1l_2}(s_1,s_2,s_3) 
\nonumber\\
&=& \frac{(s_1+s_2-s_3)^2}{2 \lambda^2} \, 
\biggl[ 
- s_1  \, B_0(s_1) - s_2  \, B_0(s_2)   
+ (2s_1 + 2s_2 - s_3) \, B_0(s_3) \nonumber\\
&-& \frac{\lambda}{s_1+s_2-s_3} 
\, \Big( 
  s_1 C_1(s_1,s_2,s_3) 
+ s_2 C_1(s_2,s_1,s_3) \Big) 
\biggr] \,. 
\en 
Now we display triangle tensor functions. 

\vspace*{.25cm}
\noindent 
\textbf{Loop integrals $C^{\mu\nu;012}$ and $C^{\mu\nu;017}$} 

\noindent 
Sets of the scalar functions: 
\eq
& &C_{2g}^{012}(s)\,,
   C_{2PP}^{012}(s)\,,
   C_{2RR}^{012}(s)\,,
   C_{2TT}^{012}(s)\,,
   C_{2PR}^{012}(s)\,,
   C_{2PT}^{012}(s)\,,
   C_{2RT}^{012}(s)\,,
\nonumber\\
& &C_{2g}^{017}(s)\,,
   C_{2PP}^{017}(s)\,,
   C_{2RR}^{017}(s)\,,
   C_{2TT}^{017}(s)\,,
   C_{2PR}^{017}(s)\,,
   C_{2PT}^{017}(s)\,,
   C_{2RT}^{017}(s)\,.
\en
\noindent 
Relations:
\eq 
\left\{   
\begin{array}{lccl}
 C_{2g}^{012}(s) &=& +&  C_{2g}^{017}(s)\,, \nonumber\\
C_{2PP}^{012}(s) &=& +& C_{2PP}^{017}(s) - \frac{1}{4} C_0(s,0)\,, \nonumber\\
C_{2RR}^{012}(s) &=& +& C_{2RR}^{017}(s) - F(s,0) - \frac{1}{4} C_0(s,0)\,, \nonumber\\
C_{2TT}^{012}(s) &=& +& C_{2TT}^{017}(s)\,, \nonumber\\
C_{2PR}^{012}(s) &=& +& C_{2PR}^{017}(s) 
- \frac{1}{2} F(s,0) - \frac{1}{4} C_0(s,0)\,, \nonumber\\
C_{2PT}^{012}(s) &=& +& C_{2PT}^{017}(s)\,, \nonumber\\
C_{2RT}^{012}(s) &=& +& C_{2RT}^{017}(s)\,. \nonumber
\end{array}
\right.
\en
Explicit results: 
\eq 
C_{2g}^{012}(s)  &=& C_{21}(0,0,s) 
=  \frac{F(s,0)+G(s,0)}{D-2} \, \frac{s}{4} 
\,, \nonumber\\[1mm]
C_{2PP}^{012}(s) &=& \frac{C_{22}(0,0,s)-C_{23}(0,0,s)}{2} 
= \frac{G(s,0)-F(s,0)}{8}
\,, \nonumber\\[1mm]
C_{2RR}^{012}(s) &=& \frac{C_{22}(0,0,s)+C_{23}(0,0,s)}{2} 
= - \frac{G(s,0)+3F(s,0)}{8}
\,, \nonumber\\[1mm]
C_{2TT}^{012}(s) &=& - \frac{s}{u t} \, C_{2g}^{012}(s) 
\,, \nonumber\\[1mm]
C_{2PR}^{012}(s) &=& C_{2PT}^{012}(s) \,=\, C_{2RT}^{012}(s) \,=\, 0 
\,. 
\en 
\vspace*{.25cm}
\noindent 
\textbf{Loop integrals $C^{\mu\nu;015}$ and $C^{\mu\nu;026}$} 

\noindent 
Sets of the scalar functions: 
\eq
& & C_{2g}^{015}(s,u,Q^2)\,,
   C_{2PP}^{015}(s,u,Q^2)\,,
   C_{2RR}^{015}(s,u,Q^2)\,,
   C_{2TT}^{015}(s,u,Q^2)\,,
\nonumber\\
& &C_{2PR}^{015}(s,u,Q^2)\,,
   C_{2PT}^{015}(s,u,Q^2)\,,
   C_{2RT}^{015}(s,u,Q^2)\,,
\nonumber\\
& &\\
& & C_{2g}^{026}(s,t,Q^2)\,,
   C_{2PP}^{026}(s,t,Q^2)\,,
   C_{2RR}^{026}(s,t,Q^2)\,,
   C_{2TT}^{026}(s,t,Q^2)\,,
\nonumber\\
& &C_{2PR}^{026}(s,t,Q^2)\,,
   C_{2PT}^{026}(s,t,Q^2)\,,
   C_{2RT}^{026}(s,t,Q^2)\,.
\nonumber
\en
\noindent 
Relations:
\eq 
\left\{   
\begin{array}{lccl}
 C_{2g}^{015}(s,u,Q^2) &=& +&  C_{2g}^{026}(s,u,Q^2)\,, \nonumber\\
C_{2PP}^{015}(s,u,Q^2) &=& +& C_{2PP}^{026}(s,u,Q^2)\,, \nonumber\\
C_{2RR}^{015}(s,u,Q^2) &=& +& C_{2RR}^{026}(s,u,Q^2)\,, \nonumber\\
C_{2TT}^{015}(s,u,Q^2) &=& +& C_{2TT}^{026}(s,u,Q^2)\,, \nonumber\\
C_{2PR}^{015}(s,u,Q^2) &=& -& C_{2PR}^{026}(s,u,Q^2)\,, \nonumber\\
C_{2PT}^{015}(s,u,Q^2) &=& +& C_{2PT}^{026}(s,u,Q^2)\,, \nonumber\\
C_{2RT}^{015}(s,u,Q^2) &=& -& C_{2RT}^{026}(s,u,Q^2)\,. \nonumber
\end{array}
\right.
\en
Explicit results: 
\eq 
C_{2g}^{015}(s,u,Q^2) 
&=& C_{21}(0,u,0) 
= \frac{F(u,0)+G(u,0)}{D-2} \, \frac{u}{4} 
\,, \nonumber\\[1mm]
C_{2PP}^{015}(s,u,Q^2)  &=& 
- \frac{t}{s u}             \, C_{21}(0,u,0) 
+ \frac{C_{22}(0,u,0)}{4}               
+ \frac{(Q^2 - 2s)^2}{4s^2} \, C_{22}(u,0,0) 
\nonumber\\[1mm]
&+& \frac{Q^2-2s}{2s}         \, C_{23}(0,u,0) 
=  
- \frac{F(u,0)}{2} \, \biggl[ \frac{1}{D-4} + \frac{Q^4}{4 s^2} \biggr]
- \frac{F(u,0) + G(u,0)}{D-2} \, \frac{t}{4s}
\nonumber\\[1mm]
&+& \Big[G(u,0) - F(u,0)\Big] \,  \frac{Q^2 - s}{8 s} 
\,, \nonumber 
\en 
\eq 
C_{2RR}^{015}(s,u,Q^2)  &=& 
-  \frac{t}{s u} \, C_{21}(0,u,0) 
+ \frac{C_{22}(0,u,0)}{4}   
+ \frac{(Q^2 - 2t)^2}{4s^2} \, C_{22}(u,0,0) 
\nonumber\\[1mm]
&-& \frac{Q^2-2t}{2s}   \, C_{23}(0,u,0) 
\nonumber\\[1mm]
&=&  
C_{2PP}^{015}(s,u,Q^2) + \frac{u}{4 s} \, 
\, \biggl[ 2 F(u,0) \, \frac{Q^2-u}{s} + F(u,0) - G(u,0) \biggr] 
\,, \nonumber\\[1mm]
C_{2TT}^{015}(s,u,Q^2)  &=& 
-  \frac{s}{t u} \, C_{21}(0,u,0) 
                  + C_{22}(u,0,0)  
=  - \frac{F(u,0) + G(u,0)}{D-2} \, \frac{s}{4t}
     - \frac{F(u,0)}{2} 
\,, \nonumber\\[1mm]
C_{2PR}^{015}(s,u,Q^2)  &=& 
-  \frac{t}{s u} \, C_{21}(0,u,0) 
+ \frac{C_{22}(0,u,0)}{4}   
- \frac{(Q^2 - 2s) (Q^2 - 2t)}{4 s^2} \, C_{22}(u,0,0) 
\nonumber\\[1mm]
&+& \frac{t-s}{2 s} \, C_{23}(0,u,0)
= C_{2PP}^{015}(s,u,Q^2)
\nonumber\\[1mm]  
&+&  \frac{u}{8 s} \, 
\, \biggl[ 2 F(u,0) \, \frac{Q^2}{s} + F(u,0) - G(u,0) \biggr] 
\nonumber\\[1mm]
C_{2PT}^{015}(s,u,Q^2)  &=& 
  \frac{C_{21}(0,u,0)}{u} \, 
- \frac{Q^2 - 2s}{2s} \, C_{22}(u,0,0) 
- \frac{C_{23}(0,u,0)}{2}  
\nonumber\\[1mm]
&=&  
  \frac{F(u,0) + G(u,0)}{4 (D-2)} 
+ F(u,0) \, \frac{Q^2}{4 s} 
+ \frac{F(u,0) - G(u,0)}{8} 
\,, \nonumber\\[1mm]
C_{2RT}^{015}(s,u,Q^2)  &=& 
  \frac{C_{21}(0,u,0)}{u} \, 
+ \frac{Q^2 - 2t}{2s} \, C_{22}(u,0,0) 
- \frac{C_{23}(0,u,0)}{2}  
\nonumber\\[1mm]
&=& C_{2PT}^{015}(s,u,Q^2)  - F(u,0) \, \frac{u}{2s} 
\,. 
\en 
\vspace*{.25cm}
\noindent 
\textbf{Loop integrals $C^{\mu\nu;016}$ and $C^{\mu\nu;025}$} 

\noindent 
Sets of the scalar functions: 
\eq
& & C_{2g}^{016}(s,t,Q^2)\,,
   C_{2PP}^{016}(s,t,Q^2)\,,
   C_{2RR}^{016}(s,t,Q^2)\,,
   C_{2TT}^{016}(s,t,Q^2)\,,
\nonumber\\
& &C_{2PR}^{016}(s,t,Q^2)\,,
   C_{2PT}^{016}(s,t,Q^2)\,,
   C_{2RT}^{016}(s,t,Q^2)\,,
\nonumber\\
& &\\
& & C_{2g}^{025}(s,u,Q^2)\,,
   C_{2PP}^{025}(s,u,Q^2)\,,
   C_{2RR}^{025}(s,u,Q^2)\,,
   C_{2TT}^{025}(s,u,Q^2)\,,
\nonumber\\
& &C_{2PR}^{025}(s,u,Q^2)\,,
   C_{2PT}^{025}(s,u,Q^2)\,,
   C_{2RT}^{025}(s,u,Q^2)\,.
\nonumber
\en
\noindent 
Relations:
\eq 
\left\{   
\begin{array}{lccl}
 C_{2g}^{016}(s,t,Q^2) &=& +&  C_{2g}^{025}(s,t,Q^2)\,, \nonumber\\
C_{2PP}^{016}(s,t,Q^2) &=& +& C_{2PP}^{025}(s,t,Q^2)\,, \nonumber\\
C_{2RR}^{016}(s,t,Q^2) &=& +& C_{2RR}^{025}(s,t,Q^2)\,, \nonumber\\
C_{2TT}^{016}(s,t,Q^2) &=& +& C_{2TT}^{025}(s,t,Q^2)\,, \nonumber\\
C_{2PR}^{016}(s,t,Q^2) &=& -& C_{2PR}^{025}(s,t,Q^2)\,, \nonumber\\
C_{2PT}^{016}(s,t,Q^2) &=& +& C_{2PT}^{025}(s,t,Q^2)\,, \nonumber\\
C_{2RT}^{016}(s,t,Q^2) &=& -& C_{2RT}^{025}(s,t,Q^2)\,. \nonumber
\end{array}
\right.
\en
Explicit results: 
\eq 
C_{2g}^{016}(s,t,Q^2) &=& C_{21}(0,t,Q^2) = 
 \frac{H(Q^2,t)}{D-2} \, \frac{Q^2-t}{4} 
\,, \nonumber
\en 
\eq 
C_{2PP}^{016}(s,t,Q^2) &=& 
- \frac{ut \, C_{21}(0,t,Q^2)}{s (Q^2-t)^2} 
+ \frac{C_{22}(0,t,Q^2)}{4} 
+ \frac{\gamma_-(t,u,s)^2}{4} \, C_{22}(t,0,Q^2)  
\nonumber\\[1mm]
&-& \frac{\gamma_-(t,u,s)}{2} \, C_{23}(0,t,Q^2) 
\nonumber\\[1mm]
&=& 
- \frac{F(Q^2,t)}{2} \, \biggl[ \frac{t^2}{(Q^2-t)^2} \frac{1}{D-4} 
+ \frac{Q^4}{4 s^2} \biggr]  
- \frac{H(Q^2,t)}{D-2} \, 
\frac{u t}{4s (Q^2-t)} 
\nonumber\\[1mm]
&+& \frac{Q^2}{8 s} \, 
 \biggl[G(Q^2,0) \frac{(Q^2-t)^2+ut}{(Q^2-t)^2} 
+ F(Q^2,0) \frac{(Q^2-s)(Q^2-t)+st}{(Q^2-t)^2} \biggr] 
\,, \nonumber\\
C_{2RR}^{016}(s,t,Q^2) &=& 
- \frac{ut \, C_{21}(0,t,Q^2)}{s (Q^2-t)^2} 
+ \frac{C_{22}(0,t,Q^2)}{4} 
+ \frac{\gamma_-(t,s,u)^2}{4} \, C_{22}(t,0,Q^2)  
\nonumber\\[1mm]
&+& \frac{\gamma_-(t,s,u)}{2} \, C_{23}(0,t,Q^2) 
\nonumber\\[1mm]
&=& C_{2PP}^{016}(s,t,Q^2) - \biggl[ F(Q^2,t) (s-2t) + G(Q^2,t) s \biggr] 
\, \frac{Q^2-t}{4 s^2}
\,, \nonumber\\[1mm]
C_{2TT}^{016}(s,t,Q^2) &=& 
- \frac{s}{u t} \, C_{21}(0,t,Q^2) 
+ \, C_{22}(t,0,Q^2) =
- \frac{F(Q^2,t)}{2} 
- \frac{H(Q^2,t)}{D-2} \, \frac{s (Q^2-t)}{4 u t} 
\,, \nonumber\\
C_{2PR}^{016}(s,t,Q^2) &=& 
-  \frac{u t}{s (Q^2-t)^2} \, C_{21}(0,t,Q^2) 
+ \frac{C_{22}(0,t,Q^2)}{4} 
\nonumber\\[1mm]
&-& \frac{\gamma_-(t,s,u) \, \gamma_-(t,u,s)}{4} 
\, C_{22}(t,0,Q^2) + \frac{t (s-u)}{2 s (Q^2-t)} \, C_{23}(0,t,Q^2) 
\nonumber\\[1mm]
&=& C_{2PP}^{016}(s,t,Q^2) + \biggl[ F(Q^2,t) (2Q^2-s) - G(Q^2,t) s \biggr] 
\, \frac{Q^2-t}{8 s^2}
\,, \nonumber\\
C_{2PT}^{016}(s,t,Q^2) &=& 
  \frac{C_{21}(0,t,Q^2)}{Q^2-t} \, 
- \frac{\gamma_-(t,u,s)}{2} \, C_{22}(t,0,Q^2) 
+ \frac{C_{23}(0,t,Q^2)}{2}  \nonumber\\[1mm]
&=& 
\frac{H(Q^2,t)}{4 (D-2)} 
+ F(Q^2,t) \biggl[ \frac{Q^2}{4s} - \frac{1}{8} \biggr] 
- \frac{G(Q^2,t)}{8} 
\,, \nonumber\\[1mm]
C_{2RT}^{016}(s,t,Q^2) &=& 
  \frac{C_{21}(0,t,Q^2)}{Q^2-t} \, 
+ \frac{\gamma_-(t,s,u)}{2} \, C_{22}(t,0,Q^2) 
+ \frac{C_{23}(0,t,Q^2)}{2} 
\nonumber\\[1mm]
&=& C_{2PT}^{016}(s,t,Q^2) - F(Q^2,t) \, \frac{Q^2-t}{2s}
\,. 
\en

\vspace*{.25cm}
\noindent 
\textbf{Loop integrals $C^{\mu\nu;018}$ and $C^{\mu\nu;024}$} 

\noindent 
Sets of the scalar functions: 
\eq
& & C_{2g}^{018}(s,t,Q^2)\,,
   C_{2PP}^{018}(s,t,Q^2)\,,
   C_{2RR}^{018}(s,t,Q^2)\,,
   C_{2TT}^{018}(s,t,Q^2)\,,
\nonumber\\
& &C_{2PR}^{018}(s,t,Q^2)\,,
   C_{2PT}^{018}(s,t,Q^2)\,,
   C_{2RT}^{018}(s,t,Q^2)\,,
\nonumber\\
& &\\
& & C_{2g}^{024}(s,u,Q^2)\,,
   C_{2PP}^{024}(s,u,Q^2)\,,
   C_{2RR}^{024}(s,u,Q^2)\,,
   C_{2TT}^{024}(s,u,Q^2)\,,
\nonumber\\
& &C_{2PR}^{024}(s,u,Q^2)\,,
   C_{2PT}^{024}(s,u,Q^2)\,,
   C_{2RT}^{024}(s,u,Q^2)\,.
\nonumber
\en
\noindent 
Relations:
\eq 
\left\{   
\begin{array}{lccl}
 C_{2g}^{018}(s,t,Q^2) &=& +&  C_{2g}^{024}(s,t,Q^2)\,, \nonumber\\
C_{2PP}^{018}(s,t,Q^2) &=& +& C_{2PP}^{024}(s,t,Q^2)\,, \nonumber\\
C_{2RR}^{018}(s,t,Q^2) &=& +& C_{2RR}^{024}(s,t,Q^2)\,, \nonumber\\
C_{2TT}^{018}(s,t,Q^2) &=& +& C_{2TT}^{024}(s,t,Q^2)\,, \nonumber\\
C_{2PR}^{018}(s,t,Q^2) &=& -& C_{2PR}^{024}(s,t,Q^2)\,, \nonumber\\
C_{2PT}^{018}(s,t,Q^2) &=& +& C_{2PT}^{024}(s,t,Q^2)\,, \nonumber\\
C_{2RT}^{018}(s,t,Q^2) &=& -& C_{2RT}^{024}(s,t,Q^2)\,. \nonumber
\end{array}
\right.
\en 
Explicit results: 
\eq 
C_{2g}^{018}(s,t,Q^2) &=& C_{21}(0,Q^2,t) 
= 
\frac{H(Q^2,t)}{D-2} \, \frac{Q^2-t}{4} 
\,, \nonumber\\
C_{2PP}^{018}(s,t,Q^2) &=& 
- \frac{u t}{s (Q^2-t)^2} \, C_{21}(0,Q^2,t)
+ \frac{C_{22}(0,Q^2,t)}{4} 
+ \frac{\alpha^2_-(t,s,u)}{4} \, C_{22}(Q^2,0,t)
\nonumber\\ 
&+&\frac{\alpha_-(t,s,u)}{2} \, C_{23}(0,Q^2,t)
= - \frac{F(Q^2,t)}{2} \, \biggl[ 
\frac{Q^4}{(Q^2-t)^2 (D-4)} + \frac{(Q^2+2s)^2}{4s^2} \biggr] 
\nonumber\\
&-& \frac{H(Q^2,t)}{D-2} \, \frac{u t}{4 s (Q^2-t)} 
 +  F(Q^2,0) \, \frac{Q^2 (3Q^2-t)}{8 (Q^2-t)^2} 
\nonumber\\
&+& \frac{F(Q^2,0) + G(Q^2,0)}{8} \, \frac{Q^2 ((Q^2-t)^2+ut)}{s (Q^2-t)^2} 
\,, \nonumber\\
C_{2RR}^{018}(s,t,Q^2) &=& 
- \frac{u t}{s (Q^2-t)^2} \, C_{21}(0,Q^2,t)
+ \frac{C_{22}(0,Q^2,t)}{4} 
+ \frac{\beta^2_-(t,s,u)}{4} \, C_{22}(Q^2,0,t)
\nonumber\\
&-& \frac{\beta_-(t,s,u)}{2} \, C_{23}(0,Q^2,t)
\nonumber\\
&=& C_{2PP}^{018}(s,t,Q^2) + 
\biggl[ F(Q^2,t) \Big(2 (Q^2-u)+s\Big) - G(Q^2,t) s \biggr] \, 
\frac{Q^2-t}{4 s^2}
\,, 
\nonumber\\
C_{2TT}^{018}(s,t,Q^2) &=& 
- \frac{s}{u t} \, C_{21}(0,Q^2,t) 
+ C_{22}(Q^2,0,t) 
= - \frac{H(Q^2,t)}{D-2} \, \frac{s (Q^2- t)}{4 u t} 
- \frac{F(Q^2,t)}{2} 
\,, \nonumber\\
C_{2PR}^{018}(s,t,Q^2)  &=& 
- \frac{u t}{s (Q^2-t)^2} \, C_{21}(0,Q^2,t)
+ \frac{C_{22}(0,Q^2,t)}{4} 
\nonumber\\
&-& \frac{\alpha_-(t,s,u) \, \beta_-(t,s,u)}{4} 
\, C_{22}(Q^2,0,t)
- \frac{\alpha_-(t,u,s)}{2} \, C_{23}(0,Q^2,t)
\nonumber\\
&=& C_{2PP}^{018}(s,t,Q^2) + 
\biggl[ F(Q^2,t) (2Q^2+3s) - G(Q^2,t) s \biggr] \, 
\frac{Q^2-t}{8 s^2} \,,
\nonumber\\
C_{2PT}^{018}(s,t,Q^2)  &=& \frac{C_{21}(0,Q^2,t)}{Q^2-t} 
- \frac{\alpha_-(t,s,u)}{2} C_{22}(Q^2,0,t)
- \frac{C_{23}(0,Q^2,t)}{2} 
\nonumber\\
&=& 
  \frac{H(Q^2,t)}{4 (D-2)} 
+ \frac{F (Q^2,t) (2Q^2 + 3s) - G(Q^2,t) s}{8s} 
\,,\nonumber\\
C_{2RT}^{018}(s,t,Q^2)  &=& \frac{C_{21}(0,Q^2,t)}{Q^2-t} 
+ \frac{\beta_-(t,s,u)}{2} C_{22}(Q^2,0,t)
- \frac{C_{23}(0,Q^2,t)}{2}
\nonumber\\
&=& 
C_{2PT}^{018}(s,t,Q^2)  - F(Q^2,t) \, \frac{Q^2-t}{2s} 
\,.  
\en 

\vspace*{.25cm}
\noindent 
\textbf{Loop integrals $C^{\mu\nu;019}$ and $C^{\mu\nu;023}$} 

\noindent 
Sets of the scalar functions:
\eq
& & C_{2g}^{019}(s,u,Q^2)\,,
   C_{2PP}^{019}(s,u,Q^2)\,,
   C_{2RR}^{019}(s,u,Q^2)\,,
   C_{2TT}^{019}(s,u,Q^2)\,,
\nonumber\\
& &C_{2PR}^{019}(s,u,Q^2)\,,
   C_{2PT}^{019}(s,u,Q^2)\,,
   C_{2RT}^{019}(s,u,Q^2)\,,
\nonumber\\
& &\\
& & C_{2g}^{023}(s,t,Q^2)\,,
   C_{2PP}^{023}(s,t,Q^2)\,,
   C_{2RR}^{023}(s,t,Q^2)\,,
   C_{2TT}^{023}(s,t,Q^2)\,,
\nonumber\\
& &C_{2PR}^{023}(s,t,Q^2)\,,
   C_{2PT}^{023}(s,t,Q^2)\,,
   C_{2RT}^{023}(s,t,Q^2)\,.
\nonumber
\en
\noindent 
Relations:
\eq 
\left\{   
\begin{array}{lccl}
 C_{2g}^{019}(s,u,Q^2) &=& +&  C_{2g}^{023}(s,u,Q^2)\,, \nonumber\\
C_{2PP}^{019}(s,u,Q^2) &=& +& C_{2PP}^{023}(s,u,Q^2)\,, \nonumber\\
C_{2RR}^{019}(s,u,Q^2) &=& +& C_{2RR}^{023}(s,u,Q^2)\,, \nonumber\\
C_{2TT}^{019}(s,u,Q^2) &=& +& C_{2TT}^{023}(s,u,Q^2)\,, \nonumber\\
C_{2PR}^{019}(s,u,Q^2) &=& -& C_{2PR}^{023}(s,u,Q^2)\,, \nonumber\\
C_{2PT}^{019}(s,u,Q^2) &=& +& C_{2PT}^{023}(s,u,Q^2)\,, \nonumber\\
C_{2RT}^{019}(s,u,Q^2) &=& -& C_{2RT}^{023}(s,u,Q^2)\,. \nonumber
\end{array}
\right.
\en 
\noindent 
Explicit results: 
\eq 
C_{2g}^{019}(s,u,Q^2) &=& C_{21}(0,0,u) = 
\frac{F(u,0)+G_0(u,0)}{D-2} \, \frac{u}{4} 
\,, \nonumber\\
C_{2PP}^{019}(s,u,Q^2) &=& 
- \frac{t}{s u} \, C_{21}(0,0,u) 
+ \frac{s^2 + (Q^2-s)^2}{4 s^2} \, C_{22}(0,0,u) 
- \frac{Q^2-s}{2 s} \, C_{23}(0,0,u)  \nonumber\\
&=& 
- \frac{F(u,0)+G(u,0)}{D-2} \, \frac{t}{4s} 
- F(u,0) \, \frac{(Q^2-2s)^2}{8s^2} 
\nonumber\\
&+& \biggl[G(u,0) - F(u,0)\biggr] \, \frac{Q^2-s}{8s} \,,  
\nonumber\\
C_{2RR}^{019}(s,u,Q^2) &=& 
- \frac{t}{s u} \, C_{21}(0,0,u) 
+ \frac{s^2 + (u-t)^2}{4 s^2} \, C_{22}(0,0,u) 
+ \frac{u-t}{2 s} \, C_{23}(0,0,u) 
\nonumber\\ 
&=& C_{2PP}^{019}(s,u,Q^2) - \Big[ F(u,0) + G(u,0) \Big] \, \frac{u}{4s} 
 + F(u,0) \, \frac{u t}{2s^2} \,,
\nonumber\\ 
C_{2TT}^{019}(s,u,Q^2) &=& 
- \frac{s}{u t} \, C_{21}(0,0,u) 
+ C_{22}(0,0,u) = 
- \frac{F(u,0)+G(u,0)}{D-2} \, \frac{s}{4t} 
- \frac{F(u,0)}{2} 
\,, \nonumber\\
C_{2PR}^{019}(s,u,Q^2) &=& 
- \frac{t}{s u} \, C_{21}(0,0,u) 
+ \frac{s^2 + t^2 - u^2}{4 s u} \, C_{22}(0,0,u) 
- \frac{t}{2 s} \, C_{23}(0,0,u) \nonumber\\ 
&=& 
C_{2PP}^{019}(s,u,Q^2) - \Big[F(u,0)+G(u,0)\Big] \, \frac{u}{8s} 
\nonumber\\
&-& F(u,0) \, \biggl[ \frac{(Q^2-s)^2 + s^2}{8su} 
- \frac{Q^4}{8s^2} \biggr] 
\,, \nonumber\\
C_{2PT}^{019}(s,u,Q^2) &=&   
  \frac{1}{u} \, C_{21}(0,0,u) 
- \frac{Q^2-s}{2 s} \, C_{22}(0,0,u) 
+ \frac{1}{2} \, C_{23}(0,0,u) \nonumber\\ 
&=& 
     \frac{G(u,0)+F(u,0)}{8} \, \frac{4-D}{D-2}  
  +  F(u,0) \, \frac{Q^2-s}{4s} 
\,, \nonumber
\en
\eq
C_{2RT}^{019}(s,u,Q^2) &=&   
  \frac{1}{u} \, C_{21}(0,0,u) 
+ \frac{u-t}{2 s} \, C_{22}(0,0,u) 
+ \frac{1}{2} \, C_{23}(0,0,u) 
 \nonumber\\ 
&=& C_{2PT}^{019}(s,u,Q^2) - F(u,0) \, \frac{u}{2s}
\,. 
\en 
\vspace*{.25cm}
\noindent 
\textbf{Loop integrals $C^{\mu\nu;035}$ and $C^{\mu\nu;069}$} 

\noindent 
Sets of the scalar functions:
\eq
& & C_{2g}^{035}(s,u,Q^2)\,,
   C_{2PP}^{035}(s,u,Q^2)\,,
   C_{2RR}^{035}(s,u,Q^2)\,,
   C_{2TT}^{035}(s,u,Q^2)\,,
\nonumber\\
& &C_{2PR}^{035}(s,u,Q^2)\,,
   C_{2PT}^{035}(s,u,Q^2)\,,
   C_{2RT}^{035}(s,u,Q^2)\,,
\nonumber\\
& &\\
& & C_{2g}^{069}(s,t,Q^2)\,,
   C_{2PP}^{069}(s,t,Q^2)\,,
   C_{2RR}^{069}(s,t,Q^2)\,,
   C_{2TT}^{069}(s,t,Q^2)\,,
\nonumber\\
& &C_{2PR}^{069}(s,t,Q^2)\,,
   C_{2PT}^{069}(s,t,Q^2)\,,
   C_{2RT}^{069}(s,t,Q^2)\,.
\nonumber
\en
\noindent 
Relations:
\eq 
\left\{   
\begin{array}{lccl}
 C_{2g}^{035}(s,u,Q^2) &=& +&  C_{2g}^{069}(s,u,Q^2)\,, \nonumber\\
C_{2PP}^{035}(s,u,Q^2) &=& +& C_{2PP}^{069}(s,u,Q^2)\,, \nonumber\\
C_{2RR}^{035}(s,u,Q^2) &=& +& C_{2RR}^{069}(s,u,Q^2)\,, \nonumber\\
C_{2TT}^{035}(s,u,Q^2) &=& +& C_{2TT}^{069}(s,u,Q^2)\,, \nonumber\\
C_{2PR}^{035}(s,u,Q^2) &=& -& C_{2PR}^{069}(s,u,Q^2)\,, \nonumber\\
C_{2PT}^{035}(s,u,Q^2) &=& +& C_{2PT}^{069}(s,u,Q^2)\,, \nonumber\\
C_{2RT}^{035}(s,u,Q^2) &=& -& C_{2RT}^{069}(s,u,Q^2)\,. \nonumber
\end{array}
\right.
\en
\noindent 
Explicit results: 
\eq 
C_{2g}^{035}(s,u,Q^2) &=& C_{21}(0,u,0) 
= \frac{F(u,0)+G(u,0)}{D-2} \, \frac{u}{4} 
\,, \nonumber\\
C_{2PP}^{035}(s,u,Q^2) &=& 
- \frac{t}{s u} \, C_{21}(0,u,0)
+ \frac{(Q^2-s)^2}{4  s^2} \, C_{22}(0,u,0) 
+ \frac{(Q^2-2s)^2}{4 s^2} \, C_{22}(u,0,0) 
\nonumber\\
&-& \frac{(Q^2-s) (Q^2-2s)}{2 s^2} \, C_{23}(0,u,0) 
\nonumber\\
&=& - \frac{F(u,0)}{2s^2} \, 
\biggl[ \frac{(Q^2-s)^2}{D-4} + \frac{Q^4}{4} \biggr]  
- \frac{F(u,0)+G(u,0)}{D-2} \, \frac{t}{4s} 
\nonumber\\
&+& \Big[G(u,0)-F(u,0)\Big] \, \frac{Q^2-s}{8s} \,, 
\nonumber\\
C_{2RR}^{035}(s,u,Q^2) &=& 
- \frac{t}{s u} \, C_{21}(0,u,0)
+ \frac{(u-t)^2}{4 s^2}   \, C_{22}(0,u,0) 
+ \frac{(Q^2-2t)^2}{4 s^2} \, C_{22}(u,0,0) 
\nonumber\\
&-& \frac{(Q^2-2t) (u-z)}{2 s^2} \, C_{23}(0,u,0) \nonumber\\[2mm]
&=& C_{2PP}^{035}(s,u,Q^2) + F(u,0) \, \frac{2 u t}{s^2} \, 
\biggl[ \frac{1}{D-4} + \frac{Q^2-u}{4t} \biggr] 
\nonumber\\[1mm]
&+& \biggl[ F(u,0) - G(u,0) \biggr] \, \frac{u}{4s} \,, 
\nonumber
\en
\eq 
C_{2TT}^{035}(s,u,Q^2) &=& 
- \frac{s}{u t} \, C_{21}(0,u,0) 
+ C_{22}(0,u,0) + C_{22}(u,0,0) 
- 2 C_{23}(0,u,0) \nonumber\\[1mm] 
&=& - F(u,0) \, \frac{D}{2 (D-4)} - \frac{F(u,0) + G(u,0)}{D-2} 
\, \frac{s}{4t} \,, 
\nonumber\\
C_{2PR}^{035}(s,u,Q^2) &=& 
- \frac{t}{s u} \, C_{21}(0,u,0) 
- \frac{(u-t) (Q^2-s)}{4 s^2} \, C_{22}(0,u,0) 
\nonumber\\[1mm]
&-& \frac{(Q^2-2s) \, (Q^2-2t)}{4 s^2} \, C_{22}(u,0,0) 
 + \frac{(s-t) t + u^2}{2 s^2} \, C_{23}(0,u,0) \nonumber\\
&=& C_{2PP}^{035}(s,u,Q^2) + F(u,0) \, \frac{u}{s^2} \, 
\biggl[ \frac{Q^2-s}{D-4} + \frac{Q^2}{4} \biggr]
\nonumber\\
&+& \Big[F(u,0)-G(u,0)\Big] \, \frac{u}{8s}
\,, 
\nonumber\\
C_{2PT}^{035}(s,u,Q^2) &=&   
  \frac{1}{u} \, C_{21}(0,u,0) 
- \frac{Q^2-s}{2 s} \, C_{22}(0,u,0) 
- \frac{Q^2-2s}{2 s} \, C_{22}(u,0,0) 
\nonumber\\[1mm]
&+& \frac{2Q^2-3s}{2 s} \, C_{23}(0,u,0) \nonumber\\
&=& F(u,0) \frac{1}{s} 
\, \biggl[ \frac{Q^2-s}{D-4} + \frac{Q^2+s}{4} \biggr] 
+ \frac{F(u,0) + G(u,0)}{8} \, \frac{4-D}{D-2} 
\,, \nonumber\\[1mm]
C_{2RT}^{035}(s,u,Q^2) &=&   
  \frac{1}{u} \, C_{21}(0,u,0) 
+ \frac{u-t}{2 s} \, C_{22}(0,u,0) 
+ \frac{Q^2-2t}{2s} \, C_{22}(u,0,0) 
\nonumber\\[1mm]
&-& \frac{2(u-t)+s}{2s} \, C_{23}(0,u,0) \nonumber\\
&=& C_{2PT}^{035}(s,u,Q^2) - F(u,0) \, \frac{u}{2s} \, \frac{D}{D-4} 
\,.   
\en 
\vspace*{.25cm}
\noindent 
\textbf{Loop integrals $C^{\mu\nu;038}$ and $C^{\mu\nu;049}$} 

\noindent 
Sets of the scalar functions:
\eq
& & C_{2g}^{038}(s,u,Q^2)\,,
   C_{2PP}^{038}(s,u,Q^2)\,,
   C_{2RR}^{038}(s,u,Q^2)\,,
   C_{2TT}^{038}(s,u,Q^2)\,,
\nonumber\\
& &C_{2PR}^{038}(s,u,Q^2)\,,
   C_{2PT}^{038}(s,u,Q^2)\,,
   C_{2RT}^{038}(s,u,Q^2)\,,
\nonumber\\
& &\\
& & C_{2g}^{049}(s,t,Q^2)\,,
   C_{2PP}^{049}(s,t,Q^2)\,,
   C_{2RR}^{049}(s,t,Q^2)\,,
   C_{2TT}^{049}(s,t,Q^2)\,,
\nonumber\\
& &C_{2PR}^{049}(s,t,Q^2)\,,
   C_{2PT}^{049}(s,t,Q^2)\,,
   C_{2RT}^{049}(s,t,Q^2)\,.
\nonumber
\en 
\noindent 
Relations:
\eq 
\left\{   
\begin{array}{lccl}
 C_{2g}^{038}(s,u,Q^2) &=& +&  C_{2g}^{049}(s,u,Q^2)\,, \nonumber\\
C_{2PP}^{039}(s,u,Q^2) &=& +& C_{2PP}^{049}(s,u,Q^2)\,, \nonumber\\
C_{2RR}^{038}(s,u,Q^2) &=& +& C_{2RR}^{049}(s,u,Q^2)\,, \nonumber\\
C_{2TT}^{038}(s,u,Q^2) &=& +& C_{2TT}^{049}(s,u,Q^2)\,, \nonumber\\
C_{2PR}^{038}(s,u,Q^2) &=& +& C_{2PR}^{049}(s,u,Q^2)\,, \nonumber\\
C_{2PT}^{038}(s,u,Q^2) &=& +& C_{2PT}^{049}(s,u,Q^2)\,, \nonumber\\
C_{2RT}^{038}(s,u,Q^2) &=& +& C_{2RT}^{049}(s,u,Q^2)\,. \nonumber
\end{array}
\right.
\en
\noindent 
Explicit results: 
\eq
C_{2g}^{038}(s,u,Q^2) &=& 
C_{21}(0,Q^2,s)(s,u,Q^2) = 
\frac{H(Q^2,s)}{D-2} \, \frac{Q^2-s}{4} 
\,, \nonumber\\
C_{2PP}^{038}(s,u,Q^2) &=& 
\frac{(Q^2-s)^2}{4  s^2} \, \biggl[ 
C_{22}(0,Q^2,s) + C_{22}(Q^2,0,s) 
- 2 C_{23}(0,Q^2,s)\biggr] \nonumber\\
&=& - F(Q^2,s) \, \frac{Q^4}{2s^2} \, \frac{D-3}{D-4} 
+ \Big[ G(Q^2,0) s + 3 F(Q^2,0) Q^2 \Big] \, \frac{Q^2}{8s^2}  
\,, \nonumber\\ 
C_{2RR}^{038}(s,u,Q^2) &=& 
- \frac{4 t u}{s (Q^2-s)^2 } \, C_{21}(0,Q^2,s)
+ \frac{(t-u)^2}{4 s^2} \, C_{22}(0,Q^2,s)
\nonumber\\
&+& \frac{(t-u)^2 (Q^2+s)^2}{4 s^2 (Q^2-s)^2}   \, C_{22}(Q^2,0,s) 
 -  \frac{(t-u)^2 (Q^2+s)}{2 s^2 (Q^2-s)} \, C_{23}(0,Q^2,s) 
\nonumber\\
&=& - F(Q^2,s) \, \frac{(u-t)^2}{2 s^2 (Q^2-s)^2} \, 
\biggl[ \frac{Q^4}{D-4} + (Q^2-s)^2 \biggr] 
- \frac{H(Q^2,s)}{D-2} \, \frac{u t}{s (Q^2-s)} \nonumber\\
&-& \biggl[ G(Q^2,0) s + F(Q^2,0) (2s - 3Q^2) \biggr] 
\, \frac{(u-t)^2 Q^2}{8 s^2 (Q^2-s)^2}
\,, \nonumber\\
C_{2TT}^{038}(s,u,Q^2) &=& 
- \frac{s (u-t)^2}{u t (Q^2-s)^2} \, C_{21}(0,Q^2,s) 
+ C_{22}(0,Q^2,s) 
+ \frac{(Q^2+s)^2}{(Q^2-s)^2} \, C_{22}(Q^2,0,s) \nonumber\\
&-& 2 \, \frac{Q^2+s}{Q^2-s} \, C_{23}(0,Q^2,s) 
 = \frac{4s^2}{(u-t)^2} \, 
\biggl[ C_{2RR}^{038} 
+ \frac{4ut - (u-t)^2}{4stu} \, C_{2g}^{038} 
\biggr]
\,, \nonumber\\ 
C_{2PR}^{038}(s,u,Q^2) &=& 
  \frac{t-u}{4s^2} \, \biggl[ 
  (Q^2-s) \, C_{22}(0,Q^2,s) 
+ (Q^2+s) \, C_{22}(Q^2,0,s) 
\nonumber\\[1mm]
&-& 2 Q^2   \, C_{23}(0,Q^2,s) \biggr] 
\nonumber\\ 
&=& \frac{t-u}{Q^2-s} \, 
\biggl[ C_{2PP}^{038}(s,u,Q^2) 
- \frac{(3Q^2+s) F(Q^2,s) - (Q^2-s) G(Q^2,s)}{8s} 
\biggr]
\,, \nonumber\\
C_{2PT}^{038}(s,u,Q^2) &=& \frac{2s}{u-t} \, 
C_{2PR}^{038}(s,u,Q^2)  
\,, \nonumber\\
C_{2RT}^{038}(s,u,Q^2) &=& \frac{u-t}{2s} \, \biggl[ 
         \frac{4s}{(Q^2-s)^2} \, C_{21}(0,Q^2,s)
+                                C_{22}(0,Q^2,s) 
\nonumber\\
&+& \frac{(Q^2+s)^2}{(Q^2-s)^2} \, C_{22}(Q^2,0,s) 
- \frac{2 (Q^2+s)}{Q^2-s}     \, C_{23}(0,Q^2,s) 
\biggr] \nonumber\\
&=& \frac{2s}{u-t} \, 
\biggl[ C_{2PR}^{038}(s,u,Q^2) + \frac{1}{s} \, C_{2g}^{038}
\biggr]
\,. 
\en
\vspace*{.25cm} 
\noindent 
\textbf{Loop integrals $C^{\mu\nu;046}$ and $C^{\mu\nu;058}$} 

\noindent 
Sets of the scalar functions:
Sets of the scalar functions:
\eq
& & C_{2g}^{046}(s,t,Q^2)\,,
   C_{2PP}^{046}(s,t,Q^2)\,,
   C_{2RR}^{046}(s,t,Q^2)\,,
   C_{2TT}^{046}(s,t,Q^2)\,,
\nonumber\\
& &C_{2PR}^{046}(s,t,Q^2)\,,
   C_{2PT}^{046}(s,t,Q^2)\,,
   C_{2RT}^{046}(s,t,Q^2)\,,
\nonumber\\
& &\\
& & C_{2g}^{058}(s,u,Q^2)\,,
   C_{2PP}^{058}(s,u,Q^2)\,,
   C_{2RR}^{058}(s,u,Q^2)\,,
   C_{2TT}^{058}(s,u,Q^2)\,,
\nonumber\\
& &C_{2PR}^{058}(s,u,Q^2)\,,
   C_{2PT}^{058}(s,u,Q^2)\,,
   C_{2RT}^{058}(s,u,Q^2)\,.
\nonumber
\en
\noindent 
Relations:
\eq 
\left\{   
\begin{array}{lccl}
 C_{2g}^{046}(s,t,Q^2) &=& +&  C_{2g}^{058}(s,t,Q^2)\,, \nonumber\\
C_{2PP}^{046}(s,t,Q^2) &=& +& C_{2PP}^{058}(s,t,Q^2)\,, \nonumber\\
C_{2RR}^{046}(s,t,Q^2) &=& +& C_{2RR}^{058}(s,t,Q^2)\,, \nonumber\\
C_{2TT}^{046}(s,t,Q^2) &=& +& C_{2TT}^{058}(s,t,Q^2)\,, \nonumber\\
C_{2PR}^{046}(s,t,Q^2) &=& -& C_{2PR}^{058}(s,t,Q^2)\,, \nonumber\\
C_{2PT}^{046}(s,t,Q^2) &=& +& C_{2PT}^{058}(s,t,Q^2)\,, \nonumber\\
C_{2RT}^{046}(s,t,Q^2) &=& -& C_{2RT}^{058}(s,t,Q^2)\,. \nonumber
\end{array}
\right.
\en
\noindent 
Explicit results: 
\eq 
C_{2g}^{046}(s,t,Q^2) &=& C_{21}(Q^2,t,0) 
= \frac{H(Q^2,t)}{D-2} \, \frac{Q^2-t}{4} 
\,, \nonumber\\
C_{2PP}^{046}(s,t,Q^2) &=& 
-\frac{u t}{s (Q^2-t)^2}       \, C_{21}(Q^2,t,0) 
+ \frac{\alpha^2_+(t,s,u)}{4}  \, C_{22}(Q^2,t,0) 
\nonumber\\
&+& \frac{\gamma^2_+(t,u,s)}{4}  \, C_{22}(t,Q^2,0) 
 - \frac{\alpha_+(t,s,u) \, \gamma_+(t,u,s)}{2} \, C_{23}(Q^2,t,0) 
\nonumber\\
&=& - \frac{F(Q^2,t)}{2s^2} 
\, \biggl[\frac{((Q^2-t)^2+ut)^2}{(Q^2-t)^2 (D-4)} + \frac{Q^4}{4} \biggr] 
    - \frac{H(Q^2,t)}{D-2} \, \frac{u t}{4 s (Q^2-t)} 
\nonumber\\
&-& F(Q^2,t) \, \frac{Q^2 (2Q^2+s)}{8 s (Q^2-t)}  
    + \Big[ G(Q^2,0) - F(Q^2,0) \Big] \, \frac{Q^2 ((Q^2-t)^2+ut)}{8 s (Q^2-t)^2}
\,, \nonumber\\
C_{2RR}^{046}(s,t,Q^2) &=& 
-  \frac{u t}{s (Q^2-t)^2}     \, C_{21}(Q^2,t,0) 
+ \frac{\beta^2_+(t,s,u)}{4}   \, C_{22}(Q^2,t,0)
\nonumber\\
&+& \frac{\gamma^2_+(t,s,u)}{4}  \, C_{22}(t,Q^2,0) 
 - \frac{\beta_+(t,s,u) \, \gamma_+(t,s,u)}{2} \, C_{23}(Q^2,t,0) 
\nonumber\\
&=& C_{2PP}^{046}(s,t,Q^2) + 2F(Q^2,t)\, \frac{u t}{s^2} 
\, \biggl[ \frac{1}{D-4} + \frac{Q^2-t}{4u} \biggr]
\nonumber\\
&+& \frac{Q^2}{4s} \, \Big[ 3F(Q^2,0) - G(Q^2,0) \Big]
\,, \nonumber\\
C_{2TT}^{046}(s,t,Q^2) &=& 
- \frac{s}{u t} \, C_{21}(Q^2,t,0) 
+ \frac{(Q^2-t)^2}{(Q^2+t)^2} \, 
\Big[ C_{22}(Q^2,t,0) + C_{22}(t,Q^2,0) 
\nonumber\\
&-& 2 C_{23}(Q^2,t,0) \Big] 
= - F(Q^2,t) \, \frac{D}{2 (D-4)} - \frac{H(Q^2,t)}{D-2} \, \frac{s (Q^2-t)}{4 u t} 
\,, \nonumber\\
C_{2PR}^{046}(s,t,Q^2) &=& 
- \frac{u t}{s (Q^2-t)^2} \, C_{21}(Q^2,t,0) 
- \frac{\alpha_+(t,s,u)   \, \beta_+(t,s,u)}{4}  \, C_{22}(Q^2,t,0) 
\nonumber\\
&-& \frac{\gamma_+(t,s,u)   \, \gamma_+(t,u,s)}{2} \, C_{22}(t,Q^2,0) 
+ \frac{\sigma_+(t,u,s)}{2} \, C_{23}(Q^2,t,0) 
\nonumber\\
&=& C_{2PP}^{046}(s,t,Q^2) + \frac{F(Q^2,t)}{s^2} \, \biggl[ 
\frac{(Q^2-t)^2+ut}{(Q^2-t)^2 (D-4)} 
+ \frac{Q^2 (Q^2-2t)}{4} \biggr] 
\nonumber\\
&+& - \frac{H(Q^2,t)}{D-2} \, \frac{u t}{4 s (Q^2-t)} 
+ \biggl[3F(Q^2,0) - G(Q^2,0)\biggr] \, \frac{Q^2}{8s}  
\,, \nonumber
\en 
\eq 
C_{2PT}^{046}(s,t,Q^2) &=&   
  \frac{1}{Q^2-t}     \, C_{21}(Q^2,t,0) 
\nonumber\\
&-& \frac{Q^2-t}{Q^2+t} \,  \biggl[ 
  \frac{\alpha_+(t,s,u)}{2} \, \Big(C_{22}(Q^2,t,0) - C_{23}(Q^2,t,0)\Big) 
\nonumber\\
&+& \frac{\gamma_+(t,u,s)}{2} \, \Big(C_{22}(t,Q^2,0) - C_{23}(Q^2,t,0)\Big) 
\biggr] \nonumber\\
&=& F(Q^2,t) \, \frac{Q^2}{s} \,  
\biggl[ \frac{(Q^2-t)^2 + ut}{Q^2 (Q^2-t) (D-4)} 
+ \frac{1}{4} \biggr]
\nonumber\\
&+& \frac{H(Q^2,t)}{4 (D-2)} 
 +  \Big[3F(Q^2,0) - G(Q^2,0)\Big] \, \frac{Q^2}{8 (Q^2-t)}
\,, \nonumber\\
C_{2RT}^{046}(s,t,Q^2) &=&   
  \frac{1}{Q^2-t}         \, C_{21}(Q^2,t,0) 
\nonumber\\
&+& \frac{Q^2-t}{Q^2+t} \,  \biggl[ 
  \frac{\beta_+(t,s,u)}{2}  \, \Big(C_{22}(Q^2,t,0) - C_{23}(Q^2,t,0)\Big) 
\nonumber\\
&+& \frac{\gamma_+(t,s,u)}{2} \, \Big(C_{22}(t,Q^2,0) - C_{23}(Q^2,t,0)\Big) 
\biggr] \nonumber\\
&=& C_{2PT}^{046}(s,t,Q^2) - F(Q^2,t) \, \frac{Q^2-t}{2s} \, \frac{D}{D-4}
\,. 
\en 

\vspace*{0.25cm}
\noindent 
\textbf{Loop integrals $C^{\mu\nu;078}$ and $C^{\mu\nu;079}$}

\noindent 
Sets of the scalar functions:
\eq
& & C_{2g}^{078}(s,u,Q^2)\,,
   C_{2PP}^{068}(s,u,Q^2)\,,
   C_{2RR}^{078}(s,u,Q^2)\,,
   C_{2TT}^{078}(s,u,Q^2)\,,
\nonumber\\
& &C_{2PR}^{078}(s,u,Q^2)\,,
   C_{2PT}^{078}(s,u,^2)\,,
  C_{2RT}^{078}(s,u,Q^2)\,,
\nonumber\\
& &\\
& & C_{2g}^{079}(s,u,Q^2)\,,
   C_{2PP}^{079}(s,u,Q^2)\,,
   C_{2RR}^{079}(s,u,Q^2)\,,
   C_{2TT}^{079}(s,u,Q^2)\,,
\nonumber\\
& &C_{2PR}^{079}(s,u,Q^2)\,,
   C_{2PT}^{079}(s,u,Q^2)\,,
   C_{2RT}^{079}(s,u,Q^2)\,.
\nonumber
\en
\noindent 
Relations:
\eq 
\left\{   
\begin{array}{lccl}
C_{2K}^{078}(s,u,Q^2) &=& &  C_{2K}^{079}(s,u,Q^2)\,, \qquad
K = g, RR, TT, RT \,, \nonumber\\
C_{2PP}^{078}(s,u,Q^2) + C_{1P}^{078}(s,u,Q^2) 
&=& & 
C_{2PP}^{078}(s,u,Q^2) + C_{1P}^{079}(s,u,Q^2)   \,,  \nonumber\\
C_{2PR}^{078}(s,u,Q^2) + \frac{1}{2} C_{1R}^{078}(s,u,Q^2)
&=& & 
C_{2PR}^{079}(s,u,Q^2) + \frac{1}{2} C_{1R}^{079}(s,u,Q^2)
\,, \nonumber\\
C_{2PT}^{078}(s,u,Q^2) + \frac{1}{2} C_{1T}^{078}(s,u,Q^2) 
&=& & 
C_{2PT}^{079}(s,u,Q^2) + \frac{1}{2} C_{1T}^{079}(s,u,Q^2)
\,. \nonumber
\end{array}
\right.
\en 
\noindent 
Explicit results: 
\eq 
C_{2g}^{078}(s,u,Q^2)  &=& C_{21}(s,Q^2,0) 
= \frac{H(Q^2,s)}{D-2} \, \frac{(Q^2-s)}{4} 
\,, \nonumber\\ 
C_{2PP}^{078}(s,u,Q^2)  &=& 
\frac{(Q^2-s)^4}{4  s^2 (Q^2+s)^2} \, C_{22}(Q^2,s,0) \nonumber\\[1mm]
&=& - \frac{F(Q^2,s)}{2} \, \frac{D-3}{D-4} 
    + \frac{Q^2}{8 s} \biggl[ G(Q^2,0) - \frac{Q^2+4s}{s} F(Q^2,0) \biggr]
\,, \nonumber
\en 
\eq
C_{2RR}^{078}(s,u,Q^2)  &=& \frac{(u-t)^2}{4s^2} \, 
\biggl[
- \frac{16 sut}{(Q^2-s)^2 (u-t)^2} \, C_{21}(s,Q^2,0)
\nonumber\\ 
&+& \frac{4s^2}{(Q^2+s)^2}         \, C_{22}(s,Q^2,0)
 +                                    C_{22}(Q^2,s,0) 
 - \frac{4s}{Q^2+s}                \, C_{23}(s,Q^2,0) 
\biggr] 
\nonumber\\
&=& - \frac{F(Q^2,s)}{D-4} \, \frac{(u-t)^2}{2 (Q^2-s)^2}   
    - \frac{H(Q^2,s)}{D-2} \, \frac{u t}{s (Q^2-s)} 
\nonumber\\
&-& \frac{Q^2 (u-t)^2}{8 s^2 (Q^2-s)^2} \, 
\biggl[G(Q^2,0) s + F(Q^2,0) (Q^2-2s)\biggr] 
\,,
\nonumber\\
C_{2TT}^{078}(s,u,Q^2)  &=& 
- \frac{s (u-t)^2}{u t (Q^2-s)^2} \, C_{21}(s,Q^2,0) 
+ \frac{4 s^2}{(Q^2+s)^2}         \, C_{22}(s,Q^2,0) 
\nonumber\\
&+&                                  C_{22}(Q^2,s,0) 
- \frac{4 s}{Q^2+s}               \, C_{23}(s,Q^2,0) 
\nonumber\\
&=& \frac{4s^2}{(u-t)^2} \, 
\biggl[ C_{2RR}^{078}(s,u,Q^2)  
+ \frac{4ut - (u-t)^2}{4stu} \, C_{2g}^{078}(s,u,Q^2 
\biggr]
\,, \nonumber\\ 
C_{2PR}^{078}(s,u,Q^2)  &=& 
 \frac{(t-u) \, (Q^2-s)^2}{4 s^2 (Q^2+s)^2}   \,
\biggl[ (Q^2+s) C_{22}(Q^2,s,0) - 2 s C_{23}(s,Q^2,0)
\biggr]  \nonumber\\ 
&=& - \frac{F(Q^2,s)}{D-4} \, \frac{u-t}{2 (Q^2-s)}
    + \frac{Q^2 (Q^2+s) (u-t)}{8 (Q^2-s) s^2} \, F(Q^2,0) 
\,, \nonumber\\
C_{2PT}^{078}(s,u,Q^2)  &=& \frac{2s}{u-t} \, C_{2PR}^{078}(s,u,Q^2) 
\,, \nonumber\\
C_{2RT}^{078}(s,u,Q^2) &=& \frac{u-t}{2s} \, \biggl[ 
  \frac{4s}{(Q^2-s)^2}   \, C_{21}(s,Q^2,0) 
+ \frac{4s^2}{(Q^2+s)^2} \, C_{22}(s,Q^2,0) 
\nonumber\\ 
&+&                         C_{22}(Q^2,s,0) 
 - \frac{4s}{Q^2+s}      \, C_{23}(s,Q^2,0) \biggr]
\nonumber\\ 
&=& \frac{u-t}{2s} \, \biggl[ C_{2TT}^{078}(s,u,Q^2) + \frac{s}{u t}  
C_{2g}^{078}(s,u,Q^2) \biggr]
\,. 
\en 

\vspace*{.25cm}
\section{Box integrals}

\subsection{Scalar box integral} 
\label{Box_calculation}  

First, we discuss calculation of the scalar box integral $D_0(s,u,Q^2)$.  
After $\alpha$ parametrization and integration over loop 
momentum the integral $D_0(s,u,Q^2)$ takes 
the form 
\eq 
D_0(s,u,Q^2) &=&
(\pi\mu^{-2})^{\frac{D}{2}-2} 
\, \Gamma\Big(4-\frac{D}{2}\Big) 
\, \int\limits_0^\infty d\alpha_1 \ldots \int\limits_0^\infty d\alpha_4 
\nonumber\\
&\times&
 \frac{\delta\biggl(1-\sum\limits_{i=1}^4 \alpha_i\biggr) \, 
	\biggl(\sum\limits_{i=1}^4 \alpha_i\biggr)^{4-D}} 
{\Big(- s \alpha_3 (\alpha_2 + \alpha_4) 
	- u \alpha_4 (\alpha_1 + \alpha_3) 
	- t \alpha_3 \alpha_4\Big)^{4-\frac{D}{2}}} \,.
\en 
Next we apply the change of variables ($\alpha$ parameters)  
using the trick from~\cite{Smirnov:2006ry}: 
\eq 
\alpha_1 = \eta_1 (1-\xi_1)\,, \qquad 
\alpha_2 = \eta_2 (1-\xi_2)\,, \qquad 
\alpha_3 = \eta_1 \xi_1\,,     \qquad 
\alpha_4 = \eta_2 \xi_2\,. 
\en 
Next we take into account that the Jacobian of such 
the change of variables is $J = \eta_1 \eta_2$,  
integrate over $\eta_2$ using $\delta$ function, 
and integrate over $\eta_1$ using the integral representation 
for beta function\index{Beta function}~(\ref{Beta_int}). 

After these tricks the box integral takes the form 
\eq 
\hspace*{-1cm}
D_0(s,u,Q^2) = (\pi\mu^{-2})^{\frac{D}{2}-2} \, 
\frac{\Gamma\Big(4-\frac{D}{2}\Big) \, \Gamma^2\Big(\frac{D}{2}-2\Big)} 
{\Gamma(D-4)} \, \int\limits_0^1 \int\limits_0^1 
\frac{d\xi_1 d\xi_2}
{\Big(-s \xi_1 - u \xi_2 - t \xi_1 \xi_2\Big)^{4-\frac{D}{2}}} \,.
\en 
Integration over $\xi_2$ gives 
\eq\label{D0st_xi} 
D_0(s,u,Q^2) &=& \frac{(\pi\mu^{-2})^{\frac{D}{2}-2}}{u} \, 
\frac{\Gamma\Big(3-\frac{D}{2}\Big) \, \Gamma^2\Big(\frac{D}{2}-2\Big)} 
{\Gamma(D-4)} \, \int\limits_0^1 d\xi_1  \,
\biggl(1 + \frac{t}{u} \xi_1\biggr)^{-1} 
\nonumber\\
&\times&
\biggl[ (-u)^{\frac{D}{2}-3} \, \biggl(1 + \frac{s+t}{u} 
\xi_1\biggr)^{\frac{D}{2}-3} \,- \  
(-s)^{\frac{D}{2}-3} \, \xi_1^{\frac{D}{2}-3} 
\biggr]  \,.
\en 
Next step is to identify the box integral as a combination of 
Gauss hypergeometric functions $_2F_1$. We can proceed in two equivalent ways. 

First, using $\alpha$ parametrization for the first term in 
Eq.~(\ref{D0st_xi})  
\eq 
\frac{1}{AB^n} = n \, \int\limits_0^1 d\alpha \, 
\frac{\alpha^{n-1}}{\Big[A + (B-A) \alpha\Big]^{n+1}} \,,
\en  
and integrating over $\xi_1$ we write down the integral as 
combination of three terms 
\eq
D_0(s,u,Q^2) &=& (\pi\mu^{-2})^{\frac{D}{2}-2} \, 
\frac{\Gamma\Big(3-\frac{D}{2}\Big) \, \Gamma^2\Big(\frac{D}{2}-2\Big)} 
{\Gamma(D-4)} \, \int\limits_0^1 d\alpha \nonumber\\
&\times&  \biggl[ \frac{1}{t} \, 
\biggl(-\frac{(s+t) \, u}{t}\biggr)^{\frac{D}{2}-3}  
\, \alpha^{2-\frac{D}{2}} \, 
\biggl(1 - \alpha \, \frac{s}{s+t}\biggr)^{\frac{D}{2}-3}
\nonumber\\
&-&
\frac{1}{t} \, \biggl(-\frac{(s+t) (u+t)}{t}\biggr)^{\frac{D}{2}-3}  
\, \alpha^{2-\frac{D}{2}} \, 
\biggl(1 - \alpha \, \frac{su}{(s+t) (u+t)}\biggr)^{\frac{D}{2}-3}
\nonumber\\
&+&
\frac{(-s)^{\frac{D}{2}-2}}{su} \, \alpha^{\frac{D}{2}-3} \, \
\biggl(1 + \alpha \, \frac{t}{u}\biggr)^{-1}\, 
\biggr] \,,
\en 
Next using integral representation for the 
Gauss hypergeometric function 
\eq 
_2F_1(a,b,c,z) = \frac{\Gamma(c)}{\Gamma(b) \Gamma(c-b)} \, 
\int\limits_0^1 d\alpha \, \alpha^{b-1} \, (1-\alpha)^{c-b-1} 
\, (1-z\alpha)^{-a} \,,
\en 
we get 
\eq 
\hspace*{-.5cm}
D_0(s,u,Q^2) &=& - (\pi\mu^{-2})^{\frac{D}{2}-2} \, 
\frac{\Gamma\Big(2-\frac{D}{2}\Big) \, \Gamma^2\Big(\frac{D}{2}-2\Big)} 
{\Gamma(D-4)} 
\ \biggl[ 
\frac{(-s)^{\frac{D}{2}-2}}{su} 
\ _2F_1\Big(1,\frac{D}{2}-2,\frac{D}{2}-1,-\frac{t}{u}\Big)
\nonumber\\
\hspace*{-.5cm}
&-&\frac{2-\frac{D}{2}}{3-\frac{D}{2}} \, \frac{1}{t} \, 
\biggl(-\frac{(s+t) u}{t}\biggr)^{\frac{D}{2}-3} 
\ _2F_1\Big(3-\frac{D}{2},3-\frac{D}{2},4-\frac{D}{2},\frac{s}{s+t}\Big)
\nonumber\\
\hspace*{-.5cm}
&+&\frac{2-\frac{D}{2}}{3-\frac{D}{2}} \, \frac{1}{t} \, 
\biggl(-\frac{(s+t) (u+t)}{t}\biggr)^{\frac{D}{2}-3} 
\ _2F_1\Big(3-\frac{D}{2},3-\frac{D}{2},4-\frac{D}{2},
\frac{s u}{(s+t) (u+t)}\Big)
\biggr] \,. \nonumber\\
\en 
Further applying the identity  
\eq 
_2F_1(a,b,c,z) = (1-z)^{-a} \ _2F_1\Big(c-a,b,c,\frac{z}{z-1}\Big)\,,
\en 
for the second and third 
hypergeometric function\index{Hypergeometric function} we get:
\eq\label{D0_prefin} 
D_0(s,u,Q^2) &=& - (\pi\mu^{-2})^{\frac{D}{2}-2} \, 
\frac{\Gamma\Big(2-\frac{D}{2}\Big) \, \Gamma^2\Big(\frac{D}{2}-2\Big)} 
{\Gamma(D-4)} \ \biggl[ 
\frac{(-s)^{\frac{D}{2}-2}}{su} 
\, _2F_1\Big(1,\frac{D}{2}-2,\frac{D}{2}-1,-\frac{t}{u}\Big)
\nonumber\\
&-&\frac{2-\frac{D}{2}}{3-\frac{D}{2}} \, 
\frac{(-u)^{\frac{D}{2}-3}}{t} 
\, _2F_1\Big(1,3-\frac{D}{2},4-\frac{D}{2},-\frac{s}{t}\Big)
\nonumber\\
&+&\frac{2-\frac{D}{2}}{3-\frac{D}{2}} \, 
\frac{(-s-t-u)^{\frac{D}{2}-3}}{t} 
\, _2F_1\Big(1,3-\frac{D}{2},4-\frac{D}{2},-\frac{su}{(s + t) (u + t)}\Big)
\biggr] \,.
\en 
Note this representation can also be obtained
from Eq.~(\ref{D0st_xi}) identifying there the first term 
with the Appell function using integral representation 
\eq 
\hspace*{-.5cm}
F_1(a,b,c,d,m,n) = \frac{\Gamma(d)}{\Gamma(a) \Gamma(d-a)} \, 
\int\limits_0^1 d\alpha \, \alpha^{a-1} \, (1-\alpha)^{d-a-1} \, 
(1-m\alpha)^{-b} \, (1-n\alpha)^{-c}  \,,
\en 
and further using the identity transforming the 
Appell function\index{Appell function}  
into a combination of two Gauss functions\index{Gauss hypergeometric function}   
(see proof in the end of this section): 
\eq\label{ID_Appel_2Gauss} 
F_1(1,b,c,2,m,n) &=& \frac{1}{(1-b-c) m} \, 
\biggl[\, _2F_1\Big(1,c,b+c,1-\frac{n}{m}\Big) 
 \nonumber\\
 &-& (1-m)^{1-b} \, (1-n)^{-c} \ _2F_1\Big(1,c,b+c,\frac{m-n}{m (1-n)}\Big)  
\biggr] \,.
\en 
Now we are in the position to derive the final expression for the 
box integral applying the analytic continuation formula for 
the Gauss functions in the second and third terms 
of Eq.~(\ref{D0_prefin}): 
\eq\label{id1F2} 
_2F_1(a,b,c,z) &=& 
\frac{\Gamma(c) \Gamma(b-a)}{\Gamma(b) \Gamma(c-a)} 
\, (-z)^{-a} \, _2F_1\Big(a,1-c+a,1-b+a,\frac{1}{z}\Big) 
\nonumber\\
&+& 
\frac{\Gamma(c) \Gamma(a-b)}{\Gamma(a) \Gamma(c-b)} 
\, (-z)^{-b} \, _2F_1\Big(b,1-c+b,1-a+b,\frac{1}{z}\Big) \,.
\en  
Note that only the first term in the identity~(\ref{id1F2}) 
contributes to the expression for the scalar box integral, 
while the second term vanishes since it generates 
the equal and opposite in sign contributions in the 
transformations of the second and third term in Eq.~(\ref{D0_prefin}). 

Final expression for the box 
integral is (see also Ref.~\cite{Duplancic:2000sk});  
\eq\label{final_box} 
D_0(s,u,Q^2) &=& - \frac{1}{su} \, 
\Biggl[\pi^{\frac{D}{2}-2} \frac{\Gamma\Big(2-\frac{D}{2}\Big) \, 
	\Gamma^2\Big(\frac{D}{2}-2\Big)}{\Gamma(D-4)}\Biggr] \, 
\nonumber\\
&\times&
\Biggl[ 
\Big(-\frac{s}{\mu^2}\Big)^{\frac{D}{2}-2} 
\, _2F_1\Big(1,\frac{D}{2}-2,\frac{D}{2}-1,-\frac{t}{u}\Big) 
\nonumber\\
&+&
\Big(-\frac{u}{\mu^2}\Big)^{\frac{D}{2}-2} 
\, _2F_1\Big(1,\frac{D}{2}-2,\frac{D}{2}-1,-\frac{t}{s}\Big) \nonumber\\
&-& 
\Big(-\frac{Q^2}{\mu^2}\Big)^{\frac{D}{2}-2} 
\, _2F_1\Big(1,\frac{D}{2}-2,\frac{D}{2}-1,-\frac{Q^2 t}{s u}\Big) 
\Biggr]\,. 
\en 
Note that in the on-shell limit $Q^2 = s + t + u = 0$ we 
reproduce the result for the scalar box given in Ref.~\cite{Smirnov:2006ry}. 

The formula for the scalar box integral can be nicely written in terms 
of the scalar bubble PV function 
$B_0(p^2)$~(\ref{PV_B0}) as: 
\eq\label{D0fin} 
D_0(s,u,Q^2) &=& - \frac{4}{su} \ \frac{D-3}{D-4} \ I_0(s,u,Q^2) \,, 
\nonumber\\
I_0(s,u,Q^2) &=& 
B_0(s) \ _2F_1\Big(1,\frac{D}{2}-2,\frac{D}{2}-1,-\frac{t}{u}\Big) 
+ 
B_0(u) \ _2F_1\Big(1,\frac{D}{2}-2,\frac{D}{2}-1,-\frac{t}{s}\Big) 
\nonumber\\
&-&  
B_0(Q^2) \ _2F_1\Big(1,\frac{D}{2}-2,\frac{D}{2}-1,-\frac{Q^2 t}{s u}\Big) 
\,.
\en 
Now we make the $\eps$ expansion of the scalar box diagram. 
The $\eps$ expansion of the Gauss function reads: 
\eq\label{2F1_expand_res} 
_2F_1(1,-\eps,1-\eps,z) &=& 
\frac{1}{1-z} - \sum\limits_{n=0}^\infty \eps^n \, {\rm Li}_n(z)  \nonumber\\
&=& 1 - \sum\limits_{n=1}^\infty \eps^n \, {\rm Li}_n(z) 
= 1 + \eps \log(1-z) -  
\sum\limits_{n=2}^\infty \eps^n \, {\rm Li}_n(z)   \,.
\en 
The proof of Eq.~(\ref{2F1_expand_res}) 
is given below in this section. 
Note that the ${\cal O}(\eps)$ term in the expansion of the 
Gauss function does not contribute to the $\eps$ expansion 
of scalar box due to compensation between the three 
Gauss functions\index{Gauss hypergeometric function}   
in Eq.~(\ref{D0fin})
\eq 
& &\log\Big(1+\frac{t}{s}\Big) + 
\log\Big(1+\frac{t}{u}\Big) - 
\log\Big(1+\frac{Q^2 t}{su}\Big) \nonumber\\
&=& 
\log\Big(\frac{s+t}{s}\Big) + 
\log\Big(\frac{u+t}{u}\Big) - 
\log\Big(\frac{(s+t) (u+t)}{su}\Big) \equiv 0    \,.
\en 
Therefore, the $\eps^2$ contribution of the Gauss functions 
reads: 
\eq 
\beta(s,u,Q^2) &=& 
{\rm Li}_2\biggl(-\frac{Q^2 t}{su}\biggr) 
- {\rm Li}_2\biggl(-\frac{t}{s}\biggr) 
- {\rm Li}_2\biggl(-\frac{t}{u}\biggr) \nonumber\\ 
&-&  \log\biggl(\frac{-s}{\mu^2}\biggr) \log\Big(\frac{u+t}{u}\Big) 
-    \log\biggl(\frac{-u}{\mu^2}\biggr) \log\Big(\frac{s+t}{s}\Big) \nonumber\\
&+&    \log\biggl(\frac{-Q^2}{\mu^2}\biggr) \log\Big(\frac{(s+t) (u+t)}{s u}\Big) \,.  
\en 
The expression for $\beta(s,u,Q^2)$ can be simplified 
using the reflection identity for the dilogarithm:  
\eq 
{\rm Li}_2(z) + {\rm Li}_2(1-z) = \frac{\pi^2}{6} - 
\log(z) \log(1-z)  \,.
\en 
One gets~\cite{Duplancic:2000sk}: 
\eq 
\beta(s,u,Q^2) = 
{\rm Li}_2\biggl(1+\frac{t}{s}\biggr)
+ {\rm Li}_2\biggl(1+\frac{t}{u}\biggr)
- {\rm Li}_2\biggl(1+\frac{Q^2 t}{su}\biggr) 
- \frac{\pi^2}{6} \,.
\en 
Next using well-known identity (Abel formula) relating 5 
dilogarithms~\cite{Cachazo:2004zb,Ellis:2007qk} 
\eq 
\frac{1}{2} \, \log^2\frac{s}{u} &+&{\rm Li}_2\biggl(1+\frac{t}{s}\biggr) 
+ {\rm Li}_2\biggl(1+\frac{t}{u}\biggr) \nonumber\\
&-& {\rm Li}_2\biggl(1+\frac{Q^2 t}{s u}\biggr) 
+ {\rm Li}_2\biggl(1-\frac{Q^2}{s}\biggr) 
+ {\rm Li}_2\biggl(1-\frac{Q^2}{u}\biggr) = 0  \,,
\en 
one can represent the $\beta(s,u,Q^2)$ 
in terms of two dilogarithms~\cite{Ellis:2007qk}: 
\eq 
\beta(s,u,Q^2) = - {\rm Li}_2\biggl(1-\frac{Q^2}{s}\biggr) 
- {\rm Li}_2\biggl(1-\frac{Q^2}{u}\biggr) 
- \frac{1}{2} \, \log^2\frac{s}{u} - \frac{\pi^2}{6} \,.
\en 
Here for convenience we do not expand the $B_0$ function and expand only the 
Gauss hypergeometric functions\index{Gauss hypergeometric function}. 

Now we derive the $\eps$ expansion of the 
Gauss hypergeometric function $_2F_1(1,-\epsilon,1-\epsilon,z)$ stated in Eq.~\ref{2F1_expand_res}. 
We start with the Euler integral representation for this function: 
\begin{align}
\ghy(1,-\eps,1-\eps,x)&=\frac{\Gamma(1-\eps)}{\Gamma(-\eps)}
\int_0^1\dx t\,t^{-1-\eps}(1-xt)^{-1}
\nonumber\\
&=-\eps\int_0^1\dx t\,\frac{t^{-1-\eps}}{1-xt}\,.
\end{align} 
To get rid of the $\eps$-pole in the above integral we perform a partial 
integration.\footnote{Subtracting the pole like in Sec.~\ref{sec:All order expansion} 
would work equally well.} Afterwards, we can safely expand the integrand in the 
$\eps$ series: 
\begin{align}
\ghy(1,-\eps,1-\eps,x)&=\left.\frac{t^{-\eps}}{1-xt}\right|_{t=0}^1
-x\int_0^1\dx t\,\frac{t^{-\eps}}{(1-x t)^2}
\nonumber\\
&=\frac{1}{1-x}-x\sum_{n=0}^\infty\frac{(-\eps)^n}{n!}
\int_0^1\dx t\,\frac{\log^n t}{(1-xt)^2}\,.
\label{eq:epsilon expansion intermediate}
\end{align}
Now we use the identity 
\begin{align}
x\,\frac{(-1)^n}{n!}\int_0^1\dx t\,\frac{\log^n t}{(1-xt)^2}=\mathrm{Li}_n(x)
\label{eq:PolyLog identity}
\end{align}
which can be established by induction for all $n\geq 0$ utilizing 
the recursive definition of the polylogarithms via
\begin{align}
\mathrm{Li}_{n+1}(x)=\int_0^x\frac{\dx t}{t}\,\mathrm{Li}_n(t)\,.
\end{align}
Plugging (\ref{eq:PolyLog identity}) into 
(\ref{eq:epsilon expansion intermediate}) gives
\begin{align}
\ghy(1,-\eps,1-\eps,x)&=\frac{1}{1-x}-\sum_{n=0}^\infty\eps^n\,
\mathrm{Li}_n(x)\,.
\end{align}
With $\mathrm{Li}_0(x)=\frac{x}{1-x}$ and $\mathrm{Li}_1(x)=-\log(1-x)$ 
we arrive at the all order expansion
\begin{align}
\ghy(1,-\eps,1-\eps,x)&=1+\eps\log(1-x)-\sum_{n=2}^\infty\eps^n\,
\mathrm{Li}_n(x)\,.
\label{eq:ghy all order expansion}
\end{align}
The gamma functions in Eq.~(\ref{final_box})  generates 
a $\frac{1}{\eps^2}$-pole, as
\begin{align}
\frac{\Gamma(\eps)\Gamma^2(-\eps)}{\Gamma(-2\eps)}=
-\mathrm{e}^{-\gamma_\text{E}\eps}\left(\frac{2}{\eps^2}
-\frac{\pi^2}{6}+\mathcal{O}(\eps)\right)\,. 
\end{align} 

To conclude this section, we prove the identity~(\ref{ID_Appel_2Gauss}), which to our knowledge 
occurs for the first time in literature. Our proof is based on 
the use of the Mellin-Barnes integral technique.

We start with the integral representation 
\begin{align}
\appell(1,b_1,b_2,2,x,y)=\int_0^1\dx t\,(1-xt)^{-b_1}(1-y t)^{-b_2}\,.
\end{align}
We factorize both multipliers in the integrand applying the 
Binomi-Mellin-Newton integral\index{Binomi-Mellin-Newton integral}  
\begin{align}
(a+b)^w=\frac{1}{\Gamma(-w)}\int_{-i\infty}^{i\infty}
\frac{\dx z}{2\pi i}\,a^{w-z}b^{z}\Gamma(-z)\Gamma(-w+z)\,.
\label{eq: Binomi-Newton-Mellin integral}
\end{align}
It yields
\begin{align}
\appell(1,b_1,b_2,2,x,y)=\frac{1}{\Gamma(b_1)\Gamma(b_2)}
&\int_{-i\infty}^{i\infty}\frac{\dx z_1}{2\pi i}\int_{-i\infty}^{i\infty}
\frac{\dx z_2}{2\pi i}(-x)^{z_1} \, (-y)^{z_2} 
\nonumber\\
&\times
\Gamma(-z_1)\Gamma(b_1+z_1)\Gamma(-z_2)\Gamma(b_2+z_2)
\int_0^1\dx t\, t^{z_1+z_2}\,.
\end{align}
Evaluating the integral over $t$ and substituting $z_1$ 
by $z=z_1+z_2$ lead to
\begin{align}
\appell(1,b_1,b_2,2,x,y)=\frac{1}{\Gamma(b_1)\Gamma(b_2)}
&\int_{-i\infty}^{i\infty}\frac{\dx z}{2\pi i}\frac{(-x)^z}{z+1}
\int_{-i\infty}^{i\infty}\frac{\dx z_2}{2\pi i}
\left(-\frac{y}{x}\right)^{z_2}
\nonumber\\
&\times
\Gamma(-z+z_2)\Gamma(b_2+z_2)\Gamma(b_1+z-z_2)\Gamma(-z_2)\,.
\end{align}
The integral over the $z_2$ variable now matches the Mellin-Barnes representation  
of a hypergeometric function. It can be subsequently expressed through 
its real integral representation to obtain
\begin{align}
\appell(1,b_1,b_2,2,x,y)&=\frac{1}{\Gamma(b_1)\Gamma(b_2)}
\int_{-i\infty}^{i\infty}\frac{\dx z}{2\pi i}\frac{(-x)^z}{z+1}
\frac{\Gamma(-z)\Gamma(b_2)\Gamma(b_1)\Gamma(b_1+b_2+z)}{\Gamma(b_1+b_2)}
\nonumber\\
&\hphantom{\frac{1}{\Gamma(b_1)\Gamma(b_2)}
	\int_{-i\infty}^{i\infty}
	\frac{\dx z}{2\pi i}}\times\ghy\left(-z,b_2,b_1+b_2,1-\frac{y}{x}\right)
\nonumber\\
&=\frac{1}{\Gamma(b_1+b_2)}\int_{-i\infty}^{i\infty}
\frac{\dx z}{2\pi i}\frac{(-x)^z}{z+1}\Gamma(-z)\Gamma(b_1+b_2+z)
\nonumber\\
&\times\frac{\Gamma(b_1+b_2)}{\Gamma(b_1)\Gamma(b_2)}\int_0^1\dx t\, 
t^{b_2-1}(1-t)^{b_1-1}\left(1-t\left(1-\frac{y}{x}\right)\right)^z\,.
\end{align}
Interchanging the order of integration and rearranging the terms leads to
\begin{align}
\appell(1,b_1,b_2,2,x,y)&=\frac{1}{\Gamma(b_1)\Gamma(b_2)}
\int_0^1\dx t\, t^{b_2-1}(1-t)^{b_1-1}
\nonumber\\
&\times
\int_{-i\infty}^{i\infty}\frac{\dx z}{2\pi i}\frac{\Gamma(z+1)
	\Gamma(b_1+b_2+z)\Gamma(-z)}{\Gamma(z+2)} \, \Big(-(1-t)x-ty\Big)^z\,.
\end{align}
We recognize in the latter formula the $z$ integral as another 
Mellin-Barnes representation\index{Mellin-Barnes integral} 
of the Gauss function $_2F_1$, thus
\begin{align}
\appell(1,b_1,b_2,2,x,y)&=\frac{\Gamma(b_1+b_2)}{\Gamma(b_1)
	\Gamma(b_2)}\int_0^1\dx t\, t^{b_2-1}(1-t)^{b_1-1} \, 
\ghy\left(1,b_1+b_2,2,(1-t)x+ty\right)\,.
\label{EqMB}
\end{align}
The Gauss function occurring in Eq.~(\ref{EqMB}) can be further simplified 
using symmetry under interchange of the first and second argument: 
\begin{align}
\ghy(1,b,2,x)=\ghy(b,1,2,x)=\int_0^1\dx t\, (1-xt)^{-b}
=\frac{1-\left(1-x\right)^{1-b}}{(1-b) x}\,.
\end{align}
Therefore
\eq 
\appell(1,b_1,b_2,2,x,y)&=&\frac{\Gamma(b_1+b_2)}{\Gamma(b_1)\Gamma(b_2)}
\int_0^1\dx t\, t^{b_2-1}(1-t)^{b_1-1}
\frac{1-(1-(1-t)x-ty))^{1-b_1-b_2}}{(1-b_1-b_2)[(1-t)x+ty]}
\nonumber\\[2mm]
&=&\frac{\Gamma(b_1+b_2)}{(1-b_1-b_2)\Gamma(b_1)\Gamma(b_2)}\biggl[
\frac{1}{x}\int_0^1\dx t\, t^{b_2-1}(1-t)^{b_1-1}
\left(1-\frac{x-y}{x}\,t\right)^{-1}
\nonumber\\[2mm]
&-&\frac{(1-x)^{1-b_1-b_2}}{x}\int_0^1\dx t\, t^{b_2-1}(1-t)^{b_1-1}
\, \left(1-\frac{x-y}{x}t\,\right)^{-1} 
\nonumber\\[2mm]
&\times&\left(1+\frac{x-y}{1-x}\,t\right)^{1-b_1-b_2}
\biggr] = \frac{1}{(1-b_1-b_2)x}
\biggl[\ghy\left(1,b_2,b_1+b_2,\frac{x-y}{x}\right) 
\nonumber\\[2mm]
&-&(1-x)^{1-b_1-b_2}\appell\left(b_2,1,b_1+b_2-1,b_1+b_2,
\frac{x-y}{x},\frac{y-x}{1-x}\right)
\biggr]\,,
\label{eq:Appell_fun_unred}
\en
where in the last line we recognized again the integral representations 
of Gauss and Appell hypergeometric functions.

The Appell function in~(\ref{eq:Appell_fun_unred}) matches the reduction 
formula (see Eq.~(1) in section 5.10 in Ref.~\cite{Erdelyi:1981ea}); 
\begin{align}
\appell(a,b_1,b_2,b_1+b_2,x,y)=(1-y)^{-a}\,\ghy
\left(a,b_1,b_1+b_2,\frac{x-y}{1-y}\right)\,.
\label{eq:Appell_red_f}
\end{align}
Using~(\ref{eq:Appell_red_f}) in (\ref{eq:Appell_fun_unred}) 
yields our desired result
\begin{align}
\appell(1,b_1,b_2,2,x,y)&=\frac{1}{(1-b_1-b_2)x}
\left[\ghy\left(1,b_2,b_1+b_2,\frac{x-y}{x}\right)\right.
\nonumber\\
&\quad\left.-(1-x)^{1-b_1-b_2}\left(1-\frac{y-x}{1-x}\right)^{-b_2}
\,\ghy\left(b_2,1,b_1+b_2,\frac{\frac{x-y}{x}
	-\frac{y-x}{1-x}}{1-\frac{y-x}{1-x}}\right)\right]
\nonumber\\
&=\frac{1}{(1-b_1-b_2)x}\left[\ghy\left(1,b_2,b_1+b_2,
\frac{x-y}{x}\right)\right.
\nonumber\\
&\quad\left.-(1-x)^{1-b_1}\left(1-y\right)^{-b_2}
\,\ghy\left(1,b_2,b_1+b_2,\frac{x-y}{x(1-y)}\right)\right]\,,
\end{align}
which we used in the main text (compare Eq.~\ref{ID_Appel_2Gauss}).

\hspace*{.25cm}
\subsection{Vector box integrals}
\label{Vector_Box_Functions}

Here we list the scalar functions occurring in the expansion 
of the vector box integrals in the orthogonal basis $(P,R,T)$. 
First, we specify fractioning identities needed to 
reduce tensorial box diagrams: 

Set of identities useful for simplifying the 
$D_1(s,u,Q^2)$ integral 
\eq 
2kP &=& \Delta_1 - \Delta_2 \,, \nonumber\\
2kR &=& - 2 \Delta_0 + \Delta_1 + \Delta_2\,, \\
2kT &=& \Delta_1 - \Delta_5 + u + kP \, \frac{u+t}{s} 
+ kR \, \frac{t-u}{s} \,. \nonumber
\en 

Set of identities useful for simplifying the 
$D_1(s,t,Q^2)$ integral 
\eq 
2kP &=& \Delta_1 - \Delta_2 \,, \nonumber\\
2kR &=& - 2 \Delta_0 + \Delta_1 + \Delta_2\,, \\ 
2kT &=& - \Delta_2 + \Delta_6 - t + kP \, \frac{u+t}{s} 
+ kR \, \frac{t-u}{s} \,. \nonumber
\en 

Set of the identities useful for the simplifying the 
$D_1(t,u,Q^2)$ integral 
\eq 
2kP &=&  \Delta_0 - \Delta_2 - \Delta_3 + \Delta_5 - u \,, 
\nonumber\\  
2kR &=& -\Delta_0 + \Delta_2 - \Delta_3 + \Delta_5 - u \,, \\ 
2kT &=& \Delta_0 - \Delta_3 + kP \, \frac{u+t}{s} 
+ kR \, \frac{t-u}{s} \,. \nonumber
\en 
Here we use $p_1^2 = p_2^2 = k_1^2 = 0$. 

The scalar functions occurring in the expansion of vector box 
integrals read 
\eq
D_{1P}^{0125}(s,u,Q^2) 
&=& \frac{P_\mu }{P^2} \, 
D_1^{\mu; 0125}(p_1,p_2,k_1) 
\nonumber\\
&=&  \frac{1}{2 s} \, 
\biggl[C_0(u,Q^2) - C_0(u)\biggr] 
\,, \nonumber\\
D_{1R}^{0125}(s,u,Q^2) 
&=& \frac{R_\mu}{R^2} \, 
D_1^{\mu; 0125}(p_1,p_2,k_1) 
\nonumber\\
&=& \frac{1}{2 s} \, 
\biggl[- C_0(u,Q^2) - C_0(u) + 2 C_0(s,Q^2)\biggr] 
\,, \nonumber\\
D_{1T}^{0125}(s,u,Q^2) 
&=& \frac{T_\mu}{T^2} \, 
D_1^{\mu; 0125}(p_1,p_2,k_1) 
\nonumber\\
&=& 
\frac{s}{2 u t} \, 
\biggl[ C_0(s) - C_0(u,Q^2)  
- u \, D_0(s,u,Q^2)    
\nonumber\\
&-&   (u+t) \, D_{1P}^{0125}(s,u,Q^2) 
    - (u-t) \, D_{1R}^{0125}(s,u,Q^2)  
\biggr] 
\nonumber\\
&=& \frac{1}{2 u t} \, 
\biggl[ 
    (t-u) \, C_0(s,Q^2) 
- (Q^2-u) \, C_0(u,Q^2) 
\nonumber\\
&+& s \, C_0(s) 
  + u \, C_0(u) 
- s u \, D_0(s,u,Q^2) 
\biggr] \,. 
\en 
\eq 
D_{1P}^{0126}(s,t,Q^2) &=& \frac{P_\mu}{P^2} \, 
D_1^{\mu; 0126}(p_1,p_2,k_1) = \frac{1}{2 s} \, 
\biggl[C_0(t) - C_0(t,Q^2)\biggr] = - D_{1P}^{0125}(s,t,Q^2)
\,, \nonumber\\
D_{1R}^{0126}(s,t,Q^2) &=& \frac{R_\mu}{R^2} \, 
D_1^{\mu; 0126}(p_1,p_2,k_1) = \frac{1}{2 s} \, 
\biggl[- C_0(t) - C_0(t,Q^2) + 2 C_0(s,Q^2) \biggr] 
\nonumber\\
&=& +  D_{1R}^{0125}(s,t,Q^2)
\,, \nonumber\\
D_{1T}^{0126}(s,t,Q^2) &=& \frac{T_\mu}{T^2} \, D_1^\mu(s,t,Q^2) 
\nonumber\\
&=& \frac{s}{2 u t} \, 
\biggl[- C_0(s) + C_0(t,Q^2) 
+ t \, D_0(s,t,Q^2) 
\nonumber\\
&+& (t+u) \, D_{1P}^{0125}(s,t,Q^2) 
  + (t-u) \, D_{1R}^{0125}(s,u,Q^2) \biggr] 
\nonumber\\
&=& - \frac{1}{2 u t} \, 
\biggl[ 
  (u-t)   \, C_0(s,Q^2) 
- (Q^2-t) \, C_0(t,Q^2) 
\nonumber\\
&+& s \, C_0(s) 
  + t \, C_0(t) 
  - s t \,  D_0(s,t,Q^2) 
\biggr] = - D_{1T}^{0125}(s,t,Q^2)
\,. 
\en 
\eq 
D_{1P}^{0235}(t,u,Q^2) 
&=& \frac{P_\mu}{P^2} \, 
D_1^{\mu; 0235}(p_1,p_2,k_1) 
\nonumber\\
&=& \frac{1}{2 s} \, 
\biggl[ C_0(t,Q^2) 
      - C_0(u,Q^2) 
      - C_0(u) 
      + C_0(t) 
      - u \, D_0(t,u,Q^2) \biggr] 
\,, \nonumber\\
D_{1R}^{0235}(t,u,Q^2) 
&=& \frac{R_\mu}{R^2} \, 
D_1^{\mu; 0235}(p_1,p_2,k_1) 
\nonumber\\
&=& \frac{1}{2 s} \, 
\biggl[ C_0(t,Q^2)
      + C_0(u,Q^2)  
      - C_0(u) 
      - C_0(t) 
+ u \, D_0(t,u,Q^2) \biggr] \,, 
\nonumber
\en 
\eq 
D_{1T}^{0235}(t,u,Q^2) 
&=& \frac{T_\mu}{T^2} \, 
D_1^{\mu; 0235}(p_1,p_2,k_1) 
\nonumber\\
&=& \frac{s}{2 u t} \, 
\biggl[ - C_0(t,Q^2) 
        + C_0(u,Q^2) 
        - (u+t) \, D_{1P}^{0235}(t,u,Q^2) 
\nonumber\\
&-&       (u-t) \, D_{1R}^{0235}(t,u,Q^2) 
\biggr] = \frac{1}{2 u t} \, 
\biggl[ 
  (Q^2-u) \, C_0(u,Q^2) 
\nonumber\\
&-& (Q^2-t) \, C_0(t,Q^2) 
  + u \, C_0(u) 
  - t \, C_0(t) 
  + u t \, D_0(t,u,Q^2) 
\biggr]
\,. 
\en  
\noindent
Relations: 
\eq 
\left\{   
\begin{array}{lccl}
D_{1P/1T}^{0125}(s,u,Q^2) &=& -& D_{1P/1T}^{0126}(s,u,Q^2) \,, \nonumber\\[1mm]
D_{1R}^{0125}(s,u,Q^2)    &=& +& D_{1R}^{0126}(s,u,Q^2) \,, \nonumber\\[1mm]
D_{1P}^{0125/0126}(s,u,Q^2)    &=& \pm & D_{1R}^{0125/0126}(s,u,Q^2) 
\mp \dfrac{C_0(s,Q^2) - C_0(u,Q^2)}{s} 
\,, \nonumber\\[1mm]
D_{1P}^{0235}(t,u,Q^2) &=& -& D_{1R}^{0235}(t,u,Q^2) 
+ \dfrac{C_0(t,Q^2) - C_0(u)}{s} \, 
\,. \nonumber
\end{array}
\right.
\en 

\vspace*{.25cm}
\subsection{Tensor box integrals}
\label{Tensor_rank23_Box} 

Here we list scalar functions occurring in  
expansion of tensor box integrals. 

\vspace*{.2cm}
\noindent 
\textbf{Integral $D_2^{\mu\nu; 0125}(p_1,p_2,k_1)$}
\eq\label{D2suQ2_list} 
D_{2PP}^{0125}(s,u,Q^2) &=& 
\frac{P_\mu P_\nu}{P^4} \, 
D_2^{\mu\nu; 0125}(p_1,p_2,k_1) 
= \frac{1}{2s} \, \biggl[
  C_{1P}^{025}(s,u,Q^2)  
- C_{1P}^{015}(s,u,Q^2)\biggr]\,, 
\nonumber\\
D_{2PR}^{0125}(s,u,Q^2)  &=& 
\frac{P_\mu R_\nu}{P^2 R^2} \, 
D_2^{\mu\nu; 0125}(p_1,p_2,k_1) 
= \frac{1}{2s} \, \biggl[
  C_{1R}^{025}(s,u,Q^2) 
- C_{1R}^{015}(s,u,Q^2) \biggr]\,, 
\nonumber\\
D_{2PT}^{0125}(s,u,Q^2)  &=& 
\frac{P_\mu T_\nu}{P^2 T^2} \, 
D_2^{\mu\nu; 0125}(p_1,p_2,k_1) 
= \frac{1}{2s} \, \biggl[
  C_{1T}^{025}(u,Q^2) 
- C_{1T}^{015}(u)\biggr]\,, 
\nonumber\\
D_{2RR}^{0125}(s,u,Q^2)  &=& 
\frac{R_\mu R_\nu}{R^4} \, 
D_2^{\mu\nu; 0125}(p_1,p_2,k_1) 
= \frac{1}{2s} \, \biggl[
- C_{1R}^{025}(s,u,Q^2)  
- C_{1R}^{015}(s,u,Q^2)  
\nonumber\\
&+& 2 C_{1R}^{078}(s,u,Q^2)  
    - C_0(s,Q^2) \biggr]\,, 
\nonumber\\
D_{2RT}^{0125}(s,u,Q^2)  &=& 
\frac{R_\mu T_\nu}{R^2 T^2} \, 
D_2^{\mu\nu; 0125}(p_1,p_2,k_1) 
= \frac{1}{2s} \, \biggl[
- C_{1T}^{025}(u,Q^2) 
- C_{1T}^{015}(u)  
\nonumber\\
&+& 2 C_{1T}^{078}(s,Q^2) \biggr]\,, 
\nonumber   
\en 
\eq 
D_{2TT}^{0125}(s,u,Q^2)  &=& 
\frac{T_\mu T_\nu}{T^4} \, 
D_2^{\mu\nu; 0125}(p_1,p_2,k_1) 
= \frac{s}{2ut} \, \biggl[ 
  \frac{u}{s} \, C_{1T}^{015}(u) 
- \frac{s+t}{s} \, C_{1T}^{025}(u,Q^2) 
\nonumber\\
&-& \frac{u-t}{s} \, C_{1T}^{078}(s,Q^2) 
 - u \, D_{1T}^{0125}(s,u,Q^2)  
\biggr]\,, 
\nonumber\\
D_{2g}^{0125}(s,u,Q^2)  &=& 
\frac{g_{\mu\nu}^\perp}{D-3} \, 
D_2^{\mu\nu; 0125}(p_1,p_2,k_1) 
= \frac{1}{D-3} \, \biggl[C_0(s,Q^2) 
- s \, D_{2PP}^{0125}(s,u,Q^2) 
\nonumber\\
&+& s \, D_{2RR}^{0125}(s,u,Q^2) 
+ \frac{ut}{s} \, D_{2TT}^{0125}(s,u,Q^2)  
\biggr] \,.
\en 

\vspace*{.25cm} 
\noindent 
\textbf{Integral  $D_2^{\mu\nu; 0126}(p_1,p_2,k_1)$} 
\eq\label{D2stQ2_list} 
D_{2PP}^{0126}(s,t,Q^2)  &=& 
\frac{P_\mu P_\nu}{P^4} \, 
D_2^{\mu\nu; 0126}(p_1,p_2,k_1) 
= \frac{1}{2s} \, \biggl[
  C_{1P}^{026}(s,t,Q^2) 
- C_{1P}^{016}(s,t,Q^2) 
\biggr] 
\nonumber\\
&=& + D_{2PP}^{0125}(s,t,Q^2) \,, 
\nonumber\\
D_{2PR}^{0126}(s,t,Q^2)   &=& 
\frac{P_\mu R_\nu}{P^2 R^2} \, 
D_2^{\mu\nu; 0126}(p_1,p_2,k_1) 
= \frac{1}{2s} \, \biggl[
  C_{1R}^{026}(s,t,Q^2)  
- C_{1R}^{016}(s,t,Q^2) \biggr]
\nonumber\\
&=& - D_{2PP}^{0125}(s,t,Q^2) \,, 
\nonumber\\
D_{2PT}^{0126}(s,t,Q^2)   &=& 
\frac{P_\mu T_\nu}{P^2 T^2} \, 
D_2^{\mu\nu; 0126}(p_1,p_2,k_1) 
= \frac{1}{2s} \, \biggl[
  C_{1T}^{026}(t)
- C_{1T}^{016}(t,Q^2) \biggr]
\nonumber\\
&=& + D_{2PT}^{0125}(s,t,Q^2) \,, 
\nonumber\\
D_{2RR}^{0126}(s,t,Q^2)   &=& 
\frac{R_\mu R_\nu}{R^4} \, 
D_2^{\mu\nu; 0126}(p_1,p_2,k_1) 
= \frac{1}{2s} \, \biggl[
- C_{1R}^{026}(s,t,Q^2)  
- C_{1R}^{016}(s,t,Q^2)   
\nonumber\\
&+& 2 C_{1R}^{079}(s,t,Q^2)   
- C_0(s,Q^2) \biggr] 
= + D_{2RR}^{0125}(s,t,Q^2) \,, 
\nonumber\\
D_{2RT}^{0126}(s,t,Q^2)   &=& 
\frac{R_\mu T_\nu}{R^2 T^2} \, 
D_2^{\mu\nu; 0126}(p_1,p_2,k_1) 
= \frac{1}{2s} \, \biggl[
- C_{1T}^{026}(t,Q^2) 
- C_{1T}^{016}(t) 
\nonumber\\
&+& 2 C_{1T}^{079}(s,Q^2) \biggr] 
= - D_{2RT}^{0125}(s,t,Q^2) \,, 
\nonumber\\
D_{2TT}^{0126}(s,t,Q^2)   &=& 
\frac{T_\mu T_\nu}{T^4} \, 
D_2^{\mu\nu; 0126}(p_1,p_2,k_1) 
= \frac{s}{2ut} \, \biggl[
- \frac{t}{s} \, C_{1T}^{026}(t) 
+ \frac{s+u}{s} \, C_{1T}^{016}(t,Q^2) 
\nonumber\\
&-& \frac{u-t}{s} \, C_{1T}^{079}(s,Q^2) 
+ t \, D_{1T}^{0126}(s,t,Q^2) 
\biggr] = + D_{2TT}^{0125}(s,t,Q^2) \,, 
\nonumber\\
D_{2g}^{0126}(s,t,Q^2)   &=& 
\frac{g_{\mu\nu}^\perp}{D-3} \, 
D_2^{\mu\nu; 0126}(p_1,p_2,k_1) 
\nonumber\\
&=& \frac{1}{D-3} \, \biggl[C_0(s,Q^2)  
- s \, D_{2PP}^{0126}(s,t,Q^2) 
+ s \, D_{2RR}^{0126}(s,t,Q^2) 
\nonumber\\
&+& \frac{ut}{s} \, D_{2TT}^{0126}(s,t,Q^2)  
\biggr] = + D_{2g}^{0125}(s,t,Q^2) \,. 
\en 

\noindent 
\textbf{Integral $D_2^{\mu\nu; 0235}(p_1,p_2,k_1)$} 
\eq
D_{2PP}^{0235}(t,u,Q^2) &=& 
\frac{P_\mu P_\nu}{P^4} \, 
D_2^{\mu\nu; 0235}(p_1,p_2,k_1) 
= \frac{1}{2s} \, 
\biggl[ \frac{C_0(t,Q^2)}{2} 
            - C_{1P}^{046}(s,t,Q^2) 
\nonumber\\
&+& C_{1P}^{023}(s,t,Q^2) 
- C_{1P}^{025}(s,u,Q^2) 
- C_{1P}^{035}(s,u,Q^2) - u D_{1P}^{0235}(t,u,Q^2)
\biggr]\,, 
\nonumber\\
D_{2PR}^{0235}(t,u,Q^2)  &=& 
\frac{P_\mu R_\nu}{P^2 R^2} \, 
D_2^{\mu\nu; 0235}(p_1,p_2,k_1) 
= \frac{1}{2s} \, 
\biggl[ - \frac{C_0(t,Q^2)}{2} 
              - C_{1R}^{046}(s,t,Q^2) 
\nonumber\\
&+& C_{1R}^{023}(s,t,Q^2) 
- C_{1R}^{025}(s,u,Q^2) 
- C_{1R}^{035}(s,u,Q^2) 
- u D_{1R}^{0235}(t,u,Q^2) 
\biggr]\,, 
\nonumber
\en
\eq
D_{2PT}^{0235}(t,u,Q^2)  &=& 
\frac{P_\mu T_\nu}{P^2 T^2} \, 
D_2^{\mu\nu; 0235}(p_1,p_2,k_1) 
= \frac{1}{2s} \, 
\biggl[ - C_{1T}^{046}(t,Q^2) 
+ C_{1T}^{023}(t) 
\nonumber\\
&-& C_{1T}^{025}(u,Q^2) 
- C_{1T}^{035}(u) 
- u D_{1T}^{0235}(t,u,Q^2) 
\biggr]\,, 
\nonumber\\
D_{2RR}^{0235}(t,u,Q^2)  &=& 
\frac{R_\mu R_\nu}{R^4} \, 
D_2^{\mu\nu; 0235}(p_1,p_2,k_1) 
= \frac{1}{2s} \, \biggl[ 
- \frac{C_0(t,Q^2)}{2} 
- C_{1R}^{046}(s,t,Q^2)
\nonumber\\
&-& C_{1R}^{023}(s,t,Q^2) 
+ C_{1R}^{025}(s,u,Q^2) 
- C_{1R}^{035}(s,u,Q^2) 
+ u D_{1R}^{0235}(t,u,Q^2) 
\biggr]\,, 
\nonumber\\
D_{2RT}^{0235}(t,u,Q^2)  &=& 
\frac{R_\mu T_\nu}{R^2 T^2} \, 
D_2^{\mu\nu; 0235}(p_1,p_2,k_1) 
= \frac{1}{2s} \, \biggl[
- C_{1T}^{046}(t,Q^2) 
- C_{1T}^{023}(t) 
+ C_{1T}^{025}(u,Q^2) 
\nonumber\\
&-& C_{1T}^{035}(u) 
+ u D_{1T}^{0235}(t,u,Q^2) 
\biggr]\,, 
\nonumber\\
D_{2TT}^{0235}(t,u,Q^2)  &=& 
\frac{T_\mu T_\nu}{T^4} \, 
D_2^{\mu\nu; 0235}(p_1,p_2,k_1) 
= \frac{s}{2ut} \, \biggl[ 
  \frac{s+u}{s} \, C_{1T}^{046}(t,Q^2) 
+ \frac{s+t}{s} \, C_{1T}^{025}(u,Q^2) 
\nonumber\\
&+& \frac{u}{s} \, C_{1T}^{035}(u) 
  - \frac{t}{s} \, C_{1T}^{023}(t) 
+ \frac{ut}{s} D_{1T}^{0235}(t,u,Q^2)
\biggr]\,, 
\nonumber\\
D_{2g}^{0235}(t,u,Q^2)  &=& 
\frac{g_{\mu\nu}^\perp}{D-3} \, 
D_2^{\mu\nu; 0235}(p_1,p_2,k_1) 
= \frac{1}{D-3} \, \biggl[C_0^{046}(t,Q^2)  
- s \, D_{2PP}^{0235}(t,u,Q^2) 
\nonumber\\
&+& s \, D_{2RR}^{0235}(t,u,Q^2) 
+ \frac{ut}{s} \, D_{2TT}^{0235}(t,u,Q^2)  
\biggr] \,.
\label{D2tuQ2_list}   
\en 

\noindent 
Relations: 
\eq 
\left\{   
\begin{array}{lccl}
D_{i}^{0125}(s,u,Q^2) &=& +& D_{i}^{0126}(s,u,Q^2) \,, 
\quad i = 2g, 2PP, 2RR, 2TT, 2PT \nonumber\\[1mm]
D_{i}^{0125}(s,u,Q^2) &=& -& D_{i}^{0126}(s,u,Q^2) \,, 
\quad i = 2PR, 2TR \nonumber\\
\end{array}
\right.
\en 

\vspace*{.25cm}
Functions occurring in the 
expansion of tensor rank-3 box 
integrals~(\ref{Box_Tensor3_1})-(\ref{Box_Tensor3_3}) read: 

\vspace*{.5cm}
\noindent 
\textbf{Integral $D_3^{\mu\nu\alpha}(s,u,Q^2)$} 
\eq
D_{3PPP}^{0125}(s,u,Q^2) 
&=& 
\frac{P_\mu P_\nu P_\alpha}{P^6} \, 
D_3^{\mu\nu\alpha}(s,u,Q^2)  
= \frac{1}{2s} \, \biggl[
  C_{2PP}^{025}(s,u,Q^2) 
- C_{2PP}^{015}(s,u,Q^2) \biggr]\,, 
\nonumber\\[1mm]
D_{3PPR}^{0125}(s,u,Q^2)  &=& 
\frac{P_\mu P_\nu R_\alpha}{P^4 R^2} \, 
D_3^{\mu\nu\alpha}(s,u,Q^2)  
= \frac{1}{2s} \, \biggl[
  C_{2PR}^{025}(s,u,Q^2)  
- C_{2PR}^{015}(s,u,Q^2) \biggr]\,, 
\nonumber\\[1mm]
D_{3PPT}^{0125}(s,u,Q^2)  &=& 
\frac{P_\mu P_\nu T_\alpha}{P^4 T^2} \, 
D_3^{\mu\nu\alpha}(s,u,Q^2)  
= \frac{1}{2s} \, \biggl[
  C_{2PT}^{025}(s,u,Q^2) 
- C_{2PT}^{015}(s,u,Q^2) \biggr]\,, 
\nonumber\\[1mm]
D_{3RRP}^{0125}(s,u,Q^2)  &=& 
\frac{P_\mu R_\nu R_\alpha}{R^4 P^2} \, 
D_3^{\mu\nu\alpha}(s,u,Q^2)  
= \frac{1}{2s} \, \biggl[
  C_{2RR}^{025}(s,u,Q^2)  
- C_{2RR}^{015}(s,u,Q^2)  \biggr]\,, 
\nonumber\\[1mm]
D_{3TTP}^{0125}(s,u,Q^2)  &=& 
\frac{P_\mu T_\nu T_\alpha}{T^4 P^2} \, 
D_3^{\mu\nu\alpha}(s,u,Q^2)  
= \frac{1}{2s} \, \biggl[
  C_{2TT}^{025}(s,u,Q^2)  
- C_{2TT}^{015}(s,u,Q^2)  
\biggr]\,, 
\nonumber\\[1mm]
D_{3PRT}^{0125}(s,u,Q^2)  &=& 
\frac{P_\mu R_\nu T_\alpha}{P^2 R^2 T^2} \, 
D_3^{\mu\nu\alpha}(s,u,Q^2)  
= \frac{1}{2s} \, \biggl[
  C_{2RT}^{025}(s,u,Q^2)  
- C_{2RT}^{015}(s,u,Q^2)  
\biggr]\,, 
\nonumber
\en
\eq
D_{3RRR}^{0125}(s,u,Q^2)  &=& 
\frac{R_\mu R_\nu R_\alpha}{R^6} \, 
D_3^{\mu\nu\alpha}(s,u,Q^2)
\nonumber\\[1mm]  
&=& \frac{1}{2s} \, \biggl[
- C_{2RR}^{025}(s,u,Q^2)  
- C_{2RR}^{015}(s,u,Q^2)  
+ \frac{C_0^{078}(s,Q^2)}{2} 
\nonumber\\[1mm]
&-& 2C_{1R}^{078}(s,u,Q^2)  
+ 2C_{2RR}^{078}(s,u,Q^2)   
\biggr]\,, 
\nonumber\\[1mm]
D_{3RRT}^{0125}(s,u,Q^2)  &=& 
\frac{R_\mu R_\nu T_\alpha}{R^4 T^2} \, 
D_3^{\mu\nu\alpha}(s,u,Q^2)  
\nonumber\\[1mm]  
&=& \frac{1}{2s} \, \biggl[
- C_{2RT}^{025}(s,u,Q^2)  
- C_{2RT}^{015}(s,u,Q^2)   
\nonumber\\
&+& 2 C_{2RT}^{078}(s,u,Q^2)   
- C_{1T}^{078}(s,Q^2)   
\biggr]\,, \nonumber\\[1mm]
D_{3TTR}^{0125}(s,u,Q^2)  &=& 
\frac{R_\mu T_\nu T_\alpha}{T^4 R^2} \, 
D_3^{\mu\nu\alpha}(s,u,Q^2)  
\nonumber\\[1mm]
&=& \frac{1}{2s} \, \biggl[
- C_{2TT}^{025}(s,u,Q^2)   
- C_{2TT}^{015}(s,u,Q^2)   
+ 2 C_{2TT}^{078}(s,u,Q^2)    
\biggr]\,, 
\nonumber\\[1mm]
D_{3TTT}^{0125}(s,u,Q^2)  &=& 
\frac{T_\mu T_\nu T_\alpha}{T^6} \, 
D_3^{\mu\nu\alpha}(s,u,Q^2)  
\nonumber\\
&=& \frac{s}{2ut} \, \biggl[ 
- u D_{2TT}^{0125}(s,u,Q^2)   
+ C_{2TT}^{012}(s) 
+ \frac{u}{s}   \, C_{2TT}^{015}(s,u,Q^2)   
\nonumber\\[1mm]
&-& \frac{s+t}{s} \, C_{2TT}^{025}(s,u,Q^2)   
- \frac{u-t}{s} \, C_{2TT}^{078}(s,u,Q^2)   
\biggr]\,, 
\nonumber\\[1mm]
D_{3gP}^{0125}(s,u,Q^2)  &=& 
\frac{g_{\mu\nu}^\perp P_\alpha}{(D-3) P^2} \, 
D_3^{\mu\nu\alpha}(s,u,Q^2)  
\nonumber\\[1mm]
&=& \frac{1}{D-3} \, \biggl[ 
  \frac{C_0(s,Q^2)}{2} 
+ C_{1P}^{078}(s,u,Q^2) 
+ s \, D_{3P}^{0125}(s,u,Q^2)  
\biggr] \,,
\nonumber\\[1mm]
D_{3gR}^{0125}(s,u,Q^2)  &=& 
\frac{g_{\mu\nu}^\perp R_\alpha}{(D-3) R^2} \, 
D_3^{\mu\nu\alpha}(s,u,Q^2)  
\nonumber\\[1mm]
&=& \frac{1}{D-3} \, \biggl[ 
- \frac{C_0(s,Q^2)}{2} 
+ C_{1R}^{078}(s,u,Q^2) 
+ s \, D_{3R}^{0125}(s,u,Q^2)  
\biggr] \,, 
\nonumber\\[1mm]
D_{3gT}^{0125}(s,u,Q^2)  &=& 
\frac{g_{\mu\nu}^\perp T_\alpha}{(D-3) T^2} \, 
D_3^{\mu\nu\alpha}(s,u,Q^2)  
\nonumber\\[1mm]
&=&\frac{1}{D-3} \, \biggl[ 
  C_{1T}^{078}(s,Q^2)  
+ s \, D_{3T}^{0125}(s,u,Q^2)  
\biggr] \,. 
\label{D3suQ2_list} 
\en 
Here and in the following 
\eq 
D_{3L}^{0ijk}(s,u,Q^2)  &=& 
- D_{3LPP}^{0ijk}(s,u,Q^2)  
+ D_{3LRR}^{0ijk}(s,u,Q^2)  
+ \frac{ut}{s^2}D_{3LTT}^{0ijk}(s,u,Q^2)  \,, 
\nonumber\\
L &=& P, R, T \,.
\en 

\clearpage 

\noindent 
\textbf{Integral $D_3^{\mu\nu\alpha}(s,t,Q^2)$} 
\eq
D_{3PPP}^{0126}(s,t,Q^2)  &=& 
\frac{P_\mu P_\nu P_\alpha}{P^6} \, 
D_3^{\mu\nu\alpha}(s,t,Q^2)  
= \frac{1}{2s} \, \biggl[
  C_{2PP}^{026}(s,t,Q^2) 
- C_{2PP}^{016}(s,t,Q^2) \biggr]\,, 
\nonumber\\[1mm]
D_{3PPR}^{0126}(s,t,Q^2) &=& 
\frac{P_\mu P_\nu R_\alpha}{P^4 R^2} \, 
D_3^{\mu\nu\alpha}(s,t,Q^2)  
= \frac{1}{2s} \, \biggl[
  C_{2PR}^{026}(s,t,Q^2)  
- C_{2PR}^{016}(s,t,Q^2) \biggr]\,, 
\nonumber\\[1mm]
D_{3PPT}^{0126}(s,t,Q^2) &=& 
\frac{P_\mu P_\nu T_\alpha}{P^4 T^2} \, 
D_3^{\mu\nu\alpha}(s,t,Q^2)  
= \frac{1}{2s} \, \biggl[
  C_{2PT}^{026}(s,t,Q^2)  
- C_{2PT}^{016}(s,t,Q^2) \biggr]\,, 
\nonumber\\[1mm]
D_{3RRP}^{0126}(s,t,Q^2) &=& 
\frac{P_\mu R_\nu R_\alpha}{R^4 P^2} \, 
D_3^{\mu\nu\alpha}(s,t,Q^2)  
= \frac{1}{2s} \, \biggl[
  C_{2RR}^{026}(s,t,Q^2)  
- C_{2RR}^{016}(s,t,Q^2)  \biggr]\,, 
\nonumber\\[1mm]
D_{3TTP}^{0126}(s,t,Q^2) &=& 
\frac{P_\mu T_\nu T_\alpha}{T^4 P^2} \, 
D_3^{\mu\nu\alpha}(s,t,Q^2)  
= \frac{1}{2s} \, \biggl[
  C_{2TT}^{026}(s,t,Q^2)  
- C_{2TT}^{016}(s,t,Q^2)  
\biggr]\,, 
\nonumber\\[1mm]
D_{3PRT}^{0126}(s,t,Q^2) &=& 
\frac{P_\mu R_\nu T_\alpha}{P^2 R^2 T^2} \, 
D_3^{\mu\nu\alpha}(s,z,Q^2)  
= \frac{1}{2s} \, \biggl[
  C_{2RT}^{026}(s,t,Q^2)  
- C_{2RT}^{016}(s,t,Q^2)  
\biggr]\,, 
\nonumber\\[1mm]
D_{3RRR}^{0126}(s,t,Q^2) &=& 
\frac{R_\mu R_\nu R_\alpha}{R^6} \, 
D_3^{\mu\nu\alpha}(s,t,Q^2)  
= \frac{1}{2s} \, \biggl[
- C_{2RR}^{026}(s,t,Q^2)  
- C_{2RR}^{016}(s,t,Q^2)  
\nonumber\\[1mm]
&+&  \frac{C_0(s,Q^2)}{2}
- 2 C_{1R}^{079}(s,t,Q^2) 
+ 2 C_{2RR}^{079}(s,t,Q^2)  
\biggr]\,, 
\nonumber\\[1mm]
D_{3RRT}^{0126}(s,t,Q^2) &=& 
\frac{R_\mu R_\nu T_\alpha}{R^4 T^2} \, 
D_3^{\mu\nu\alpha}(s,t,Q^2)  
\nonumber\\[1mm]
&=& \frac{1}{2s} \, \biggl[ 
- C_{2RT}^{026}(s,t,Q^2)  
- C_{2RT}^{016}(s,t,Q^2)  
\nonumber\\
&+& 2 C_{2RT}^{079}(s,t,Q^2) 
- C_{1T}^{079}(t,Q^2)  
\biggr]\,, \nonumber\\[1mm]
D_{3TTR}^{0126}(s,t,Q^2) &=& 
\frac{R_\mu T_\nu T_\alpha}{T^4 R^2} \, 
D_3^{\mu\nu\alpha}(s,t,Q^2)  
\nonumber\\[1mm]
&=& \frac{1}{2s} \, \biggl[
- C_{2TT}^{026}(s,t,Q^2) 
- C_{2TT}^{016}(s,t,Q^2)  
+ 2 C_{2TT}^{079}(s,t,Q^2)   
\biggr]\,, 
\nonumber\\[1mm]
D_{3TTT}^{0126}(s,t,Q^2) &=& 
\frac{T_\mu T_\nu T_\alpha}{T^6} \, 
D_3^{\mu\nu\alpha}(s,t,Q^2)  
\nonumber\\[1mm]
&=& \frac{s}{2ut} \, \biggl[ 
  t \, D_{2TT}^{0126}(s,t,Q^2)   
- C_{2TT}^{012}(s)   
- \frac{t}{s}   \, C_{2TT}^{026}(s,t,Q^2)    
\nonumber\\[1mm]
&+& \frac{s+u}{s} \, C_{2TT}^{016}(s,t,Q^2)    
  + \frac{t-u}{s} \, C_{2TT}^{079}(s,t,Q^2)    
\biggr]\,, 
\nonumber\\
D_{3gP}^{0126}(s,t,Q^2) &=& 
\frac{g_{\mu\nu}^\perp P_\alpha}{(D-3) P^2} \, 
D_3^{\mu\nu\alpha}(s,t,Q^2) 
\nonumber\\[1mm]
&=& \frac{1}{D-3} \, \biggl[ 
  \frac{C_0(s,Q^2)}{2}   
+ C_{1P}^{079}(s,t,Q^2)    
+ s \, D_{3P}^{0126}(s,t,Q^2)    
\biggr] \,,
\nonumber
\en

\clearpage 

\eq
D_{3gR}^{0126}(s,t,Q^2) &=& 
\frac{g_{\mu\nu}^\perp R_\alpha}{(D-3) R^2} \, 
D_3^{\mu\nu\alpha}(s,t,Q^2) 
\nonumber\\[1mm]
&=& \frac{1}{D-3} \, \biggl[ 
- \frac{C_0(s,Q^2)}{2}   
+ C_{1R}^{079}(s,t,Q^2)    
+ s \, D_{3R}^{0126}(s,t,Q^2)    
\biggr] \,, 
\nonumber\\  
D_{3gT}^{0126}(s,t,Q^2) &=& 
\frac{g_{\mu\nu}^\perp T_\alpha}{(D-3) T^2} \, 
D_3^{\mu\nu\alpha}(s,t,Q^2) 
\nonumber\\[1mm]
&=& \frac{1}{D-3} \, \biggl[ 
  C_{1T}^{079}(t,Q^2)    
+ s \, D_{3T}^{0126}(s,t,Q^2)    
\biggr] \,.
\label{D3stQ2_list} 
\en
\noindent 
Relations: 
\eq 
\left\{   
\begin{array}{l}
D_{i}^{0125}(s,u,Q^2) = + D_{i}^{0126}(s,u,Q^2)\,, \\  
i = 3gR, 3PPR, 3TTR, 3PRT, 3RRR                    \\\nonumber\\
D_{i}^{0125}(s,u,Q^2) = - D_{i}^{0126}(s,u,Q^2)\,, \\ 
i = 3gP, 3gT, 3PPP, 3TTT, 3PPT, 3TTP, 3RRP, 3RRT   \\
\end{array}
\right.
\en 

\noindent 
\textbf{Integral $D_3^{\mu\nu\alpha}(t,u,Q^2)$} 
\eq 
D_{3PPP}^{0235}(t,u,Q^2) &=& 
\frac{P_\mu P_\nu P_\alpha}{P^6} \, 
D_3^{\mu\nu\alpha}(t,u,Q^2)  
\nonumber\\ 
&=& \frac{1}{2s} \, \biggl[
  C_{2PP}^{023}(s,t,Q^2) 
- C_{2PP}^{025}(s,u,Q^2)  
- C_{2PP}^{035}(s,u,Q^2)   
\nonumber\\
&+& \frac{C_0(t,Q^2)}{4}
- C_{1P}^{046}(s,t,Q^2)  
+ C_{2PP}^{046}(s,t,Q^2)  
- u D_{2PP}^{0235}(u,t,Q^2)   
\biggr]\,, 
\nonumber\\
D_{3PPR}^{0235}(t,u,Q^2)  &=& 
\frac{P_\mu P_\nu R_\alpha}{P^4 R^2} \, 
D_3^{\mu\nu\alpha}(t,u,Q^2)  \nonumber\\ 
&=& \frac{1}{2s} \, \biggl[
   C_{2PR}^{023}(s,t,Q^2)  
 - C_{2PR}^{025}(s,u,Q^2)   
 - C_{2PR}^{035}(s,t,Q^2)  
\nonumber\\
&-& \frac{C_0(t,Q^2)}{4} 
 + \frac{C_{1P}^{046}(s,t,Q^2)}{2} 
 - \frac{C_{1R}^{046}(s,t,Q^2)}{2} 
 + C_{2PR}^{046}(s,t,Q^2) 
\nonumber\\
&-& u D_{2PR}^{0235}(t,u,Q^2) 
\biggr]\,, 
\nonumber\\
D_{3PPT}^{0235}(t,u,Q^2)  &=& 
\frac{P_\mu P_\nu T_\alpha}{P^4 T^2} 
\, D_3^{\mu\nu\alpha}(t,u,Q^2)  
\nonumber\\
&=& \frac{1}{2s} \, \biggl[
  C_{2PT}^{023}(s,t,Q^2) 
- C_{2PT}^{025}(s,u,Q^2) 
- C_{2PT}^{035}(s,u,Q^2)  
\nonumber\\
&-& \frac{C_{1T}^{046}(s,t,Q^2)}{2} 
+ C_{2PT}^{046}(s,t,Q^2) 
- u D_{2PT}^{0235}(t,u,Q^2) 
\biggr]\,, 
\nonumber
\en 
\eq 
D_{3RRP}^{0235}(t,u,Q^2)  &=& 
\frac{P_\mu R_\nu R_\alpha}{R^4 P^2} \, D_3^{\mu\nu\alpha}(t,u,Q^2)  
\nonumber\\
&=& \frac{1}{2s} \, 
  \biggl[C_{2RR}^{023}(s,t,Q^2)  
       - C_{2RR}^{025}(s,u,Q^2) 
       - C_{2RR}^{035}(s,u,Q^2)  
\nonumber\\
&+& \frac{C_0(t,Q^2)}{4}  
      + C_{1R}^{046}(s,t,Q^2) 
      + C_{2RR}^{046}(s,t,Q^2) 
- u D_{2RR}^{0235}(t,u,Q^2)  
\biggr]\,, 
\nonumber\\
D_{3TTP}^{0235}(t,u,Q^2)  &=& 
\frac{P_\mu T_\nu T_\alpha}{T^4 P^2}  
 \, D_3^{\mu\nu\alpha}(t,u,Q^2)  
\nonumber\\
&=& \frac{1}{2s} \, \biggl[
  C_{2TT}^{023}(s,t,Q^2)  
- C_{2TT}^{025}(s,u,Q^2)  
- C_{2TT}^{035}(s,u,Q^2)  
\nonumber\\
&+& C_{2TT}^{046}(s,t,Q^2)  
- u D_{2TT}^{0235}(t,u,Q^2)   
\biggr]\,, 
\nonumber\\
D_{3PRT}^{0235}(t,u,Q^2)  &=& 
\frac{P_\mu R_\nu T_\alpha}{P^2 R^2 T^2} \, 
D_3^{\mu\nu\alpha}(t,u,Q^2) \nonumber\\  
&=& \frac{1}{2s} \, \biggl[
  C_{2RT}^{023}(s,t,Q^2)  
- C_{2RT}^{025}(s,u,Q^2)  
- C_{2RT}^{035}(s,u,Q^2)  
\nonumber\\
&+& \frac{C_{1T}^{046}(t,Q^2)}{2} 
+ C_{2RT}^{046}(s,t,Q^2)  
- u D_{2RT}^{0235}(t,u,Q^2)  
\biggr]\,, 
\nonumber\\ 
D_{3RRR}^{0235}(t,u,Q^2)  &=& 
\frac{R_\mu R_\nu R_\alpha}{R^6} \, 
D_3^{\mu\nu\alpha}(t,u,Q^2)  \nonumber\\
&=& \frac{1}{2s} \, \biggl[
- C_{2RR}^{023}(s,t,Q^2)  
+ C_{2RR}^{025}(s,u,Q^2)  
- C_{2RR}^{035}(s,u,Q^2)   
\nonumber\\
&+& \frac{C_0^{046}(t,Q^2)}{4} 
 + C_{1R}^{046}(s,t,Q^2) 
 + C_{2RR}^{046}(s,t,Q^2)  
+ u D_{2RR}^{0235}(t,u,Q^2)   
\biggr]\,, 
\nonumber\\
D_{3RRT}^{0235}(t,u,Q^2)  &=& 
\frac{R_\mu R_\nu T_\alpha}{R^4 T^2} \, 
D_3^{\mu\nu\alpha}(t,u,Q^2)  
\nonumber\\
&=& \frac{1}{2s} 
\, \biggl[- C_{2RT}^{023}(s,t,Q^2) 
 + C_{2RT}^{025}(s,u,Q^2)  
 - C_{2RT}^{035}(s,u,Q^2)   
\nonumber\\
&+& \frac{C_{1T}^{046}(t,Q^2)}{2} + 
 + C_{2RT}^{046}(s,t,Q^2)    
+ u D_{2RT}^{0235}(u,t,Q^2)    
\biggr]\,, 
\nonumber\\
D_{3TTR}^{0235}(t,u,Q^2)  &=& 
\frac{R_\mu T_\nu T_\alpha}{T^4 R^2} \, 
D_3^{\mu\nu\alpha}(t,u,Q^2)  
\nonumber\\
&=& \frac{1}{2s} \, 
\, \biggl[- C_{2TT}^{023}(s,t,Q^2)   
 + C_{2TT}^{025}(s,u,Q^2)    
- C_{2TT}^{035}(s,u,Q^2)    
\nonumber\\
&+& C_{2TT}^{046}(s,t,Q^2)    
+ u D_{2TT}^{0235}(u,t,Q^2)    
\biggr]\,, 
\nonumber\\
D_{3TTT}^{0235}(t,u,Q^2)  &=& 
\frac{T_\mu T_\nu T_\alpha}{T^6} \, 
D_3^{\mu\nu\alpha}(t,u,Q^2)  \nonumber\\
&=& \frac{1}{2 ut} \, \biggl[ 
- t C_{2TT}^{023}(s,t,Q^2)    
+ (s+t) \, C_{2TT}^{025}(s,u,Q^2)     
+ u \, C_{2TT}^{035}(s,u,Q^2)    
\nonumber\\
&-& (s+u) \, C_{2TT}^{046}(s,t,Q^2)    
+ D_{2TT}^{0235}(u,t,Q^2)     
\biggr]\,, 
\nonumber\\ 
D_{3gP}^{0235}(t,u,Q^2)  &=& 
\frac{g_{\mu\nu}^\perp P_\alpha}{(D-3) P^2} 
\, D_3^{\mu\nu\alpha}(t,u,Q^2) 
\nonumber\\
&=& \frac{1}{D-3} \, \biggl[ 
  \frac{C_0(t,Q^2)}{2}    
- C_{1P}^{046}(s,t,Q^2)     
+ s \, D_{3P}^{0235}(u,t,Q^2)      
\biggr] \,,
\nonumber
\en 
\eq 
D_{3gR}^{0235}(t,u,Q^2)  &=& 
\frac{g_{\mu\nu}^\perp R_\alpha}{(D-3) R^2} \, 
D_3^{\mu\nu\alpha}(s,t,Q^2)  
\nonumber\\
&=& \frac{1}{D-3} \, \biggl[ 
- \frac{C_0(t,Q^2)}{2} 
+ C_{1R}^{046}(s,t,Q^2)     
+ s \, D_{3R}^{0235}(u,t,Q^2)     
\biggr] \,,
\nonumber\\
D_{3gT}^{0235}(t,u,Q^2)  &=& 
\frac{g_{\mu\nu}^\perp T_\alpha}{(D-3) T^2} \, 
D_3^{\mu\nu\alpha}(s,t,Q^2)  
\nonumber\\
&=& \frac{1}{D-3} \, \biggl[ 
- C_{1T}^{046}(t,Q^2)     
+ s \, D_{3T}^{0235}(u,t,Q^2)      
\biggr] \,.
\label{D3tuQ2_list}
\en 
 
\vspace*{.25cm}

\section{Spherical integration in dimensional regularization}
\label{app:spher_int} 

The key tool to handle PSIs in dimensional regularization is a generalization of spherical 
integration to arbitrary dimension. In our consideration we follow handbook~\cite{Sterman:1994ce}. 
For any integer $n$-dimension we can introduce spherical coordinates.  For $n=2$ dimension we have
\begin{align}
\vec{x\,}^{(2)}=r_2 \vec{u\,}^{(2)}(\theta_1)
\quad\text{with}\quad
\vec{u\,}^{(2)}(\theta_1)=(\sin\theta_1,\,\,\cos\theta_1) \, ,
\end{align}
with module of vector $0\leq r_2<\infty$ and angle $0 \leq \theta_1 \leq  2\pi$. 
Here and in the following, we denote the $n$-dimensional length of a vector as $r_n$ and 
$\vec{u\,}^{(n)}(\theta_1,\ldots,\theta_{n-1})$ is 
a unit vector depending on $n-1$ angular coordinates $\theta_i$. 

Suppose that we know how to express the unit vector $\vec{u\,}^{(n-1)}(\theta_1,\ldots,\theta_{n-2})$ 
in terms of $n-2$ angular variables. 
Then we can write $\vec{x\,}^{(n)}$ in cylindrical coordinates
\begin{align}
\vec{x\,}^{(n)}=(r_{n-1}\vec{u\,}^{(n-1)}(\theta_1,\ldots,\theta_{n-2}),x_n)\,.
\end{align}
Changing variables to spherical ones by the relations
\eq 
r_{n-1}=r_n\sin\theta_{n-1}\,, \qquad 
x_n=r_n\cos\theta_{n-1}\,,
\label{eq:sphcoord cov}
\en 
where $0\leq r_n\leq\infty$ and $0\leq \theta_{n-1}<\pi$ gives
\begin{align}
\vec{x\,}^{(n)}
&=r_n \, (\sin\theta_{n-1}\vec{u\,}^{(n-1)}(\theta_1,\ldots,\theta_{n-2}),\cos\theta_{n-1}) 
\nonumber\\
&=r_n \, \vec{u\,}^{(n)}(\theta_1,\ldots,\theta_{n-1})\,.
\end{align}
With this procedure we have constructed spherical coordinates for arbitrary integer dimension 
$n\geq 2$.

Now we are facing the issue of how to evaluate integrals in $n$ dimensions without specifying~$n$. 
This will enable us to formally extend the resulting formula to non-integer~$n$. 
The central idea to do so is to first examine on which set of angular variables the integrand depends. 
For our purposes, it suffices to consider scalar integrands where the integration variable 
$\vec{x\,}^{(n)}$ only appears in scalar products of the form $\vec{x\,}^{(n)}\cdot\vec{y\,}^{(n)}$ 
with a fixed $n$-dimensional vector of the specific form $\vec{y\,}^{(n)}=(\vec{0\,}_{n-2},y_2,y_1)$ 
or as an absolute square $\vec{x\,}^{(n)}\cdot\vec{x\,}^{(n)}=r_n^2$.
We can expand $\vec{x\,}^{(n)}$ with our recursion in terms of angular variables, obtaining
\begin{align}
\vec{x\,}^{(n)}
&=r_n \vec{u\,}^{(n)}(\theta_1,\ldots,\theta_{n-1}) 
\nonumber\\
&=r_n \, (\sin\theta_{n-1}\vec{u}^{(n-1)}(\theta_1,\ldots,\theta_{n-2}),\cos\theta_{n-1}) 
\nonumber\\
&=r_n \, (\sin\theta_{n-1}\sin\theta_{n-2}\vec{u}^{(n-2)}(\theta_1,\ldots,\theta_{n-3}),
\sin\theta_{n-1}\cos\theta_{n-2},\cos\theta_{n-1})\,.
\end{align}
Therefore,
\begin{align}
\vec{x\,}^{(n)}\cdot\vec{y\,}^{(n)}
=r_n \, (y_2\sin\theta_{n-1}\cos\theta_{n-2}+y_1\cos\theta_{n-1})\,.
\end{align}
This implies that we can integrate over all angular variables except for $\theta_{n-2}$ and $\theta_{n-1}$ 
as we have integrands of the form $f(r_n,\theta_{n-1},\theta_{n-2})$.
The next step is to express the $\int\dx^n x$ integral in terms of our spherical coordinates with explicit 
$\theta_{n-2}$ and $\theta_{n-1}$ integrations. 

The integral in spherical coordinates in $n$-dimensions has the form 
\begin{align}
\int\dx^n x=\int_0^\infty\dx r_n r_n^{n-1}\int\dx\Omega_{n-1}\,.
\end{align}
Constructing $\int\dx\Omega_{n-1}$ iteratively using the spherical coordinates defined above yields
\begin{align}
\int\dx\Omega_{n-1}= \int_0^{2\pi} \dx\theta_1 \, 
\prod_{k=2}^{n-1} \int_0^\pi \dx\theta_k \, \sin^{k-1}\theta_{k} \,. 
\label{eq:n dim spherical integral}
\end{align}
The second part is
\eq 
\int_0^\pi \dx\theta \, \sin^k\theta = B\Big(\frac{1}{2},\frac{k+1}{2}\Big) \, , 
\en 
and integrating out all angles yields
\eq 
\Omega_{n} = \int\dx\Omega_n
= 2\pi \prod_{k=1}^{n-1} \, B\Big(\frac{1}{2},\frac{k+1}{2}\Big) 
= \frac{2\pi^\frac{n+1}{2}}{\Gamma\left(\frac{n+1}{2}\right)}\,. 
\label{eq:Omega n}
\en 
Formula~(\ref{eq:Omega n}) allows for continuation to arbitrary complex $n$ 
(except for poles). This observation takes us a step closer to formally evaluate integrals 
in non-integer dimensions.

The idea for calculating the integrals of interest $\int\dx\Omega_{n}$ with $n$ non-integer is to split 
it in two parts. One part is the integral over $k$ angles, where $k$ 
is a fixed integer. In our case this part contains the integrand 
$f(r_n,\theta_{n-1},\theta_{n-2})$ with  $k=2$. 
The remaining part is an $n-k$ dimensional spherically symmetric integral, which can be also evaluated 
for non-integer $n-k$ using the identity~(\ref{eq:Omega n}).
The splitting we need is
\begin{align}
\int\dx\Omega_{n-1}&=\int_0^\pi\dx\theta_{n-1}\sin^{n-2}\theta_{n-1}\int\dx\Omega_{n-2} 
\nonumber\\
&=\int_0^\pi\dx\theta_{n-1}\sin^{n-2}\theta_{n-1}\int_0^\pi\dx\theta_{n-3}\sin^{n-2}\theta_{n-2}
\int\dx\Omega_{n-3}\,.
\end{align}
Thus we obtain for our integral
\eq 
\hspace*{-1.2cm}
& &\int\dx^n x\,f(r_n,\theta_{n-1},\theta_{n-2})=\int_0^\infty\dx r_n\, r_n^{n-1}
\int\dx\Omega_{n-1}\,f(r_n,\theta_{n-1},\theta_{n-2}) 
\nonumber\\
\hspace*{-1.2cm}
&=&\Omega_{n-3}\int_0^\infty\dx r_n\, r_n^{n-1}\int_0^\pi\dx\theta_{n-1}\sin^{n-2}\theta_{n-1}
\int_0^\pi\dx\theta_{n-2}\sin^{n-3}\theta_{n-2}
\, f(r_n,\theta_{n-1},\theta_{n-2}) \,. 
\en
Commonly the angles are relabeled $\theta_{n-1}\rightarrow\theta_1$ 
and $\theta_{n-2}\rightarrow\theta_2$ at this stage.

\section{Partial fraction decomposition of linear propagators}
\label{sec:Partial Fractioning}

A common phrase to read about PSI is that it involves a tedious partial 
fraction decomposition. Algorithms of how to perform them are known. 
However, to the best of the authors knowledge, the partial fraction 
decomposition was never viewed in a wider context. Especially the question under which circumstances one 
can reduce the number of angular integrals to be calculated often lacks an explanation.
In this section we want to discuss a simple geometric representation of the occurring propagators, which 
is straightforwardly connected with Minkowski momentum space. We will see that any partial fractioning 
has its root in just one basic formula~(\ref{eq:Two-Point Splitting Lemma}). 
This formula can then be extended to not only a general picture of 
partial fraction decompositions, but also to a nice and useful formula for the splitting 
of double-massive propagators to single-massive ones used 
in Sec.~\ref{sec:Double massive integral with two denominators} to calculate 
the double massive integral in $D$ dimensions.

To set the starting stage, we consider Euclidean vectors $\mathbf{v_i}\in\mathbb{R}^{D-1}$ and 
define there \textit{scaled linear propagator} 
\begin{align}
\prop{v_i}=\frac{1}{1-\mathbf{v_i}\cdot\mathbf{k}}\,, 
\end{align}
where $\mathbf{k}$ is the PSI integration momentum choosen for convenience as an arbitrary unit vector. 
How $\prop{v_i}$ is connected with physical propagators will be clarified 
in Sec.~\ref{sec:Connection to physical propagators}.
For convenience in our consideration we introduce the shorten notation for the product 
of $N$ propagators: 
\begin{align}\label{prop_prod}
\propN{v_1}{v_2}{v_N}\equiv\prop{v_1}\prop{v_2}\cdots\prop{v_N}\,.
\end{align}

\subsection{Two-point splitting lemma}

We start by proving the so-called \textit{two-point splitting lemma} allowing 
to factorize two variables in two-propagator product $\propd{v_1}{v_2}$~[see definition in 
Eq.~\ref{prop_prod}] by introducing two-propagator products of the vectors $v_1$ and $v_2$ 
with another (third) arbitrary Euclidean vector $v_3$. 
In particular, we are interested in deriving a decomposition of the form 
\begin{align}
\propd{v_1}{v_2}=\lambda\,\propd{v_1}{v_3}+\mu\,\propd{v_2}{v_3}\,,
\label{eq:ansatz splitting}
\end{align} 
where $\lambda$ and $\mu$ are two real parameters. Such identity will be useful for 
manipulations of products of propagators containing $\mathbf{v_1}$ and $\mathbf{v_2}$ in PSIs. 
Starting from r.h.s. of Eq.~(\ref{eq:ansatz splitting}), we arrive at 
\eq 
& &\lambda\,\propd{v_1}{v_3}+\mu\,\propd{v_2}{v_3}
\nonumber\\
&=&\Big(\lambda\,\invprop{v_2}+\mu\,\invprop{v_1}\Big) \propt{v_1}{v_2}{v_3}
\nonumber\\
&=&
\Big[ \Bigl(\lambda\,\invprop{v_2}+\mu\,\invprop{v_1}\Big) \, \prop{v_3} \Big] 
\, \propd{v_1}{v_2} \,,
\en
Comparing the latter with l.h.s.of~Eq.~(\ref{eq:ansatz splitting}) 
we conclude that the condition 
\eq 
\Big(\lambda\,\invprop{v_2}+\mu\,\invprop{v_1}\Big) \, \prop{v_3} \equiv 1 \,,
\en 
must hold. It gives a system of two equations on two arbitrary real parameters $\lambda$ 
and $\mu$ and vector $\mathbf{v_3}$:   
\eq\label{SystEqs}
\left\{ 
\begin{array}{c}
\lambda + \mu = 1 \\
\mu \mathbf{v_1} + \lambda \mathbf{v_2} = \mathbf{v_3} \,,
\end{array}
\right. 
\en 
From Eq.~(\ref{SystEqs}) it follows that the parameter $\lambda$ can be chosen arbitrarily, 
while the second parameter $\mu$ and vector $\mathbf{v_3}$ are then fixed as 
$\mu=1-\lambda$ and $\mathbf{v_3}=(1-\lambda)\mathbf{v_1}+\lambda\mathbf{v_2}$. 

Therefore, we are now in the position to formulate the 
\textit{Two-Point Splitting Lemma}\index{Two-point splitting lemma}: 
\eq\label{eq:Two-Point Splitting Lemma} 
\propd{v_1}{v_2}=\lambda\,\propd{v_1}{v_3}
+(1-\lambda)\,\propd{v_2}{v_3}\,,
\en 
where $\mathbf{v_3}=(1-\lambda)\,\mathbf{v_1}+\lambda\,\mathbf{v_2}$. 

\begin{figure}[hb]
\begin{center}
\epsfig{figure=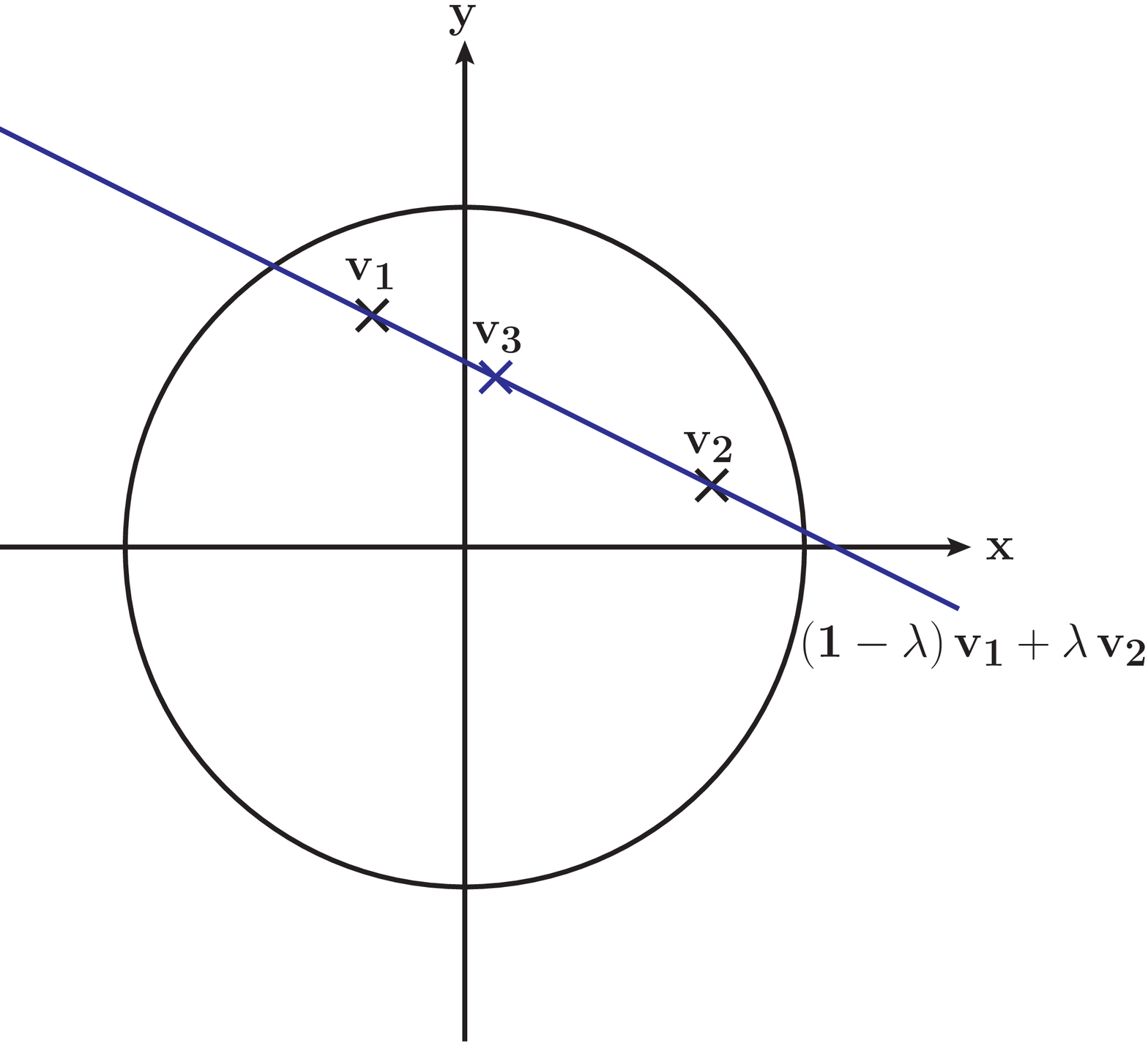,scale=.31}
\end{center}
\caption{A graphical representation of the two-point splitting lemma. 
The two-dimensional $(xy)$ plane is spanned by two arbitrary linearly independent 
vectors $\mathbf{v_1}$ and $\mathbf{v_2}$. The head of the third vector $\mathbf{v_3}$ (we 
show one typical example) lies 
in the straight line connecting the heads of $\mathbf{v_1}$ and $\mathbf{v_2}$. 
The circle represents all vectors with $\mathbf{v}\cdot\mathbf{v}=1$.} 
\label{fig2_jhep}
\end{figure}

Geometrical interpretation of the two-point splitting lemma is that the 
heads of the vector $\mathbf{v_3}$ defined as 
$\mathbf{v_3}=(1-\lambda)\,\mathbf{v_1}+\lambda\,\mathbf{v_2}$ with $\lambda$ 
being arbitrary real parameter 
represents the line passing through the heads of $\mathbf{v_1}$ and $\mathbf{v_2}$.  
In other words, it is a definition of a line by two linear independent vectors. 
All three vectors $\mathbf{v_1}$, $\mathbf{v_2}$, and $\mathbf{v_3}$ obviously lie 
in the two-dimensional plane spanned by $\mathbf{v_1}$ and $\mathbf{v_2}$. 
This is depicted in Fig.~\ref{fig2_jhep}.

For $j,l\in\mathbb{N}$ the two-point splitting lemma can be generalized to
\begin{align}
\powprop{v_1}{j}\powprop{v_2}{l}&=\sum_{n=0}^{j-1}\binom{l-1+n}{l-1}
\lambda^l(1-\lambda)^n\powprop{v_1}{j-n}\powprop{v_3}{l+n}
\nonumber\\
&+\sum_{n=0}^{l-1}\binom{j-1+n}{j-1}\lambda^n(1-\lambda)^j\powprop{v_2}{l-n}\powprop{v_3}{j+n}\,.
\label{eq:generalized Two-Point Splitting Lemma}
\end{align}

To prove Eq.~(\ref{eq:generalized Two-Point Splitting Lemma}) 
we employ the binomial summation formula 
\begin{align}
\sum_{n=0}^N \binom{j-1+n}{j-1}=\binom{j+N+1}{j} \,,
\label{eq:Binomial summation}
\end{align}
and use mathematical induction. 

\subsection{Two-mass splitting and two-point partial fractioning}

\begin{figure}[hb]
	\begin{center}
        \epsfig{figure=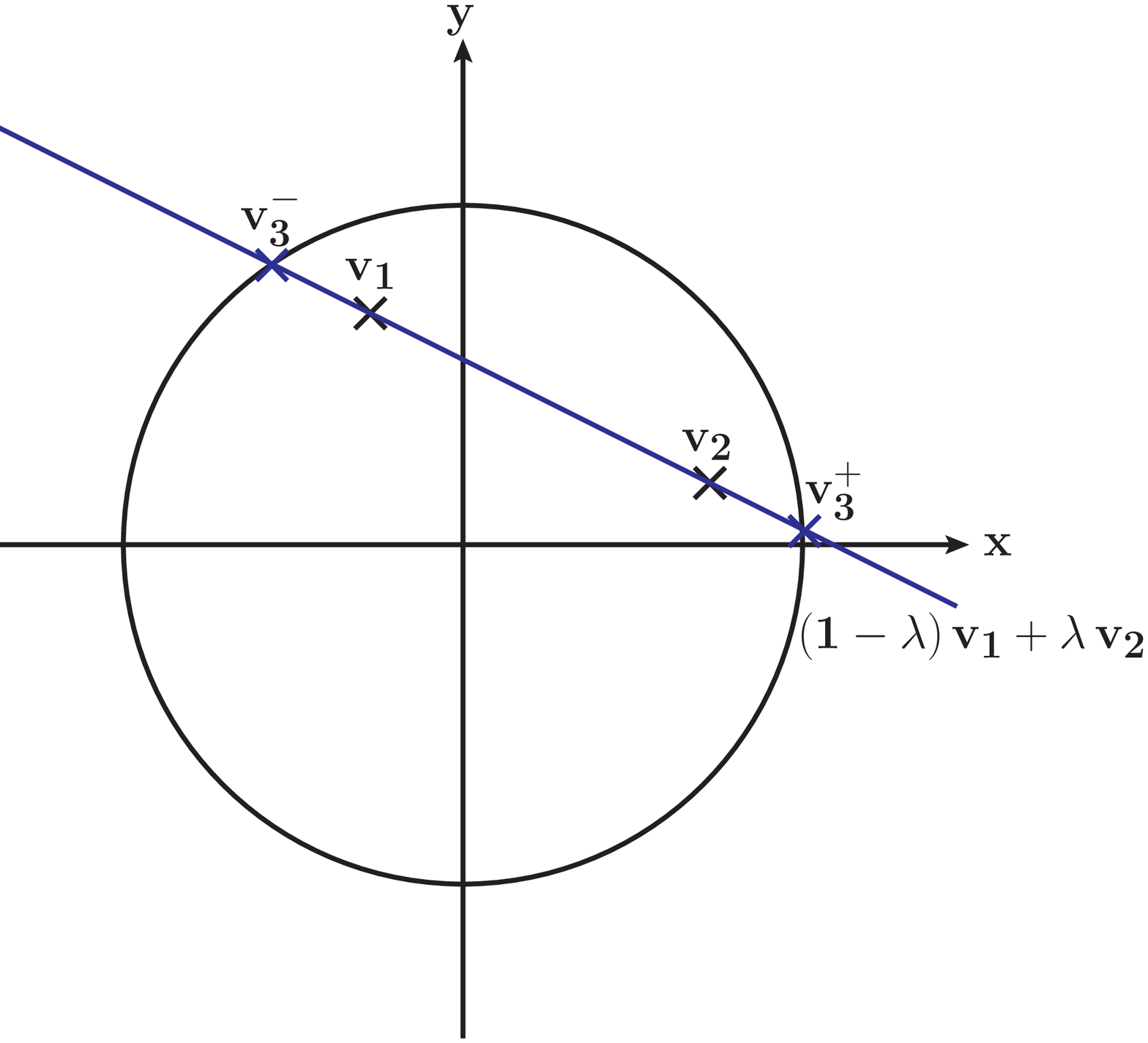,scale=.31}
	\end{center}
\caption{Graphical representation of double-to-single-massive splitting. 
$\mathbf{v_3^\pm}$ is chosen as one of the intersections of the unit circle with the line through 
$\mathbf{v_1}$ and $\mathbf{v_2}$.}
\label{fig3_jhep} 
\end{figure}

The two-mass splitting now becomes a simple corollary of the two-point splitting lemma. 
We just have to choose $\mathbf{v_3}$ such that $\mathbf{v_3}\cdot\mathbf{v_3}=1$. 
To achieve this we have to take $\mathbf{v_3}$ as the intersection of the line 
$(1-\lambda)\,\mathbf{v_1}+\lambda\,\mathbf{v_2}$ with the unit circle.
The condition 
\begin{align}
\left|(1-\lambda)\,\mathbf{v_1}+\lambda\,\mathbf{v_2}\right|^2\overset{!}{=}&1 \,,
\end{align}
leads to the quadratic equation 
(see Fig.~\ref{fig3_jhep}) 
\begin{align}
\lambda^2 \, (\mathbf{v_2}-\mathbf{v_1})^2 + 2 \, \lambda 
\,\mathbf{v_1}\cdot(\mathbf{v_2}-\mathbf{v_1})
+\mathbf{v_1}\cdot\mathbf{v_1}-1\overset{!}{=}0\,.
\label{eq:Line Circle Intersection}
\end{align}
Defining the \textit{scaled Mandelstam variables}
\begin{align}
v_{11}=1-\mathbf{v_1}\cdot\mathbf{v_1}\,, \quad  
v_{12}=1-\mathbf{v_1}\cdot\mathbf{v_2}\,, \quad  
v_{22}=1-\mathbf{v_2}\cdot\mathbf{v_2} \,,
\end{align} 
we write down Eq.~(\ref{eq:Line Circle Intersection}) as 
\eq 
\lambda^2 \, (2 v_{12}-v_{11}-v_{22}) + 2 \lambda \, (v_{11}-v_{12}) - v_{11} = 0 \,. 
\en 
The solution of the latter equation is 
\begin{align}
\lambda_{\pm}=\frac{v_{12}-v_{11}\pm\sqrt{v_{12}^2-v_{11} v_{22}}}{2 v_{12}-v_{11}-v_{22}}\,.
\label{eq:double-massive lambda}
\end{align}

The massless vectors associated with $\mathbf{v_1}$ and $\mathbf{v_2}$ thus are
\begin{align}
\mathbf{v_3^\pm}=\frac{v_{12}-v_{22}\pm\sqrt{v_{12}^2-v_{11} v_{22}}}{2 v_{12}-v_{11}-v_{22}}\,\mathbf{v_1}
+\frac{v_{12}-v_{11}\mp\sqrt{v_{12}^2-v_{11} v_{22}}}{2 v_{12}-v_{11}-v_{22}}\,\mathbf{v_2}\,.
\label{eq:double-massive associated massless vector}
\end{align}
Using the $\lambda$ from Eq.~(\ref{eq:double-massive lambda}) and 
the associated vector from~(\ref{eq:double-massive associated massless vector}) 
we have the \textit{Double-to-Single-Massive Splitting} (or simply \textit{Two-mass splitting})
\begin{align}
\propd{v_1}{v_2}=\lambda_\pm\,\propd{v_1}{v_3^\pm}+(1-\lambda_\pm)\propd{v_2}{v_3^\pm}\,.
\label{eq:Double-to-Single-Massive splitting}
\end{align}
It splits the double-massive propagator on the left into two single-massive propagators on the right. 
This proves very useful for the calculation of the double-massive integral with integer coefficients.

The two mass-splitting becomes even more powerful in combination with the generalized form of 
the two point splitting lemma. For $j,l\in\mathbb{N}$ it holds
\begin{align}
\powprop{v_1}{j}\powprop{v_2}{l}&=\sum_{n=0}^{j-1}\binom{l-1+n}{l-1}
\lambda_\pm^l(1-\lambda_\pm)^n\powprop{v_1}{j-n}\powprop{v^\pm_3}{l+n}
\nonumber\\
&+\sum_{n=0}^{l-1}\binom{j-1+n}{j-1}\lambda_\pm^n(1-\lambda_\pm)^j\powprop{v_2}{l-n}\powprop{v^\pm_3}{j+n}\,.
\label{eq:Generalized Two-mass splitting}
\end{align}

\begin{figure}[hb]
        \begin{center}
	        \epsfig{figure=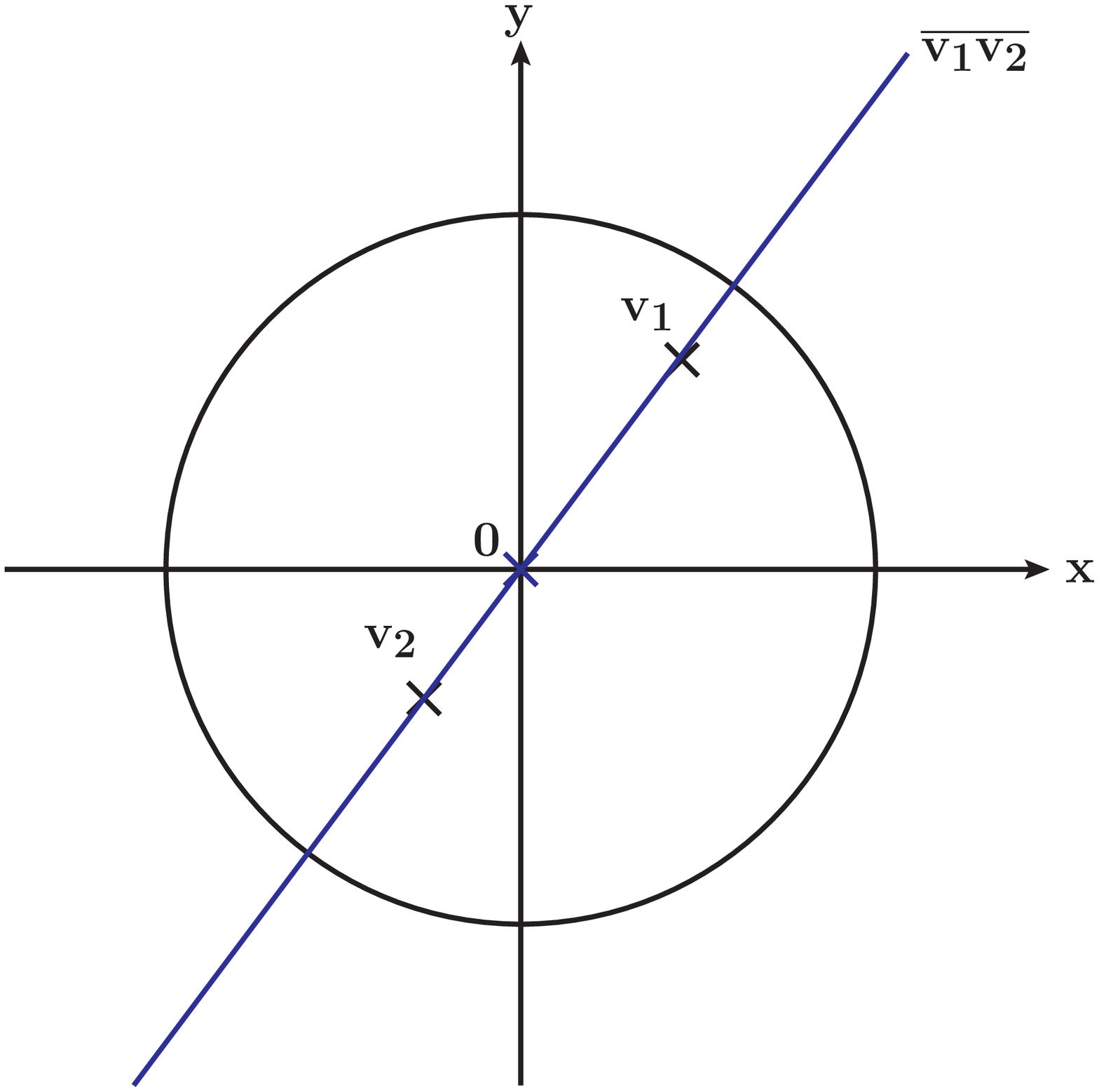,scale=.31}
        \end{center}
\caption{Graphical illustration of two-point partial fractioning. 
The origin lies on the line $\overline{\mathbf{v_1}\mathbf{v_2}}$, 
we can use $\prop{0}$ to split $\prop{v_1}$ and $\prop{v_2}$.} 
\label{fig4_jhep} 
\end{figure}

Another direct consequence of the two-point splitting lemma is the partial fractioning of the propagator 
of two different linear dependent vectors. If $\lambda_1\mathbf{v_1}+\lambda_2\mathbf{v_2} = \mathbf{0}$ with 
$\lambda_1+\lambda_2 \neq 0$, then the line connecting $\mathbf{v_1}$ and $\mathbf{v_2}$ passes through the 
origin. Thus $\mathbf{0}$ is associated with them. As $\prop{\mathbf{0}}=1$ the two-point splitting lemma 
can be used to perform a reduction of the number of propagators. Thus we have 
the \textit{two-point partial fractioning}\index{Two-point partial fractioning}~(see Fig.~\ref{fig4_jhep}) 
\eq\label{eq:two-point partial fractioning}  
\propd{v_1}{v_2}=\frac{1}{\lambda_1+\lambda_2}\left[\lambda_2\,\prop{v_1}+\lambda_1\,\prop{v_2}\right] \,,
\en
with two constraints $\lambda_1\mathbf{v_1}+\lambda_2\mathbf{v_2} = \mathbf{0}$ 
and $\lambda_1+\lambda_2 \neq 0$. The most important special case of 
the two point partial fractioning is $\mathbf{v_2}=-\mathbf{v_1}$. 
Then one can write Eq.~(\ref{eq:two-point partial fractioning}) as 
\begin{align}
\propd{v}{-v}=\frac{1}{2}\left[\prop{v}+\prop{-v}\right]\,.
\end{align}

\subsection{Three-point partial fractioning\index{Three-point partial fractioning} 
and general denominator reduction}

Eq.~(\ref{eq:two-point partial fractioning}) gives a way to reduce the propagator of 
two linear dependent vectors to just one. Together with the two-point splitting 
lemma~(\ref{fig8_jhep}) we will geometrically construct the formula for 
the reduction of three propagators with linear dependent vectors to the sum of two propagators. 
The derived formula will be at the heart of the partial fraction decomposition for the NNLO real 
correction of the Drell-Yan angular distribution (which we plan to investigate in the near future in the context of small $Q_T$ resummation, compare Ref.~\cite{Boer:2006eq} and \cite{Berger:2007jw}). 

\begin{figure}[hb]
	\begin{center}
        \epsfig{figure=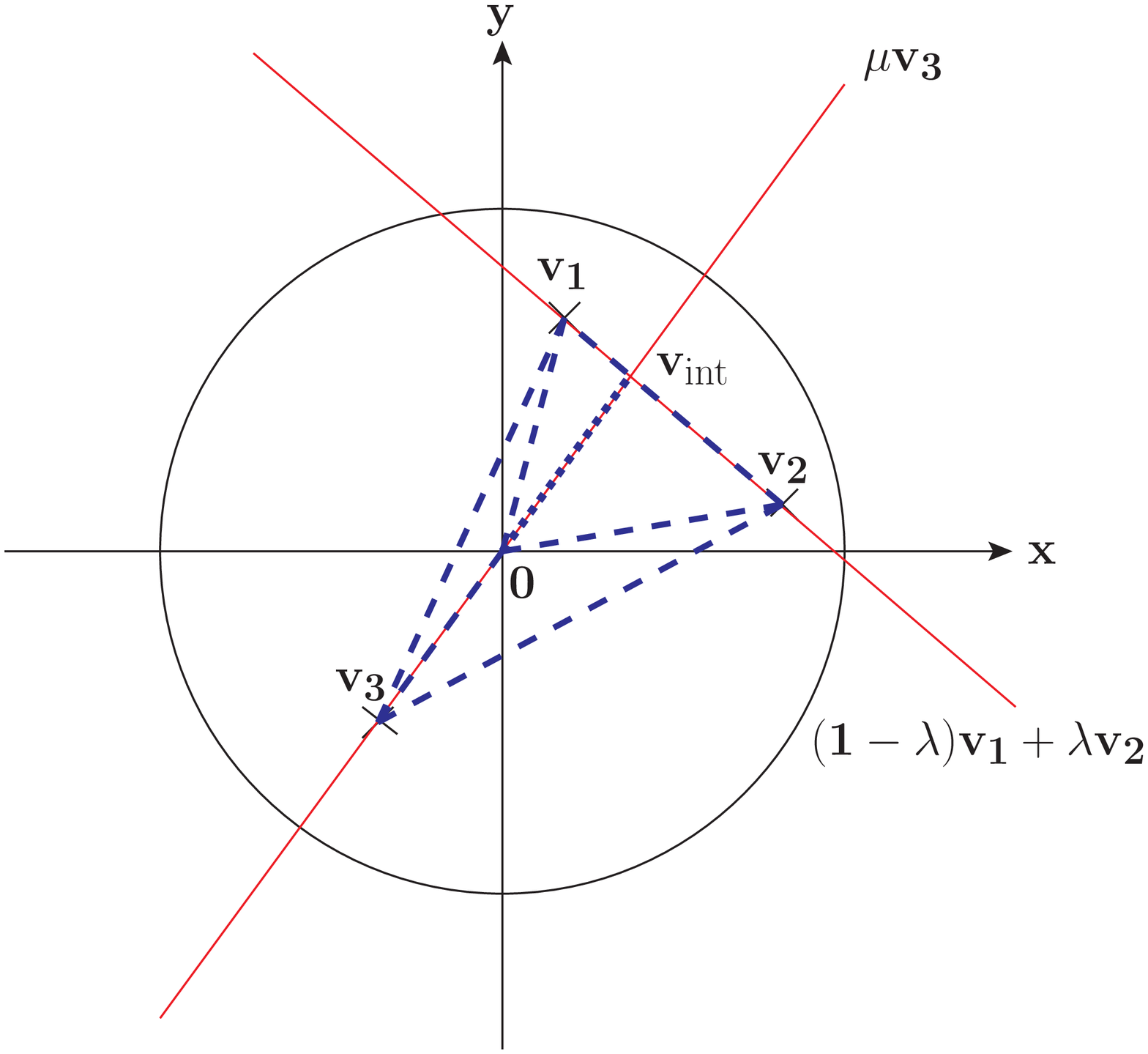,scale=.31}
	\end{center}
\caption{Graphical illustration of three-point partial fractioning.} 
\label{fig5_jhep}
\end{figure}

Let us start with three linear dependent and pairwise linear independent vectors 
$\mathbf{v_1}$, $\mathbf{v_2}$ and $\mathbf{v_3}$ satisfying 
\eq\label{v1v2v3_id}
\lambda_1\mathbf{v_1}+\lambda_2\mathbf{v_2}+\lambda_3\mathbf{v_3} = \mathbf{0}  \,,
\en  
with $\lambda_1+\lambda_2+\lambda_3 \neq 0$. 
We want to derive an expression of the form
\begin{align}
\propt{v_1}{v_2}{v_3}=a_1\propd{v_1}{v_2}+a_2\propd{v_1}{v_3}+a_3\propd{v_2}{v_3} \,,
\end{align}
with parameters $a_{1,2,3}\equiv a_{1,2,3}(\lambda_1,\lambda_2,\lambda_3)$ to be determined. 
In our graphical representation $\mathbf{v_1}$, $\mathbf{v_2}$ and $\mathbf{v_3}$ form a triangle. 
We proceed in three steps depicted in Fig.~\ref{fig5_jhep}.

First, we find the intersection of $\overline{\mathbf{v_1}\mathbf{v_2}}$ and 
$\overline{\mathbf{0}\mathbf{v_3}}$ and name it $\mathbf{v_\text{int}}$\footnote{In the case, where these 
lines are parallel such that no intersection exists, change the roles of $\mathbf{v_1}$, $\mathbf{v_2}$,  
and $\mathbf{v_3}$ appropriately.}. $\mathbf{v_\text{int}}$ is associated with $\mathbf{v_1}$ 
by the two-point splitting lemma and therefore there exists a $\lambda$ such that
\begin{align}
\propd{v_1}{v_2}=(1-\lambda)\propd{v_1}{v_\text{int}}+\lambda\propd{v_2}{v_\text{int}}\,.
\end{align}
Therefore, upon multiplication of $\propd{v_1}{v_2}$ with $\prop{v_3}$ it follows 
\begin{align}
\propt{v_1}{v_2}{v_3}=(1-\lambda)\propt{v_1}{v_\text{int}}{v_3}+\lambda\propt{v_2}{v_\text{int}}{v_3}\,.
\end{align}
Secondly, we observe that $\mathbf{v_\text{int}}$ is linear dependent on $\mathbf{v_3}$, 
thus $\mathbf{v_\text{int}}=\mu \mathbf{v_3}$ for some $\mu$. 
Hence, according to Eq.~(\ref{eq:two-point partial fractioning}) we can use 
the two-point partial fractioning\index{Two-point partial fractioning} 
\begin{align}
\propd{v_\text{int}}{v_3}=\frac{1}{1-\mu}\prop{v_3}-\frac{\mu}{1-\mu}\prop{v_\text{int}}.
\end{align}
Thus, 
\begin{align}
\propt{v_1}{v_2}{v_3}&=\frac{1-\lambda}{1-\mu}\propd{v_1}{v_3}
-\frac{(1-\lambda)\,\mu}{1-\mu}\propd{v_1}{v_\text{int}}
\nonumber\\
&+\frac{\lambda}{1-\mu}\propd{v_2}{v_3}-\frac{\lambda\,\mu}{1-\mu}\propd{v_2}{v_\text{int}}\,,
\end{align}
with each of the summands corresponding to one of the four triangles with the corners 
$(\mathbf{0},\mathbf{v_1},\mathbf{v_3})$, $(\mathbf{0},\mathbf{v_1},\mathbf{v_\text{int}})$, 
$(\mathbf{0},\mathbf{v_2},\mathbf{v_3})$, $(\mathbf{0},\mathbf{v_2},\mathbf{v_\text{int}})$.

In a third step we get rid of $\mathbf{v_\text{int}}$ by combining the triangles 
$(\mathbf{0},\mathbf{v_1},\mathbf{v_\text{int}})$ and $(\mathbf{0},\mathbf{v_2},\mathbf{v_\text{int}})$ 
with two-point splitting in reverse via Eq~(\ref{eq:Two-Point Splitting Lemma}) such that
\begin{align}
\propt{v_1}{v_2}{v_3}=-\frac{\mu}{1-\mu}\propd{v_1}{v_2}+\frac{1-\lambda}{1-\mu}\propd{v_1}{v_3}
+\frac{\lambda}{1-\mu}\propd{v_2}{v_3}\,. 
\label{eq:Three-point partial fractioning prelim}
\end{align}

Solving the intersection equation
\begin{align}
\mu\mathbf{v_3}\overset{!}{=}\lambda\mathbf{v_1}+(1-\lambda)\mathbf{v_2} \,,
\end{align} 
with boundary condition~(\ref{v1v2v3_id}) yields for $\lambda$ and $\mu$ 
\begin{align}\label{solution_lambda_mu}
\lambda=\frac{\lambda_1}{\lambda_1+\lambda_2}
\quad\text{and}\quad
\mu=-\frac{\lambda_3}{\lambda_1+\lambda_2}\,.
\end{align}
Plugging solution~(\ref{solution_lambda_mu}) 
into Eq.~(\ref{eq:Three-point partial fractioning prelim}) 
we have the \textit{three-point partial fractioning}\index{Three-point partial fractioning} 
\eq 
\hspace*{-.75cm}
\propt{v_1}{v_2}{v_3}=\frac{1}{\lambda_1+\lambda_2+\lambda_3}\left[\lambda_3\propd{v_1}{v_2}
+\lambda_2\propd{v_1}{v_3}+\lambda_1\propd{v_2}{v_3}\right]\,,
\en 
with boundary conditions $\lambda_1\mathbf{v_1}+\lambda_2\mathbf{v_2}+\lambda_3\mathbf{v_3} = \mathbf{0}$ 
and $\lambda_1+\lambda_2+\lambda_3 \neq 0$. 

After the illustration of the application of two-point splitting lemma to 
partial-fractioning of products of two and three propagators, 
we are in the position to extend it to the general case of the product of 
$n$ propagators with $n$ being arbitrary. Therefore, we formulate the generalized  
\textit{reduction of products of linear propagators}\index{Reduction of products of linear propagators}: 

If $\sum_{i=1}^n \lambda_i \mathbf{v_i}= \mathbf{0}$ with $\lambda\equiv\sum_{i=1}^n\lambda_i\neq 0$, 
$n\in\mathbb{N}$, $\lambda_i\in\mathbb{C}$, then it holds
\begin{align}
\prod_{i=1}^n\prop{v_i}
=\frac{1}{\lambda} \, \sum_{j=1}^n \lambda_j \, \prod_{i=1}^{n \neq j} \, \prop{v_i} \,. 
\label{eq:linear propagator reduction}
\end{align}
The proof of the identity~(\ref{eq:linear propagator reduction}) is 
based on the linearity of the propagator and is straightforward. 
In particular, using ${\bf 0} = \sum_{j=1}^n \lambda_j \mathbf{v_j}$ 
one gets 
\eq 
\lambda = \sum\limits_{j=1}^n \ \lambda_j 
= \sum\limits_{j=1}^n \ \lambda_j \, \Big(1 - {\bf v_j} {\bf k}\Big) 
= \sum\limits_{j=1}^n \ \lambda_j \, \invprop{v_j} \,. 
\en  
Therefore, 
\eq 
\propN{v_1}{v_2}{v_n} 
&=& \prod_{i=1}^n \prop{v_i} = 
\frac{1}{\lambda} \,\sum_{j=1}^n\lambda_j \, 
\prod_{i=1}^n \prop{v_i} 
\nonumber\\
&=& 
\frac{1}{\lambda} \, 
\sum_{j=1}^n \lambda_j \, \prod_{i=1}^n \, \invprop{v_j} \, \prop{v_i} 
= \frac{1}{\lambda} \, 
\sum_{j=1}^n \lambda_j \, \prod_{i=1}^{n \neq j} \, \prop{v_i}  
\nonumber\\
&=& \frac{1}{\lambda} \, \biggl[ 
\lambda_1 \, \propN{v_2}{v_3}{v_n} \,+\, 
\lambda_2 \, \propN{v_1}{v_3}{v_n} 
\nonumber\\ 
&+& \ldots \,+\,  
\lambda_n \, \propN{v_1}{v_2}{v_{n-1}} 
\biggr] \,. 
\en
Geometrically Eq.~(\ref{eq:linear propagator reduction}) follows 
from the fact that $\mathbf{0}$ 
is in the $n-1$-dimensional hyperplane spanned by the vectors $\mathbf{v_i}$ and, thus 
it can be associated with them repeatedly using the two-point splitting lemma. 
Of course, the general reduction of linear 
propagators contains two- and three-point partial fractioning as mere special cases.  
However, we think that the presented inductive approach is more pedagogically suited to build up 
an geometric intuition. 

\subsection{Propagators to negative integer powers} 

In practical calculations one is also confronted
with negative integer powers $j\in\mathbb{Z}_{\leq 0}$ of massive propagators. 
Those can always be made massless by simple scaling. It holds
\begin{align}
\invprop{v_1}&=1-\mathbf{v_1}\cdot\mathbf{k}=1-|\mathbf{v_1}|+|\mathbf{v_1}|\,
\Delta^{-1}_\mathbf{k}\left({\frac{\mathbf{v_1}}{|\mathbf{v_1}|}}\right)
\nonumber\\
&=1-\sqrt{1-v_{11}}+\sqrt{1-v_{11}}\,\invprop{\hat{v}_1} \,,
\label{eq:PropagatorNegativePower}
\end{align}
with $\mathbf{\hat{v}_1}=\frac{\mathbf{v_1}}{|\mathbf{v_1}|}=\frac{\mathbf{v_1}}{\sqrt{1-v_{11}}}$.
Contracting with another vector $\mathbf{v_2}$ gives the scalar product
\begin{align}
\hat{v}_{12}=1-\mathbf{\hat{v}_1}\cdot\mathbf{v_2}
=1-\frac{\mathbf{v_1}\cdot\mathbf{v_2}}{\sqrt{1-v_{11}}}
=1-\frac{1-v_{12}}{\sqrt{1-v_{11}}}\,.
\end{align}
Eq.~(\ref{eq:PropagatorNegativePower}) is straightforwardly generalized to $j\in\mathbb{Z}_{\leq 0}$
by application of the Binomial theorem 
\begin{align}
\powprop{v_1}{j}&=\left(1- \sqrt{1-v_{11}}+\sqrt{1-v_{11}}\,\invprop{\hat{v}_1}\right)^{-j}
\nonumber\\
&=\sum_{n=0}^{-j}\binom{-j}{n}\left(1-\sqrt{1-v_{11}}\,\right)^{-j-n} \, 
\left(1-v_{11}\right)^{n/2} \, \powprop{\hat{v}_1}{-n}\,.
\label{eq:PropagatorNegativePowerBinomial}
\end{align}
This equation allows to generalize the two-mass splitting to all $j,l\in\mathbb{Z}$. 
Another practical application is that one can reduce two-mass angular integral 
to one-mass one in case of $j\in\mathbb{Z}_{\leq 0}$. 
In particular, for that case it holds
\begin{align}\label{2M_to_1M_red_a} 
I_{j,l}^{(2)}(v_{12},v_{11},v_{22};\eps)&
=\int\dx\Omega_{k_1 k_2}\powprop{v_1}{j}\powprop{v_2}{l}
\nonumber\\
&\hspace*{-4cm}
=\sum_{n=0}^{-j}\binom{-j}{n}
\left(1-\sqrt{1-v_{11}}\,\right)^{-j-n} \left(1-v_{11}\right)^{n/2} \, 
\int\dx\Omega_{k_1 k_2}\powprop{\hat{v}_1}{-n}\powprop{v_2}{l} 
\nonumber\\
&\hspace*{-4cm}
=\sum_{n=0}^{-j}\binom{-j}{n}
\left(1-\sqrt{1-v_{11}}\,\right)^{-j-n} \left(1-v_{11}\right)^{n/2} \, 
I_{l,-n}^{(1)}\left(1-\frac{1-v_{12}}{\sqrt{1-v_{11}}},v_{22};\eps\right) \,. 
\end{align} 
Equivalently for $l\in\mathbb{Z}_{\leq 0}$
\begin{align}\label{2M_to_1M_red_b} 
I_{j,l}^{(2)}(v_{12},v_{11},v_{22};\eps)&=\sum_{n=0}^{-l}\binom{-l}{n}  
\left(1-\sqrt{1-v_{22}}\,\right)^{-l-n} \left(1-v_{22}\right)^{n/2}
\nonumber\\
&\qquad\times  I_{j,-n}^{(1)}\left(1-\frac{1-v_{12}}{\sqrt{1-v_{22}}},v_{11};\eps\right)\,.
\end{align}

\subsection{Connection to physical propagators}
\label{sec:Connection to physical propagators}

Let us consider a situation with fixed momenta $p_1$, $p_2$, $\dots$, $p_n$ and a PSI 
over the external momenta $k_1$ and $k_2$, where $\sum_{j=1}^n p_j=k_1+k_2$ and 
establish the connection of physical propagators with 
the scaled propagators occurring in the derived partial fractioning formulas. 
We use the CMF for the angular integration momenta $(\mathbf{k_1} + \mathbf{k_2} = 0)$ and
scale all momenta by their corresponding energy component: 
\begin{align}
k_1=E_k \, \left(1, \mathbf{k}\right)\,, \qquad 
k_2=E_k \, \left(1,-\mathbf{k}\right)\,, \qquad 
p_i=E_i \, \left(1, \mathbf{v_i}\right)\,. 
\end{align}
Also we suppose that $k_{1,2}$ momenta are on their mass-shell and for simplicity 
we assume that they correspond to the same mass: $k_{1,2}^2=m_k^2$. 
Considering generic physical propagator depending 
on external $p_i$ and angular loop $(k_1,k_2)$ momenta 
\eq 
\Delta_{\rm phys}(k_{1,2},p_i,m) = \frac{1}{(k_{1,2}-p_i)^2-m^2}  \,,
\en
we can always express it in terms of scaled propagator\index{Scaled propagator} 
$\prop{\pm\bar{v}_i}$: 
\eq\label{eq:propagator scaling} 
\Delta_{\rm phys}\left(k_{1,2},p_i,m\right) = 
\frac{1}{p_i^2-m^2+m_k^2-2E_iE_k} \,\,\prop{\pm\bar{v}_i} \,,
\en  
where we introduced the scaled vector\index{Scaled vector} 
\begin{align}\label{eq:vector scaling}  
{\bf\bar{v}_i} = \frac{{\bf v_i}}{\frac{p_i^2-m^2+m_k^2}{2E_iE_k}-1} \,.
\end{align}
Note the two vectors $v_i = (1, {\bf v_i})$ and 
$\bar{v}_i = (1, {\bf\bar{v}_i})$ are differed only by spatial part 
[see Eq.~(\ref{eq:vector scaling})]. 
So, up to an angular independent factor we can express any physical propagator 
of the discussed form as a scaled propagator with a suitably chosen scaled vector.

\subsection{Example: Drell-Yan double real corrections kinematics}
\label{DY_example}

\begin{figure}[ht]
        \begin{center}
        \epsfig{figure=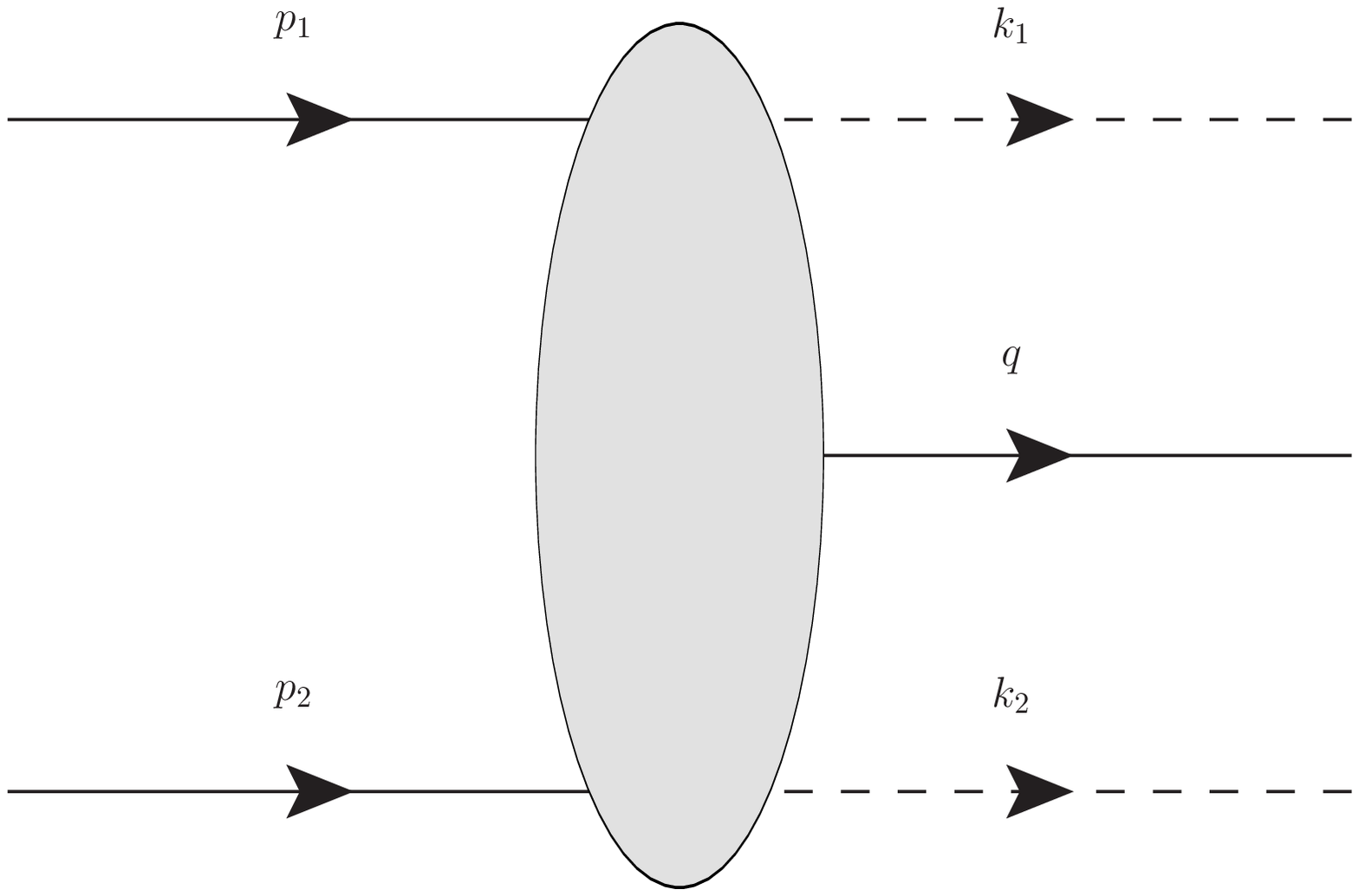,scale=.5}
        \end{center}
\caption{Kinematics for double real corrections of the Drell-Yan process.
$p_1$ and $p_2$ are incoming momenta, $q$ is outgoing.
The PSI is performed over the dashed momenta $k_1$ and $k_2$.}
\label{fig6_jhep}

\vspace*{.5cm}
        \begin{center}
        \epsfig{figure=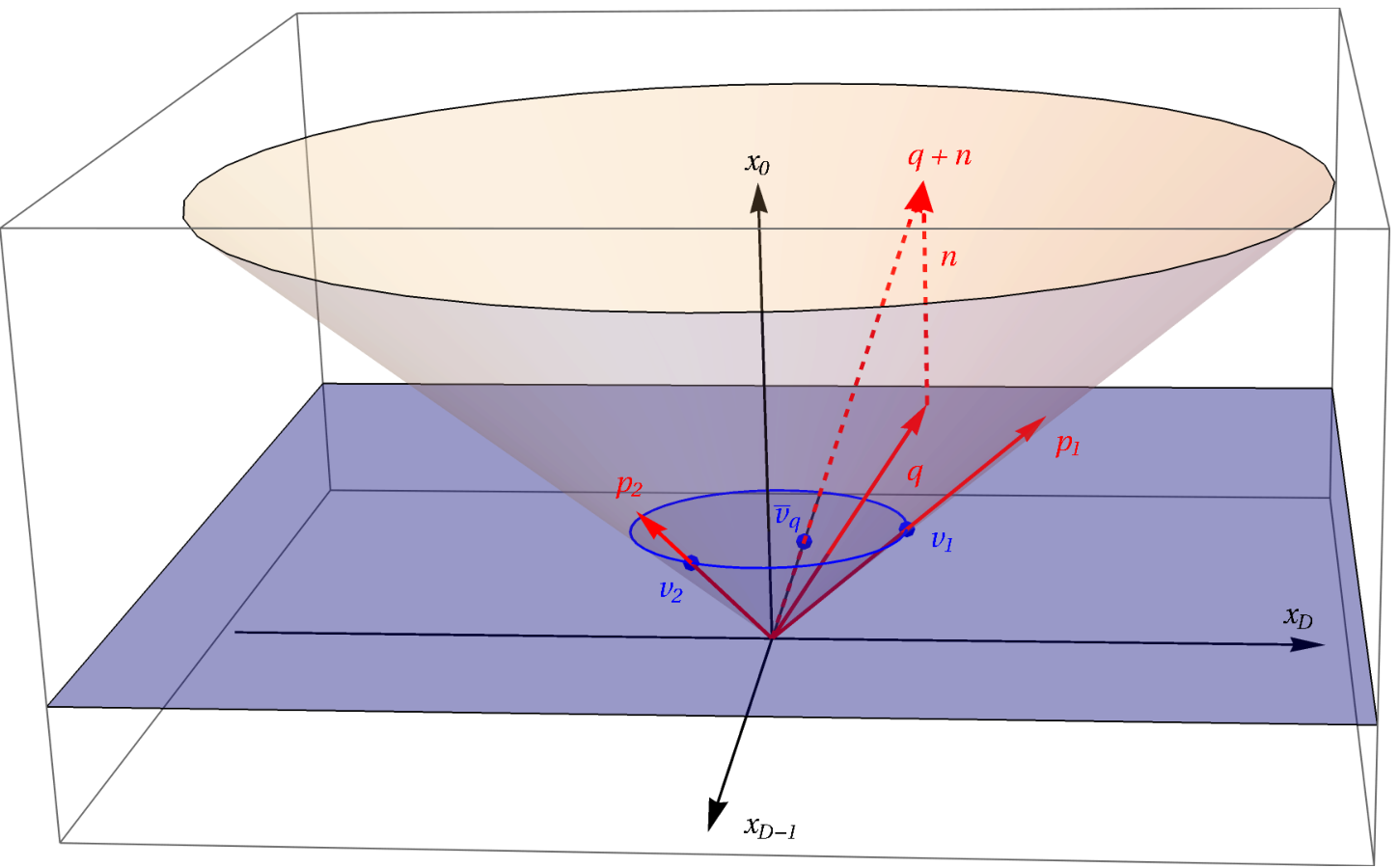,scale=.9}
        \end{center}
\caption{The picture shows the $0\equiv x_1\equiv\dots\equiv x_{D-2}$ slice of Minkowski space. 
The lightcone is shown in light yellow. The external momenta are depicted as red arrows, 
$p_1$ and $p_2$ are massless and hence on the light-cone, $q$ has mass $Q^2$ and is inside the light-cone. 
The scaled momenta $v$, marked by blue dots, are in the blue $x_0\equiv 1$ plane. The blue circle marks 
the intersection of the light-cone with the $x_0\equiv 1$ plane.  
The dashed red momenta illustrate the construction of $\bar{v}_q$.}
\label{fig7_jhep}
\end{figure}

To demonstrate how we can bring our general considerations about partial fractioning to good 
use in a physical situation we consider the kinematics schematically shown 
in Fig.~\ref{fig6_jhep}. 
This is the kinematics of the double real corrections for the Drell-Yan process. 
Conservation of momentum implies that $p_1+p_2=q+k_1+k_2$. 
For the PSI $P_\text{ext}=p_1+p_2-q$. Hence, the angular PSI 
is defined in the CMF of $p_1$, $p_2$, and $-q$ (or equivalently in the CMF of $k_1$ and $k_2$): 
\begin{align}
\mathbf{p}_1+\mathbf{p}_2-\mathbf{q}=0\,.
\label{eq:CMS condition}
\end{align}

\begin{figure}[htb]
	\begin{center}
        \epsfig{figure=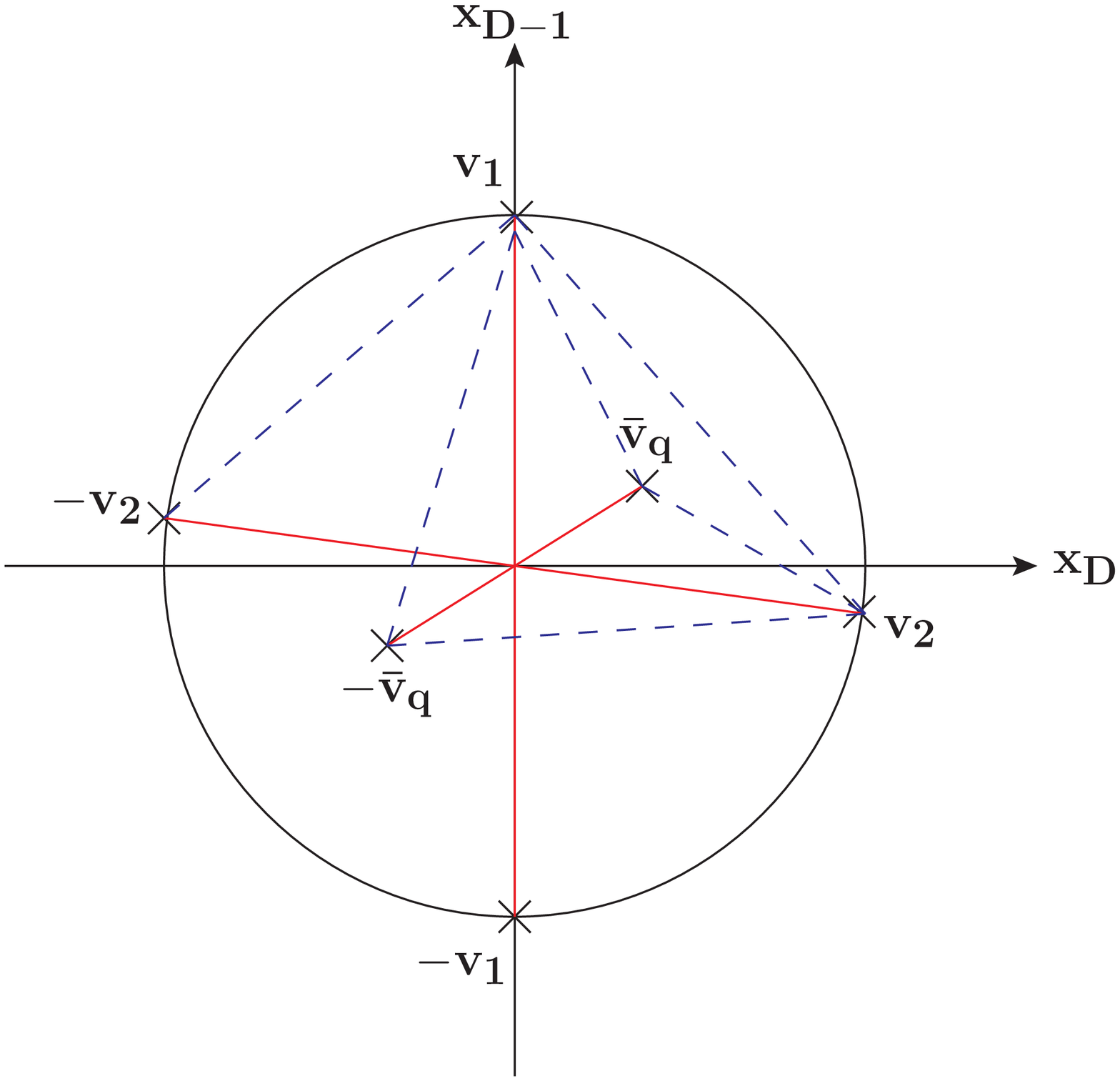,scale=.31}
	\end{center}
\caption{Drell-Yan kinematics in terms of scaled vectors.
Red lines indicate pairs of linear dependent vectors. The dashed blue lines mark contributions
to genuinely different PSIs.}
\label{fig8_jhep}
\end{figure}

These three vectors are linear dependent and thus lie in a common plane, defining the $(x_{D-1},x_D)$-plane. 
We can then depict the kinematics in a three dimensional slice of $D$ dimensional space-time 
(see Fig.~\ref{fig7_jhep}). 
Since all involved particles are massless, we have propagators of the general form
\begin{align}
\frac{1}{(k_{1,2}-p)^2+i0}\,,
\end{align}
where $p$ is one of the fixed momenta. In the rest frame of $p_1$, $p_2$, and $-q$ 
with the $x_D$-axis chosen in the direction of $\mathbf{p_1}$ the vectors can be 
parametrized as
\begin{align}
k_1&=
\frac{\sqrt{s_2}}{2}\left(1,{\bf v_k}\right)= 
\frac{\sqrt{s_2}}{2}\left(1,\dots,\sin\theta_1\cos\theta_2,\cos\theta_1\right)
\,,\nonumber\\
k_2&=\frac{\sqrt{s_2}}{2}\left(1,{\bf v_k}\right)
=\frac{\sqrt{s_2}}{2}\left(1,\dots,-\sin\theta_1\cos\theta_2,-\cos\theta_1\right)
\,,\nonumber\\
p_1&=\frac{s_2-u}{2\sqrt{s_2}}(1,{\bf v_1}) 
=\frac{s_2-u}{2\sqrt{s_2}}\left(1,\mathbf{0}_{D-3},0,1\right)
\,,
\nonumber\\
p_2&=\frac{s_2-t}{2\sqrt{s_2}}(1,{\bf v_2}) 
=\frac{s_2-t}{2\sqrt{s_2}}\left(1,\mathbf{0}_{D-3},\sin\vartheta,\cos\vartheta\right)\,,
\nonumber\\
q&=-\frac{t+u}{2\sqrt{s_2}}\left(1,{\bf v_q}\right)
=-\frac{t+u}{2\sqrt{s_2}}\left(1,\mathbf{0}_{D-3},\frac{t-s_2}{t+u}\sin\vartheta,
\frac{u-s_2}{t+u}+\frac{t-s_2}{t+u}\cos\vartheta\right)\,,
\end{align}
with
\begin{align}
\cos\vartheta=1-\frac{2 s s_2}{(s_2-u)(s_2-t)} \,,
\end{align}
where 
\eq
s=(p_1+p_2)^2\,, \quad  
t=(p_1-q)^2  \,, \quad 
u=(p_2-q)^2  \,, \quad
s_2=(k_1+k_2)^2 \,.
\en 
By conservation of four momentum it holds $Q^2=q^2=s+t+u-s_2$. 
It is also useful to write the relations between Mandelstam variables 
and components of transverse momentum in light-cone frame $q=(Q^+,Q^-,Q_T)$, 
where 
\eq
Q^\pm &=& \frac{Q^0 \pm Q^3}{\sqrt{2}}\,, \quad q^2 \, = \, Q^2 = 2 Q^+ Q^- - Q^2_T \,, 
\nonumber\\
Q_T^2 &=& \frac{\lambda(s,s_2,Q^2)}{4 s_2} \, = \, \frac{ut - s_2Q^2}{s}
\,, \nonumber \\
\cos\vartheta &=& \frac{Q_T^2-s_2}{Q_T^2+s_2}
\,, \quad 
\sin\vartheta \ = \ \frac{2 Q_T \sqrt{s_2}}{Q_T^2+s_2}
\,, \quad 
\sqrt{s_2} \ = \ Q_T \, {\rm tan}\frac{\vartheta}{2} \,.
\en 
Introducing scaled momenta
\eq 
& & v_1=\frac{p_1}{E_1}=(1,{\bf v_1})\,, \qquad 
    v_2=\frac{p_2}{E_2}=(1,{\bf v_2})\,, \nonumber\\
& & v_q=\frac{q}{E_q}=(1,{\bf v_q})  \,, \qquad 
\bar{v}_q=(1,{\bf\bar{v}_q})   \,, 
\en 
where 
\eq
{\bf\bar{v}_q} = \frac{{\bf v_q}}{1+\frac{Q^2}{2E_kE_q}} 
 = {\bf v_q} \ \frac{t+u}{t+u-2 Q^2} \,,
\en
the propagators become
\begin{align}
\frac{1}{t_{1,2}}\equiv\frac{1}{(k_{1,2}-p_1)^2}&=\frac{1}{-2k_{1,2}\cdot p_1}
=-\frac{2}{s_2-u}\prop{\pm v_1}\,,
\\
\frac{1}{u_{1,2}}\equiv\frac{1}{(k_{1,2}-p_2)^2}&=\frac{1}{-2k_{1,2}\cdot p_2}
=-\frac{2}{s_2-t}\prop{\pm v_2}\,,
\\
\frac{1}{w_{1,2}}\equiv\frac{1}{(k_{1,2}+q)^2}&=\frac{1}{2k_{1,2}\cdot q+Q^2}
=\frac{2}{2Q^2-t-u}\prop{\pm \bar{v}_q}\,.
\end{align}
The construction of the scaled momenta is depicted in Fig.~\ref{fig7_jhep}. 
The general denominators appearing in our calculations will have the form
\begin{align}
\frac{1}{t_1^{n_1}t_2^{n_2}u_1^{n_3}u_2^{n_4}w_1^{n_5}w_2^{n_6}}
=&\left(-\frac{2}{s_2-u}\right)^{n_1+n_2}
\left(-\frac{2}{s_2-t}\right)^{n_3+n_4}\left(\frac{2}{2Q^2-t-u}\right)^{n_5+n_6}
\nonumber\\
\times &
\powprop{v_1}{n_1}\powprop{-v_1}{n_2}\powprop{v_2}{n_3}\powprop{-v_2}{n_4}
\powprop{\bar{v}_q}{n_5}\powprop{-\bar{v}_q}{n_6}\,.
\label{eq:general DY propagator}
\end{align}
In terms of the pictures from our general considerations, 
the situation in the $x_0\equiv 1$ plane looks 
as Fig.~\ref{fig8_jhep}. 

We can decompose Eq.~(\ref{eq:general DY propagator}) into a sum of terms were each one contains 
a pair of linearly independent vectors by: 
\begin{itemize}
\item[1.] performing 
two-point partial fractioning between the pairs of linearly dependent vectors;
\item[2.]doing three-point partial fractioning between the remaining three vectors.
\end{itemize}
The $\lambda$ used for three point partial fractioning can be read of using 
the condition 
\begin{align}
{\bf 0}=(s_2-u)\mathbf{v_1}+(s_2-t)\mathbf{v_2}+(2Q^2-t-u)\mathbf{\bar{v}_q} = 2 \sqrt{s_2} \, 
\Big(\mathbf{p}_1+\mathbf{p}_2-\mathbf{q}\Big) 
\,,
\end{align} 
which is the direct consequence of Eq.~(\ref{eq:CMS condition}).
 
Expressing the numerator originating from some Dirac trace as a polynomial of the 
two corresponding Mandelstam variables we can simplify our angular integrals. 
In particular, the numerator simply decreases the powers of propagators in 
denominator. 
All that we need is to complete the calculation of the following set of integrals
\eq 
\hspace*{-1cm}
I^{(0)}_{j,l}(u_{12}^\pm;\eps)
&=&\int\dx\Omega_{k_1,k_2}\powprop{v_1}{j}\powprop{\mp v_2}{l}
=\int\dx\Omega_{k_1,k_2}\powprop{-v_1}{j}\powprop{\pm v_2}{l}
\,, \nonumber\\
\hspace*{-1cm}
I^{(1)}_{j,l}(u_{i\bar{q}}^\pm,v_{\bar{q}\bar{q}};\eps)
&=&\int\dx\Omega_{k_1,k_2}\, \powprop{\bar{v}_q}{j}\powprop{\mp v_i}{l}
=\int\dx\Omega_{k_1,k_2}\, \powprop{-\bar{v}_q}{j}\powprop{\pm v_i}{l}
\,,  
\en
where $i=1,2$ and 
\eq 
\hspace*{-1cm}
& &
u_{12}^\pm = 1 \pm (1-v_{12})\,, \qquad 
u_{i\bar{q}}^\pm = 1 \pm (1-v_{i\bar{q}})\,, \nonumber\\
\hspace*{-1cm}
& &
v_{12}=1-\cos\vartheta=\frac{2 s s_2}{(s_2-u)(s_2-t)} \,, 
\qquad
v_{\bar{q}\bar{q}}=\frac{4s Q^2}{(s+Q^2-s_2)^2}\,, \nonumber\\
\hspace*{-1cm}
& &
v_{1\bar{q}}=2 \biggl[1 + \frac{su}{(s_2-u)(s+Q^2-s_2)}\biggr]\,,
\qquad 
v_{2\bar{q}}=2 \biggl[1 + \frac{st}{(s_2-t)(s+Q^2-s_2)}\biggr] \,,
\en 
for integer valued $j$ and $l$. These are exactly the massless and single-massive two-denomi\-nator 
integrals presented in the main text. Note that the integrals depending on 
the arguments $v_{1\bar{q}}$ and $v_{2\bar{q}}$ 
are related by a crossing symmetry, i.e.\,by interchange $t \leftrightarrow u$. 
All in all, we were able to write down all potentially 
appearing integrals for the process without any concrete calulation.

\section{Consistency of the hypergeometric representation of angular integrals}
\label{sec:Consistency of the hypergeometric representation of angular integrals} 

In this appendix we present further details regarding hypergeometric 
representation of the general two denominator angular integrals in $D$ dimensions. 
First, we check consistency of the master formula for the two denominator integral in terms 
of the Lauricella function derived in Eq.~(\ref{eq:Hypergeometric representation of Ijl}) 
with known special results, e.g. derived in Ref.~\cite{Somogyi:2011ir}.  

\subsection{Double massive integral in four dimensions}

To perform the check of Eq.~(\ref{eq:Hypergeometric representation of Ijl}) 
we start with the double massive integral in four 
dimensions. In particular, for the special case $j=l=1$ and $\eps=0$ we recover the well-known 
four-dimensional expression for $I_{1,1}$ using the chain of the reduction identities
\begin{align}
&\lauricella(a_1,a_2,a_3,b_1,b_2,0,c,x_1,x_2,x_3)=\appellthree(a_1,a_2,b_1,b_2,c,x_1,x_2)\,,
\label{eq:lauricellatoappellthree}
\\
&\appellthree(a_1,a_2,b_1,b_2,a_1+a_2,x_1,x_2)=(1-x_2)^{-b_2} \, 
\appell\left(a_1,b_1,b_2,a_1+a_2,x_1,\frac{x_2}{x_2-1}\right)\,,
\label{eq:appellthreetoappell}
\\
&\appell(a,b_1,b_2,b_1+b_2,x_1,x_2)=(1-x_2)^{-a} \, 
\ghy\left(a,b_1,b_1+b_2,\frac{x_1-x_2}{1-x_2}\right)\,,
\label{eq:appelltoghy}
\end{align}
the quadratic hypergeometric transformation
\begin{align}
\ghy\left(\frac{a}{2},\frac{a+1}{2},b+\frac{1}{2},x\right)=(1+\sqrt{x})^{-a}
\,\ghy\left(a,b,2b,2-\frac{2}{1+\sqrt{x}}\right) \,,
\label{eq:ghy quadratic transformation}
\end{align}
and 
\begin{align}
\ghy(1,1,2,x)=\sum_{n=0}^\infty\frac{\pochhammer{1}{n}\pochhammer{1}{n}}{\pochhammer{2}{n}}
\,\frac{x^n}{n!}=\sum_{n=0}^\infty\frac{x^n}{n+1}=-\frac{\log(1-x)}{x}\,.
\label{eq:ghytolog}
\end{align}
Starting with formula~(\ref{eq:Hypergeometric representation of Ijl}), where we interchange 
the symmetric parameters $a_1$ and $b_1$, we straightforwardly obtain
\begin{align}
I_{1,1}(v_{12},v_{11},v_{22};0)&\overset{(\ref{eq:Hypergeometric representation of Ijl})}{=}
\frac{2\pi}{v_{12}}\,\lauricella\left(1,\frac{1}{2},\frac{1}{2},
\frac{1}{2},1,0,\frac{3}{2};1-\frac{v_{11}}{v_{12}},1-\frac{v_{22}}{v_{12}},1-v_{12}\right)
\nonumber\\
&\overset{(\ref{eq:lauricellatoappellthree})}{=}\frac{2\pi}{v_{12}}\,
\appellthree\left(1,\frac{1}{2},\frac{1}{2},1,\frac{3}{2},1-\frac{v_{11}}{v_{12}},1-\frac{v_{22}}{v_{12}}\right)
\nonumber\\
&\overset{(\ref{eq:appellthreetoappell})}{=}\frac{2\pi}{v_{12}}\,\frac{v_{12}}{v_{22}}\,
\appell\left(1,\frac{1}{2},1,\frac{3}{2},1-\frac{v_{11}}{v_{12}},1-\frac{v_{12}}{v_{22}}\right)
\nonumber\\
&\overset{(\ref{eq:appelltoghy})}{=}\frac{2\pi}{v_{12}}\,\ghy\left(1,\frac{1}{2},
\frac{3}{2};1-\frac{v_{11}v_{22}}{v_{12}^2}\right)
\nonumber\\
&\overset{(\ref{eq:ghy quadratic transformation})}{=}\frac{2\pi}{v_{12}+\sqrt{v_{12}^2-v_{11}v_{22}}}
\,\ghy\left(1,1,2,1-\frac{v_{12}-\sqrt{v_{12}^2-v_{11}v_{22}}}{v_{12}+\sqrt{v_{12}^2-v_{11}v_{22}}}\right)
\nonumber\\
&\overset{(\ref{eq:ghytolog})}{=}\frac{\pi}{\sqrt{v_{12}^2-v_{11}v_{22}}}
\,\log\left(\frac{v_{12}+\sqrt{v_{12}^2-v_{11}v_{22}}}{v_{12}-\sqrt{v_{12}^2-v_{11}v_{22}}}\right)\,.
\end{align}

\subsection{Single massive integral}

In consideration of single massive integral we use two important indentities 
\begin{align}
&\appell\left(a,b,b,c,\frac{y+\sqrt{y+(1-y)x}}{y-1},\frac{y-\sqrt{y+(1-y)x}}{y-1}\right)
\nonumber\\
&= \frac{(1-y)^b}{B(a,c-a)} \ 
\int_0^1\dx t\,t^{a-1}(1-t)^{c-a-1}(1-x t^2-y (1-t)^2)^{-b}
\nonumber\\
&=(1-y)^b \sum_{m,n=0}^\infty\frac{\pochhammer{a}{2m}\pochhammer{c-a}{2n}
\pochhammer{b}{m+n}}{\pochhammer{c}{2m+2n}}\frac{x^{m}}{m!} \frac{y^n}{n!} \,, 
\label{eq:Appell function bb sum}
\end{align}
and 
\eq 
\sum_{n=0}^\infty \frac{\pochhammer{l}{2n}4^{-n}}{\pochhammer{\frac{3}{2}-\eps}{m+n+p}n!}
= \ghy\left(\frac{l}{2},\frac{l+1}{2},\frac{3}{2}-\eps,1\right)
\frac{\pochhammer{1-l-\eps}{m+p}}{\pochhammer{2-l-2\eps}{2m+2p}}\,4^{m+p}\,. 
\label{eq:sum_id_Poch}
\en 
The identity~(\ref{eq:Appell function bb sum}) 
is derived using integral representations for the Appell and gamma functions 
and the bivariate generating function 
\eq 
\Big(1-x t^2-y (1-t)^2\Big)^{-b} = 
\sum_{m,n=0}^\infty \, \pochhammer{b}{m+n} \, 
\frac{x^{m}}{m!} \frac{y^n}{n!} \, t^{2m} \, (1-t)^{2n} \,.
\en 
and then using duplication formula for 
the Pochhammer symbols\index{Pochhammer symbol}     
\eq\label{eq:pochhammer doubling}
(x)_n \, \biggl(x+\frac{1}{2}\biggr)_n = 2^{-2n} \, (2x)_{2n} \,. 
\en 
The second identity~(\ref{eq:sum_id_Poch}) is proved using the 
Pochhammer shift identities\index{Pochhammer shift identity}   
\eq
(x)_{m+n} &=& (x+m)_{n} \, (x)_{m}\,, 
\label{eq:pochhammer shift} \\ 
(x)_{m+n} \, (x-n)_n &=& (x+m)_{n} \, (x-n)_{m+n} \,, 
\label{eq:pochhammer shift2}
\en 
duplication identity~(\ref{eq:pochhammer doubling}), 
and the Pochhammer-Gauss summation formula 
\eq\label{Pochhamer_Gauss_Theorem} 
\sum\limits_{n=0}^\infty \, \frac{(a)_n \, (b)_n}{(c)_n n!} 
= \ghy(a,b,c,1) = \frac{(c-a)_a}{(c-a-b)_a} \,.
\en 
Here, Eq.~(\ref{eq:pochhammer shift2}) is a generalization of Eq.~(\ref{eq:pochhammer shift}) 
derived by multiplying of l.h.s. and r.h.s. of Eq.~(\ref{eq:pochhammer shift}) by $(x-n)_n$: 
\eq 
(x)_{m+n} \, (x-n)_n = (x+m)_{n} 
\, \underbrace{(x)_{m} \, (x-n)_{n}}_{= (x-n)_{m+n}} = (x+m)_n \, (x-n)_{m+n}\,. 
\en  
Eq.~(\ref{Pochhamer_Gauss_Theorem}) is the combination of two formulas --- power expansion 
of Gauss function~(\ref{Gauss_Pochhamer}) and Gauss summation theorem~(\ref{Gauss_theorem}).  

In particular, Eq.~(\ref{eq:Appell function bb sum}) is derived as: 
\begin{align}\label{sum_proof}
\sum_{n=0}^\infty \frac{\pochhammer{l}{2n}4^{-n}}{\pochhammer{\frac{3}{2}-\eps}{m+n+p}n!}
&\overset{\rm(\ref{eq:pochhammer doubling}),\rm(\ref{eq:pochhammer shift})}{=} 
\frac{1}{\pochhammer{\frac{3}{2}-\eps}{m+p}} \ 
\sum_{n=0}^\infty\frac{\pochhammer{\frac{l}{2}}{n}
\pochhammer{\frac{l+1}{2}}{n}}{\pochhammer{\frac{3}{2}-\eps+m+p}{n} n!}
\nonumber\\
&\overset{\rm(\ref{Pochhamer_Gauss_Theorem})}{=}
\frac{1}{\pochhammer{\frac{3}{2}-\eps}{m+p}} \ 
\frac{\left(\frac{3}{2}-\frac{l}{2}+m+p-\eps\right)_{\frac{l}{2}}}  
     {\left(1-l+m+p-\eps\right)_{\frac{l}{2}}}  
\nonumber\\
&\overset{\rm(\ref{eq:pochhammer shift2})} 
{=}\frac{\left(\frac{3}{2}-\frac{l}{2}-\eps\right)_\frac{l}{2}}
{\left(1-l-\eps\right)_\frac{l}{2}} \  
\frac{\pochhammer{1-l-\eps}{m+p}}{\pochhammer{\frac{3}{2}
-\frac{l}{2}-\eps}{m+p}\pochhammer{1-\frac{l}{2}-\eps}{m+p}}
\nonumber\\
&\overset{\rm(\ref{eq:pochhammer doubling})}{=}\ghy\left(\frac{l}{2},\frac{l+1}{2},
\frac{3}{2}-\eps,1\right) \  
\frac{\pochhammer{1-l-\eps}{m+p}}{\pochhammer{2-l-2\eps}{2m+2p}}\,4^{m+p}\,.
\end{align}
Note, in the step from the second to the third line of Eq.~(\ref{sum_proof}) we apply 
the Pochhammer shift identity~(\ref{eq:pochhammer shift2}) twice. 

Now we are in the position to derive the formula for the angular 
single massive integral 
\begin{align}\label{single_mass}
&I_{j,l}(v_{12},v_{11},0,\eps)=\frac{2\pi}{1-2\eps}\, v_{12}^{1-j-l-\eps}
\nonumber\\
&\times\lauricella\left(\frac{j}{2},\frac{l}{2},\frac{3-j-l-2\eps}{2},\frac{j+1}{2},\frac{l+1}{2},
\frac{2-j-l-2\eps}{2},\frac{3}{2}-\eps,1-\frac{v_{11}}{v_{12}},1,1-v_{12}\right)
\nonumber\\
&=\frac{2\pi}{1-2\eps}\, v_{12}^{1-j-l-\eps}\sum_{m,n,p=0}^\infty
\frac{\pochhammer{j}{2m}\pochhammer{l}{2n}\pochhammer{2-j-l-2\eps}{2p}}
{\pochhammer{\frac{3}{2}-\eps}{m+n+p}m!n!p!}\,4^{-m-n-p}
\left(1-\frac{v_{11}}{v_{12}}\right)^m(1-v_{12})^p
\nonumber\\
&=\frac{2\pi}{1-2\eps}\, v_{12}^{1-j-l-\eps}
\,\ghy\left(\frac{l}{2},\frac{l+1}{2},\frac{3}{2}-\eps,1\right)
\nonumber\\
&\times\sum_{m,p=0}^\infty\frac{\pochhammer{j}{2m}\pochhammer{2-j-l-2\eps}{2p}
\pochhammer{1-l-\eps}{m+p}}{\pochhammer{2-l-2\eps}{2m+2p}m!p!}
\,\left(1-\frac{v_{11}}{v_{12}}\right)^m(1-v_{12})^p
\nonumber\\
&\overset{(\ref{eq:Appell function bb sum})}{=}I_{l}^{(0)}(\eps)\,v_{12}^{-j}
\appell\left(j,1-l-\eps,1-l-\eps,2-l-2\eps,1-\frac{1+\sqrt{1-v_{11}}}{v_{12}},
1-\frac{1-\sqrt{1-v_{11}}}{v_{12}}\right)
\nonumber\\
&=I_{j,l}^{(1)}(v_{12},v_{11},\eps)\,,  
\end{align}
which is in the exact agreement with the known analytic result 
for the single massive integral in $D$ dimensions~\cite{Somogyi:2011ir}. 

\subsection{Massless integral}

We perform the check of the massless integral starting from the single massive 
result~(\ref{single_mass}), where we put $v_{11}=0$. 
Using the identities 
\begin{align}
\appell(a,b_1,b_2,c,x,1)&=\ghy(a,b_2,c,1)\,\ghy(a,b_1,c-b_2,x)\,,
\\
\ghy(a,b,c,x)&=(1-x)^{-a}\,\ghy\left(a,c-b,c,\frac{x}{x-1}\right)\,, 
\end{align}
the Gauss theorem\index{Gauss theorem}~(\ref{Gauss_theorem})  
\eq 
\ghy(j,1-l-\eps,2-l-2\eps,1) &=& 
\frac{\Gamma(2-l-2\eps) \Gamma(1-j-\eps)}{\Gamma(2-l-j-2\eps) \Gamma(1-\eps)} 
= \frac{\pochhammer{2-j-l-2\eps}{j}}{\pochhammer{1-j-\eps}{j}} 
\nonumber\\
&=& \frac{\pochhammer{2-j-l-2\eps}{j+l}}{\pochhammer{1-j-\eps}{j} \, 
\pochhammer{2-l-2\eps}{l}}  \,,
\en 
and Eq.~(\ref{Ijl_v12_eps}) 
we obtain the formula 
\begin{align}
I_{j,l}(v_{12},0,0;\eps)&=I_{l}^{(0)}(\eps)\,v_{12}^{-j}
\, \appell\left(j,1-l-\eps,1-l-\eps,2-l-2\eps,1-\frac{2}{v_{12}},1\right)
\nonumber\\
&=I_{l}^{(0)}(\eps)\,v_{12}^{-j}\ghy(j,1-l-\eps,2-l-2\eps,1)
\, \ghy\left(j,1-l-\eps,1-\eps,1-\frac{2}{v_{12}}\right)
\nonumber\\
&=\frac{I_{l}^{(0)}(\eps)}{2^j}\ghy(j,1-l-\eps,2-l-2\eps,1)
\, \ghy\left(j,l,1-\eps,1-\frac{v_{12}}{2}\right)
\nonumber\\
&=\frac{I^{(0)}(\eps)}{2^{j+l}} \, 
\frac{\pochhammer{2-j-l-2\eps}{j+l}}{\pochhammer{1-j-\eps}{j}\pochhammer{1-l-\eps}{l}}
\, \ghy\left(j,l,1-\eps,1-\frac{v_{12}}{2}\right)
\nonumber\\
&=I^{(0)}_{j,l}(v_{12};\eps)\,,
\end{align}
which is again in complete agreement with the result of Ref.~\cite{Somogyi:2011ir}. 

\section{Double Nielsen polylogarithms}
\label{app:Nielsen}   

\subsection{Basic properties of double Nielsen polylogarithms}

Basic properties of the \textit{double Nielsen polylogarithm}\index{Nielsen polylogarithm}
\begin{align}
S_{n,p_1,p_2}(x,y)\equiv\frac{(-1)^{n+p_1+p_2-1}}{(n-1)!p_1!p_2!}
\int_0^1\frac{\dx t}{t}\,\log^{n-1} t\log^{p_1}(1-xt)\log^{p_2}(1-yt)\,.
\label{eq:double Nielsen polylog definition}
\end{align}
are: 

(1) symmetry $S_{n,p_1,p_2}(x,y)=S_{n,p_2,p_1}(y,x)$, 

(2) reduction to single Nielsen polylogarithm 
$S_{n,p_1,0}(x,y)=S_{n,p_1}(x)$ and $S_{n,0,p_2}(x,y)=S_{n,p_2}(y)$. 

With expansion
\begin{align}
\log^m(1+x)=m!\sum_{k=m}^\infty S_k^{(m)}\frac{x^k}{k!}\quad\text{for }|x|<1\,,
\end{align}
where  
\begin{align}
S_k^{(m)}=\sum_{i=0}^{k-m}\frac{1}{i!}\binom{k-1+i}{k-m+i}
\binom{2k-m}{k-m-i}\sum_{j=0}^i(-1)^i \binom{i}{j}^{k-m+i}\,,
\end{align}
are Stirling numbers\index{Stirling number} 
of the 1st kind and integral 
\eq
\int\limits_0^1\dx t t^{m-1}(\log t)^{n-1} = -(-m)^{-n}(n-1)! \,,
\en 
one obtains the series 
\begin{align}
S_{n,p_1,p_2}(x,y)=\sum_{k,l=0}^\infty\frac{(-1)^{k+l}
S_{p_1+k}^{(p_1)}S_{p_2+l}^{(p_2)}}{(p_1+k)!(p_2+l)!(p_1+p_2+k+l)^n}\,x^{p_1+k} 
y^{p_2+l}\quad\text{for } |x|\leq 1\,,\,|y|\leq 1\,.
\end{align}

\subsection{Relation of Goncharov and Nielsen polylogarithms}

Goncharov polylogarithms\index{Goncharov polylogarithm}  
are a powerful tool for working with a wide range of generalized
logarithms~\cite{Goncharov:2001iea}. They are recursively defined via the iterated integral
\begin{align}
G(a_1,\dots,a_n;z)\equiv\int_0^1\frac{\dx t}{t-a_1}\,G(a_2,\dots,a_n;t)
\end{align}
with $G(z)=1$.
It holds  with $\vec{a}_n=(\underbrace{a,\dots,a}_{n\text{-times}})$:
\eq
& &G(\vec{0}_n;z)=\frac{1}{n!}\log^n z\,, \quad
   G(\vec{a}_n;z)=\frac{1}{n!}\log^n \left(1-\frac{z}{a}\right)\,, \nonumber\\
& &G(\vec{0}_{n-1},a;z)=-\mathrm{Li}_n\left(\frac{z}{a}\right), \quad
   G(\vec{0}_{n-k},\vec{a}_k;z)=(-1)^k S_{n-k,k}\left(\frac{z}{a}\right) \,.
\en
Goncharov polylogarithms are related to other polylogarithms by the relations:

(1) with  Nielsen polylogarithms\index{Nielsen polylogarithm} by
$S_{n,p}(z)=(-1)^p G(\overbrace{0,\dots,0}^n,\overbrace{1,\dots,1}^p;z)$;

(2) with Harmonic polylogarithms\index{Harmonic polylogarithm} by 
$H(a_1,\dots a_n;z)=(-1)^p G(a_1,\dots,a_n;z)$, 

where $a_i\in\lbrace 0,1\rbrace$

(3) with Multiple polylogarithms\index{Multiple polylogarithm} by
$$G(\vec{0}_{m_1-1},a_1,\dots,\vec{0}_{m_k-1},a_k;z)=(-1)^k \mathrm{Li}_{m_k,\dots,m_1} 
\left(\frac{a_{k-1}}{a_k},\dots,\frac{a_{1}}{a_2},\frac{z}{a_1}\right)\,,$$ where
\begin{align}
\mathrm{Li}_{m_1,\dots,m_k}=\sum_{0<n_1<\dots<n_k}\frac{z_1^{n_1} z_2^{n_2} \dots
z_k^{n_k}}{n_1^{m_1}n_2^{m_2}\dots n_k^{m_k}}\,.
\end{align}
Goncharov polylogarithms satisfy the \textit{shuffle algebra}\index{Shuffle algebra} 
\begin{align}
G(\vec{a};z)G(\vec{b},z)=\sum_{\vec{c}\in\vec{a}\shuffle\vec{b}}G(\vec{c};z)\,,
\end{align}
where $\vec{a}\shuffle\vec{b}$ denotes the set of all shuffles of the vectors $\vec{a}$
and $\vec{b}$, i.e.\,all the ways of interlacing them while keeping the order 
within $\vec{a}$ and $\vec{b}$.
Goncharov polylogarithm has the integration by parts property
\begin{align}
G(z_1,z_2,\dots,z_n;1)&=G(z_1;1)G(z_2,\dots,z_n;1)-G(z_2,z_1;1)G(z_3,\dots,z_n;1)
\nonumber\\
&+\dots-(-1)^n G(z_n,\dots,z_2,z_1;1)\,.
\end{align}
Another useful identity is the logarithmic integral
\begin{align}
G(\vec{0}_m,z_1,z_2,\dots,z_n;1)=\frac{(-1)^m}{m!}\int_0^1\frac{\dx t\,\log^m t}{t-z_1}\,G(z_2,\dots,z_n;t)\,.
\label{eq:Goncharov log integral}
\end{align}

\subsection{Double Nielsen polylogarithms in terms of multiple polylogarithms}

Using the logarithmic integral Eq.~(\ref{eq:Goncharov log integral}) and the shuffle algebra
we can represent the double Nielsen polylogarithm in terms of multiple polylogarithms. Starting
from Eq.~(\ref{eq:double Nielsen polylog definition}) we find
\begin{align}
S_{n,p_1,p_2}(x,y)&=\frac{(-1)^{n+p_1+p_2-1}}{(n-1)!p_1!p_2!}\int_0^1\frac{\dx t}{t}\,\log^{n-1}
t\log^{p_1}(1-xt)\log^{p_2}(1-yt)
\nonumber\\
&=\frac{(-1)^{n+p_1+p_2-1}}{(n-1)!}\int_0^1\frac{\dx t}{t}\,\log^{n-1}t\,
G\left(\vec{\left(\frac{1}{x}\right)}_{p_1};t\right)
\, G\left(\vec{\left(\frac{1}{y}\right)}_{p_2};t\right)
\nonumber\\
&=\frac{(-1)^{n+p_1+p_2-1}}{(n-1)!}\int_0^1\frac{\dx t}{t}\,\log^{n-1}t\,
\sum_{\vec{z}\in\vec{\frac{1}{x}}\shuffle\vec{\frac{1}{y}}} G(z_1,\dots,z_{p_1+p_2};t)
\nonumber\\
&=(-1)^{p_1+p_2}\sum_{\vec{z}\in\vec{\frac{1}{x}}\shuffle\vec{\frac{1}{y}}}
G(\vec{0}_{n-1},0,z_1,\dots,z_{p_1+p_2};1)
\nonumber\\
&=\sum_{\vec{z}\in\vec{x}_{p_1}\shuffle\vec{y}_{p_2}} \mathrm{Li}_{1,1,\dots,1,n+1}
\left(\frac{z_{p_1+p_2}}{z_{p_1+p_2-1}},\dots,\frac{z_2}{z_1},z_1\right)\,.
\end{align}
Therefore, double Nielsen polylogarithms can be calculated from multiple polylogarithms 
using the identity
\begin{equation}
S_{n,p_1,p_2}(x,y)=\sum_{\vec{z}\in\vec{x}_{p_1}\shuffle\vec{y}_{p_2}}
\mathrm{Li}_{1,1,\dots,1,n+1}\left(\frac{z_{p_1+p_2}}{z_{p_1+p_2-1}},\dots,\frac{z_2}{z_1},z_1\right)\,.
\end{equation}
For the reduction and calculation of multiple polylogarithms there exist publicly available
tools~\cite{Frellesvig:2016ske,Duhr:2019tlz}.
Some explicit examples for the shuffle sums are
\begin{align}
S_{n,1,1}(x,y)&=\sum_{\vec{z}\in x\shuffle y} \mathrm{Li}_{1,n+1}\left(\frac{z_2}{z_1},z_1\right)
=\mathrm{Li}_{1,n+1}\left(\frac{y}{x},x\right)+\mathrm{Li}_{1,n+1}\left(\frac{x}{y},y\right)\,,
\\
S_{n,2,1}(x,y)&=\sum_{\vec{z}\in \vec{x}_2\shuffle y}
\mathrm{Li}_{1,1,n+1}\left(\frac{z_3}{z_2},\frac{z_2}{z_1},z_1\right)
\nonumber\\
&=\mathrm{Li}_{1,1,n+1}\left(\frac{y}{x},1,x\right)+\mathrm{Li}_{1,1,n+1}
\left(\frac{x}{y},\frac{y}{x},x\right)
+\mathrm{Li}_{1,1,n+1}\left(1,\frac{x}{y},y\right)\,,
\\
\\
S_{n,2,2}(x,y)&=\sum_{\vec{z}\in \vec{x}_2\shuffle \vec{y}_2} \mathrm{Li}_{1,1,1,n+1}
\left(\frac{z_4}{z_3},\frac{z_3}{z_2},\frac{z_2}{z_1},z_1\right)
\nonumber\\
&=\mathrm{Li}_{1,1,1,n+1}\left(1,\frac{y}{x},1,x\right)
+\mathrm{Li}_{1,1,1,n+1}\left(\frac{y}{x},\frac{x}{y},\frac{y}{x},x\right)
\nonumber\\
&
+\mathrm{Li}_{1,1,1,n+1}\left(\frac{x}{y},1,\frac{y}{x},1,x\right)
+\mathrm{Li}_{1,1,1,n+1}\left(\frac{y}{x},1,\frac{x}{y},y\right)
\nonumber\\
&
+\mathrm{Li}_{1,1,1,n+1}\left(\frac{x}{y},\frac{y}{x},\frac{x}{y},y\right)
+\mathrm{Li}_{1,1,1,n+1}\left(1,\frac{x}{y},1,y\right)\,.
\end{align}

\clearpage 

\section{Tables of angular integrals}
\label{sec:tables_AI}

\subsection{Massless integral with one denominator}

\begin{align}
I^{(0)}_{-4}(\eps)&= 4I^{(0)}(\eps) \, 
\frac{(3-\eps) (4-\eps)}{(3-2\eps) (5-2\eps)} 
\nonumber\\
&=
\frac{32 \pi }{5}+\frac{1192 \pi  \eps}{75}+\frac{38152 \pi  \eps^2}{1125}
+\frac{1168912 \pi  \eps^3}{16875}+\frac{35313472 \pi  \eps^4}{253125}+\mathcal{O}\left(\eps^5\right)\,,
\\
I^{(0)}_{-3}(\eps)&= 2I^{(0)}(\eps) \, 
\frac{3-\eps}{3-2\eps} 
\nonumber\\
&= 
4 \pi +\frac{28 \pi  \eps }{3}+\frac{176 \pi  \eps^2}{9}
+\frac{1072 \pi  \eps^3}{27}+\frac{6464 \pi  \eps^4}{81}+\mathcal{O}\left(\eps^5\right)\,,
\\
I^{(0)}_{-2}(\eps)&= 2I^{(0)}(\eps) \, \frac{2-\eps}{3-2\eps} 
\nonumber\\
&=\frac{8 \pi }{3}+\frac{52 \pi  \eps }{9}+\frac{320 \pi  \eps^2}{27}
+\frac{1936 \pi  \eps^3}{81}+\frac{11648 \pi  \eps^4}{243}+\mathcal{O}\left(\eps^5\right)\,,
\\
I^{(0)}_{0}(\eps)&=I^{(0)}_{-1}(\eps) = I^{(0)}(\eps) 
= \frac{2 \pi}{1 - 2 \eps} = 
  2 \pi +4 \pi  \eps +8 \pi  \eps^2+16 \pi  \eps^3
+32 \pi  \eps^4+\mathcal{O}\left(\eps^5\right)\,,
\\
I^{(0)}_{1}(\eps)&=- I^{(0)}(\eps) \, \frac{1-2\eps}{2 \eps} = -\frac{\pi }{\eps }\,,
\\
I^{(0)}_{2}(\eps)&= - I^{(0)}(\eps) \, \frac{1-2\eps}{2 (1+\eps)} = - \frac{\pi}{1+\eps} = 
-\pi +\pi  \eps -\pi  \eps^2+\pi  \eps^3-\pi  \eps^4+\mathcal{O}\left(\eps^5\right)
\,.
\end{align}

\vspace*{.25cm}
\subsection{Massive integral with one denominator}
\begin{align}
I^{(1)}_{-4}(v_{11};\eps)&= I^{(0)}(\eps) \, \frac{2 (6-3v_{11}-2\eps) (4-\eps) 
- 3 v_{11} (4-v_{11}-2\eps)}{(3-2\eps) (5-2\eps)} \nonumber\\
&=
 \frac{2}{5} \pi  \left(16-12 v_{11}+v_{11}^2\right)
+\frac{4}{75} \pi  \left(298-246 v_{11}+23 v_{11}^2\right) \eps 
+\mathcal{O} \left(\eps^2\right)\,,
\\
I^{(1)}_{-3}(v_{11};\eps)&= I^{(0)}(\eps) \, \frac{6-3v_{11}-2\eps}{3-2\eps} \nonumber\\
&= 
2 \pi (2-v_{11})+\frac{4}{3} \pi  (7-4 v_{11}) \eps
+\mathcal{O}\left(\eps^2\right)\,,
\\
I^{(1)}_{-2}(v_{11};\eps)&= I^{(0)}(\eps) \, \frac{4-v_{11}-2\eps}{3-2\eps} \nonumber\\
&= 
\frac{2}{3} \pi  (4-v_{11})+\frac{4}{9} \pi  (13-4 v_{11}) \eps
+\mathcal{O}\left(\eps^2\right)\,,
\\
I^{(1)}_{0}(v_{11};\eps)&=I^{(1)}_{-1}(v_{11};\eps) = I^{(0)}(\eps) 
= \frac{2 \pi}{1 - 2 \eps} = 
  2 \pi +4 \pi  \eps +\mathcal{O}\left(\eps^2\right)\,,\\
I^{(1)}_{1}(v_{11};\eps)&
=\frac{\pi  \log \left(\frac{1+\sqrt{1-v_{11}}}{1-\sqrt{1-v_{11}}}\right)}{\sqrt{1-v_{11}}}
-\frac{\pi  \eps \left(\log ^2\left(\frac{1+\sqrt{1-v_{11}}}{1-\sqrt{1-v_{11}}}\right)
+4 \dilog\left(\frac{2 \sqrt{1-v_{11}}}{\sqrt{1-v_{11}}-1}\right)\right)}{2 \sqrt{1-v_{11}}}
\nonumber\\ 
&+\mathcal{O}\left(\eps^2\right)\,,
\\
I^{(1)}_{2}(v_{11};\eps)
&=\frac{2\eps}{v_{11}} \, I^{(1)}_{1}(v_{11};\eps) 
+ \frac{1-2\eps}{v_{11}} \, I^{(0)} = 
\frac{2 \pi }{v_{11}} \, 
\biggl[1 + \frac{\eps \log \left(\frac{1+\sqrt{1-v_{11}}}{1-\sqrt{1-v_{11}}}\right)}{\sqrt{1-v_{11}}} 
\biggr] 
\nonumber\\
&+ \mathcal{O}\left(\eps^2\right)\,.
\end{align}

\subsection{Massless integral with two denominators}

\begin{align}
I^{(0)}_{-4,0}(v_{12};\eps)&=I^{(0)}_{-4}(\eps)\,, 
\\
I^{(0)}_{-4,-1}(v_{12};\eps)
&= \frac{1}{4 (3-\eps)} \,
\biggl[
  (7-2\eps)  I^{(0)}_{-4}(\eps)
+ (13-2\eps)  I^{(0)}_{-3,-1}(v_{12};\eps)
- 5 v_{12}   I^{(0)}_{-3}(\eps) \biggr]
\nonumber\\
&=\frac{32\pi}{15} \left(5-2 v_{12}\right)
 +\frac{8\pi}{225} \left(785 - 338 v_{12}\right) \eps
+\mathcal{O}\left(\eps^2\right)\,,
\\
I^{(0)}_{-4,-2}(v_{12};\eps)
&= \frac{1}{7-2\eps} \,
\biggl[
  (5-\eps)   I^{(0)}_{-4,-1}(v_{12};\eps)
+ (7-\eps)   I^{(0)}_{-3,-2}(v_{12};\eps)
- 3 v_{12}   I^{(0)}_{-3,-1}(v_{12};\eps) \biggr]
\nonumber\\
&=\frac{64\pi}{105} \left(30-20 v_{12}+3 v_{12}^2\right)
+\frac{16 \pi}{11025} \left(34470-24380 v_{12}+3909 v_{12}^2\right) \eps 
\nonumber\\
&+\mathcal{O}\left(\eps^2\right)\,,
\\
I^{(0)}_{-4,-3}(v_{12};\eps)
&= \frac{1}{4 (4-\eps)} \,
\biggl[
  (13-2\eps) I^{(0)}_{-4,-2}(v_{12};\eps)
+ (15-2\eps) I^{(0)}_{-3,-3}(v_{12};\eps)
\nonumber\\
&- 7 v_{12}   I^{(0)}_{-3,-2}(v_{12};\eps) \biggr]
=\frac{16\pi}{35} \left(70-60 v_{12}+15 v_{12}^2-v_{12}^3\right)
\nonumber\\
&+\frac{16 \pi}{3675}  \left(20895-18810 v_{12}
+4965 v_{12}^2-352 v_{12}^3\right) \eps 
+\mathcal{O}\left(\eps^2\right)\,,
\\
I^{(0)}_{-4,-4}(v_{12};\eps)
&= \frac{1}{9-\eps} \,
\biggl[
  (8-\eps) \Big(I^{(0)}_{-4,-3}(v_{12};\eps)+
                I^{(0)}_{-3,-4}(v_{12};\eps)\Big)
     - 4 v_{12} I^{(0)}_{-3,-3}(v_{12};\eps) \biggr]
\nonumber\\
&=\frac{16\pi}{315} \left(
1120-1120 v_{12}+360 v_{12}^2-40 v_{12}^3+v_{12}^4
\right)
\nonumber\\
&+\frac{32 \pi}{99225}  \left(518630-540680 v_{12} 
+181890 v_{12}^2-21260 v_{12}^3+563 v_{12}^4 
\right) \eps 
\nonumber\\
&+\mathcal{O}\left(\eps^2\right)\,,
\\
I^{(0)}_{-4,1}(v_{12};\eps)
&=(1-v_{12}) I^{(0)}_{-3}(\eps)
\nonumber\\
&+ \frac{v_{12}}{2 (2-\eps)} \,
\biggl[
  3 I^{(0)}_{-2}(\eps)
+ (7-2\eps) I^{(0)}_{-3,1}(v_{12};\eps)
- 3 v_{12}  I^{(0)}_{-2,1}(v_{12};\eps)
\biggr]
\nonumber\\
&=-\frac{\pi}{\eps} v_{12}^4
+\frac{\pi}{6} \left(
24+16 v_{12}+12 v_{12}^2+12 v_{12}^3-25 v_{12}^4 
\right)
\nonumber\\
&+\frac{\pi}{36}  \left(
336+256 v_{12}+228 v_{12}^2 
+300 v_{12}^3-415 v_{12}^4
\right) \eps+\mathcal{O}\left(\eps^2\right)\,,
\\
I^{(0)}_{-4,2}(v_{12};\eps)
&=(1-v_{12}) I^{(0)}_{-3,1}(v_{12};\eps)
\nonumber\\
&+ \frac{v_{12}}{3-2\eps} \,
\biggl[
          3 I^{(0)}_{-2,1}(v_{12};\eps)
+ (7-2\eps) I^{(0)}_{-3,2}(v_{12};\eps)
- 3 v_{12}  I^{(0)}_{-2,2}(v_{12};\eps)
\biggr]
\nonumber\\
&=-\frac{2 \pi}{\eps}  \left(8-5 v_{12}\right) v_{12}^3 
+\frac{4\pi}{3} \left(2+4v_{12}+9 v_{12}^2
-44 v_{12}^3+20 v_{12}^4 
\right)
\nonumber\\
&+\frac{2\pi}{9}
\, \left(26+64 v_{12}+198 v_{12}^2-680 v_{12}^3+305 v_{12}^4\right) \, 
\eps+\mathcal{O}\left(\eps^2\right)\,,
\\
I^{(0)}_{-3,0}(v_{12};\eps)&=I^{(0)}_{-3}(\eps)
\,, 
\\
I^{(0)}_{-3,-1}(v_{12};\eps)
&= \frac{1}{(5-2\eps)} \,
\biggl[
  (3-\eps) I^{(0)}_{-3}(\eps)
+ (5-\eps) I^{(0)}_{-2,-1}(v_{12};\eps)
- 2 v_{12} I^{(0)}_{-2}(\eps) \biggr]
\nonumber\\
&=\frac{4\pi}{5} \left(8-3 v_{12}\right)
+\frac{4\pi}{75} \left(298-123 v_{12}\right) \eps
+\mathcal{O}\left(\eps^2\right)\,, 
\end{align}

\begin{align} 
I^{(0)}_{-3,-2}(v_{12};\eps)
&= \frac{1}{4 (3-\eps)} \,
\biggl[
  (9-2\eps) I^{(0)}_{-3,-1}(v_{12};\eps)
+ (11-2\eps) I^{(0)}_{-2,-2}(v_{12};\eps)
\nonumber\\
&- 5 v_{12} I^{(0)}_{-2,-1}(v_{12};\eps) \biggr]
\nonumber\\
&=\frac{4\pi}{15} \left(
40-24 v_{12}+3 v_{12}^2
\right)+\frac{8\pi}{225} 
\left(785-507 v_{12}+69 v_{12}^2
\right) \eps
+\mathcal{O}\left(\eps^2\right)\,,
\\
I^{(0)}_{-3,-3}(v_{12};\eps)
&= \frac{1}{7-2\eps} \,
\biggl[
  (6-\eps) \Big(I^{(0)}_{-3,-2}(v_{12};\eps)+
                I^{(0)}_{-2,-3}(v_{12};\eps)\Big)
     - 3 v_{12} I^{(0)}_{-2,-2}(v_{12};\eps) \biggr]
\nonumber\\
&=\frac{4\pi}{35} \left(
160-120 v_{12}+24 v_{12}^2-v_{12}^3
\right)
\nonumber\\
&+\frac{8 \pi}{3675}  \left(
22980-18285 v_{12}+3909 v_{12}^2-176 v_{12}^3
\right) \eps 
+\mathcal{O}\left(\eps^2\right)\,,
\\
I^{(0)}_{-3,1}(v_{12};\eps)
&=(1-v_{12}) I^{(0)}_{-2}(\eps)
\nonumber\\
&+ \frac{v_{12}}{(3-2\eps)} \,
\biggl[
  2 I^{(0)}_{-1}(\eps)
+ (5-2\eps) I^{(0)}_{-2,1}(v_{12};\eps)
- 2 v_{12}  I^{(0)}_{-1,1}(v_{12};\eps)
\biggr]
\nonumber\\
&=-\frac{\pi}{\eps}  v_{12}^3 + \frac{\pi}{3}
\left(8+6 v_{12}+6 v_{12}^2-11 v_{12}^3\right)
\nonumber\\
&+\frac{\pi}{9} \left(
52+48 v_{12}+66 v_{12}^2-85 v_{12}^3
\right) \eps
+\mathcal{O}\left(\eps^2\right)\,,
\\
I^{(0)}_{-3,2}(v_{12};\eps)
&=(1-v_{12}) I^{(0)}_{-2,1}(v_{12};\eps)
\nonumber\\
&+ \frac{v_{12}}{(1-\eps)} \,
\biggl[
  I^{(0)}_{-1,1}(v_{12};\eps)
+ (2-\eps) I^{(0)}_{-2,2}(v_{12};\eps)
- v_{12}  I^{(0)}_{-1,2}(v_{12};\eps)
\biggr]
\nonumber\\
&=-\frac{3 \pi}{\eps}  \left(3-2 v_{12}\right) v_{12}^2 
+\pi \left(2+6 v_{12}-27 v_{12}^2+12 v_{12}^3\right)
\nonumber\\
&+\pi
\left(4+18 v_{12}-63 v_{12}^2+30 v_{12}^3\right) \eps 
+\mathcal{O}\left(\eps^2\right)\,,
\\
I^{(0)}_{-2,0}(v_{12};\eps)&=I^{(0)}_{-2}(\eps)
\,,
\\
I^{(0)}_{-2,-1}(v_{12};\eps)
&= \frac{1}{4 (2-\eps)} \,
\biggl[
  (7-2\eps) I^{(0)}_{-2}(\eps)
+ (5-2\eps) I^{(0)}_{-1,-1}(v_{12};\eps) 
- 3 v_{12}  I^{(0)}(\eps) \biggr]
\nonumber\\
&=\frac{4\pi}{3} \left(3-v_{12}\right)
+\frac{4\pi}{9} \left(21-8 v_{12}\right) \eps
+\mathcal{O}\left(\eps^2\right)\,,
\\
I^{(0)}_{-2,-2}(v_{12};\eps)
&= \frac{1}{5-2\eps} \,
\biggl[
  (4-\eps) \Big(I^{(0)}_{-2,-1}(v_{12};\eps)+
                I^{(0)}_{-1,-2}(v_{12};\eps)\Big)
     - 2 v_{12} I^{(0)}_{-1,-1}(v_{12};\eps) \biggr]
\nonumber\\
&=\frac{4\pi}{15} \left(24-12 v_{12}+v_{12}^2\right)
+\frac{8\pi}{225} \left(447-246 v_{12}+23 v_{12}^2\right) \eps
+\mathcal{O}\left(\eps^2\right)\,,
\\
I^{(0)}_{-2,1}(v_{12};\eps)
&=(1-v_{12}) I^{(0)}(\eps)
\nonumber\\
&+ \frac{v_{12}}{2 (1-\eps)} \,
\biggl[
  I^{(0)}(\eps)
+ (3-2\eps) I^{(0)}_{-1,1}(v_{12};\eps)
- v_{12}  I^{(0)}_{1}(\eps)
\biggr]
\nonumber\\
&=-\frac{\pi}{\eps}  v_{12}^2 
+\pi  \left(2+2 v_{12}-3 v_{12}^2\right)
+\pi  \left(4+6 v_{12}-7 v_{12}^2\right) \eps
+\mathcal{O}\left(\eps^2\right)\,,
\\
I^{(0)}_{-2,2}(v_{12};\eps)
&=(1-v_{12}) I^{(0)}_{-1,1}(v_{12};\eps)
\nonumber\\
&+ \frac{v_{12}}{(1-2\eps)} \,
\biggl[
  I^{(0)}_{1}(\eps)
+ 2 (1-\eps) I^{(0)}_{-1,2}(v_{12};\eps)
- v_{12} I^{(0)}_{2}(\eps)
\biggr]
\nonumber\\
&=-\frac{\pi}{\eps}  v_{12} \left(4-3 v_{12}\right) 
+\pi \left(2-8 v_{12}+3 v_{12}^2\right)
+\pi \left(4-16 v_{12}+9 v_{12}^2\right) \eps
\nonumber\\
&+\mathcal{O}\left(\eps^2\right)\,,
\end{align}

\clearpage 

\begin{align}
I^{(0)}_{-1,0}(v_{12};\eps)&=I^{(0)}(\eps)\,,
\\
I^{(0)}_{-1,-1}(v_{12};\eps)
&= \frac{4-v_{12}-2\eps}{3-2\eps} \, I^{(0)}
\nonumber\\
&=\frac{2\pi}{3}
\left(4-v_{12}\right)+\frac{4\pi}{9} \left(13-4 v_{12}\right) \eps
+\mathcal{O}\left(\eps^2\right)\,,
\\
I^{(0)}_{-1,1}(v_{12};\eps)
&=(1-v_{12}) I^{(0)}(\eps) + v_{12} I^{(0)}_{1}(\eps)
\nonumber\\
&=-\frac{\pi}{\eps}  v_{12}+2 \pi  \left(1-v_{12}\right)
+4 \pi  \left(1-v_{12}\right) \eps
+\mathcal{O}\left(\eps^2\right)\,,
\\
I^{(0)}_{-1,2}(v_{12};\eps)
&=(1-v_{12}) I^{(0)}_1(\eps) + v_{12} I^{(0)}_{2}(\eps)
\nonumber\\
&=
-\frac{\pi}{\eps}  \left(1-v_{12}\right) - \pi  v_{12}+\pi  v_{12} \eps
+\mathcal{O}\left(\eps^2\right)\,,
\\
I^{(0)}_{1,1}(v_{12};\eps)&=\frac{2\pi}{v_{12}} \, 
\biggl[ -\frac{1}{\eps}
+\log\left(\frac{v_{12}}{2}\right) 
- \eps \biggl( \frac{1}{2} \log^2\left(\frac{v_{12}}{2}\right)
+ \dilog\left(1-\frac{v_{12}}{2}\right)\biggr) 
\biggr] 
\nonumber\\
&+\mathcal{O}\left(\eps^2\right)\,,
\\
I^{(0)}_{1,2}(v_{12};\eps)
&=\frac{1}{v_{12}} \, 
\biggl[ 
(1+2\eps) I^{(0)}_{1,1}(v_{12};\eps) 
- 2\eps I^{(0)}_1(\eps) + I^{(0)}_2(\eps) 
\biggr] 
\nonumber\\  
&=\frac{\pi}{v_{12}^2} \, 
\biggl[ 
- \frac{2}{\eps} - 4 + v_{12} + 2 \log\left(\frac{v_{12}}{2}\right) 
\nonumber\\
&+\eps \biggl(v_{12}+4 \log\left(\frac{v_{12}}{2}\right) 
-\log^2\left(\frac{v_{12}}{2}\right)
-2 \dilog\left(1-\frac{v_{12}}{2}\right) \biggr)\biggr] 
+\mathcal{O}\left(\eps^2\right)\,,
\\
I^{(0)}_{2,2}(v_{12};\eps)
&=\frac{1}{v_{12}^2} \, 
\biggl[ 
(1+2\eps) (1-v_{12}) I^{(0)}_{1,1}(v_{12};\eps) 
- (1+2\eps) I^{(0)}_2(\eps) + (3+2\eps) v_{12} 
I^{(0)}_{1,2}(v_{12};\eps) 
\biggr] 
\nonumber\\   
&=- \frac{2 \pi}{v_{12}^3} \, \biggl[  
\frac{4-v_{12}}{\eps}
+ 10 - 4 v_{12} - (4 - v_{12}) \log\left(\frac{v_{12}}{2}\right) 
\nonumber\\
&+\eps \biggl(4-3 v_{12}
-2 \left(5-v_{12}\right) 
\log\left(\frac{v_{12}}{2}\right) 
+\left(2-\frac{v_{12}}{2}\right)
\log^2\left(\frac{v_{12}}{2}\right)
\nonumber\\
&+\left(4-v_{12}\right)
   \dilog\left(1-\frac{v_{12}}{2}\right)
   \biggr)\biggr]+\mathcal{O}\left(\eps^2\right)\,.
\end{align}

\vspace*{.25cm}
\subsection{Single massive integral with two denominators}

\begin{align}
I^{(1)}_{-4,-1}(v_{12},v_{11};\eps)
&=\frac{1}{4 (3-\eps)} \, 
\biggl[ 
   (7-2\eps)  I^{(1)}_{-4}(v_{11};\eps)
 + (13-2\eps) I^{(1)}_{-3,-1}(v_{12},v_{11};\eps) 
\nonumber\\
&- 3 v_{11}   I^{(1)}_{-2,-1}(v_{12},v_{11};\eps) 
 - 5 v_{12}   I^{(1)}_{-3}(v_{11};\eps)
\biggr] 
\nonumber\\
&=\frac{2\pi}{15} 
\Big(80-32 v_{12}-48 v_{11}+3 v_{11}^2+12 v_{12} v_{11}
\Big)
\nonumber\\
&+\frac{4\pi}{225}  
\Big(1570-676 v_{12}-1014 v_{11}+69 v_{11}^2+276 v_{12} v_{11}
\Big) \eps 
\nonumber\\
&+\mathcal{O}\left(\eps^2\right)\,,
\end{align}

\begin{align}
I^{(1)}_{-4,-2}(v_{12},v_{11};\eps)
&=\frac{1}{2 (7-2\eps)} \, 
\biggl[ 
   2 (5-\eps) I^{(1)}_{-4,-1}(v_{12},v_{11};\eps) 
 + 2 (7-\eps) I^{(1)}_{-3,-2}(v_{12},v_{11};\eps) 
\nonumber\\
&- 3 v_{11}   I^{(1)}_{-2,-2}(v_{12},v_{11};\eps) 
 - 6 v_{12}   I^{(1)}_{-3,-1}(v_{12},v_{11};\eps) 
\biggr]
\nonumber\\
&=\frac{8\pi}{105} 
   \Big(240 - 160 v_{12} - 120 v_{11}
   + 6  v_{11}^2 + 48 v_{12} v_{11}
   + 24 v_{12}^2 - 3  v_{12}^2 v_{11}   
\Big)
   \nonumber\\
   &+\frac{4 \pi}{11025}  
   \Big(137880 - 97520 v_{12} - 73140 v_{11}
   + 3909  v_{11}^2 + 31272 v_{12} v_{11}
   \nonumber\\
&  + 15636 v_{12}^2 - 2112  v_{12}^2 v_{11}
\Big) \eps
+\mathcal{O}\left(\eps^2\right)\,, 
\\[3mm]
I^{(1)}_{-4,-3}(v_{12},v_{11};\eps)
&=\frac{1}{4 (4-\eps)} \, 
\biggl[ 
   (13-2\eps) I^{(1)}_{-4,-2}(v_{12},v_{11};\eps) 
 + (15-2\eps) I^{(1)}_{-3,-3}(v_{12},v_{11};\eps) 
\nonumber\\
&- 3 v_{11}   I^{(1)}_{-2,-3}(v_{12},v_{11};\eps) 
 - 7 v_{12}   I^{(1)}_{-3,-2}(v_{12},v_{11};\eps) 
\biggr]
\nonumber\\
&=
\frac{4\pi}{35}
   \Big(280 - 240 v_{12} - 120 v_{11} 
         + 5 v_{11}^2 + 60 v_{11} v_{12} + 60 v_{12}^2
\nonumber\\
&       - 4 v_{12}^3 - 6 v_{11} v_{12}^2\Big) 
+ \frac{4 \pi}{3675}  \Big(83580 - 75240 v_{12} - 37620 v_{11} 
+ 1655 v_{11}^2 
\nonumber\\
& + 19860 v_{11} v_{12} + 19860 v_{12}^2
- 1408 v_{12}^3 - 2112 v_{11} v_{12}^2  
\Big) \eps + \mathcal{O}\left(\eps^2\right)\,, 
\\[3mm]
I^{(1)}_{-4,-4}(v_{12},v_{11};\eps)
&=\frac{1}{2 (9-2\eps)} \, 
\biggl[ 
   2 (8-\eps) I^{(1)}_{-4,-3}(v_{12},v_{11};\eps) 
 + 2 (8-\eps) I^{(1)}_{-3,-4}(v_{12},v_{11};\eps) 
\nonumber\\
&- 3 v_{11}   I^{(1)}_{-2,-4}(v_{12},v_{11};\eps) 
 - 8 v_{12}   I^{(1)}_{-3,-3}(v_{12},v_{11};\eps) 
\biggr]
\nonumber\\
&=\frac{16 \pi}{315} \Big(
1120 - 1120 v_{12} - 420 v_{11} + 
15 v_{11}^2 + 240 v_{11} v_{12} + 360 v_{12}^2 
\nonumber\\
&- 30 v_{11} v_{12}^2 - 40 v_{12}^3 + v_{12}^4
\Big)
+ \frac{8 \pi}{99225} \Big(
2074520 - 2162720 v_{12} 
\nonumber\\
&- 811020 v_{11} 
+ 30315  v_{11}^2 + 727560 v_{12}^2 + 485040 v_{11} v_{12} 
- 63780 v_{11} v_{12}^2 
\nonumber\\
&- 85040 v_{12}^3 + 2252 v_{12}^4
\Big) \eps   
+\mathcal{O}\left(\eps^2\right)\,, 
\\[3mm]
I^{(1)}_{-4,1}(v_{12},v_{11};\eps)
&=(1-v_{12}) I^{(1)}_{-3}(v_{11};\eps) 
+ \frac{1}{2 (2-\eps)} \, 
\biggl[ 3 (v_{12}-v_{11}) I^{(1)}_{-2}(v_{11};\eps) 
\nonumber\\
&- 3 v_{12}^2 I^{(1)}_{-2,1}(v_{11},v_{12};\eps) 
 + (7-2\eps) I^{(1)}_{-3,1}(v_{12},v_{11};\eps) 
\biggr]
\nonumber\\
&=-\frac{\pi}{\eps}  v_{12}^4
+  \frac{\pi}{6} \Big(24 + 16 v_{12} - 24 v_{11} 
+  3 v_{11}^2 - 12 v_{11} v_{12} + 12 v_{12}^2 
\nonumber\\
&- 6 v_{11} v_{12}^2 + 12 v_{12}^3 - 25 v_{12}^4
\Big)+\frac{\pi}{36} \Big(
  336 
+ 256 v_{12} 
- 384 v_{11}
+ 57 v_{11}^2 
\nonumber\\  
&
- 228 v_{11} v_{12}
+ 228 v_{12}^2
- 150 v_{11} v_{12}^2
+ 300 v_{12}^3 
- 415 v_{12}^4 
\Big) \eps 
\nonumber\\  
&+\mathcal{O}\left(\eps^2\right)\,,
\end{align}
\begin{align}
I^{(1)}_{-4,2}(v_{12},v_{11};\eps)
&=(1-v_{12}) I^{(1)}_{-3,1}(v_{12},v_{11};\eps) 
+ \frac{1}{(3-2\eps)} \, 
\biggl[ 3 (v_{12}-v_{11}) I^{(1)}_{-2,1}(v_{11},v_{12};\eps)  
\nonumber\\
&- 3 v_{12}^2 I^{(1)}_{-2,2}(v_{11},v_{12};\eps) 
 + 2 (3-\eps) I^{(1)}_{-3,2}(v_{12},v_{11};\eps) 
\biggr]
\nonumber\\
&=
\frac{2 \pi}{\eps}  v_{12}^2 \Big(3v_{11} - 8 v_{12} + 5 v_{12}^2\Big) 
+\frac{\pi}{3} 
\Big(
  8 
+ 16  v_{12} 
- 12  v_{11} 
+ 3   v_{11}^2
- 24  v_{11} v_{12} 
\nonumber\\ 
&
+ 36  v_{12}^2
+ 66  v_{11} v_{12}^2 
- 176 v_{12}^3 
+ 80  v_{12}^4 
\Big)
+\frac{\pi}{9} 
    \Big(
  52 
+ 128 v_{12}
- 96 v_{11}
\nonumber\\
&+ 33 v_{11}^2
- 264 v_{11} v_{12}
+ 396 v_{12}^2
+ 510 v_{11} v_{12}^2 
- 1360 v_{12}^3
+ 610 v_{12}^4
\Big) \eps 
\nonumber\\
&+\mathcal{O}\left(\eps^2\right)\,,
\\[3mm]
I^{(1)}_{-3,-1}(v_{12},v_{11};\eps)
&=\frac{1}{5-2\eps} \, 
\biggl[ 
  (3-\eps) I^{(1)}_{-3}(v_{11};\eps)  
+ (5-\eps) I^{(1)}_{-2,-1}(v_{11},v_{12};\eps)  
\nonumber\\
&- v_{11}   I^{(1)}_{-1,-1}(v_{11},v_{12};\eps) 
 - 2 v_{12} I^{(1)}_{-2}(v_{11};\eps) 
\biggr]
\nonumber\\
&=\frac{2\pi}{5}   
\Big(
  16
- 6 v_{12}
- 6 v_{11}
+ v_{12} v_{11}
\Big)
\nonumber\\
&+\frac{4\pi}{75} 
\Big(
  298
- 123 v_{12}
- 123 v_{11} 
+ 23  v_{11} v_{12}
\Big) \eps
\nonumber\\
&+\mathcal{O}\left(\eps^2\right)\,,
\\[3mm]
I^{(1)}_{-3,-2}(v_{12},v_{11};\eps)
&=\frac{1}{4 (3-\eps)} \, 
\biggl[ 
  (9-2\eps)  I^{(1)}_{-3,-1}(v_{11},v_{12};\eps)  
+ (11-2\eps) I^{(1)}_{-2,-2}(v_{11},v_{12};\eps)  
\nonumber\\
&- 2 v_{11} I^{(1)}_{-1,-2}(v_{11},v_{12};\eps) 
 - 5 v_{12} I^{(1)}_{-2,-1}(v_{11},v_{12};\eps) 
\biggr]
\nonumber\\
&=\frac{4\pi}{15} 
\Big(
  40
- 24 v_{12}
- 12 v_{11}
+ 3  v_{12}^2 
+ 3  v_{11} v_{12}
\Big)
   \nonumber\\
&+\frac{4\pi}{225} 
\Big(
  1570 
- 1014 v_{12}
- 507  v_{11} 
+ 138  v_{12} v_{11}
+ 138  v_{12}^2
\Big) \eps 
\nonumber\\  
&+\mathcal{O}\left(\eps^2\right)\,,
\\[3mm]
I^{(1)}_{-3,-3}(v_{12},v_{11};\eps)
&=\frac{1}{7-2\eps} \, 
\biggl[ 
   (6-\eps) \Big(I^{(1)}_{-3,-2}(v_{12},v_{11};\eps) 
             +   I^{(1)}_{-2,-3}(v_{12},v_{11};\eps)\Big)  
\nonumber\\
&- v_{11}   I^{(1)}_{-1,-3}(v_{12},v_{11};\eps) 
 - 3 v_{12} I^{(1)}_{-2,-2}(v_{12},v_{11};\eps) 
\biggr] 
\nonumber\\
&=\frac{4\pi}{35} 
\Big(
  160
- 120 v_{12}
- 40  v_{11}
+ 12 v_{11} v_{12}
+ 24 v_{12}^2
- v_{12}^3
\Big) 
\nonumber\\
&+\frac{4\pi}{3675}  \Big( 
  45960 
- 36570 v_{12} 
- 12190 v_{11}
+ 3909 v_{11} v_{12} 
\nonumber\\
&+ 7818 v_{12}^2 - 352 v_{12}^3 
\Big) \eps +\mathcal{O}\left(\eps^2\right)\,,
\end{align}

\begin{align}
I^{(1)}_{-3,-4}(v_{12},v_{11};\eps)
&=\frac{1}{4 (4-\eps)} \, 
\biggl[ 
   (15-2\eps) I^{(1)}_{-3,-3}(v_{12},v_{11};\eps) 
 + (13-2\eps) I^{(1)}_{-2,-4}(v_{12},v_{11};\eps) 
\nonumber\\
&- 2 v_{11}   I^{(1)}_{-1,-4}(v_{12},v_{11};\eps) 
 - 7 v_{12}   I^{(1)}_{-2,-3}(v_{12},v_{11};\eps) 
\biggr] 
\nonumber\\
&=\frac{16\pi}{35} \Big(
  70 
- 60 v_{12} 
- 15 v_{11}
+ 5  v_{11} v_{12}
+ 15 v_{12}^2
- v_{12}^3
\Big)
\nonumber\\
&+\frac{8 \pi}{3675}  \Big(
  41790
- 37620 v_{12}
- 9405  v_{11}
+ 3310  v_{11} v_{12}
+ 9930  v_{12}^2
\nonumber\\
&- 704   v_{12}^3
\Big) \eps +\mathcal{O}\left(\eps^2\right)\,,
\\[3mm]
I^{(1)}_{-3,1}(v_{12},v_{11};\eps)
&=(1-v_{12}) I^{(1)}_{-2}(v_{11};\eps)  
 + \frac{1}{3-2\eps} \, 
\biggl[ 2 (v_{12}-v_{11}) 
I^{(0)}(\eps) 
\nonumber\\
&- 2 v_{12}^2 I^{(1)}_{-1,-1}(v_{11},v_{12};\eps)   
+ (5-2\eps) v_{12} I^{(1)}_{-2,1}(v_{11},v_{12};\eps)   
\biggr]
\nonumber\\
&=-\frac{\pi}{\eps} v_{12}^3 
+\frac{\pi}{3} 
\Big(
  8 
+ 6 v_{12}
- 6 v_{11}
- 3  v_{11} v_{12}
+ 6  v_{12}^2
- 11 v_{12}^3
\Big)
\nonumber\\
&+\frac{\pi}{9} 
\Big(
   52
+  48 v_{12}
-  48 v_{11}
-  33 v_{11} v_{12}
+  66 v_{12}^2
-  85 v_{12}^3
\Big) \eps
\nonumber\\
&+\mathcal{O}\left(\eps^2\right)\,,
\\[3mm]
I^{(1)}_{-3,2}(v_{12},v_{11};\eps)
&=(1-v_{12}) I^{(1)}_{-2,1}(v_{11},v_{12};\eps)   
 + \frac{1}{1-\eps} \, 
\biggl[ (v_{12}-v_{11}) 
  I^{(1)}_{-1,1}(v_{11},v_{12};\eps)   
\nonumber\\
&- v_{12}^2 I^{(1)}_{-1,2}(v_{11},v_{12};\eps)   
+ (2-\eps) v_{12} I^{(1)}_{-2,2}(v_{11},v_{12};\eps)   
\biggr]
\nonumber\\
&=\frac{3 \pi}{\eps}  v_{12} \Big(v_{11} - 3 v_{12} + 2 v_{12}^2\Big) 
\nonumber\\
&+\pi 
\Big(
  2  
+ 6 v_{12}
- 3 v_{11}
+ 9 v_{11} v_{12}
- 27 v_{12}^2
+ 12 v_{12}^3
\Big)
\nonumber\\
&+\pi  
\Big(
  4 
+ 18 v_{12}
- 9 v_{11}
+ 21 v_{11} v_{12}
- 63 v_{12}^2
+ 30 v_{12}^3
\Big) \eps
+\mathcal{O}\left(\eps^2\right)\,, 
\\[3mm]
I^{(1)}_{-2,-1}(v_{12},v_{11};\eps)
&=\frac{1}{4 (2-\eps)} \, 
\biggl[ 
   (5-2\eps) I^{(1)}_{-2}(v_{11};\eps) 
 + (7-2\eps) I^{(1)}_{-1,-1}(v_{12},v_{11};\eps) 
\nonumber\\
&- (v_{11} + 3 v_{12})  I^{(0)}(\eps) 
\biggr] 
\nonumber\\
&=\frac{2\pi}{3} 
\Big(6-2 v_{12}-v_{11}\Big)
+\frac{4\pi}{9} \pi  
\Big(21-8 v_{12}-4 v_{11}\Big) \eps 
+\mathcal{O}\left(\eps^2\right)\,,
\\[3mm]
I^{(1)}_{-2,-2}(v_{12},v_{11};\eps)
&=\frac{1}{2 (5-2\eps)} \, 
\biggl[ 
   2 (4-\eps) (I^{(1)}_{-2,-1}(v_{12},v_{11};\eps) 
             + I^{(1)}_{-1,-2}(v_{12},v_{11};\eps)) 
\nonumber\\
&- v_{11}   I^{(0)}_{-2}(\eps) 
 - 4 v_{12} I^{(1)}_{-1,-1}(v_{12},v_{11};\eps) 
\biggr] 
\nonumber\\
&= \frac{4\pi}{15}
   \Big(
  24
- 12 v_{12}
- 3 v_{11}
+ v_{12}^2
\Big)
\nonumber\\
&+
\frac{4\pi}{225}  
\Big(
  894
- 492 v_{12} 
- 123 v_{11}
+ 46 v_{12}^2
\Big) \eps
+\mathcal{O}\left(\eps^2\right)\,,
\end{align}

\begin{align}
I^{(1)}_{-2,-3}(v_{12},v_{11};\eps) 
&=\frac{1}{4 (3-\eps)} \, 
\biggl[ 
   (11-2\eps) I^{(1)}_{-2,-2}(v_{12},v_{11};\eps) 
 + (9-2\eps)  I^{(1)}_{-1,-3}(v_{12},v_{11};\eps)) 
\nonumber\\
&- v_{11}   I^{(0)}_{-3}(\eps) 
 - 5 v_{12} I^{(1)}_{-1,-2}(v_{12},v_{11};\eps) 
\biggr] 
\nonumber\\
&=\frac{4\pi}{15} 
\Big(
  40 
- 24 v_{12}
- 4  v_{11}
+ 3  v_{12}^2
\Big)
\nonumber\\
&+\frac{4\pi}{225} 
\Big(
  1570
- 1014 v_{12} 
- 169  v_{11}
+ 138  v_{12}^2
\Big) \eps
+\mathcal{O}\left(\eps^2\right)\,,
\\[3mm]
I^{(1)}_{-2,-4}(v_{12},v_{11};\eps) 
&=\frac{1}{2 (7-2\eps)} \, 
\biggl[ 
     2 (7-\eps) I^{(1)}_{-2,-3}(v_{12},v_{11};\eps) 
   + 2 (5-\eps) I^{(1)}_{-1,-4}(v_{12},v_{11};\eps)) 
\nonumber\\
&- v_{11}   I^{(0)}_{-4}(\eps) 
 - 6 v_{12} I^{(1)}_{-1,-3}(v_{12},v_{11};\eps) 
\biggr] 
\nonumber\\
&=\frac{32\pi}{105} 
\Big(
  60
- 40 v_{12}
-5 v_{11}
+ 6 v_{12}^2
\Big)
\nonumber\\
&+\frac{8 \pi}{11025}  
\Big(
  68940
- 48760 v_{12}
- 6095  v_{11}
+ 7818  v_{12}^2
\Big) \eps
+\mathcal{O}\left(\eps^2\right)\,,
\\[3mm]
I^{(1)}_{-2,1}(v_{12},v_{11};\eps)
&=(1-v_{12}) I^{(0)}(\eps)   
 + \frac{1}{2 (1-\eps)} \, 
\biggl[ (v_{12} - v_{11}) I^{(0)}(\eps)  
\nonumber\\
&- v_{12}^2 I^{(0)}_{1}(\eps) 
+ (3-2\eps) v_{12} I^{(1)}_{-1,1}(v_{11},v_{12};\eps)   
\biggr]
\nonumber\\
&=-\frac{\pi}{\eps}   v_{12}^2
+ \pi  
\Big(2-v_{11}+2 v_{12}-3 v_{12}^2\Big) 
\nonumber\\
&+\pi  
\Big(4-3 v_{11}+6 v_{12}-7 v_{12}^2\Big) \eps
+\mathcal{O}\left(\eps^2\right)\,,
\\[3mm]
I^{(1)}_{-2,2}(v_{12},v_{11};\eps)
&=(1-v_{12}) I^{(1)}_{-1,1}(v_{11},v_{12};\eps)   
 + \frac{1}{2 (1-\eps)} \, 
\biggl[ (v_{12} - v_{11}) I^{(0)}_{1}(\eps)  
\nonumber\\
&- v_{12}^2 I^{(0)}_{2}(\eps)   
+ (3-2\eps) v_{12} I^{(1)}_{-1,2}(v_{11},v_{12};\eps)   
\biggr]
\nonumber\\
&=\frac{\pi  \left(v_{11}-4 v_{12}+3 v_{12}^2\right)}{\eps } 
+\pi  \left(2+2 v_{11}-8 v_{12}+3 v_{12}^2\right)
\nonumber\\
&+\pi  \left(4+4 v_{11}-16 v_{12}+9 v_{12}^2\right) \eps
+\mathcal{O}\left(\eps^2\right)\,.
\\[3mm]
I^{(1)}_{-1,-1}(v_{12},v_{11};\eps)
&=\frac{4-v_{12}-2\eps}{3-2\eps} \, I^{(0)}(\eps) 
\nonumber\\
&=\frac{2}{3} \pi  \left(4-v_{12}\right)
+\frac{4}{9} \pi  \left(13-4 v_{12}\right) \eps
+\mathcal{O}\left(\eps^2\right)\,,
\\[3mm]
I^{(1)}_{-1,-2}(v_{12},v_{11};\eps)
&=\frac{1}{4 (2-\eps)} \, 
\biggl[ 
     (7-2\eps) I^{(1)}_{-1,-1}(v_{12},v_{11};\eps) 
   + (5-2\eps) I^{(0)}_{-2}(\eps) 
   - 3 v_{12} I^{(0)}(\eps) 
\biggr] 
\nonumber\\
&=\frac{4\pi}{3} \left(3-v_{12}\right)
+\frac{4\pi}{9} \left(21-8 v_{12}\right) \eps 
+\mathcal{O}\left(\eps^2\right)\,,
\end{align}

\begin{align}
I^{(1)}_{-1,-3}(v_{12},v_{11};\eps) 
&=\frac{1}{5-2\eps} \, 
\biggl[ 
     (5-\eps) I^{(1)}_{-1,-2}(v_{12},v_{11};\eps) 
   + (3-\eps) I^{(0)}_{-3}(\eps) 
   - 2 v_{12} I^{(0)}_{-2}(\eps) 
\biggr] 
\nonumber\\
&= \frac{4\pi}{5}  
\left(8-3 v_{12}\right)+\frac{4\pi}{75} \left(298-123 v_{12}\right) \eps
+\mathcal{O}\left(\eps^2\right)\,,
\\[3mm]
I^{(1)}_{-1,-4}(v_{12},v_{11};\eps) 
&=\frac{1}{4 (3-\eps)} \, 
\biggl[ 
     (13-2\eps) I^{(1)}_{-1,-3}(v_{12},v_{11};\eps) 
   + (7-2\eps)  I^{(0)}_{-4}(\eps) 
   - 5 v_{12}   I^{(0)}_{-3}(\eps) 
\biggr] 
\nonumber\\
&=\frac{32\pi}{15} \left(5-2 v_{12}\right)+\frac{8\pi}{225} 
\left(785-338 v_{12}\right)
   \eps
+\mathcal{O}\left(\eps^2\right)\,,
\\[3mm]
I^{(1)}_{-1,1}(v_{12},v_{11};\eps)
&=(1-v_{12}) I^{(0)}(\eps) + v_{12} I^{(0)}_1(\eps) 
\nonumber\\
&=-\frac{\pi}{\eps} v_{12}
  + 2 \pi \left(1-v_{12}\right) (1+2\eps) 
  + \mathcal{O}\left(\eps^2\right)\,,
\\[3mm]
I^{(1)}_{-1,2}(v_{12},v_{11};\eps)
&=(1-v_{12}) I^{(0)}_1(\eps) + v_{12} I^{(0)}_2(\eps) \, 
\nonumber\\
&=- \frac{\pi  \left(1-v_{12}\right)}{\eps} - \pi v_{12} \left(1-\eps\right)
+\mathcal{O}\left(\eps^2\right)\,,
\\[3mm]
I^{(1)}_{1,1}(v_{12},v_{11};\eps)
&=\frac{\pi}{v_{12}}\biggl[-\frac{1}{\eps}-\log\frac{v_{11}}{v_{12}^2}
- \frac{\eps}{2} \, L(v_{11},v_{12}) +\mathcal{O}(\eps^2)\biggr]\,,
\\[3mm]
I^{(1)}_{1,2}(v_{12},v_{11};\eps)&=\frac{\pi}{v_{12}^3}
\biggl[\frac{v_{11}-v_{12}}{\eps}+2 v_{11}-4v_{12}+v_{12}^2+(v_{11}-v_{12})
\log\frac{v_{11}}{v_{12}^2}
\nonumber\\
&+\frac{\eps}{2}\biggl(2 v_{12}^2+4(v_{11}-v_{12})
\log\frac{v_{11}}{v_{12}^2} 
-4 v_{12}\sqrt{1-v_{11}}\log\biggl(\frac{1+\sqrt{1-v_{11}}}{1-\sqrt{1-v_{11}}}\biggr)
\nonumber\\
&+(v_{11}-v_{12}) L(v_{11},v_{12})
\biggr)\biggr]+\mathcal{O}(\eps^2)\,,
\\[3mm]
I^{(1)}_{2,1}(v_{12},v_{11};\eps)
&=\frac{\pi}{v_{12}^2 v_{11}}
\biggl[-\frac{v_{11}}{\eps}+2v_{12}-2v_{11}
-v_{11} \log\left(\frac{v_{11}}{v_{12}^2}\right)
\nonumber\\
&-\eps\biggl(2v_{11}\log\frac{v_{11}}{v_{12}^2}
-2v_{12}\sqrt{1-v_{11}}
\log\left(\frac{1+\sqrt{1-v_{11}}}{1-\sqrt{1-v_{11}}}\right)
\nonumber\\
&+\frac{v_{11}}{2}\, L(v_{11},v_{12}) \biggr)\biggr] 
+\mathcal{O}(\eps^2)\,,
\\[3mm]
I^{(1)}_{2,2}(v_{12},v_{11};\eps)&=\frac{\pi}{v_{12}^4v_{11}}
\biggl[ 
\frac{1}{\eps} \left(3v_{11}^2-4v_{11}v_{12}+v_{11}v_{12}^2\right) 
+8v_{11}^2+5v_{11}v_{12}^2-16v_{11}v_{12}+2v_{12}^2
\nonumber\\
&+(3v_{11}^2-4v_{11}v_{12}+v_{11}v_{12}^2)
\log\frac{v_{11}}{v_{12}^2}+\eps\biggl(4v_{11}^2-8v_{11}v_{12}+5v_{11}v_{12}^2
\nonumber\\
&+(8v_{11}^2-10v_{11}v_{12}+2v_{11}v_{12}^2)\log\frac{v_{11}}{v_{12}^2}
\nonumber\\
&-(6v_{11}v_{12}-2v_{12}^2)\sqrt{1-v_{11}}
\log\left(\frac{1+\sqrt{1-v_{11}}}{1-\sqrt{1-v_{11}}}\right)
\nonumber\\
&+\frac{1}{2} (3v_{11}^2-4v_{11}v_{12}+v_{11}v_{12}^2)\, L(v_{11},v_{12})
\biggr)\biggr]+\mathcal{O}(\eps^2)\,, 
\end{align}

\begin{align}
I^{(1)}_{1,3}(v_{12},v_{11};\eps)
&=\frac{\pi}{2 v_{12}^5} \biggl[
\frac{1}{\eps} \left(-3v_{11}^2+6v_{11}v_{12}-v_{11}v_{12}^2-2 v_{12}^2\right) 
-8v_{11}^2+22 v_{11}v_{12}-5v_{11}v_{12}^2 
\nonumber\\
&-12 v_{12}^2+2v_{12}^3+\frac{v_{12}^4}{2}-(3v_{11}^2
-6v_{11}v_{12}+v_{11}v_{12}^2+2 v_{12}^2)\log\frac{v_{11}}{v_{12}^2}
\nonumber\\
&+\eps \biggl(-4v_{11}^2+12 v_{11}v_{12}\vphantom{\frac{v_{11}}{v_{12}^2}} 
-5v_{11}v_{12}^2-8v_{12}^2+4 v_{12}^3+\frac{3 v_{12}^4}{4}
\nonumber\\
&-(8v_{11}^2+6 v_{12}^2-16v_{11}v_{12}+2v_{11}v_{12}^2)\log\frac{v_{11}}{v_{12}^2}
\nonumber\\
&+6v_{12} (v_{11}-v_{12}) 
\sqrt{1-v_{11}}\log\left(\frac{1+\sqrt{1-v_{11}}}{1-\sqrt{1-v_{11}}}\right)
\nonumber\\
&-\frac{1}{2} \, (3v_{11}^2-6 v_{11}v_{12}+v_{11}v_{12}^2+2v_{12}^2) \, L(v_{11},v_{12}) 
\biggr)\biggr]+\mathcal{O}(\eps^2)\,,
\\
I^{(1)}_{3,1}(v_{12},v_{11};\eps)
&=\frac{\pi}{v_{12}^3 v_{11}^2} \bigg[ 
- \frac{v_{11}^2}{\eps} 
- 3 v_{11}^2 + 2 v_{11} v_{12} + 2 v_{12}^2 
- v_{11} v_{12}^2 - v_{11}^2 \log\frac{v_{11}}{v_{12}^2}
\nonumber\\
&+ \frac{\eps}{2} \biggl(
  4 v_{11} v_{12} + 4 v_{12}^2 - 4 v_{11}^2 - 4 v_{11} v_{12}^2 
- 6 v_{11}^2 \log\frac{v_{11}}{v_{12}^2} 
+ 4 v_{12} (v_{11} + v_{12}) 
\nonumber\\
&\times \sqrt{1-v_{11}} 
\log\left(\frac{1+\sqrt{1-v_{11}}}{1-\sqrt{1-v_{11}}}\right)
- v_{11}^2 \, L(v_{11},v_{12})
\biggr)\biggr]
+\mathcal{O}(\eps^2)\,,
\\
I^{(1)}_{2,3}(v_{12},v_{11};\eps)
&=\frac{\pi}{8 v_{12}^6 v_{11}} \biggl[
\frac{12}{\eps} \, 
v_{11} \left(-5v_{11}^2+12v_{11}v_{12}-6 v_{12}^2-3v_{11}v_{12}^2
+2 v_{12}^3\right) 
\nonumber\\
&- 184 v_{11}^3 + 552 v_{11}^2 v_{12} 
- 384 v_{11} v_{12}^2 - 156 v_{11}^2 v_{12}^2
+ 16  v_{12}^3 + 160 v_{11} v_{12}^3 
\nonumber\\
&- 6 v_{11} v_{12}^4
 -12 v_{11} \left(
  5  v_{11}^2 
- 12 v_{11} v_{12}
+ 6  v_{12}^2 
+ 3  v_{11} v_{12}^2 
- 2  v_{12}^3 \right)  
\log\frac{v_{11}}{v_{12}^2}
\nonumber\\
&+\eps \biggl(
- 144 v_{11}^3
+ 448 v_{11}^2 v_{12}
- 320 v_{11}   v_{12}^2 
- 172 v_{11}^2 v_{12}^2 
+ 200 v_{11}   v_{12}^3 
- 13  v_{11}   v_{12}^4
\nonumber\\
&+8 v_{11} 
\left(
- 23  v_{11}^2
+ 54  v_{11}   v_{12}
- 27  v_{12}^2 
- 12  v_{11}   v_{12}^2 
+ 8   v_{12}^3 
\right) \log\frac{v_{11}}{v_{12}^2}
\nonumber\\
&+8 v_{12} 
\left(
  15  v_{11}^2
- 21  v_{11}   v_{12}
+ 2   v_{12}^2 
+ 4   v_{11}   v_{12}^2 
\right) 
\sqrt{1-v_{11}} 
\log\left(\frac{1+\sqrt{1-v_{11}}}{1-\sqrt{1-v_{11}}}\right)
\nonumber\\
&-6 v_{11} 
\left(
  5   v_{11}^2
- 12  v_{11}   v_{12}
+ 6   v_{12}^2 
+ 3   v_{11}   v_{12}^2 
- 2   v_{12}^3 
\right) 
\, L(v_{11},v_{12}) 
\biggr)\biggr]
\nonumber\\
&+\mathcal{O}(\eps^2)\,,
\\
I^{(1)}_{3,2}(v_{12},v_{11};\eps)
&=\frac{\pi}{v_{12}^5 v_{11}^2} \biggl[
\frac{3}{\eps} \, 
v_{11}^2 \left(
  2 v_{11}
- 3 v_{12}
+ v_{12}^2
\right) 
+ 19 v_{11}^3 
+ 6  v_{11}   v_{12}^2 
- 39 v_{11}^2 v_{12}
\nonumber\\
&+ 2  v_{12}^3 
- 3  v_{11}   v_{12}^3 
+ 14 v_{11}^2 v_{12}^2
+ 3  v_{11}^2 
  \left(
  2  v_{11} 
- 3  v_{12}
+    v_{12}^2 
  \right)  
\log\frac{v_{11}}{v_{12}^2}
\nonumber\\
&+\frac{\eps}{2} 
  \biggl(
  32  v_{11}^3
+ 4   v_{12}^3
+ 8   v_{11}   v_{12}^2
- 68  v_{11}^2 v_{12}  
- 8   v_{11}   v_{12}^3 
+ 34  v_{11}^2 v_{12}^2 
\nonumber\\
&+2 v_{11}^2 
\left(
  19  v_{11}
- 27  v_{12}
+ 8   v_{12}^2 
\right) \log\frac{v_{11}}{v_{12}^2}
\nonumber\\
&-4 v_{12} 
\left(
  6  v_{11}^2
- 3  v_{11}   v_{12}
-    v_{12}^2 
+    v_{11}   v_{12}^2 
\right) 
\sqrt{1-v_{11}} 
\log\left(\frac{1+\sqrt{1-v_{11}}}{1-\sqrt{1-v_{11}}}\right)
\nonumber\\
&+ 3 v_{11}^2 
\left(
  2   v_{11}
- 3   v_{12} 
+     v_{12}^2 
\right) 
\, L(v_{11},v_{12}) 
\biggr)\biggr]
+\mathcal{O}(\eps^2)\,,
\end{align}

\begin{align}
I^{(1)}_{3,3}(v_{12},v_{11};\eps)
&=\frac{\pi}{4 v_{12}^7 v_{11}^2} \biggl[
- \frac{6}{\eps} \, 
  v_{11}^2 \left(
  15 v_{11}^2
- 40 v_{11} v_{12}
+ 24 v_{12}^2 
+ 12 v_{11} v_{12}^2
- 12 v_{12}^3 
+    v_{12}^4
\right) 
\nonumber\\
&- 306 v_{11}^4 
+ 976 v_{11}^3  v_{12} 
- 768 v_{11}^2  v_{12}^2 
- 312 v_{11}^3  v_{12}^2
+ 48  v_{11}    v_{12}^3 
+ 420 v_{11}^2  v_{12}^3
\nonumber\\
&+ 8  v_{12}^4 
- 24  v_{11}    v_{12}^4  
- 43  v_{11}^2  v_{12}^4  
- 6   v_{11}^2  \Big(
  15  v_{11}^2  
- 40  v_{11}    v_{12}
+ 12  v_{11}    v_{12}^2 
\nonumber\\
&+ 24 v_{12}^2 
- 12  v_{12}^3 
+     v_{12}^4  
  \Big)  
\log\frac{v_{11}}{v_{12}^2}
+\frac{\eps}{2} 
  \biggl(
- 616  v_{11}^4
+ 2024 v_{11}^3 v_{12} 
- 1616 v_{11}^2 v_{12}^2
\nonumber\\
&- 784  v_{11}^3 v_{12}^2
+ 48   v_{11}   v_{12}^3  
+ 1120 v_{11}^2 v_{12}^3 
+ 16   v_{12}^4 
- 48   v_{11}   v_{12}^4
- 145  v_{11}^2 v_{12}^4
\nonumber\\
&+ 2 v_{11}^2 
  \left(
- 306 v_{11}^2
+ 796 v_{11} v_{12}
- 468 v_{12}^2  
- 222 v_{11} v_{12}^2 
+ 216 v_{12}^3 
- 16  v_{12}^4 
\right) \log\frac{v_{11}}{v_{12}^2}
\nonumber\\
&+ 8 v_{12}
\left(
  21  v_{11}^2 v_{12}^2
+ 12  v_{11}   v_{12}^2
+ 2   v_{12}^3 
+ 45  v_{11}^3
- 75  v_{11}^2 v_{12} 
- 5   v_{11}   v_{12}^3 
\right) 
\nonumber\\
&\times 
\sqrt{1-v_{11}} 
\log\left(\frac{1+\sqrt{1-v_{11}}}{1-\sqrt{1-v_{11}}}\right)
-6 v_{11}^2 
\Big(
  15 v_{11}^2
- 40 v_{11}   v_{12}
+ 24 v_{12}^2 
\nonumber\\
&+ 12 v_{11}   v_{12}^2 
- 12 v_{12}^3 
+    v_{12}^4 
\Big) 
\, L(v_{11},v_{12}) 
\biggr)\biggr]
+\mathcal{O}(\eps^2)\,,
\end{align} 
where 
\eq
\hspace*{-.5cm}
L(v_{11},v_{12})= 
\log^2\frac{v_{11}}{v_{12}^2}
+4 \dilog\left(1-\frac{v_{12}}{1-\sqrt{1-v_{11}}}\right) 
+4 \dilog\left(1-\frac{v_{12}}{1+\sqrt{1-v_{11}}}\right)\,. 
\en

\vspace*{.25cm}
\subsection{Double massive integral with two denominators}

In this section we briefly review how we proceed with 
double massive integrals with two denominators. 
As we showed in Sec.~\ref{sec:Double_massive}, the $\eps$ expansion 
of the double massive master integral with two denominators 
$I_{1,1}^{(2)}(v_{12},v_{11},v_{22};\eps)$ 
is generated by analytically using Eq.~(\ref{Eq_I11_2mass}). 
Eq.~(\ref{dm_eps_2nd}) explicitly illustrates 
the $\eps$ expansion of this integral up to second order. 
All other double massive integrals are obtained using 
recursive relations (\ref{recursions_negative}),
(\ref{Req_jl1}), (\ref{Req_jl2}), and (\ref{Main_Rec_5})
and $\eps$-expansion of the master integrals 
$I^{(0)}$, $I_{1,0}^{(1)}(v_{11};\eps)$, and  
$I_{1,1}^{(2)}(v_{12},v_{11},v_{22};\eps)$. 

\vspace*{.25cm}
\subsection{Mirkes type III integrals}
\label{sec:tables_Mirkes}

Mirkes type III integrals are defined as 
\begin{align}\label{Mirkes1app}
I^{j,l}_D(A,B,C) = 
\int\dx\Omega_{k_1k_2}
\frac{(-\cos\theta_1)^l}{\left(A+B\cos\theta_1+C\sin\theta_1\cos\theta_2\right)^j} 
= A^{-j} \, I^{j,l}_D(b,\rho^2) \,,  
\end{align} 
where $b=B/A$\,, $c=C/A$\,, and $\rho^2 = b^2 + c^2$\,. 

By application of the recursion relations~(\ref{RecIII_1})-(\ref{RecIII_0}) we find 
the integrals with specific values of indices $j=1,2$ and $l=0,\ldots,4$: 
\eq 
I^{1,0}_D(b,\rho^2) &=&I^{(1)}_1\,,
\\
I^{1,1}_D(b,\rho^2)&=&\frac{b}{\rho^2} \, 
\Big[- I^{(0)} + I^{(1)}_1\Big]\,, 
\en
\eq 
I^{1,2}_D(b,\rho^2)&=&\frac{1}{2 \rho^4 (1-\eps)} \, 
\biggl[ I^{(0)} \, \Big(\rho^2 - b^2 (3-2\eps) \Big)
\nonumber\\
&-& I^{(1)}_1 \, \Big( \rho^2 (1-\rho^2) - b^2 (3-\rho^2-2\eps) \Big)
\biggr]\,,
\\[2mm]
I^{1,3}_D(b,\rho^2)&=&\frac{b}{2 \rho^6 \, (1-\eps)}  \, 
\biggl[ \frac{I^{(0)}}{3-2\eps} \, \Big(\rho^2 (1-\eps) (9-6\rho^2+4b^2-2\eps) 
\nonumber\\
&-&b^2 (3-2\eps) (5-2\eps)\Big) 
 - I^{(1)}_1 \, \Big( 3 \rho^2 (1-\rho^2) - b^2 (5-3\rho^2-2\eps) \Big)
\biggr] \,, 
\\[2mm]
I^{1,4}_D(b,\rho^2)&=&
\frac{1}{4 \rho^8 (1-\eps) (2-\eps)} \,
\biggl[ \frac{I^{(0)}}{3-2\eps} \, \Big(
- b^4 (3 - 2\eps) (5 - 2\eps) (7 - 2\eps)
\nonumber\\[2mm]
&-& 3 \rho^4 (3 - 2\eps)
+ 3 \rho^6 (5 - 4\eps)
+ b^4 \rho^2 (5 - 2\eps) (11 - 10\eps)
\nonumber\\[2mm]
&-& 6 b^2 \rho^4 (13 - 16\eps + 4\eps^2)
+   6 b^2 \rho^2 (3 - 2\eps) (5-2\eps)
\Big)
\nonumber\\[2mm]
&+& I^{(1)}_1 \, \Big(3 \rho^4 (1 - \rho^2 + b^2)^2
                - 6 b^2 \rho^2 (1 - \rho^2 + b^2) (5 - 2\eps) 
\nonumber\\[2mm]
&+& b^4 (5 - 2\eps) (7 - 2\eps)\Big) \biggr]\,,
\\
I^{2,0}_D(b,\rho^2)&=& \frac{1}{1-\rho^2} \biggl[
(1-2\eps) \, I^{(0)} + 2\eps \, I^{(1)}_1 \biggr]  \,, 
\\[2mm]
I^{2,1}_D(b,\rho^2)&=&\frac{b}{\rho^2 \, (1-\rho^2)} \, 
\Big[ (1-2\eps) \, I^{(0)} \,-\,  (1-\rho^2-2\eps)  \, I^{(1)}_1 \Big] \,, 
\\[2mm]
I^{2,2}_D(b,\rho^2)&=&
I^{(0)} \, \biggl[
- \frac{1}{\rho^2}
+ \frac{b^2 (3-2\rho^2-2\eps)}{\rho^4 (1-\rho^2)} \biggr]
\nonumber\\[2mm]
&+& I^{(1)}_1 \, \biggl[ \frac{\rho^2-3b^2}{\rho^4} 
+ \frac{2 b^2 \eps}{\rho^4 (1-\rho^2)}
\biggr] \,, 
\\[2mm]
I^{2,3}_D(b,\rho^2)&=& 
\frac{b}{2 \rho^6 \, (1-\rho^2) \, (1-\eps)} \, \biggl[  
I^{(0)}\, \Big(
- 3 \rho^2 (1-\rho^2) (3-2\eps) 
\nonumber\\[2mm]
&+& b^2 (3-2\eps) (5-2\eps) - b^2 \rho^2 (13-10\eps) \Big)  
\nonumber\\[2mm]
&+& I^{(1)}_1 \, \Big(3 \rho^2 (1-\rho^2) (3-\rho^2 - 2\eps)  
\nonumber\\[2mm]
&-& b^2 ( (3-2\eps) (5-2\eps) - 3 \rho^2 (6-\rho^2-4\eps))\Big)
\biggr] \,, 
\\[2mm]
I^{2,4}_D(b,\rho^2)&=& 
\frac{I^{(0)}}{2 \rho^2} \,
\biggl[
  \frac{3}{\rho^2 (1-\eps)} 
- \frac{6}{3-2\eps}
+ \frac{6 b^2}{\rho^2} \, 
\biggl(\frac{4}{3-2\eps} 
     - \frac{5-2\eps}{\rho^2 (1-\eps)}
\biggr) 
\nonumber\\[2mm]
&+& 
  \frac{b^4}{\rho^2 (1-\rho^2)} \, 
\biggl( 
  \frac{16}{3-2\eps}
+ \frac{(5-2\eps) (7-2\eps)}{\rho^4 (1-\eps)}
- \frac{(5-2\eps) (23-18\eps)}{\rho^2 (1-\eps) (3-2\eps)}
\biggr) 
\biggr]
\nonumber\\[2mm]
&+& 
\frac{I^{(1)}_1}{2 \rho^8} \, 
\biggl[ 4 b^2 (3\beta^2 - 5b^2) - \frac{3 (1-\rho^2) (b^2-\rho^2) (5b^2-\rho^2)}{1-\eps} 
+ \frac{4 b^4 \eps}{1-\rho^2} 
\biggr] \,. 
\en

\printindex

\end{document}